%% file: report_main.tex
\documentclass[11pt]{report}
\usepackage{titletoc}% http://ctan.org/pkg/titletoc
\usepackage{amsmath} %Never write a paper without using *amsmath* for
\usepackage{amssymb} %Some extra symbols
\usepackage{makeidx} %If you want to generate an index, automatically
\usepackage{graphicx} %If you want to include postscript graphics
\usepackage{caption} %If you want to include postscript graphics
\usepackage{subfigure}
\usepackage[normalem]{ulem}
\usepackage{epsfig}

\newcommand{\lsim}{\, \lower2truept\hbox{${< \atop\hbox{\raise4truept\hbox{$\sim$}}}$}\,}
\newcommand{\gsim}{\, \lower2truept\hbox{${> \atop\hbox{\raise4truept\hbox{$\sim$}}}$}\,}
\newcommand{\nid}{\noindent}
\newcommand{\truth}{\emph{truth}}
\newcommand{\no}{\emph{noise-only}}
\begin{document}
\centerline{\bf \LARGE SPIRE Map-Making Test Report}
\vskip0.3truecm
\centerline{(Version 6; \today)}
\vskip1truecm
{\centering
\begin{tabular}{|c|c|}
\hline
Coordinator & Kevin Xu (NHSC) \\
\hline
Simulators & Andreas Papageorgiou (Uni Cardiff) \\ 
           & Kevin Xu (NHSC) \\
\hline
Map-makers & Hacheme Ayasso (IAS)\\
           & Alexandre Beelen (IAS) \\
           & Lorenzo Piazzo (Uni Roma) \\ 
           & H\'{e}l\`{e}ne Roussel (IAP) \\
           & Bernhard Schulz (NHSC)\\ 
           & David Shupe (NHSC) \\
\hline
Testers & Luca Conversi (HSC) \\
        & Vera K\"onyves (IAS/CEA) \\
        & Andreas Papageorgiou (Uni Cardiff)\\
        & David Shupe (NHSC) \\
        & Kevin Xu (NHSC) \\
\hline
\end{tabular}\par
}
%
\include{executive_summary}
\tableofcontents
%\chapter{Introduction}
\include{chapter_introduction}
%%\section{First section}
\include{chapter_testcases}

\include{chapter_simulations}

\include{chapter_mapmakers}
% \section{Naive (Schulz)}
 \include{naiveMapperRep}

%\section{Destriper (Schulz)}
 \include{destriperRep}

% \section{Scanamorphos (Roussel)}
 \include{mapmaking_report_scanamorphos_v4}
\include{mapmaking_report_SANEPIC}
% %\section{Unimap (Piazzo)}
 \include{UniRep_new}

% %\section{HiRes (Shupe)} 
 \include{mapmaker_report_hires_v5_pdf}

% %\section{SUPREME (Hacheme Ayasso)}
 \include{SupremeV2_pdf}
%\chapter{Metrics and Results}
\include{chapter_metrics}
%\chapter{Metrics and Results}
% %\section{Diviation from Truth (K\"onyves/Papageorgiou)}
% %\section{Diviation from Truth (K\"onyves/Papageorgiou)}
 \include{metrics_SPIRE_DIFFmap_report_update2013sept_pdf}  % !
 \include{metric_report_fft_v2.1_pdf}

 \include{chapter5_metrics_sources_pdf} %!
 \include{metrics_report_superresolution_v4_pdf}

% %\section{Metrics for Super-Resolution Maps (Shupe)}
%%\chapter{Summary}
\include{chapter_summary}
%%
\bibliographystyle{abbrv}
\bibliography{main}
\end{document}

%% file: executive_summary.tex
% NOTE: to be included in the report_main.tex
%\documentclass[11pt]{book}

%\begin{document}
%\newcommand{\truth}{\emph{truth}}
%\newcommand{\no}{\emph{noise-only}}
\begin{abstract}

The photometer section of SPIRE
is one of the key instruments on board of Herschel.
Its legacy depends very much on how well the scanmap observations that
it carried out during the Herschel mission can be converted to high
quality maps. In order to have a comprehensive assessment on the
current status of SPIRE map-making, as well as to provide guidance for
future development of the SPIRE scan-map data reduction pipeline, we
carried out a test campaign on SPIRE map-making.  In this report, we
present results of the tests in this campaign.  The goals are: (1)
Compare the map-makers in the SPIRE pipeline with other mapmakers.
(2) In particular, identify the strengths and limitations of different
mapmakers in dealing with the known SPIRE map-making issues, such as
the cooler burp effect.  (3) Assess the resolution-enhancement
capabilities of the super-resolution mappers, as compared to the
destriper (the pipeline default), and investigate their applicability
to various kinds of data as well as caveats or pitfalls to avoid.
(4) Enable users to choose the right map-maker for their science.  (5)
Provide guidance for future development of the SPIRE scan-map data
reduction pipeline.

For these purposes, 13 test cases were generated, including data sets
obtained in different observational modes and scan speeds, with
different map sizes, source brightness, and levels of complexity of
the extended emission. They also include observations suffering from
the ``cooler burp'' effect, and those having strong large-scale
gradients in the background radiation. The input data for these test
cases are time-ordered data (TODs\footnote{In this report, TOD is used in the broad sense of a collection of samples containing time, flux density and position information. The data were not formatted as a single HIPE Tod product, but rather consisted of many FITS files, one per scan. Each file is known within HIPE as a Photometer Scan Product (PSP) and contains tables of the calibrated signal, right ascension and declination, with each row corresponding to a time sample and with separate columns for each bolometer.}). The map-making process turns the
TODs into maps. Among the test cases, 8 are simulated
and 5 are real observations.

Comparing to real observations, a simulated test case has the advantage of 
possessing the ``truth'', namely the sky model, based on which the
simulation is carried out. The truth map provides an unbiased standard against
which test maps made by different map-makers are to be compared.
Allowing for the effects of noise in a given map, deviations from the truth 
can be used as objective measures for the bias introduced by the 
map-making process. In the simulations, TODs were generated using two 
layers of data: a noise layer taken from real SPIRE observations of
dark fields (this allows the simulation to include both instrumental 
noise and confusion noise), and a truth layer which is 
a sky-model map based either on a real Spitzer 24$\mu m$
map or a map of artificial sources.

Seven map-makers participated in this test campaign, including
(1) Naive mapper (default of the SPIRE standard pipeline until HIPE~8);
(2) Destriper in two flavors: (i) Destriper-P0: Destriper with
polynomial-order~=~0
(default of SPIRE standard pipeline 
since HIPE~9), and (ii) Destriper-P1: Destriper with 
polynomial-order~=~1~;
(3) Scanamorphos; (4) SANEPIC (GLS mapmaker); (5) Unimap (GLS mapmaker); 
(6) HiRes (super-resolution map-maker); 
(7) SUPREME (super-resolution map-maker). 
Because of time constraints, not all map-makers processed all the
test cases (see Table~\ref{tbl0:mapmaker_testcases} for details).
%\begin{savenotes}
\begin{table}[ht]
\begin{minipage}{\textwidth}  % this is to have the footnote!!!
\renewcommand\footnoterule{}
\caption[Caption for LOF]{Test Cases Processed by Different Map-Makers}
\centering
\begin{tabular}{|l|l|l|l|l|l|l|l|l|}
\hline
Case & Name & \multicolumn{7}{|c|}{Map-Maker\protect\footnote{Abbreviations for map-makers: N -- Naive, D -- Destriper, Sc -- Scanamorphos, SA -- SANEPIC, U -- Unimap, H -- HiRes, SU -- SUPREME.}\label{tbl0:mapmaker_testcases}
} \\ 
\hline
          &      & N & D & Sc & SA & U & H & SU \\ 
\hline
1 & Nominal~Sources & $\surd$& $\surd$& $\surd$& $\surd$& $\surd$ &$\surd$ &$\surd$ \\
\hline
2 & Nominal~Cirrus & $\surd$& $\surd$& $\surd$& $\surd$&  & & \\
\hline
3 & Nominal~Dark  & $\surd$& $\surd$& $\surd$& $\surd$& & & \\
\hline
4 & Nominal~M51  & $\surd$& $\surd$& $\surd$& $\surd$& & & \\
\hline
5 & Fast-scan~Sources & $\surd$& $\surd$& $\surd$& $\surd$& $\surd$ & & \\
\hline
6 & Fast-scan~MK Center & $\surd$& $\surd$& $\surd$& $\surd$& $\surd$ & & \\
\hline
7 & Fast-scan~Dark & $\surd$& $\surd$& $\surd$& $\surd$&  & & \\
\hline
8 & Parallel~Sources & $\surd$& $\surd$& $\surd$& $\surd$&  & & \\
\hline
9 & Parallel~Mk Center & $\surd$& $\surd$& $\surd$& $\surd$&  & & \\
\hline
10 & Parallel~Cirrus & $\surd$& $\surd$& $\surd$& $\surd$& $\surd$& $\surd$& $\surd$ \\
\hline
11 & Parallel~Dark & $\surd$& $\surd$& $\surd$& $\surd$&  & & \\
\hline
12 & Nominal~NGC~628 & $\surd$& $\surd$& $\surd$& & $\surd$& $\surd$&$\surd$ \\
\hline
13 & Para-fast~Hi-Gal-L30 & $\surd$& $\surd$& $\surd$& & $\surd$& $\surd$& $\surd$\\
\hline
\end{tabular}
\end{minipage}
\end{table}

Results of tests are presented in the framework of four sets of metrics:
\begin{description}
\item{(1)} Deviation from the truth. These metrics include:
  (i) visual examinations of the difference map $\rm Map - Map_{true}$;
  (ii) a scatter plot of (S -- S$_{\rm true}$) vs S$_{\rm true}$ for individual pixels;
  (iii) slopes of these plots; 
  (iv) absolute deviations: mean and standard deviation of S -- S$_{\rm true}$;
  (v) relative deviations: mean and standard deviation of (S -- S$_{\rm true}$)/S$_{\rm true}$. They are applied to maps
  of 5 simulated test cases (Cases 2, 4, 6, 9, 10) 
  that are based on real MIPS 24$\mu m$ maps
  (simulated cases based on artificial sources are excluded). 
  The results clearly demonstrate the applicability and limitation of
  individual map-makers. Destriper-P0 produces the least deviations in
  most cases, but its maps show artificial stripes for the cases with
  ``cooler burp effect''. Scanamorphos, running with the ``galactic option''
  and without the ``relative gain corrections'', can minimize the
  ``cooler burp effect''. However, bright pixels in Scanamorphos maps display
  large deviations, likely due to a slight positional offset
  introduced by the mapper, and a slight change in the beam
  size. Destriper-P1, SANEPIC, and Unimap introduce different types of
  large spatial scale noise. For SANEPIC, this is likely due to
  mismatches between the assumptions made in the map-maker and the
  properties of the test data.  For example, SANEPIC assumes that data
  are circulant, which is not true for the Case 9. For Unimap, the
  large scale distortion in maps of Case 6 is triggered by the ``cooler burp
  effect'', which the map-maker does not know how to handle. For
  Naive-mapper (with simple median background removal), many maps show
  large deviations due to the over-subtraction of the background when
  extended emission is present.
\item{(2)} Spatial (2-D) power spectra. These metrics include (i)
  plots and comparisons of power spectra of maps made by different
  map-makers; (ii) for simulated cases, plots and comparisons of the
  divergence from the truth
  power spectrum of the maps by different map-makers.  Most of the
  power spectra, either coming from real or simulated data, \no\ or
  with extended emission, show very similar results.  In the ``middle
  part'' ($k = [0.1, 1]$ arcmin$^{-1}$), results among different
  map-makers vary little: $\sim 1\%$ for cases where a truth map was
  available as benchmark. At smaller scales ($k > 1$ arcmin$^{-1}$),
  the standard Naive mapper produces higher powers than other
  map-makers, presumably due to the fine-stripes (baseline removal
  errors) found in its maps.  Meanwhile, at the same scale, results of
  Destriper-P0, Destriper-P1, Unimap, and SANEPIC are always very
  close, and those of Scanamorphos are usually lower. The low power at
  high spatial frequencies in Scanamorphos maps is likely due to the
  fact that, unlike other map-makers, Scanamorphos distributes the
  signal measured at a sky position among multiple adjacent map
  pixels. This is equivalent to a map smoothing, which takes away high
  frequency powers.  At larger scales ($k < 0.1$ arcmin$^{-1}$), again
  the Naive mapper produces higher powers because of the poor baseline
  removal, while the results of other map-makers are all
  comparable. 
  In the special cases with the ``cooler burp'', the power spectra of
  Naive and Destriper-P0 maps are clearly affected, showing much higher
  power at $k < 0.1$ and a \emph{peak} at $k \sim 1.5$ in the PLW
  map. No significant effects due to the  ``cooler burp'' are found
  in results of the other map-makers\footnote{It should be noted that Naive mapper and Destriper were not designed to treat the cooler burp. Previous parts of the standard pipeline will do this in future HIPE versions.}.
\item{(3)} Point source and extended source photometry. These metrics include
     (i) astrometry of point sources;
    (ii) point source and extended source photometry;
   (iii) detection rates of faint point 
      sources, obtained using 
      Starfinder (a point source extractor);
    (iv) PSF profiles.
 They are applied to the simulated test cases with
 artificial sources (Cases 1, 5 and 8). The results show that 
 bright sources in maps made by Scanamorphos have systematically larger
 position errors ($\gsim 0.1$ pixel) than those in
  maps made by other map-makers, consistent with the results on position
  offsets in Scanamorphos maps found in
  Metrics (1) for the deviation from the truth. Photometry for bright 
  point sources in all maps has small errors, 
  indicating good energy conservation by all map-makers. On the other hand,
  photometry of extended sources in the Naive mapper are significantly affected by a known bias due to the over-subtraction of baselines, while other maps have no such issue. For faint
  point sources ($\rm f = 30\; mJy$), no significant difference
  is found among results for different map-makers on both detection
  rate and photometry. Also, there is no significant difference between beam profiles of sources in maps made by different map-makers.
\item{(4)} Metrics for super-resolution maps. These metrics are applied to
   maps made by HiRes and SUPREME, the two super-resolution mappers, and 
   compare them to maps made by the destriper (the pipeline default). They
   include: 
   (i) visual examinations of the maps;
   (ii) spatial power spectra;
   (iii) point source profiles.
   The results show that SUPREME and HiRes yield
  similar resolution enhancements (factors of 2-3) at spatial scales
  around 2 arcmin$^{-1}$ for the limited datasets tested at 250
  microns. At higher spatial frequencies corresponding to spatial
  scales smaller than the beam size, there is less power in the
  SUPREME maps (intentionally, to smooth and reduce the noise at
  scales smaller than the beam). HiRes contains more power than either
  SUPREME or Destriper-P0 maps between spatial scales of 15-20
  arcseconds. The differences in SUPREME and HiRes arise mainly
  because SUPREME is tuned to enhance extended emission features, and
  HiRes is essentially performing a deconvolution in image space.
\end{description}

\nid {\bf Summary of Results:}
\begin{itemize}
\item The Destriper with polynomial order of 0 (Destriper-P0), which is the
default map-maker in the SPIRE scanmap pipeline since HIPE 9, 
performed remarkably well and compared
favorably among all map-makers in all test cases except for those suffering
from the ``cooler burp'' effect, as it does not have a mechanism to deal with this effect. In particular, it can handle observations
with complex extended emission structures and 
with large scale background gradient very well. 
\item In contrast, the Destriper with 
the polynomial order of 1 (Destriper-P1) compared poorly among its peers,
introducing significant artificial large scale gradient in many cases. 
\item Scanamorphos showed noticeable differences in all comparisons. 
On the positive side,
its maps have the smallest deviation from the truth for faint pixels 
($\rm f < 0.2 Jy$) in nearly all cases.
Particularly, as shown in both the difference maps and in
the power-spectra, it can handle the ``cooler burp'' effect very well.
On the negative side, for bright pixels ($\rm f > 0.2 Jy$), 
its maps show significant deviations from the truth, likely due to a 
slight positional offset introduced by the mapper as well as
a slight change in the beam size. This effect is also seen
in the astrometric errors of the bright sources.
However the offset is very small ($\sim 0.1 pexel$), therefore it
does not affect the photometry of
both point sources and extended sources, and does not show up
in the comparison between beam profiles (resolution: 0.2 pixels).
The power spectrum analysis indicates some smoothing of the data compared 
to the other mapmakers.
\item The GLS mapper SANEPIC can also minimize the ``cooler burp'' 
effect. It performed quite well in most cases. However, for those cases
with strong variations in very large scales (i.e. comparable to the map size),
its maps show significant deviations from the truth. This
 is because some of its assumptions (e.g. TODs are circulant) are 
invalid for the data.
\item Unimap, another participating GLS mapper, is among the best
performers in most cases. However, because it does not include a mechanism for
handling the ``cooler burp'', its maps show significant
deviations from the truth in the cases affected by the artifact.
\item The Naive-mapper (with simple median background removal) is inferior 
among its peers in general. The most severe bias it introduces is the
over-subtraction of the background when extended emission is present. In the 
cases where the extended emission is in complex structures, this bias
cannot be avoided by simple masks in the background removal.
\item The two super-resolution mapmakers, SUPREME and HiRes, yield
  similar resolution enhancements (factors of 2-3) at spatial scales
  around 2 arcmin$^{-1}$ for the limited datasets tested at 250
  microns. At higher spatial frequencies corresponding to spatial
  scales smaller than the beam size, there is less power in the
  SUPREME maps (intentionally, to smooth and reduce the noise at
  scales smaller than the beam). HiRes contains more power than either
  SUPREME or Destriper-P0 maps between spatial scales of 15-20
  arcseconds. The differences in SUPREME and HiRes arise mainly
  because SUPREME is tuned to enhance extended emission features, and
  HiRes is essentially performing a deconvolution in image space.
\end{itemize}
\end{abstract}

%% file: chapter_introduction.tex
% NOTE: to be included in the report_main.tex
%\documentclass[11pt]{book}

%\begin{document}

\chapter{Introduction and Goals}

The current version of the standard SPIRE photometer 
scanmap data reduction pipeline (in 
HIPE 11) is doing a reasonably satisfactory job
for most observations obtained using the SPIRE Photometer AOTs.  Except for
some very challenging science goals (e.g. the CIB/CMB anisotropy
studies), maps in the Level 2 (or Level 2.5 for parallel mode observations)
products of the Standard Product Generation (SPG) pipeline (i.e. the
ones coming directly out of the HSA) are already of science
quality. In particular, after replacing the median baseline remover by
the destriper as the default baseline remover in the HIPE 9 pipeline,
there is not much left for an observer to do in order to further
improve the quality of a normal SPIRE photometer map, unless some
special problems occur for which the solutions have not yet been
developed. An example of these remaining problems is the so-called
"cooler burp" effect (affecting a few percent of SPIRE scanmap data):
After every SPIRE cooler-recycle, the first 6 hours or so see a steep
increase of the temperature  of the 300~mK sorption cooler.
This causes abnormal drifts in
detector timelines, which cannot be corrected by the standard
temperature drift correction module in the pipeline, and results in
stripes in maps observed during the “cooler burp” 
period\footnote{A correction that minimizes this artifact has been included 
in the SPIRE pipeline after this map-making test campaign was concluded.}.
Another issue with standard 
SPIRE maps is related to the relatively coarse angular resolutions
(beam size $\gsim 18''$) due to diffraction. Super-resolution
mappers that can beat the the diffraction limit and at same time
cause minimal artifacts are certainly desirable.

Given the general status of the SPIRE scanmap pipeline, and the
remaining issues in the SPIRE map-making, we set the following goals
for this SPIRE map-making test campaign:
\begin{itemize}
\item	Compare the map-makers in the SPIRE pipeline with other mapmakers
objectively and comprehensively.
\item	In particular, identify the strengths and limitations of different 
mapmakers in dealing with the known SPIRE map-making issues, such as 
the cooler burp effect.
\item Assess the resolution-enhancement capabilities of the super-resolution mappers as compared to the destriper (the pipeline default), and investigate their applicability to various kinds of data, as well as caveats or pitfalls to avoid.
\item Enable users to choose the right map-maker for their science.
\item Provide guidance for future development of 
the SPIRE scan-map data reduction pipeline.
\end{itemize}

In the next chapter, Chapter~\ref{ch:testcases}, the test cases examined in
this test campaign are presented. Chapter~\ref{ch:simu} describes
the method of the simulations based on which some test cases were generated.
Chapter~\ref{ch:mapmakers} introduces the map-makers participated in
the SPIRE map-making test. The test results are presented,
in a framework of pre-designed metrics,
in Chapter~\ref{ch:metrics}. The last chapter is dedicated to a general
summary.
%\end{document}

%% file: chapter_testcases.tex
\chapter{Test Cases}\label{ch:testcases}

In order to have comprehensive assessments for map-makers,
we requested that test cases shall cover the following parameter 
space of SPIRE scanmap observations:
(1) observation mode (nominal/parallel, scan speed, sampling rate); 
(2) source brightness; 
(3) map size; 
(4) depth; 
(5) complexity of the extended emission.
They shall also include examples of: 
(i) observations suffering from the "cooler burp" effects; 
(ii) sky regions with strong large-scale gradient. 

%\begin{savenotes}
\renewcommand\footnoterule{}
%\begin{table}[h]\scriptsize
\begin{table}[ht]\footnotesize
\begin{minipage}{\textwidth}  % this is to have the footnote!!!
\caption[Caption for LOF]{Test Cases\label{tbl:testcases}
}
\centering
\begin{tabular}{|l|l|l|l|l|l|l|l|}
\hline
Case & Method &Name & Mode & Scan & Samp &Size & Bands\protect\footnote{Abbreviations for SPIRE bands: S -- PSW, M -- PMW, L -- PLW.} \\ 
 & & & & speed & rate & & \\ 
%\hline
          & &          &           & ($''$/sec) & (Hz)       &($\circ\times \circ$) & \\
\hline
1 & simulation & Nominal~Sources     & Nominal & 30 & 16 & 0.7$\times$0.7 & S, M, L\\
\hline
2 & simulation & Nominal~Cirrus      & Nominal & 30 & 16 & 0.7$\times$0.7 & S, L\\ 
\hline
3 & real~obs   & Nominal~Dark        & Nominal & 30 & 16 & 0.7$\times$0.7 & S, M, L\\
\hline
4 & simulation & Nominal~M51         & Nominal & 30 & 16 & 0.7$\times$0.7 & S, M, L\\
\hline
5 & simulation & Fast-scan~Sources   & Fast~Scan & 60 & 16 & 3.5$\times$3.5 & S, L\\
\hline
6 & simulation & Fast-scan~MK-Center & Fast~Scan & 60 & 16 & 3.5$\times$3.5 & S, L\\
\hline
7 & real~obs   & Fast-scan~Dark      & Fast~Scan & 60 & 16 & 3.5$\times$3.5 & S, L\\
\hline
8 & simulation & Parallel~Sources    & Parallel & 20 & 10 & 1.3$\times$1.3 & S, L\\ 
\hline
9 & simulation & Parallel~MK-Center  & Parallel & 20 & 10 & 1.3$\times$1.3 & S, L\\ 
\hline
10 & simulation& Parallel~Cirrus     & Parallel & 20 & 10 & 1.3$\times$1.3 & S, L\\ 
\hline
11 & real~obs  & Parallel~Dark       & Parallel & 20 & 10 & 1.3$\times$1.3 & S, L\\ 
\hline
12 & real~obs  & Nominal~NGC~628     & Nominal & 30 & 16 & 0.4$\times$0.4 & S, L\\
\hline
13 & real~obs  & Para-fast~Hi-Gal-L30& Parallel & 60 & 10 & 1.9$\times$1.9 & S, L\\
\hline
\end{tabular}
\end{minipage}
\end{table}

In total 13 test cases were generated. The input data for these test
cases are time-ordered data, or TODs. The map-making process turns the
TODs into maps. These TODs have the format of the SPIRE Level-1
Photometer Scan Product (PSP).  As shown in Table~\ref{tbl:testcases},
the test cases include both real observations (5 cases) and simulated
observations (8 cases) in 4 observational modes. In the nominal mode
(scan speed = 30$”$/sec, sampling rate = 16 Hz), we have 5 test cases
(2 real, 3 simulated). In the fast-scan mode (scan speed = 60$”$/sec,
sampling rate = 16 Hz), there are 3 test cases (1 real, 2
simulated). In the parallel mode (sampling rate = 10 Hz), we have 4
test cases (1 real, 3 simulated) in slow scan (scan speed = 20$”$/sec)
and 1 real case in fast-scan (scan speed = 60$”$/sec). In order to save time 
but not to lose information, 3 test cases (Cases 1, 2, and 4) include all
3 SPIRE bands while others include only the PSW and PLW bands.

%\end{savenotes}

%% file: chapter_simulations.tex
\chapter{Simulations}\label{ch:simu}
Comparing to real observations, a simulated test case has the advantage of 
possessing the ``truth'', namely the sky model, based on which the
simulation is carried out. The truth map provides an unbiased standard against
which test maps made by different map-makers are to be compared.
Allowing for the effects of noise in a given map, deviations from the truth 
can be used as objective measures for the bias introduced by the 
map-making process.

In the simulations, TODs were generated using two layers of data:
(1) noise layer --- real SPIRE observations (public data) of a dark field;
this allows the simulation to include both instrumental noise and confusion noise;
(2) truth layer --- a sky-model map based either on a real MIPS 24$\mu m$
map (beam-size $\sim 6''$) or a map of artificial sources.
Sky-model maps in the ``truth layer'' are fine sampled (pixel size of
$1.25''$ -- $1.5''$), convolved with SPIRE beams (sampled with 
$1''$ pixels), and
scaled to desired brightness. For those taken from the real MIPS
24$\mu m$ maps, the scaling factors were set to make the original
noise in the MIPS observations negligible compared to that in the
noise layer.
   
The simulation procedure is as follows: 
\begin{description}
\item{1)} take Level-1 data of a real SPIRE observation of a dark field
(i.e. the noise layer); 
\item{2)} take a sky-model map (i.e. the truth layer) and replace its WCS 
by that of the noise layer; 
\item{3)} obtain the RA and Dec for every sampling point in the Level-1 timelines of the noise layer;
\item{4)} then, read the signal in the sky-model map at the RA and Dec of a 
given sampling point in the noise layer, and add this signal to the signal
of that sampling point in the corresponding Level-1 timeline of the noise layer;
\item{5)} do step 4) for all sampling points in the noise layer.
\end{description}

These simulations include instrumental
noise, confusion noise, and noise due to glitches. However, they do not include 
effects due to saturation, pointing error, 
photon noise contributed by very bright
sources. Also, the Level-1 timelines provided by the simulations are already
de-glitched using the standard SPIRE pipeline, which may not be desirable for
some map-makers (e.g. Unimap).
In total, we generated 8 simulated test cases (Table~\ref{tbl:simulations}).

\bigskip
\begin{table}[ht] %\footnotesize
%\begin{table}[h]\scriptsize
   \caption{Simulated Test Cases\label{tbl:simulations}}
\centering
\begin{tabular}{lllll}
\hline \\
Case & Mode & Truth~Layer & Noise~Layer & Depth \\
\hline \\
 1 & Nominal & artificial sources & Lockman-North   &  7 repeats \\ 
 2 & Nominal & cirrus region  & Lockman-North       &  7 repeats \\
 4 & Nominal & M51 & Lockman-North                  &  7 repeats \\
 5 & Fast-Scan & artificial sources & Lockman-SWIRE &  2 repeats \\
 6 & Fast-Scan & Galactic center  & Lockman-SWIRE   &  2 repeats \\
 8 & Parallel & artificial sources &  ELAIS N1      &  5 repeats \\
 9 & Parallel & Galactic center  &  ELAIS N1        &  5 repeats \\
10 & Parallel & cirrus region &  ELAIS N1           &  5 repeats \\
\hline \\
\end{tabular}
\end{table}

%% file: chapter_mapmakers.tex
% NOTE: to be included in the report_main.tex

\chapter{Map-makers}\label{ch:mapmakers}

Seven map-makers participated in the SPIRE map-making test, including
(1) Naive mapper (default of SPIRE SPG until HIPE 8);
(2) Destriper in two flavors: (i) Destriper-P0: Destriper with
polynomial-order~=~0 
(default of SPIRE SPG since HIPE~9) and (ii) Destriper-P1: Destriper with 
polynomial-order~=~1;
(3) Scanamorphos; (4) SANEPIC; (5) Unimap; (6) HiRes; 
(7) SUPREME. 
Detailed explanations of these map-makers, written by their authors,
are presented in the sections in this chapter.

Because of time constraints, not all map-makers processed all the
test cases (though some did). In Table~\ref{tbl:mapmaker_testcases},
the information on which map-maker processed which test cases is
provided with check marks.  It should also be noted that, for each
test case, two sets of input data (TODs, see Chapter~\ref{ch:testcases}) were
generated.  They were both corrected for the following instrumental
effects using the standard SPIRE scanmap data reduction pipeline: (1)
glitches; (2) electrical low-pass filter; (3) non-linearity; (4)
bolometer time response. Additionally, for Data Set 1, the TOD were
also corrected for the detector array temperature drift using the
standard pipeline, and the “turn-around” data between individual scans
were excluded (so there are gaps between scans).  These data were used
by the Naive, Destriper, HiRes and SUPREME.  For Data Set 2, the TOD
were not corrected for the detector array temperature drift; the
“turn-around” data were included so there are no gaps between
scans. These data were used by Scanamorphos, SANEPIC, and Unimap. 

We requested that all map-makers project their maps in the same
way. In particular, the crota2 parameter should be 0 (north-up); the
pixel sizes should be those of the SPIRE standard (pixel = $6''$,
$10''$, and $14''$ for PSW, PMW and PLW maps); and the projection
should be tangential in both the RA and Dec directions (RA -- TAN, DEC
-- TAN). These requirements also apply to the ``truth'' maps in
simulated cases (c.f. Chapter~\ref{ch:simu}), to which the test maps
are to be compared.  It should be noted that the original truth maps
are in finer grids (pixel size of $1.25''$ --
$1.5''$). Therefore, in order to facilitate the
comparisons, we choose to re-generate the truth maps from
simulated noise-free TODs using the SPIRE Naive mapper.

%\begin{savenotes}
\begin{table}[ht]
\begin{minipage}{\textwidth}  % this is to have the footnote!!!
\renewcommand\footnoterule{}
\caption[Caption for LOF]{Test Cases Processed by Different Map-Makers}
\centering
\begin{tabular}{|l|l|l|l|l|l|l|l|l|}
\hline
Case & Name & \multicolumn{7}{|c|}{Map-Maker\protect\footnote{Abbreviations for map-makers: N -- Naive, D -- Destriper, Sc -- Scanamorphos, SA -- SANEPIC, U -- Unimap, H -- HiRes, SU -- SUPREME.}\label{tbl:mapmaker_testcases}
} \\ 
\hline
          &      & N & D & Sc & SA & U & H & SU \\ 
\hline
1 & Nominal~Sources & $\surd$& $\surd$& $\surd$& $\surd$& $\surd$ &$\surd$ &$\surd$ \\
\hline
2 & Nominal~Cirrus & $\surd$& $\surd$& $\surd$& $\surd$&  & & \\
\hline
3 & Nominal~Dark  & $\surd$& $\surd$& $\surd$& $\surd$& & & \\
\hline
4 & Nominal~M51  & $\surd$& $\surd$& $\surd$& $\surd$& & & \\
\hline
5 & Fast-scan~Sources & $\surd$& $\surd$& $\surd$& $\surd$& $\surd$ & & \\
\hline
6 & Fast-scan~MK Center & $\surd$& $\surd$& $\surd$& $\surd$& $\surd$ & & \\
\hline
7 & Fast-scan~Dark & $\surd$& $\surd$& $\surd$& $\surd$&  & & \\
\hline
8 & Parallel~Sources & $\surd$& $\surd$& $\surd$& $\surd$&  & & \\
\hline
9 & Parallel~Mk Center & $\surd$& $\surd$& $\surd$& $\surd$&  & & \\
\hline
10 & Parallel~Cirrus & $\surd$& $\surd$& $\surd$& $\surd$& $\surd$& $\surd$& $\surd$ \\
\hline
11 & Parallel~Dark & $\surd$& $\surd$& $\surd$& $\surd$&  & & \\
\hline
12 & Nominal~NGC~628 & $\surd$& $\surd$& $\surd$& & $\surd$& $\surd$&$\surd$ \\
\hline
13 & Para-fast~Hi-Gal-L30 & $\surd$& $\surd$& $\surd$& & $\surd$& $\surd$& $\surd$\\
\hline
\end{tabular}
\end{minipage}
\end{table}
%\end{savenotes}

%% file: naiveMapperRep.tex
% Template for map-maker reports.
% CKX, April 7, 2013
%\usepackage{titletoc}% http://ctan.org/pkg/titletoc
%\usepackage{amsmath} %Never write a paper without using *amsmath* for
                     %its many new commands
%\usepackage{amssymb} %Some extra symbols
%\usepackage{makeidx} %If you want to generate an index, automatically
%\usepackage{graphicx} %If you want to include postscript graphics
%\usepackage{caption} %If you want to include postscript graphics
%\usepackage{epsfig}
%\setcounter{chapter}{1}
%
%\begin{document}
%
%
%\chapter{Participating Map-makers}
%  Do not change above lines.
%  Do not play with the margins or fonts.  This annoys the reviewers.
%  Unusual fonts do not render on all systems.  Don't change the
%  font size of the major headers either
% 

\section{Na\"{i}ve Mapper (Bernhard Schulz)}
% NOTE: Replace ``Map-maker Name'' with the name of the map-maker, 
% and ``Author'' with your last name. Example:
% \section{Naive (Schulz}

\subsection{Introduction}\label{introduction}

The Na\"{i}ve Mapper, for the purposes of this investigation, is considered in combination with the Median offset subtractor, since the SPIRE detector signals include arbitrary and variable offsets that can not be easily cast into calibration tables. The median subtractor component first removes the signal offsets by determining the medians of all unmasked readouts of all detectors in a given Level~1 building block and subtracting these values from their respective timelines. This works on the assumption that a map is dominated by flat sky-background, which is not true in general. Then the Na\"{i}ve Mapper proper establishes a regular grid of sky bins (map pixels) that covers all coordinates associated with the timelines, and the fluxes of individual readouts are distributed into these sky bins by their celestial positions. The contents of each sky bin are averaged, resulting in a regular rectangular numerical array, called the flux map. An error map and a coverage map of the same dimensions are generated as well, based on the distributions of the readouts within each sky bin and their total number therein. The following will give more algorithmic details and will outline the specific processing of each dataset.

\subsection{The Software}

The input data typically suffers from residual instrumental effects like residual glitches, a residual temperature drift component, and arbitrary constant offsets. In order of severity, the last effect has the strongest impact. The detector offsets change relative to each other depending on detector temperature and telescope background typically at the order of 5 Jy/beam. A simple method to remove these arbitrary offsets is based on the assumption that most of what a detector sees is sky background. The solution is then approximated by subtracting the median from each detector, effectively setting the celestial background level to zero. This works well as long as the primary assumption holds true, but breaks down in crowded fields, typically close to the Galactic Plane or in regions with strong Galactic Cirrus. This method has been used in SPIRE Standard Product Generation (SPG) until HIPE 8 together with the Na\"{i}ve Mapper and is implemented as Task baselineRemovalMedian() in HIPE.

If also temperature induced residual drifts are present, fit and subtraction of a polynomial can sometimes help. However, large scale source emission at low spatial frequencies will become unreliable as fractions of it will be removed by the method. It is implemented in HIPE as Task baselineRemovalPolynomial().

The Na\"{i}ve Mapper itself is applied after baseline subtraction. It first establishes a regular grid of sky bins (map pixels) based on the resolution parameter that covers the scanned sky field. For the SPIRE detector arrays 6$^{\prime\prime}$, 10$^{\prime\prime}$, and 14$^{\prime\prime}$ have been established as standard sky bin sizes for the detector arrays PSW, PMW, and PLW respectively, to provide near Nyquist sampled resolution. The Na\"{i}ve Mapper excludes all "bad" readouts where flags fit a given default mask. It further excludes by default all readouts that have scan speeds below 5~arcsec/sec. Both settings can be changed. It then distributes the fluxes of all remaining readouts into sky bins depending on their sky position, and calculates an average flux value and an uncertainty value for each sky bin, based on the standard distribution of the readouts within this sky bin and their total number (standard deviation of the mean). A weighting scheme is available, but is not applied by default. For details see the SPIRE Pipeline Specification Manual Version 2.2

For the purposes of this comparison we consider the baseline removal method based on median or polynomial fits to the Level~1 timelines an integral part of the map making process.
For this test campaign we have applied the "baselineRemovalMedian" task and the "naiveMapper" task sequentially to test cases 1 to 11 using their default parameters in HIPE 10.0.2155, The Cases 12 and 13 have been processed in HIPE 9.0.3063. We did additional  special processing as outlined later, using HIPE 11.0.1151 that made additional use of the tasks "baselineRemovalPolynomial" and "applyRelativeGains". It was verified that there was no difference in the essential results between these versions.

\subsection{Processing}\label{processing}
The na\"{i}ve maps were produced as part of a script that also generates destriper maps, which will be covered in a separate report.

First the script reads the FITS files that contain the individual Level~1 scan data, and combines them into a Level~1 context. It then runs the median baseline removal task in default configuration and calls the Na\"{i}ve Scan Mapper with the result. This is repeated for each of the requested detector arrays. The only additional parameter given to the Na\"{i}ve Mapper is the world coordinate system (WCS) defining the grid of sky bins. Otherwise default parameters, as described above, were used in the processing of the standard na\"{i}ve  maps.
There is a WCS definition for each of the five observations we used and each of the maps that were produced, i.e. the observations Lockman-SWIRE and ELAIS N1 HerMES have no definition for the PMW map. The WCS also incorporates limits on the field that confine the final map to the part of the data that has good coverage and is not affected by other effects possibly occurring in the turnaround region close to the edge of the map.
The resulting map is saved into a regular FITS image file using the simpleFitsWriter of HIPE. The naming format is "caseNN\_AAA\_mapCombined.fits" where NN is a number from 1 to 13, and AAA designates one of the detector arrays PSW, PMW, or PLW. The files are collected in a directory named "Naive".

As explained below the results obtained with the Na\"{i}ve Mapper can be dramatically improved over the straight pipeline processing with default parameters. Thus a second set of maps was produced for selected cases (Case 1, 4, 5, 7, 8, 11, 12).
In all cases with a strong central source (Case 1, 4, 5, 8, 12) a circular region of interest was excluded with radii of 8, 8, 71, 8.3, 4.8 arcmin respectively. In Cases 5 and 7 the polynomial baseline remover was used with polyDegree set to 1 to fix residual drifts.  
For all cases the relative gain correction factors for extended sources were applied to reduce systematic patterns that appear in all maps that are normally optimized for point sources.The files are collected in a directory named "NaiveMan".

\subsection{Comments}\label{comments}
The processing using effectively the standard pipeline configuration as it was up to HIPE~8 was deliberately chosen, to not confuse the results by arbitrary human intervention. Thus clear deficits are visible sometimes. Some of them were addressed with the special "manual" processing.

An issue that can be easily cured manually are bright extended sources that lead to dark stripes in both scan directions, due to the median being an overestimation of the background level for all scans crossing the bright source. This situation appears for Cases 1, 4, 5, 8, 12, which either include one very bright source at the center, or a central galaxy on an otherwise comparatively empty background. It can be fixed by defining an exclusion zone around the bright object, resulting in a better background estimate by the median.

This method however doesn't work anymore for a structured background, where the base assumption that most of the readouts in a scan see empty sky, is invalid. This becomes evident inspecting visually the results for Cases 2, 6, 9, 10, 13 (image and error maps), which are the non-trivial tests in the dataset we consider here.

Another issue was observed with the base data of the Lockman-SWIRE field, which is quite large on the sky, taking more time for the scan across. Case 7 has no additional layer and shows several scans in vertical N-S direction that obviously drift at a low level. In this particular case of a rather empty field on the sky the problem could be fixed manually using the polynomial baseline remover with a first order polynomial. This method works also for Case 5 with the bright source masked out.  For Case 6, however, this method is unlikely to work due to the many other flux components in the field and wasn't tried.

A residual repetitive stripe pattern is observed in most error maps, most prominently with the Lockman-North, Lockman-SWIRE, and NGC628 base data and for an unknown reason less so for the ELAIS N1 HerMES field. Part of this pattern is due to a mismatch of gains between detectors because the extended gain factors were not applied. This is supported by the missing entry in the product history of the Level~1 products. Application of these factors brings somewhat of an improvement as shown in the manually reprocessed maps.

%\bibliographystyle{abbrv}
%\bibliography{main}
%
%\end{document}

%% file: destriperRep.tex
% Template for map-maker reports.
% CKX, April 7, 2013
%\documentclass[11pt]{book}
%\usepackage{titletoc}% http://ctan.org/pkg/titletoc
%\usepackage{amsmath} %Never write a paper without using *amsmath* for
%                     %its many new commands
%\usepackage{amssymb} %Some extra symbols
%\usepackage{makeidx} %If you want to generate an index, automatically
%\usepackage{graphicx} %If you want to include postscript graphics
%\usepackage{caption} %If you want to include postscript graphics
%\usepackage{epsfig}
%\setcounter{chapter}{1}
%
%\begin{document}
%
%
%\chapter{Participating Map-makers}
%  Do not change above lines.
%  Do not play with the margins or fonts.  This annoys the reviewers.
%  Unusual fonts do not render on all systems.  Don't change the
%  font size of the major headers either
% 

\section{Destriper (Bernhard Schulz)}
% NOTE: Replace ``Map-maker Name'' with the name of the map-maker, 
% and ``Author'' with your last name. Example:
% \section{Naive (Schulz}

\subsection{Introduction}\label{introduction}

Although the destriper is designed as a stage that removes arbitrary offsets from the Level~1 signal timelines, it creates Na\"{i}ve maps in the process and eventually produces a map. Thus we consider the module a map-maker as well. The iterative process starts with timelines that have their respective median values subtracted by default. A first Na\"{i}ve map is created from these timelines, which is then re-sampled in the sky-positions of every readout of the original timelines. The difference between the original and the re-sampled timelines is fitted by an offset-function which is generally a polynomial, but by default is only a polynomial of degree zero, i.e. the average. In the next iteration another Na\"{i}ve map is constructed from the differences of the original scans and the respective new offset functions. This process converges and ends when all timelines have converged based on a suitable $\chi^2$ condition. Details about the Na\"{i}ve mapper are given in the respective chapter. The following will give more algorithmic details about the destriper and will outline the specific processing of each dataset.

\subsection{The Software}

The input data typically suffers from residual instrumental effects like residual glitches, a residual temperature drift component, and arbitrary constant offsets. In order of severity, the last effect has the strongest impact. The detector offsets change relative to each other depending on detector temperature and telescope background typically at the order of 5 Jy/beam. A simple method to remove these arbitrary offsets is based on the assumption that most of what a detector sees is sky background. The solution is then approximated by subtracting the median from each detector, effectively setting the celestial background level to zero. This works well as long as the primary assumption holds true, but breaks down in crowded fields, typically close to the Galactic Plane or in regions with strong Galactic Cirrus. 

To overcome this issue, an algorithm was developed that uses the constraints from the overlap of the detector scans that cross each other, to determine the relative signal offsets for each detector. The first working version was introduced in HIPE 7 and from HIPE 9 on this method has been used in SPIRE Standard Product Generation (SPG) and is implemented as Task destriper() in HIPE.

If also temperature induced residual drifts are present, fit and subtraction of a polynomial can sometimes help. For polynomial degrees $>0$ large scale source emission at low spatial frequencies may become unreliable if polynomials are fitted to single scans. We have tested polynomial degrees of 0 and 1 for single scan mode only. Full scan mode where the data of all scans of a given detector are fitted together has been implemented but has not been tested yet.

The destriper is implemented as a replacement for the baseline subtractor, although it fulfills the function of a mapmaker at the same time. Thus input and output are Level~1 timelines, but in addition the destriper also provides a destriped map, a diagnostic product, a TOD product if requested, and a residual signal product for debugging purposes.

The destriper takes the flux timelines with positions of a list context (Level 1 context) as input. After excluding all "bad" readouts where flags fit a given default mask, by default all readouts that have scan speeds below 5~arcsec/sec are excluded too. Both settings can be changed. A preliminary Na\"{i}ve map is then produced from this data and re-sampled. For each valid readout in the input timeline, the map-signal at the same position is taken to construct a map timeline. The algorithm will then fit an offset function that is a polynomial of user specified degree, by default zero degree, to the difference between input timeline and map timeline. In the next iteration the difference between input timeline and offset function is used to construct the next map, which is again re-sampled to provide the next estimate for the offset functions. For each timeline a $\chi^2$ is calculated as $\chi^2 = \sum_i[(mapTimeline_i - inputSignalTimeline_i + offsetFunction_i)^2]$. Each timeline is considered converged when the difference between the associated $\chi^2$ of two successive iterations sinks below a user selectable threshold. The destriper finishes when all timelines have converged, or when  the maximum number of iterations is reached. In addition the algorithm contains a suppression of bright sources, a Level 2 deglitcher, a jump detector, support for multithreading, and the use of a temporary pool if computer memory is an issue. The default pixel sizes for the SPIRE detector arrays are 6$^{\prime\prime}$, 10$^{\prime\prime}$, and 14$^{\prime\prime}$ and have been established as standard sky bin sizes for the detector arrays PSW, PMW, and PLW respectively, to provide near Nyquist sampled resolution. They can be changed too. 
Future prospects for development are 1) the introduction of weighted polynomial fits, and 2) allowing to use higher order polynomials for selected scans. 

For the purposes of this comparison we use the destriper as both, baseline remover and map maker. In this test campaign we have applied the "destriper" task to test cases 1 to 13 using default parameters in HIPE 10.0.2155. In addition a second set of maps was produced with polynomial degree set to 1. Both polynomial degree versions can be regarded as different mapmakers for the comparison.
Case 12 has been processed in HIPE 9.0.3063 and and Case 13 with HIPE 10.0.2734.

\subsection{Processing}\label{processing}

The destriped maps were produced as part of a script that also generates na\"{i}ve  maps, which are covered in a separate report. 

First the script reads in the FITS files that contain the individual Level~1 scan data, and combines them into a Level~1 context. It then runs the destriper task in default configuration except for the number of threads (nThreads), which is set to four. The
Na\"{i}ve Scan Mapper is then called with the destriped Level 1 products and the world coordinate system (WCS) defining the grid of sky bins. This is repeated for each of the requested detector arrays.
There is a WCS definition for each of the five observations and each of the maps that were produced, i.e. the observations Lockman-SWIRE and ELAIS N1 HerMES have no definition for the PMW map. The WCS also incorporates limits on the field that confine the final map to the part of the data that has good coverage and is not affected by other effects possibly occurring in the turnaround region close to the edge of the map.

\subsection{Comments}\label{comments}
The processing using effectively the standard pipeline configuration as it is used since HIPE~9 was deliberately chosen, to not confuse the results by arbitrary human intervention. Thus small deficits are visible sometimes. 

Bright central sources and structured backgrounds are well handled by the destriper without the need for any intervention. However, residual drifts in the data are not treated by the run with polynomial degree 0 (P0 runs) while destriping with a polynomial degree of 1 (P1 runs) effectively eliminates the stripes originating from the base data of the Lockman-SWIRE field, which is quite large on the sky, taking more time for the scan across. Case 7 has no additional layer and shows several scans in vertical N-S direction that obviously drift at a low level with P0 but appear stripe-free in the P1 results. The same is true for the other Cases 5 and 6 that use the same base data, although the artefact effectively disappears in Case 6 that is dominated by the bright back-projected layer. 

A residual repetitive stripe pattern is observed in most error maps, most prominently with the Lockman-North, Lockman-SWIRE, and NGC628 base data and for an unknown reason less so for the ELAIS N1 HerMES field. Part of this pattern is due to a mismatch of gains between detectors because the extended gain factors were not applied. This is supported by the missing entry in the product history of the Level~1 products. Application of these factors brings somewhat of an improvement as shown in the manually reprocessed Na\"{i}ve maps, however to maintain comparability w.r.t. the other mapmakers the Level~1 products were not modified in this way.

%\bibliographystyle{abbrv}
%\bibliography{main}
%
%\end{document}

%% file: mapmaking_report_scanamorphos_v4.tex
% Template for map-maker reports.
% CKX, April 7, 2013
%\documentclass[11pt]{book}
%\usepackage{titletoc}% http://ctan.org/pkg/titletoc
%\usepackage{amsmath} %Never write a paper without using *amsmath* for
%                     %its many new commands
%\usepackage{amssymb} %Some extra symbols
%\usepackage{makeidx} %If you want to generate an index, automatically
%\usepackage{graphicx} %If you want to include postscript graphics
%\usepackage{caption} %If you want to include postscript graphics
%\usepackage{epsfig}
%\begin{document}
%\setcounter{chapter}{1}
%\chapter{Participating Map-makers}
%  Do not change above lines.
%  Do not play with the margins or fonts.  This annoys the reviewers.
%  Unusual fonts do not render on all systems.  Don't change the
%  font size of the major headers either
% 

\section{Scanamorphos (H\'{e}l\`{e}ne Roussel)}\label{sect:scanamorphos}
% NOTE: Replace ``Map-maker Name'' with the name of the map-maker, 
% and ``Author'' with your last name. Example:
% \section{Naive (Schulz}

\subsection{Introduction}\label{introduction}

Scanamorphos is an IDL software making maps from flux- and
pointing-calibrated time series, exploiting the redundancy in the
observations to compute and subtract the total low-frequency noise
(both the thermal noise, strongly correlated among detectors, and the
uncorrelated flicker noise). The required level of redundancy is
reached in Herschel PACS and SPIRE observations; a fiducial value that
is convenient to remember is 10 samples per scan pair and per FWHM/4
pixel. Its capabilities also include the detection and masking of
glitches, and (for PACS) of brightness discontinuities caused by
either glitches or instabilities in the multiplexing circuit;
low-level interference patterns sometimes affecting PACS data are not
handled. The algorithm is described, accompanied by simulations and
illustrations, in Roussel  \cite{roussel2012a} and  \cite{roussel2012b}. 
The repository of the software and up-to-date documentation is:
{\tt http://www2.iap.fr/users/roussel/herschel}

The output consists of a FITS cube, of which the third dimension is
the plane index.  The first plane is the signal map; then come the
error map (statistical error on the weighted mean), the subtracted
drifts map and the weight map.  The weight of each sample is the
inverse square high-frequency noise (one value for each bolometer and
each scan).  Whenever present, the fifth plane is the “clean” signal
map, where the mean signal from each scan has been weighted by its
inverse variance; it is provided to ease the detection of remaining
artifacts in the map (by comparison with the first plane), not for
scientific purposes.

\subsection{Processing}\label{processing}

The processing options used for each dataset (among \/parallel,
\/galactic, \/jumps pacs) can be found in the fits file headers. The log
files, available upon request, contain a summary of the processing
steps, drifts amplitudes, observation duration and processing time.\break
{\bf Dates and code versions:} October for PACS, with v19; December for
SPIRE, with v20.

For SPIRE, mapmaking with and and without
the relative gain corrections (correcting for the different
beam areas of the detectors) were carried for the real
observations as well as the simulations.

\subsection{Comments}\label{comments}

For the SPIRE benchmarks, I have produced maps with two different WCS grids:
\begin{description}
\item{-} The common grid to be used by all map-makers, mapping only a central subfield.
\item{-} An enlarged grid with the same reference coordinates and pixel size, but a number of pixels
for each axis that is sufficient to cover to whole field of view.
\end{description}

\subsubsection{SPIRE point response functions:}
Since the projection method used by Scanamorphos is different from
that used in the pipeline (matrix projection versus nearest-neighbor
projection), the SPIRE PRFs are slightly different in Scanamorphos
maps. The PRF FWHM is 1.5\% larger, and the PSF area is 3\% larger.
Ideally, this should be taken into account in the photometry of point
sources. 

%% file: mapmaking_report_SANEPIC.tex
\section{SANEPIC (Alexandre Beelen)}

%\section{Summary}\label{summary}
\subsection{Introduction}\label{introduction}

Signal And Noise Estimation Procedure Including Correlation
(\textsc{sanepic}) is a maximum likelihood mapper capable of handling
correlated noise between receiver. It was first developped to handle
the BLAST experiment data \cite{Patanchon2008}, and then fully
rewritten, parallelized, and generalized to handle any kind of
dataset. The \textsc{sanepic} package now consists of several programs~:

\begin{itemize}
\item \textsc{saneFrameOrder} : to find the best distribution of the
  input data files over several computer, if \textsc{sanepic} is used
  on a cluster of computers;
\item \textsc{sanePre} : to distribute the data to temporary
  directories. The data are stored in a dirfile format, each computer
  receiving the data segment it will process;
\item \textsc{sanePos} : to compute map size, pixel indexes and a
  naive map. One can define a projection center or use a mask as a
  reference for projection. Users can use all projection supported by
  \textsc{wcslib} \cite{Calabretta2002, Greisen2002}. Conversion
  to/from ecliptic and galactic coordinates is also possible; A mask
  for strong source can also be define to remove crossing-constraints
  between different datasets.
\item \textsc{sanePS} : to compute noise-noise power spectra. the data
  are pre-processed and decomposed, in the Fourier domain, into
  uncorrelated and $n$-correlated components, using a mixing matrix of
  the correlated component, all components, common noise power spectra
  and mixing coefficients, are found using an expectation-maximization
  algorithm ;
\item \textsc{saneInv} : to invert the noise-noise power-spectra by
  mode, as needed for the full inversion made by sanePic;
\item \textsc{sanePic} : to iteratively compute the optimal map using
  a conjugate gradient method.
\end{itemize}

All programs take inputs from a single ini file which describe all
parameters, in particular the frequency of the high-pass filtering
needed before being able to transform the data in the Fourier domain
(see below for limitations).

\subsection{Processing}\label{processing}

All datasets, PACS or SPIRE, were processed in the same way :

\begin{itemize}
\item export the data from HIPE using export\_SpireToSanepic.py or
  export\_PacsToSanepic.py scripts;
\item define a blank mask with the requested WCS, add some margin
  pixels to accommodate for flag data on the edge, as sanePic need to
  be able to project all data : even flagged data needs to be present
  in the map (although not in the final map);
\item write the ini file for the processing, defining all directories,
  parameters file and choosing a very low frequency cut (half length
  of the time-stream allows to );
\item distribute the data segment with sanePre and compute pixel
  indexes and a naive map with sanePos;
\item for blank/deep field, compute the noise-noise power spectra
  using sanePS with the raw data, or bootstrap previously computed
  noise-noise power spectra; the number of common-mode component varies from
  1 or 2 for SPIRE to 6 for PACS Green;
\item inverse the noise-noise power spectra with sanePS and run sanePic.
\end{itemize}

The last two steps can/must be iterated using the previous iteration map of
sanePic as an input to be remove from the time-stream by sanePS. The process
converge quickly, in 3 to 4 iterations. This allow to derive noise-noise
power spectra in the case where strong or weak emissions are present in the
data. This also allow to adapt the noise component to each data segment, in
particular in case of cooler burps. In case of strong emission in the data,
noise-noise power spectra from a previous empty field can be bootstrapped as
the first iteration in the process.

\subsection{Comments}\label{comments}

\textsc{sanepic} make several assumptions on the data and noise model, which
can leads to known caveats/artifacts on the maps :

\begin{description}
\item[\emph{No gaps in the time stream} :] processing data in the Fourier
  domain, request that the time stream are contiguous, without gaps, in
  order to maintain consistency in the noise frequencies features. In
  particular, even if they are not used in the final map, turnaround of
  SPIRE and PACS data must be present in the timestream.
\item[\emph{Signal is circulant} :] this is the intrinsic hypothesis when
  doing Fourier Transforms, this implies that any signal gradient between
  the beginning and the end of the time stream will be removed : if the
  observation does not end where it started on the sky then any large scale
  gradient between those two points will be filtered out. This leads to very
  large scale filtering of bright gradient in PACS and SPIRE maps as the
  observations often start and end on the two extreme point on the map. Note
  that apart from those very large scales, that are not measurable by a
  Fourier analysis of the map, all the other scales are conserved : This
  implies that any difference to the truth map will show a large gradient,
  while any Fourier analysis of the map will show a transfer function close
  to unity.
\item[\emph{Noise is stationary} :] the noise properties
  are described by a single power spectrum over a data segment, this mean
  that over the data segment the noise must be stationary, having the same
  properties from start to end. This is very well the case for SPIRE and
  PACS receiver, with the exception of the cooler burps cases, where an
  additional noise component is needed, while the receiver noise are
  unchanged. If there are strong noise properties changes, one could split
  the data segment in several part where the noise is stationary, for
  e.g. before, during and after a cooler burp. If one still want to use the
  full data segment (to avoid complex filtering) then the frequencies of the
  cooler burp will down-weight the entire time stream of the data segment,
  the cross-scan will be necessary to recover those frequencies in the map.
\item[ \emph{Sky is constant over a pixel} :] As for all the mapmakers,
  \textsc{sanepic} assume that the sky is constant/flat over a pixel in the
  final map. This assumption could be broken in the case of (1) strong
  gradient in a single pixel, (2) astrometric mismatch, (3) gain or
  calibration mismatch between data segment. These problems, in case of
  strong sources, could lead to a wrong determination of the sky level over
  the filtering length of the data segment, thus leaving strong artifacts
  (crosses) on the maps. In order to avoid those problem, \textsc{sanepic}
  can remove the crossing constraints, between two data segment, using a
  mask for strong sources, whose flux level is determined using a simple
  mean between data segments.
\item[\emph{Bad Data/Glitches/Steps/Moving Objects/...} :] \textsc{sanepic}
  being only a mapmaker, the data must be properly flagged before being
  projected. Strong glitches, steps or any strong nonphysical gradient
  (induced from simulation which mismatched background level for e.g.), not
  well described in the Fourier domain, will need to be detected and flagged
  prior to processing. This could lead strong feature in the maps, even
  crosses, for strong glitches, or in the case of several faint unflagged
  glitches, to overestimate of the white noise level.

\end{description}

%\bibliographystyle{abbrv}
%\bibliography{mapmaking_report_SANEPIC}
%
%\end{document}

%% file: UniRep_new.tex
% Template for map-maker reports.
% CKX, April 7, 2013
%\documentclass[11pt]{book}
%usepackage{titletoc}% http://ctan.org/pkg/titletoc
%\usepackage{amsmath} %Never write a paper without using *amsmath* for
%                     %its many new commands
%\usepackage{amssymb} %Some extra symbols
%\usepackage{makeidx} %If you want to generate an index, automatically
%\usepackage{graphicx} %If you want to include postscript graphics
%\usepackage{caption} %If you want to include postscript graphics
%\usepackage{epsfig}
%
%\begin{document}
%
%
%\setcounter{chapter}{1}
%\chapter{Participating Map-makers}
%  Do not change above lines.
%  Do not play with the margins or fonts.  This annoys the reviewers.
%  Unusual fonts do not render on all systems.  Don't change the
%  font size of the major headers either
% 

\section{Unimap (Lorenzo Piazzo)}
% NOTE: Replace ``Map-maker Name'' with the name of the map-maker, 
% and ``Author'' with your last name. Example:
% \section{Naive (Schulz}

\subsection{Introduction}\label{introduction}

Unimap is a map maker based on the Generalised Least Square (GLS) approach, which is also the Maximum Likelihood (ML) method when the noise has Gaussian distribution. The method is well known, e.g. \cite{tegmark97}, and several practical implementations were proposed in the last decade. Unimap is specialised for handling Herschel data (PACS and SPIRE). 

Unimap is written in Matlab and can be compiled to run on every machine where Matlab can be installed, including Windows, Linux and Mac.

Unimap is divided into several modules, which are summarised in the following.

{\bf 1. Data loading.} The input data to Unimap is a set of fits files, each one storing an observation. The first module performs the loading of these files and the formation of the internal data structures, which also entails the projection of the data onto the pixellised sky defined by the astrometry parameters. This module also takes care of performing an intial filtering of the data, by rejecting timelines with a percentage of flagged redaouts higher than a user-specified level, and of setting the unit measure as specified by the user.

{\bf 2. Pre-processing.} This module detects signal jumps due to cosmic rays. Where jumps are detected the data are flagged and the timeline is broken into two, independent timelines. The module may also remove an initial signal tilt due to the memory of the calibration block, which can be found at the beginning of the timelines. As a last step this module linearly interpolates the flagged data.

{\bf 3. Glitch.} This module performs a high-pass filtering of the timelines and carries out a glitch search on the high-pass filtered data. A sigma-clipping approach is used where, for each pixel, the readouts falling into it are found and all the outliers (readouts with a difference from the median value larger than a user-selectable multiple of the standard deviation) are marked as glitches. After detection the marked values are reconstructed using linear interpolation.

{\bf 4. Drift.} This module estimates and removes the polynomial drift affecting the timelines. It exploits an Iterative implementation of a Subspace Least Square approach \cite{piazzo2013}. The user can select the polynomial order and if the drift is to be estimated for every single bolometer or for a whole array/subarray.

{\bf 5. Noise.} This module estimates the noise spectrum and constructs the corresponding GLS noise filters. 

{\bf 6. GLS.} This module estimates and removes the noise affecting the timelines by implementing the GLS map maker. It produces two output images in the form of fits files: the naive map and the GLS map.

{\bf 7. PGLS.} This module estimates the distortion introduced by the GLS map maker. It is based on the Post-Processing for GLS (PGLS) algorithm described in \cite{piazzo2012}. The estimated distortion is subtracted from the GLS map to produce a PGLS map, which is saved in the form of a fits file.

{\bf 8. WGLS.} This module implements the Weigthed PGLS (WGLS) described in \cite{piazzo2012} where the distortion estimated by the PGLS is analysed and subtracted from the GLS image only when it is significant. In this way the noise increase caused by PGLS is minimised.

A deeper description of the map maker can be found in the User's manual, which can be downloaded from the Unimap Home Page \cite{unipage}. In that page you also find a powerpoint presentation of the Unimap pipeline. A proper paper on Unimap is being written.

%Examples of Figures and Tables
%\begin{figure}[ht]
% \begin{center}
% \epsfig{file=figures/mapmaker_report_naive_fig1.eps, width=10cm}
%\end{center}
%\caption{Sample caption for one method of including figures.}
%\label{mapmaker_report_naive_fig1}
%\end{figure}
%
%Note: \label is for cross-referencing. e.g. 
% ``This is shown in Fig.~\ref{mapmaker_report_naive_fig1}.''

% Example for Tables:
%\bigskip
%\begin{table}[ht]
%\caption{Source Table\label{mapmaker_report_naive_tab1}}  % label it for cross-referencing
%\begin{tabular}{llcc}
%\hline \\
%Source & Coordinates & Flux Density/       & Diameter \\
%       & (J2000)     &   (mJy)             & (arcsec) \\
%\hline \\
%G153.4+0.3 & 17:18:00+59:30:00 & 50 & 5 \\
%G278.3-0.5 & 09:48:38-54:23:7.1 & 3 & 1 \\
%\hline \\
%\end{tabular}
%\end{table}

\subsection{Processing of SPIRE maps}\label{processing}

For the workshop, two real and four simulated SPIRE observations were reduced, in the PLW and PSW bands. All the processing was carried out on a laptop with 8 Giga RAM. The reduction time varies from a few minutes to one hour. 

Unimap comes with a default set of parameters' values. The processing approach was to firstly use the default parameters and inspect the results. If required, additional iterations were carried out in order to improve the quality. This process is simplified by the fact that Unimap can store the intermediate results and restart the processing from any module. 

In practice the default parameters always yield satisfactory images with the following exceptions:

- Unimap estimates the noise spectrum to compute the GLS filter impulse response, which is obtained by the IFFT of the spectrum. By default, the estimated spectrum is fit to a 1/f plus white noise spectrum model before the IFFT, in order to remove noise and spikes from the estimate. However the noise affecting the SPIRE level 1 timelines does not follow this model, because of the bolometer response compensation performed by the standard pipeline, which act as a high-pass filter. Therefore the GLS filters were computed from the raw spectrum, without any fit.

- For case 6, a strong drift due to a cooler burp was there in the timelines. This was combated by increasing the polynomial order used in the dedrift (we used 7 instead of a default value of 3). Also increasing the PGLS filter length up to a few hundreds of samples turned out to improve the results.

- The WGLS threshold was set manually in all the cases. This is due to the fact that, for Unimap releases below 5.4.0, the WGLS approach did not lend itself to an automatic threshold setting. This problem has been solved in Unimap 5.4.0.

As a general comment, we note that Unimap is not yet able to handle all the disturbances found in SPIRE data. In particular, the cooler burps are not adequaltely modelled. At the time being they can be partially compensated only by stretching to the limit the current processing (see case 6 above) which is just a patch. Nevertheless Unimap can satisfactorily handle most of the SPIRE observations and we have plans to improve the software in a reasonable time scale.

As a second comment, we note that the SPIRE level 1 data are not the best input for Unimap. The biggest problem is that the HIPE deglitching is heavy and may easily flag out more than 2 $\%$ of the data (which is deemed a too high percentage), by replacing the readouts with linearly interpolated values. Then we have two options: 1) use the linearly reconstructed values 2) keep the flagged data out of the image formation. Option 1) is unsatisfactory, because we are using artificial data in the image formation. Option 2) is better but, since many SPIRE tiles do not have a deep coverage, it causes several void pixels (pixels with no redouts) which are set to NaN in the final map, which is annoying. Also the numerical stability of GLS may suffer if too many data are flagged out.

A better option would be to switch off the HIPE flagging. However we verified that in this case the bolometer response compensation filter present in the standard pipeline will cause ringing with high spikes around the glitches. Then, we also should switch off that filter, with the additional advantage that the noise spectrum follows the 1/f plus white noise theoretical model. In this way the deglitching is moved to Unimap, which flags a substantially lower percent of the readouts, below $0.5 \%$. Indeed the best L030 image was obtained in this way. However there is still a problem, namely that some distortion may be there in the final image, since the bolometer response compensation has been suppressed.

The optimal solution would be to switch off both filtering and degitching in HIPE and move these steps to Unimap. This is a planned improvement, but is for the future. The currently used approach is option 1) above.

%\bibliographystyle{abbrv}
%\bibliography{main}
%
%\end{document}

%% file: mapmaker_report_hires_v5_pdf.tex
% Template for metrics reports.
% CKX, April 6, 2013

%\documentclass[11pt]{book}
%\usepackage{titletoc}% http://ctan.org/pkg/titletoc
%\usepackage{amsmath} %Never write a paper without using *amsmath* for
                     %its many new commands
%\usepackage{amssymb} %Some extra symbols
%\usepackage{makeidx} %If you want to generate an index, automatically
%\usepackage{graphicx} %If you want to include postscript graphics
%\usepackage{caption} %If you want to include postscript graphics
%\usepackage{epsfig}
%\begin{document}

%\setcounter{chapter}{1}
%\chapter{Participating Map-Makers}
%
%\setcounter{section}{5}

\section{HiRes (David Shupe)}\label{hires-intro}

\subsection{Introduction}
The HiRes mapmaker derives from the program of the same name, developed at the Infrared
Processing and Analysis Center (IPAC) for IRAS data. The method and its application to IRAS 
data are described in Aumann, Fowler \& Melnyk (1990)~\cite{Aumann:1990}. The algorithm is 
known as the Maximum Correlation Method (MCM). It was developed in large part to account for the great
variations in beam profiles of the IRAS detectors. In the limit of a constant beam profile, the method is
equivalent to Lucy-Richardson deconvolution.

HiRes was originally coded in FORTRAN and was recently ported to Python.
It is this Python code that has been used for the map-making workshop and in the preparation of this report.
The Python source code is available on GitHub\footnote{https://github.com/stargaser/hires.git} and a webpage
is available on the NHSC public Wiki\footnote{https://nhscsci.ipac.caltech.edu/sc/index.php/Spire/HiRes}.
The SPIRE ICC has ported the method to Java and incorporated it into the 12.0 development track of HIPE.
The Java version is already publicly available through Herschel's Continuous Integration Build 
system\footnote{http://herschel.esac.esa.int/hcss/build.php}, complete with a SPIRE Useful Script showing
how to run it.

The HiRes code takes as input the SPIRE timelines processed to Level 1 with destriping applied. The method
assumes that all un-masked data are valid values. This means that all artifacts such as glitches must be
removed from the timeline data before processing. The timelines must also be conditioned or destriped 
to remove any differences in the background between scans.

The beam profiles are input to the HiRes software as FITS files. The Python software allows for inputting a beam
profile for each SPIRE bolometer. As of this writing, we have used only the average beam profile in each
of the three SPIRE bands.

The HiRes method has also been implemented for images in the ICORE 
software\footnote{http://web.ipac.caltech.edu/staff/fmasci/home/icore.html}, based on the co-adding software
developed for the WISE mission. A manual is available \cite{Masci:2013}, and application to WISE is presented
in Jarrett et~al.~(2012)\cite{Jarrett:2012} and Jarrett et~al.~(2013)\cite{Jarrett:2013}.  Some ICORE maps were made
for the January 2013 workshop with 2-arcsecond pixel sizes, generally showing that map-level MCM processing 
works about as well as starting from the timeline data, for the simplified case of assuming a single beam for all
bolometers. 

The HiRes software begins with a flat image. An overview of the process is given by Jarrett et~al.~(2012)~\cite{Jarrett:2012}
(see in particular their Figure~2 for a graphical representation of the process). The first iteration of the algorithm is a
response-weighted coadd. We have found that the resolution of this first iteration is poorer than those produced by the naive
mapmaker.

There is provision for a BOOST parameter in the Python code. This can be used to, for example, square the correction factor for
the first few iterations. The software can output a ``beam image'' but this capability was not explored in this report.

\begin{figure}[ht]
\begin{center}
\includegraphics[width=6.0in, angle=0]{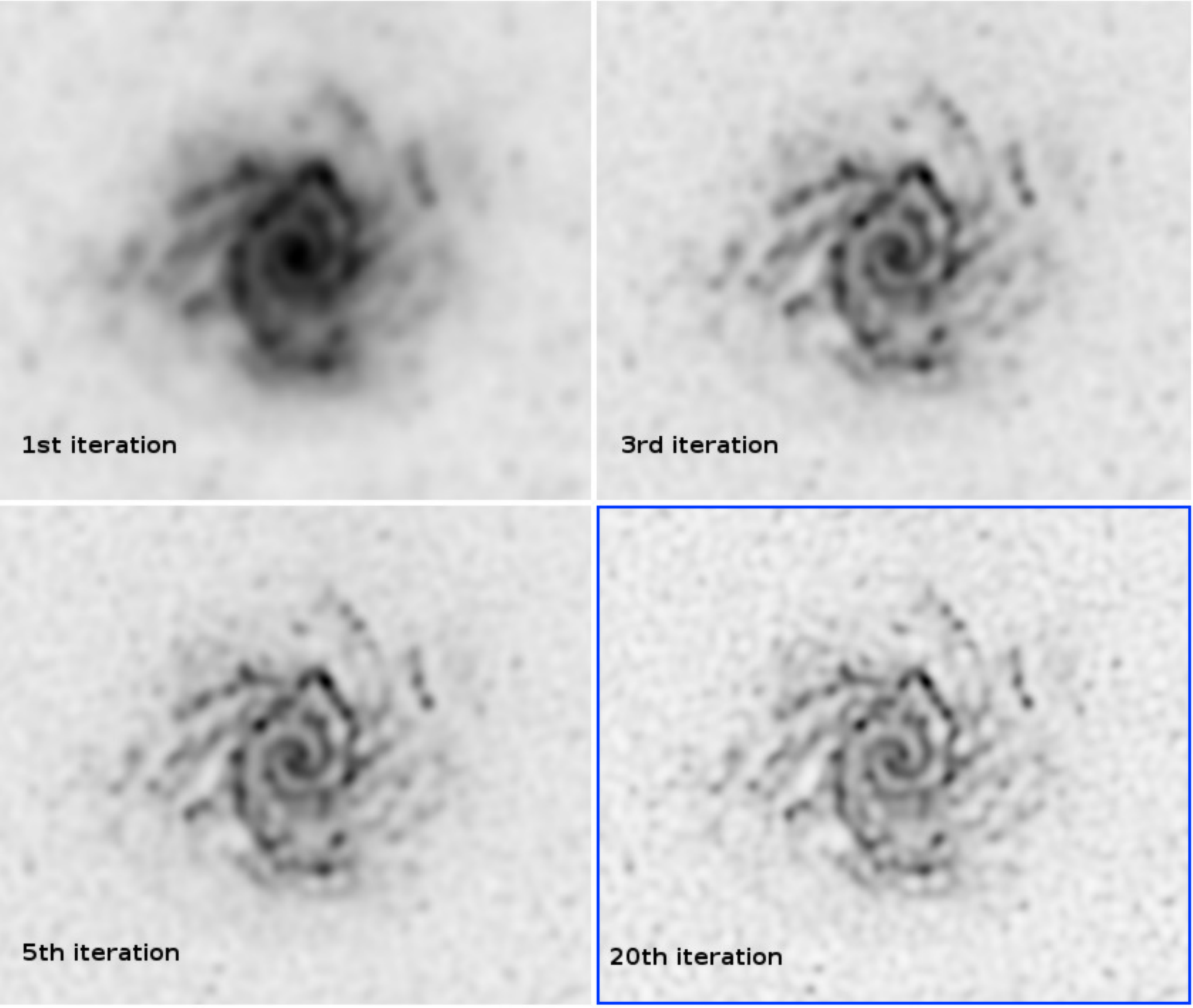}
\end{center}
\caption{The HiRes "flux" images for the PSW band (250 $\mu$m), for case 12 on NGC 628. The first, third, fifth and twentieth iterations are shown.}
\label{hires_report_example}
\end{figure}

%Examples of Figures and Tables
%\begin{figure}[ht]
% \begin{center}
% \epsfig{file=file1.eps, width=10cm}
%\end{center}
%\caption{Sample caption for one method of including figures.}
%\end{figure}

%\begin{figure*}
%    \centering
%    \includegraphics[width=7.5cm, angle=0]{file1.ps}
%    \includegraphics[width=7.5cm, angle=0]{file2.ps}
%\caption{Sample caption for another completely different but equivalent method of including figures.}
%\end{figure*}

%\bigskip
%\begin{tabular}{llcc}
%\hline \\
%Source & Coordinates & Flux Density/       & Diameter \\
%       & (J2000)     &   (mJy)             & (arcsec) \\
%\hline \\
%G153.4+0.3 & 17:18:00+59:30:00 & 50 & 5 \\
%G278.3-0.5 & 09:48:38-54:23:7.1 & 3 & 1 \\
%\hline \\
%\end{tabular}

\subsection{Processing of SPIRE maps}\label{processing}

%Describe how you processed the SPIRE maps in this test campaign. The
%goal is to make it possible for a reader to reproduce your results if
%she/he chooses to do so. Any caveats, cautions, and suggestions to a
%potential user, and comments to the SPIRE data, should go here, too.

The HiRes method requires all input data to be well-conditioned, that is, to represent real signals on the sky. There
cannot be any residual offsets in the timelines (as arise from the telescope background) and no artifacts such as
glitches can be present. We ran the SPIRE destriper on the input data to produce Level 1 products with the signal
levels normalized.

It is critical to this deconvolution method to subtract as much of the background as possible, without making 
the signals go negative. To optimize the background subtraction, the input and footprint-calculation stages 
were run in order to compute the global minimum of all the data. This minimum was then removed from all
the data before proceeding with the rest of the HiRes processing. As a precaution against negative pixels
causing problems with the deconvolution, any pixel values with negative flux were set to a value of 
$1 \times 10^{-200}$ Jy/beam.

SPIRE maps are normally made onto pixel sizes of [6, 10, 14] arcseconds for the [250, 350, 500] $\mu$m bands.
For the naive mapmaker used in the standard pipeline, smaller pixel sizes often lead to blank (NaN) values
in the output maps. With HiRes it is possible to use smaller pixel sizes. The recommended sizes are half the
standard values, that is, [3, 5, 7] arcseconds for the three SPIRE bands. For the mapmaking metrics the standard
pixel sizes were used, except for one case including simulated sources for which the half-standard sizes 
were used. The Python version was employed for all calculations for the map-making workshop as the
Java/HIPE version did not become available until early July 2013.

For each of the three bands, the same average beam profile was used as input to the method. Although the
SPIRE beam has a small ellipticity, we did not try to match the rotation of the beam with the map, instead just
assuming that the position angles were already matched. We used the average beam profiles that the SPIRE ICC
derived from Neptune observations made on a 1-arcsecond pixel grid, needing only the central region of 71x71
pixels for 500 $\mu$m and 151x151 pixels for 250 $\mu$m and 350 $\mu$m. (N.B. the inconsistency of the sizes 
is an accident but it should not affect the HiRes results very much -- the smaller size used for 500 $\mu$m was
chosen by Frank Masci when running the ICORE deconvolution on another dataset.)

In the HiRes lexicon, a ``footprint'' is the beam profile laid down on the output grid for a given detector sample. In effect
each footprint is the point response function (PRF) evaluated for a particular pixel phase.
Calculations were made with the PRF sampling at 1 arcsecond for PSW, and 2 arcseconds for both PMW and PLW.

The correction was accelerated using the BOOST SQUARED option for iterations 2 to 5. This means
that the correction factor was squared for these iterations. 

The maps were processed with a 10-pixel boundary around the final desired image sizes. The boundary was trimmed
from the 20th-iteration images to produce the final delivered maps.

%\bibliographystyle{abbrv}
%\bibliography{main}

%\end{document}

%% file: SupremeV2_pdf.tex
% !TeX spellcheck = en_US
\section{SUPREME  (Hacheme Ayasso)}
\subsection{Introduction}
SUPREME is a super-resolution mapmaker destined for extended emission with integrated destriping capacities \cite{ayasso2012b}. It is based on a realistic physical model and an unsupervised Bayesian approach which jointly estimate the map and other parameters of the model automatically from the data.  Therefore, It is easy to use since no parameter is required from the user. However, it is also possible for  the user to fix any of model parameters for fine tuning. The original code is developed in MATLAB$^{TM}$, and  a Java plugin for HIPE \cite{Hasnoun2013a} is available for the community since the beginning of 2013. This method is applied to SPIRE/Herschel, however it can be extended to other instruments. An implementation  to PACS mapmaking is being investigated.

\subsubsection{The method}

SUPREME mapmaking approach is based on physical model which links the measurements $\mathbf{y}$ to the observed sky $\mathbf{x}$. For SPIRE, the observation model is given as 
\begin{equation}
\mathbf{y} = \mathrm{H} \mathbf{x}+ \mathbf{n}
\end{equation}  
where $\mathrm{H}$ is the instrument model and $\mathbf{n}$ is the measurement noise which can be modelled by a Gaussian distribution with a unknown variance $\rho^{-1}_n$ and an offset $\mathbf{o}$ to compensate for the thermal drift in the measurement. In the current implementation of SUPREME, the instrument model is limited to telescope pointing $\mathrm{P}$ and  beam $\mathrm{C}$ \cite{Sibthorpe2011} models. Therefore, the data $\mathbf{y}$ are considered to be corrected for all other instrumental effects which is the case of Level1 data. 

Estimating the map $\mathbf{x}$ given $\mathbf{y}$ and $\mathrm{H}$ lacks a unique and a stable solution (ill posed problem) specially in a super-resolution context. Therefore, a prior information is needed for $\mathbf{x}$ to regularize the solution. SUPREME is designed for extended emission mapmaking, hence  a Markov field favorizing sky smoothness is used as a prior. The degree of sky smoothness is controlled by a correlation parameter $\rho_x$ which is supposed unknown.  

\begin{figure}[ht]
 \begin{center}
 \begin{tabular}{cc}
  \includegraphics[height=0.35\textwidth]{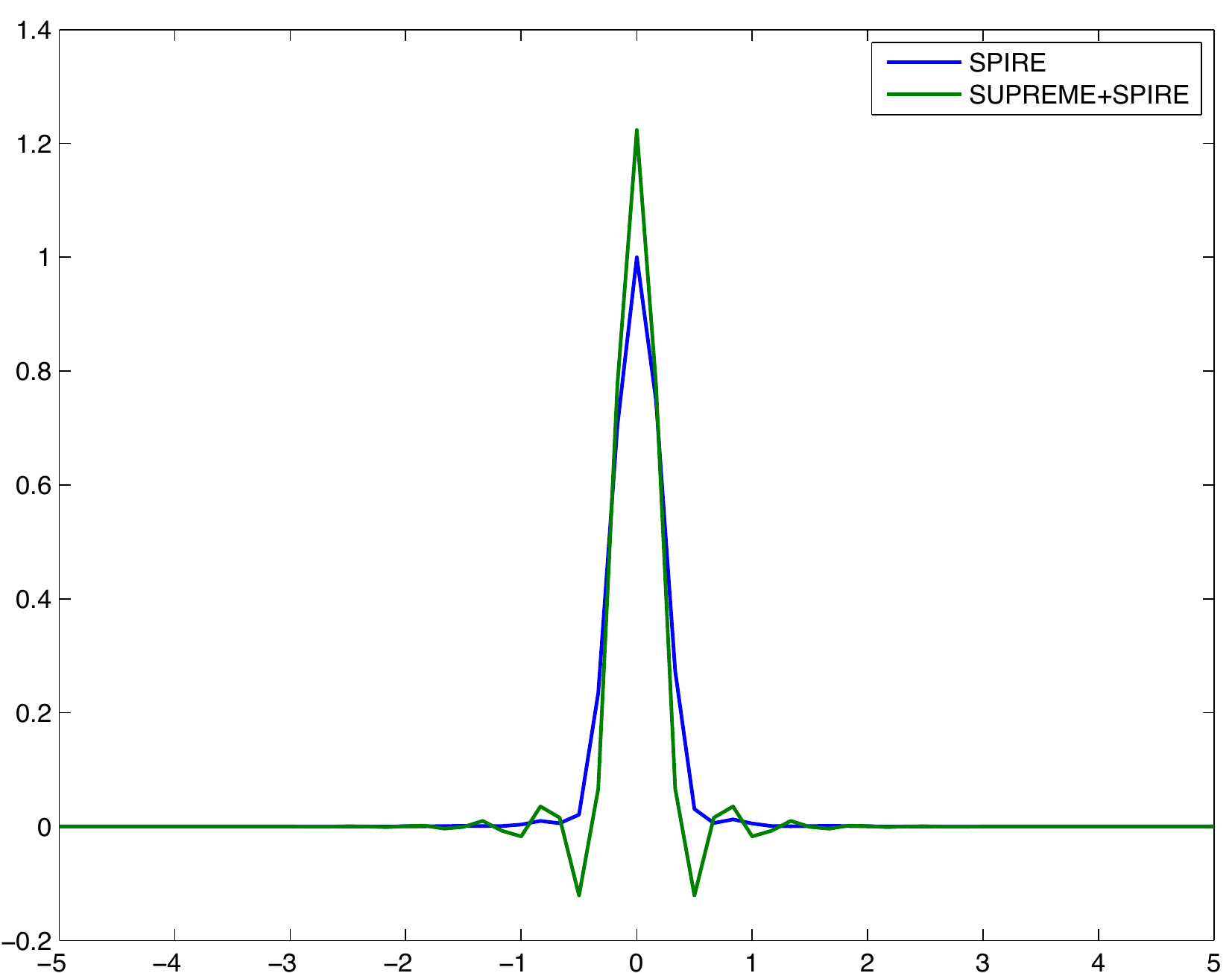}& \includegraphics[height=0.35\textwidth]{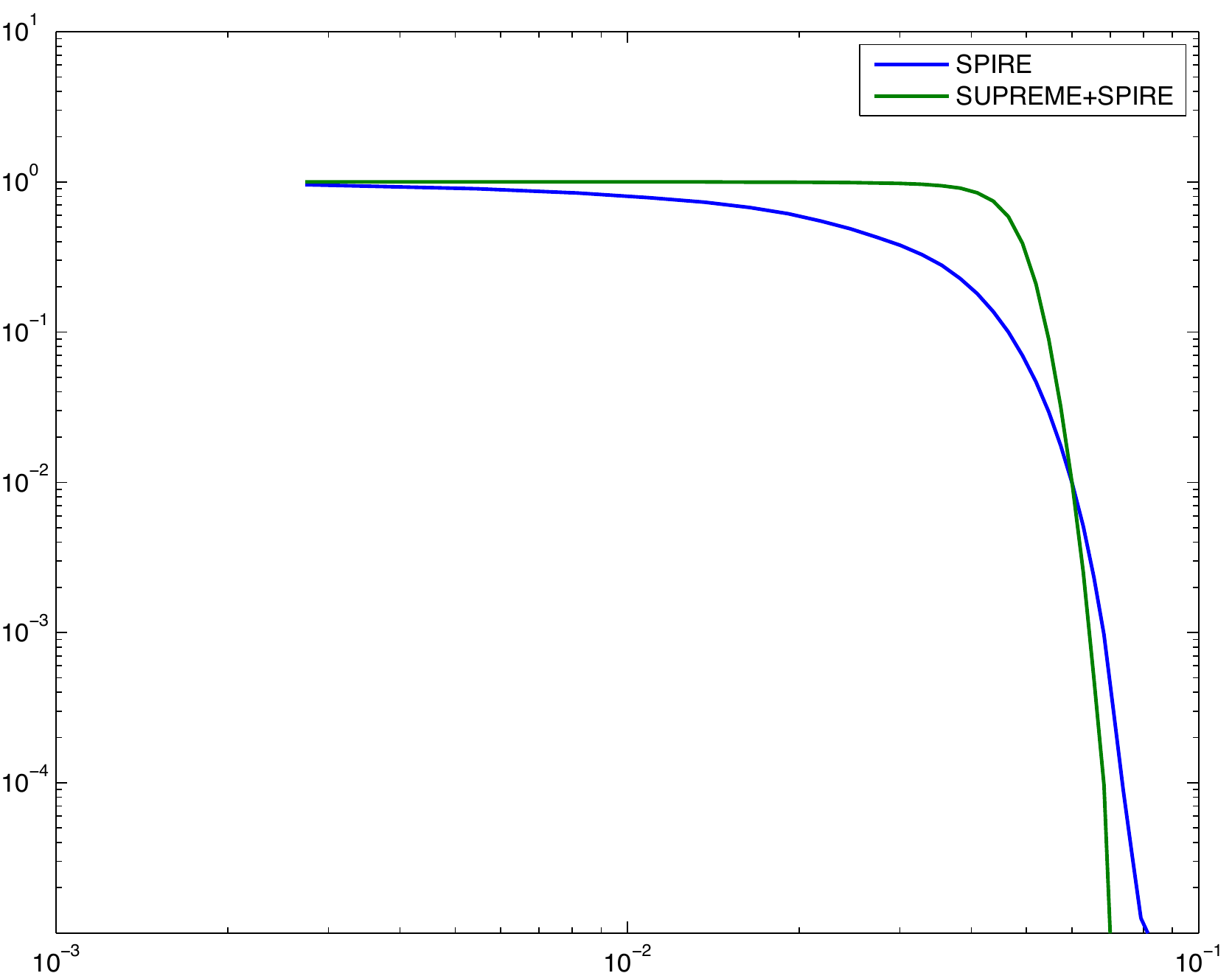} \\ 
  (a)& (b)  \\ 
 \end{tabular} 
\end{center}
\caption{Comparison between instrument beam profile (SPIRE) and the high resolution equivalent beam profile (SPIRE+SUPREME). (a) the spatial representation, (b) spectral representation }
\label{fig:HA_beam}
\end{figure}

SUPREME handles the inference in a Bayesian framework where the high resolution map $\mathbf{x}$ and the other model parameters (hyper-parameters) $\boldsymbol{\theta}=\left\{\rho_s,\rho_n,\mathbf{o} \right\}$ are chosen to maximize the joint posterior distribution. However the \textit{Joint Maximum A Posteriori} (JMAP) is intractable, the estimation is performed in an iterative scheme with an automatic stopping criteria.  After the convergence, the map and the hyper-parameters are given with their confidence intervals thanks to the probabilistic approach. Furthermore, the Bayesian theory presents a flexible framework for fixing some of the hyper-parameters within a supervised estimation situation. Moreover, the high resolution mapmaking naturally  changes the beam profile associate with the maps. The new beam profile, so called equivalent beam, is calculated directly from the instrument and the other model hyper-parameters. It has a flat spectral density for spatial scales where the signal is dominant and it decreases steeply for spatial scales where the noise is dominant (Figure.\ref{fig:HA_beam}). 

\subsubsection{Key features}
\begin{itemize}
\item High resolution maps: The resolution gain is up 3 times compared to classical Coadd mapmakers 
\item Instrument beam model is included based on the work of Sibthorpe~et~al~\cite{Sibthorpe2011}. The user gets to select between the simulated  and the empiric models.
\item 2 telescope pointing models are available: nearest neighbor model (like the model used in HIPE mapmaker) and bilinear interpolation model. The latter is more accurate compared to the former, however it is more time consuming.
\item A Gaussian noise model with variable offsets to achieve map destriping. The noise variance $\rho_n^{-1}$ and the offsets $\mathbf{o}$ can be estimated automatically or set by the user. The offsets are considered constant per leg per bolometer.
\item A markovian model for the sky with a correlation parameter $\rho_s$ which can be estimated automatically or fixed by the user.
\item The method provides also an error map and intervals of confidence for all the estimated variables as a byproduct thanks to the Bayesian framework.
\item Controlled equivalent beam model: the expression of the equivalent beam model of the high resolution map is provided for users 
\end{itemize}

\subsection{Processing}

SUPREME accepts level 1 inputs for high resolution mapmaking. Therefore, the plugin runs the standard HIPE pipeline to prepare data from level-0. However, it can accept directly level-1 data which are prepared with a non-standard pipeline.  In an automatic estimation framework, the user chooses the beam, pointing  and  thermal drift models.  

For benchmark data processing, the drift corrected data were used directly without any pre-treatment. The beam model was fixed to simulated one, the pointing model to a bilinear one and the drift model to constant per leg per bolometer. All other parameters were estimated automatically.  
 
\subsubsection{Example of performance}
The following figure (Horsehead nebula) demonstrates the high resolution capacity of our method. These parameters were used
\begin{itemize}
\item pixel size = 3'',
\item simulated beam,
\item automatic estimation for noise and field variances,
\item offset per leg per bolometer option,
\item manual stopping condition with 500 iterations
\end{itemize} 
 More results are available in \cite{ayasso2012b}.
%Examples of Figures and Tables
\begin{figure}[ht]
 \begin{center}
 \includegraphics[width=0.7\linewidth]{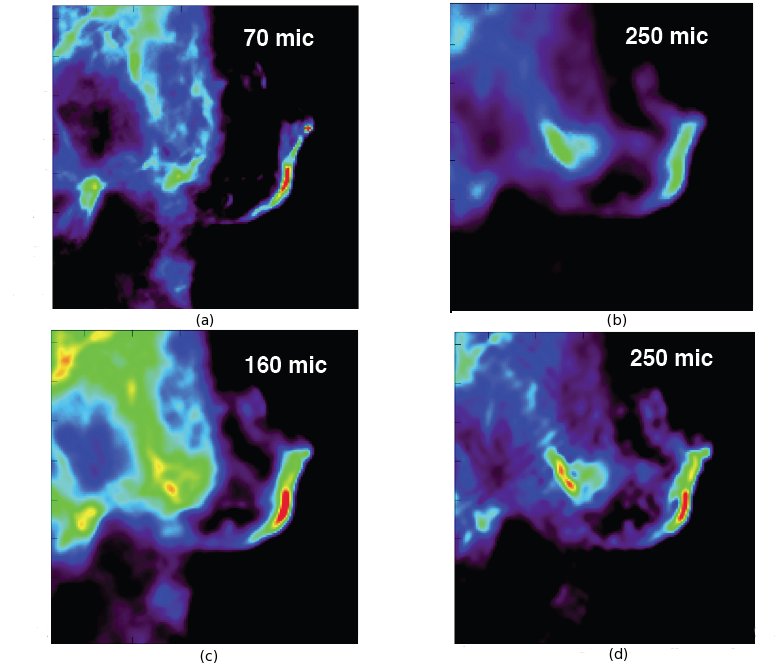}
\end{center}
\caption{Reconstruction results for Horsehead nebula (real data Herschel). (a) Coadd map form instrument PACS ($70\mu m$) , (b) Coadd map from instrument SPIRE $250 \mu m$, (c) Coadd map PACS ($160\mu m$), (d) SUPREME map from instrument SPIRE $250 \mu m$}
\label{fig:HA_HorseHead}
\end{figure}

\subsection{Comments}\label{comments}

\begin{itemize}
\item Since the method is designed for extended emission , applying it to fields with high intensities point sources might generate ring artifacts around them which corresponds to the spatial response of equivalent beam. While this ringing effect is normal, it may make SUPREME maps less useful. Therefore, another method (\textit{SUPREMEX}) was developed for joint mapmaking and source extraction. The scientific description of the method is given in Ayasso~et~al, \cite{Ayasso2012}. It will be implemented soon into HIPE plugin.
\item In certain fields where the detector offset variation is  strong the maps might still contain some stripes. Therefore, an enhanced noise model is under development for a better destriping. 
\end{itemize}

%% file: chapter_metrics.tex
% NOTE: to be included in the report_main.tex
%\documentclass[11pt]{book}

%\begin{document}

\chapter{Metrics and Results}\label{ch:metrics}

The maps made by different map-makers are examined and compared in the framework of four sets of metrics:
\begin{description}
\item{(1)} {\bf Deviation from the truth.} These metrics are applied to maps
  of simulated test cases that are based on real MIPS 24$\mu m$ maps
  (simulated cases based on artificial sources are excluded).  For
  these maps, deviations from the truth are the most direct and
  objective measures of the bias introduced by the map-making
  process. These metrics include:
 \begin{itemize}               
  \item visual examinations of the difference maps ($\rm Map - Map_{true}$);
  \item scatter plots of (S -- S$_{\rm true}$) vs S$_{true}$ for individual pixels;
  \item slopes of these plots; 
  \item absolute deviations: mean and standard deviation of S -- S$_{\rm true}$;
  \item relative deviations: mean and standard deviation of (S -- S$_{\rm true}$)/S$_{\rm true}$.
 \end{itemize}
\item{(2)} {\bf Spatial (2-D) power spectra.} 
      These are powerful tools to characterize
      the spatial distributions of maps in a general and abstract manner.
      Comparisons between the power spectra of maps made by different
      map-makers and those of the truth maps can reveal biases introduced by
      map-making processes and by SPIRE instrumental noise (e.g. the 
      ``cooler burp''). These metrics include:
 \begin{itemize}               
    \item power spectra plots
    \item plots of the divergence from the truth power spectrum,
      only for simulated cases.
  \end{itemize}
\item{(3)} {\bf Point source and extended source photometry.}
 These metrics, applying to the simulated test cases
 based on artificial sources (Cases 1, 5 and 8), are to test map-makers on their 
 ability in handling point sources and extended source, in particular on
 how well they can preserve the fluxes in the map-making process. The metrics 
 include:
  \begin{itemize}
    \item astrometry of point sources;
    \item point source and 
     extended source photometry;
    \item detection rates of faint point 
      sources, obtained using 
      Starfinder (a point source extractor);
    \item point source beam profiles. 
  \end{itemize}
\item{(4)} {\bf Metrics for super-resolution maps.} These metrics are applied to
   maps made by HiRes and SUPREME, the two super-resolution mappers, and 
   compare them to maps made by the destriper (the pipeline default). They
   include:
  \begin{itemize}
   \item visual examinations of the maps;
   \item spatial power spectra;
   \item point source beam profiles.
  \end{itemize}
\end{description}

Results of investigations using these metrics are presented in the following
sections in this chapter, written by individuals who carried out these
analyses.

%\end{document}

%% file: metrics_SPIRE_DIFFmap_report_update2013sept_pdf.tex
%% Template for report of SPIRE map-making test metrics results 
%% (adopted and modified from a similar template circulated by 
%% Roberta Paladini).
%
%  Contribution to the SPIRE map-making report.  
%
%  No style file is required.
%
%  Version 1.0    Feb 26, 2013
% 
%
% 
%%%%%%%%%%%%%%%%%%%%%%%%%%%%%%%%%%%%%%%%%%%%%%%%%%%%%%%%%%%%%%%%%%%%%%%%
%
%  The template begins here.  The font must be 12 point and the margins
%  must be at least 1-inch on all sides. 
%  Don't override this.  
%
% If you compile this and find that the text is "mushed up against
% the top of the page", the default paper size for your installation 
% of latex is A4.  In order to override this, do:
% $>$ latex texfile # where the manuscript is in a file named texfile.tex
% $>$ dvips -Ppdf -t letter -o texfile.ps texfile
% finally, to get nice (non-blurry, searchable) pdf do:
% $>$ ps2pdf13 texfile.ps  texfile.pdf
%
%
%  Do not play with the margins or fonts.  This annoys the reviewers.
%  Unusual fonts do not render on all systems.  Don't change the
%  font size of the major headers either

% NOTE: Replace ``XXX'' with the metrics name (eg. Deviation From The Truth)
%\title{Metrics: Deviation from the truth -- Difference maps}
\section{Deviation from the truth -- Difference maps (Vera K\"onyves \& Andreas Papageorgiou)}\label{sect:metrics_diff}
%\title{Metrics: Deviation from the truth -- Difference maps}
%\author{
%        Vera K\"onyves$^1$ \& Andreas Papageorgiou$^2$\\
%        $^1$IAS/Orsay -- CEA/Saclay, France\\
%        $^2$School of Physics and Astronomy, Cardiff University, UK\\
%}

%\date{\today}
%
%\documentclass[letterpaper,12pt, ]{article}
%
%\usepackage{epsfig}
%\usepackage{graphicx}
%\usepackage{graphics}
%\usepackage{url}
%\usepackage{color}
%
%\textwidth=6.5in
%\textheight=9.5in
%\topmargin=-0.75in
%\oddsidemargin=0.0in
%\evensidemargin=0.0in
%
%\pagestyle{myheadings}
%
% Please update the following line with the title of your paper 
% and your Author name (with "et al." if more than two authors).

%\markright{Metrics: Deviation from the truth\ -- K\"onyves \& Papageorgiou}
%\pagenumbering{arabic}

%\begin{document}
%
%\maketitle

\subsection{Test Data}\label{data}

Only simulated test cases (case 2, 4, 6, 9, and 10) were used in this metrics as we need a "truth" map
from which the difference will be derived. This way we can measure the biases, introduced by the various
map-making methods. In these tests we excluded the cases containing only artificial point/extended sources.  
The following table \ref{tab:simu} shows the simulated data sets for this metrics. For simplicity, but to still 
give representative results, only the PSW/250~$\mu$m and PLW/500~$\mu$m maps were tested. 
\begin{table}[h]\scriptsize
\begin{center}
%\small
\begin{tabular}{l l l l l l l}
Case		&Mode		&Truth Layer	&Noise Layer	&Map Size		&Depth		&Maps		\\
\hline
2		&Nominal	&cirrus region	&Lockman-North	&0.7d $\times$	0.7d	&7 repeats	&PSW, PMW, PLW	\\		
4		&Nominal	&M51		&Lockman-North	&0.7d $\times$	0.7d	&7 repeats	&PSW, PMW, PLW	\\	
6		&Fast-Scan	&Galactic center&Lockman-SWIRE	&3.5d $\times$	3.5d	&7 repeats	&PSW, PLW	\\	
9		&Parallel	&Galactic center&ELAIS N1	&1.3d $\times$	1.3d	&7 repeats	&PSW, PLW	\\	
10		&Parallel	&cirrus region	&ELAIS N1	&1.3d $\times$	1.3d	&7 repeats	&PSW, PLW	\\	 
\end{tabular}
\caption{Simulated data sets used in the "Difference map metrics".}
\label{tab:simu}
\end{center}
\end{table}
\normalsize

I also list here the availability of reprocessed maps which were compared to the above truth maps:
\begin{itemize}
  \item {\bf Case 2, and 9:} Naive, destriper/P0, destriper/P1, Scanamorphos, Sanepic
  \item {\bf Case 4:} Naive, Naive with human intervention, destriper/P0, destriper/P1, Scanamorphos, Sanepic
  \item {\bf Case 6, and 10:} Naive, destriper/P0, destriper/P1, Scanamorphos, Sanepic, Unimap
\end{itemize}
'P0' and 'P1' denotes the zero-, and first-order polynomial baseline removal in the HIPE destriper method.   

The maps were on the same WCS grid with the same units, therefore no further preparation was needed before making 
the difference maps. 

%More details on the simulations, mapper methods, and other metrics can be found at: 
%{\tt http://herschel.esac.esa.int/twiki/bin/view/Pacs/Map\_maker\_2013\_spire\_data}

\subsection{Analyses and Results}\label{results}

In this "Deviation from the truth -- Difference maps" metrics we present for each mapper method:
\begin{itemize}               
  \item a scatter plot of (S -- S$_{\rm true}$) vs S$_{\rm true}$ for individual pixels;
  \item slopes of these plots; 
  \item absolute deviations: mean and standard deviation of S -- S$_{\rm true}$;
  \item relative deviations: mean and standard deviation of (S -- S$_{\rm true}$)/S$_{\rm true}$;
\end{itemize}
IDL scripts were used to make difference maps, plots, and obtain statistics over the maps.

\subsubsection{Case 2 (nominal cirrus)}
\begin{figure*}[bhhh]
    \centering
    \includegraphics[width=16.0cm, angle=0]{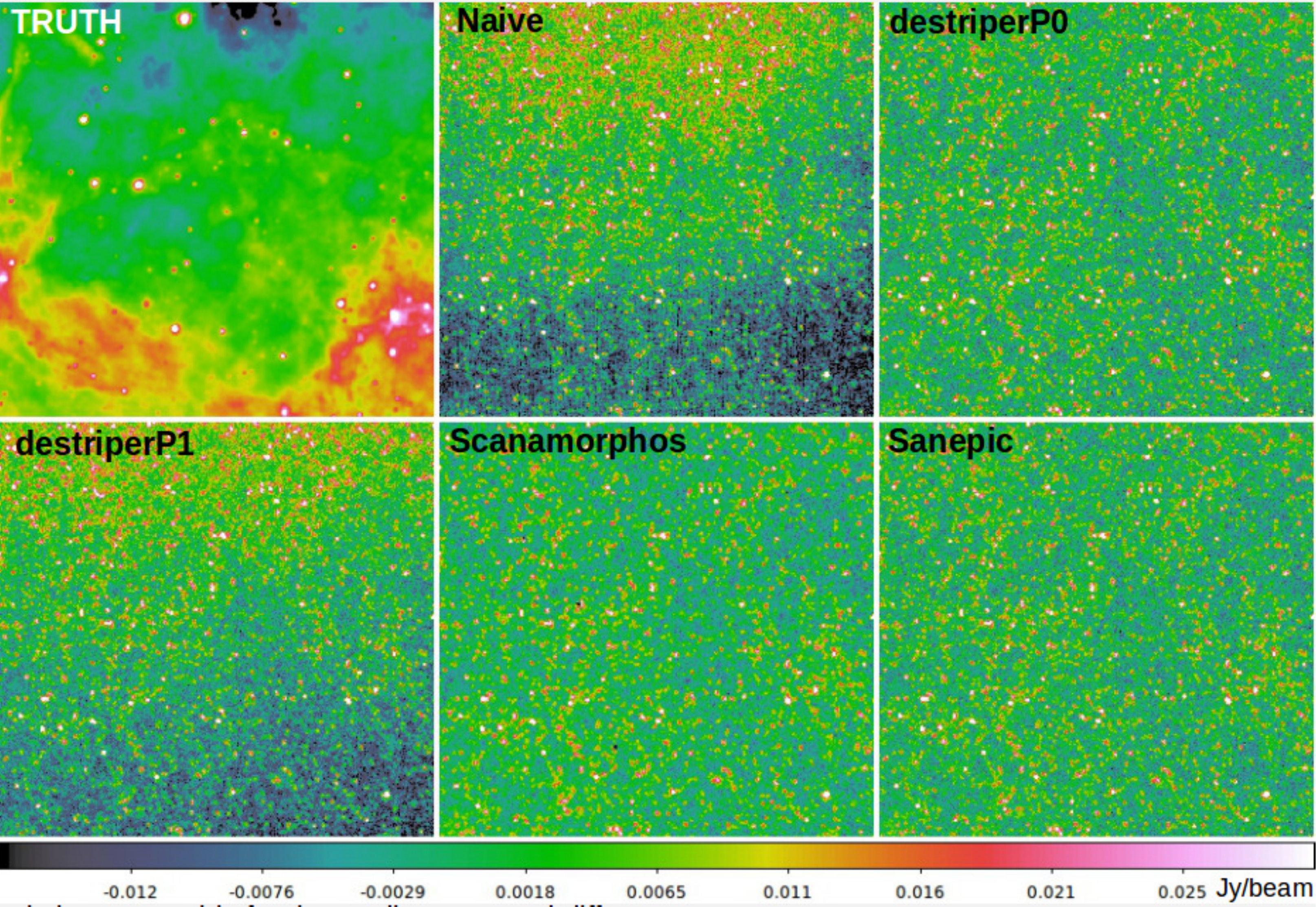}
\caption{Truth and Difference maps for case 2, PSW (nominal cirrus). For the latter ones, (Diff -- median(Diff)) is shown, 
         therefore the Jy/beam scale is comparable for the median-removed difference maps.}
\end{figure*}
\begin{figure*}
    \centering
    \includegraphics[width=17.0cm, angle=0]{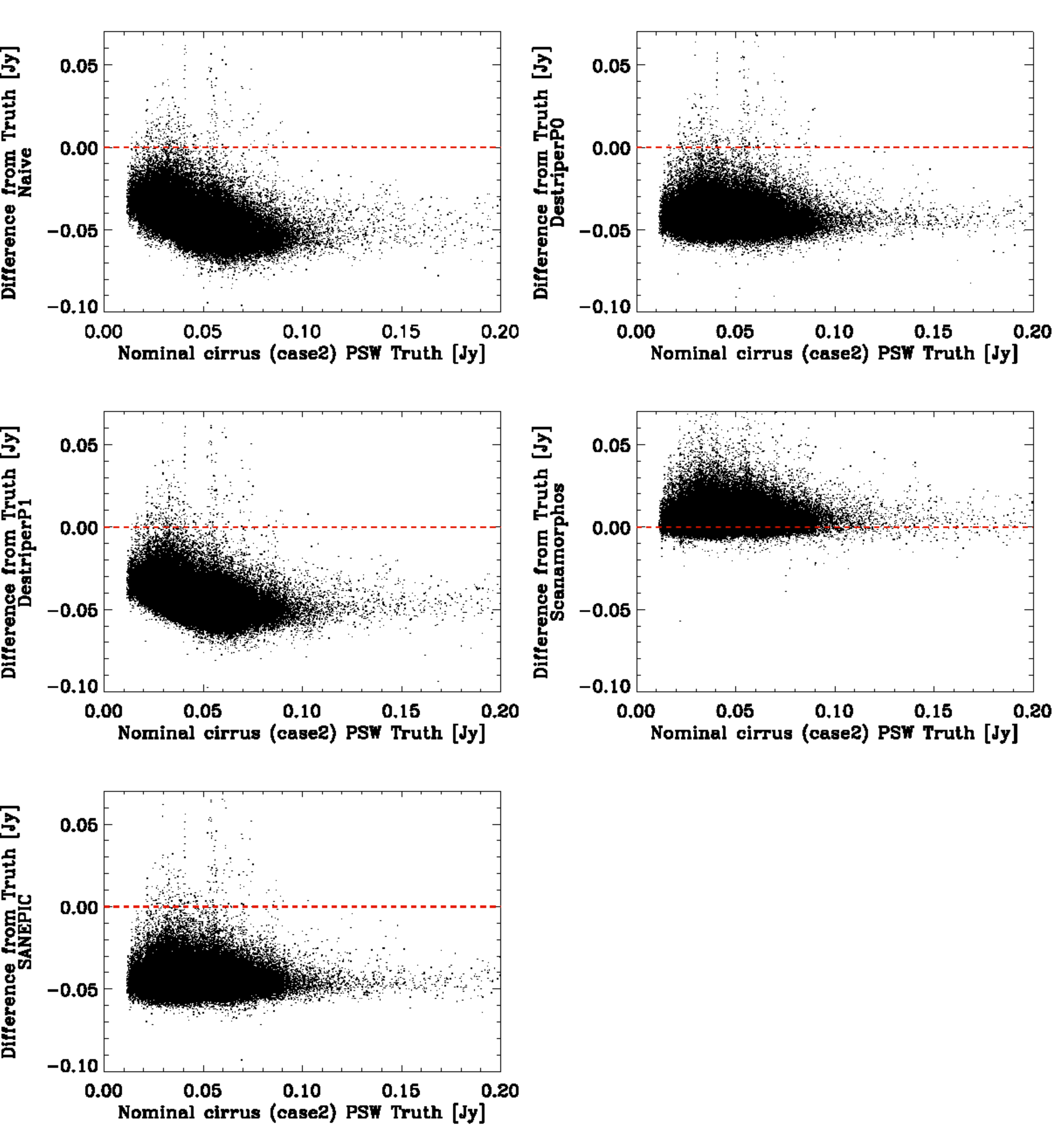}
\caption{Scatter plots of the difference maps for case 2, PSW; (S -- S$_{\rm true}$) vs S$_{\rm true}$.}
\end{figure*}
\begin{figure*}
    \centering
    \includegraphics[width=17.0cm, angle=0]{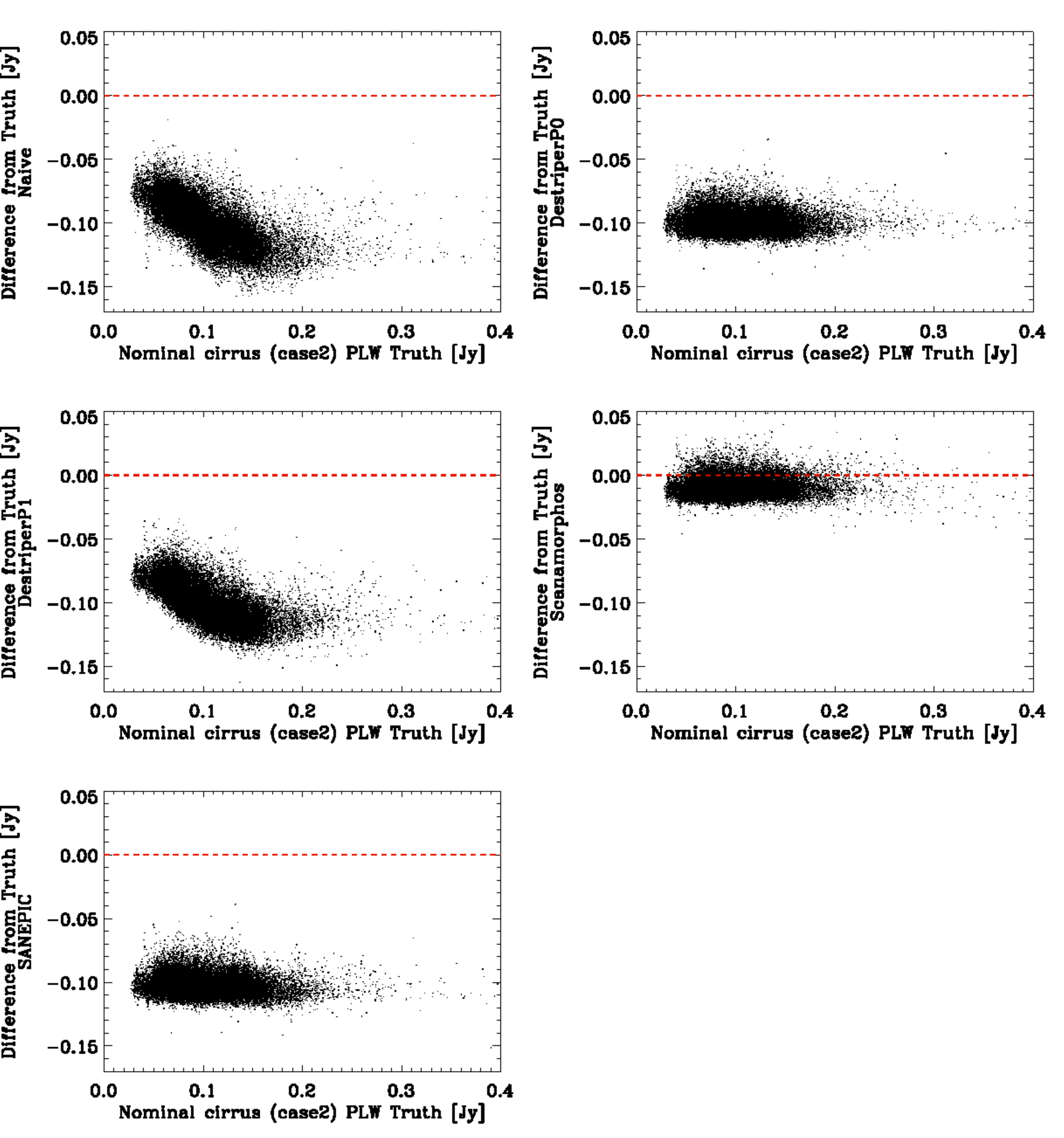}
\caption{Scatter plots of the difference maps for case 2, PLW; (S -- S$_{\rm true}$) vs S$_{\rm true}$.}
\end{figure*}
\begin{figure*}
    \centering
    \includegraphics[width=16.0cm, angle=0]{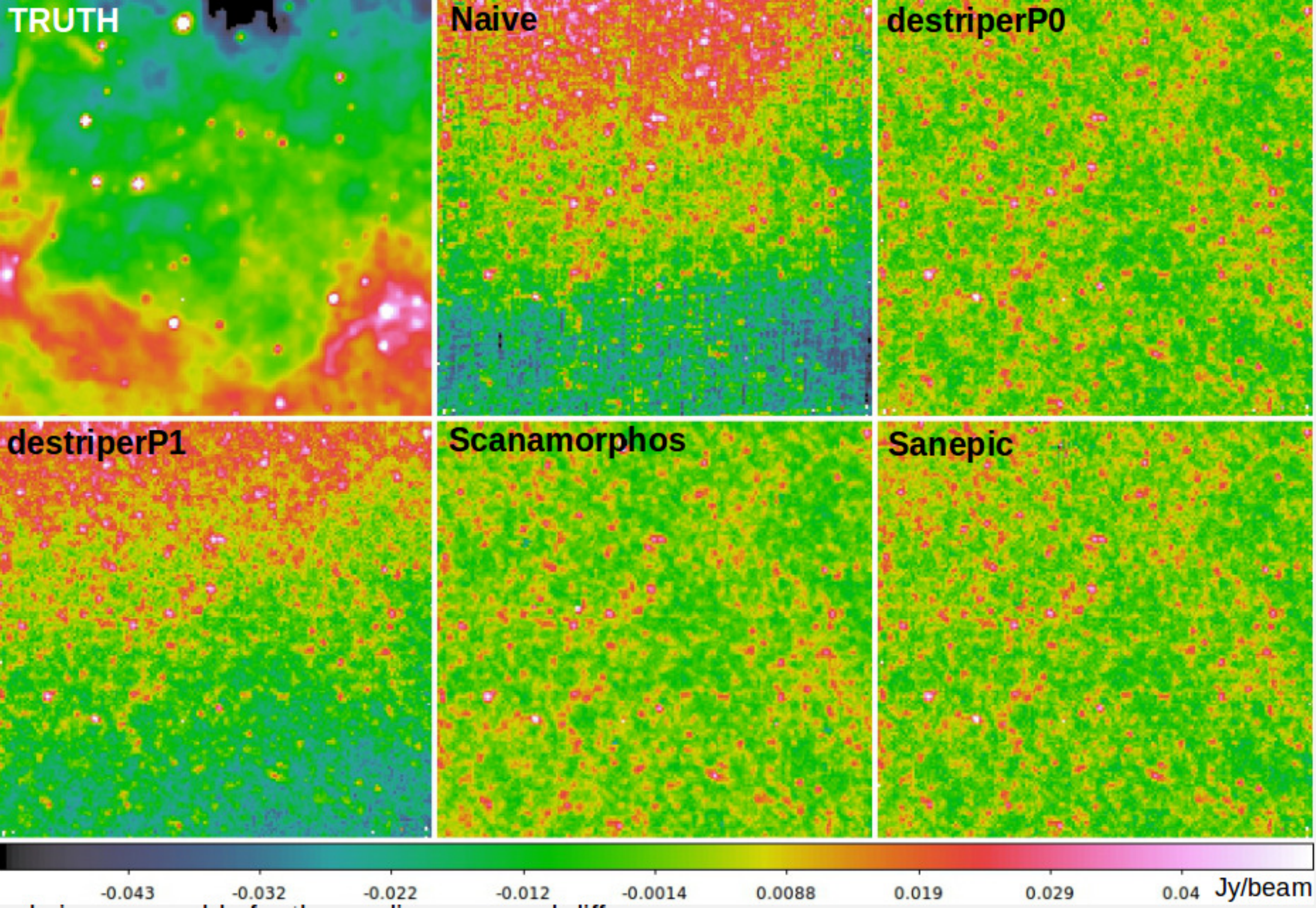}
\caption{Truth and Difference maps for case 2, PLW (nominal cirrus). For the latter ones, (Diff -- median(Diff)) is shown, 
         therefore the Jy/beam scale is comparable for the median-removed difference maps.}
\end{figure*}
\begin{figure*}
    \centering
    \includegraphics[width=8.0cm, angle=0]{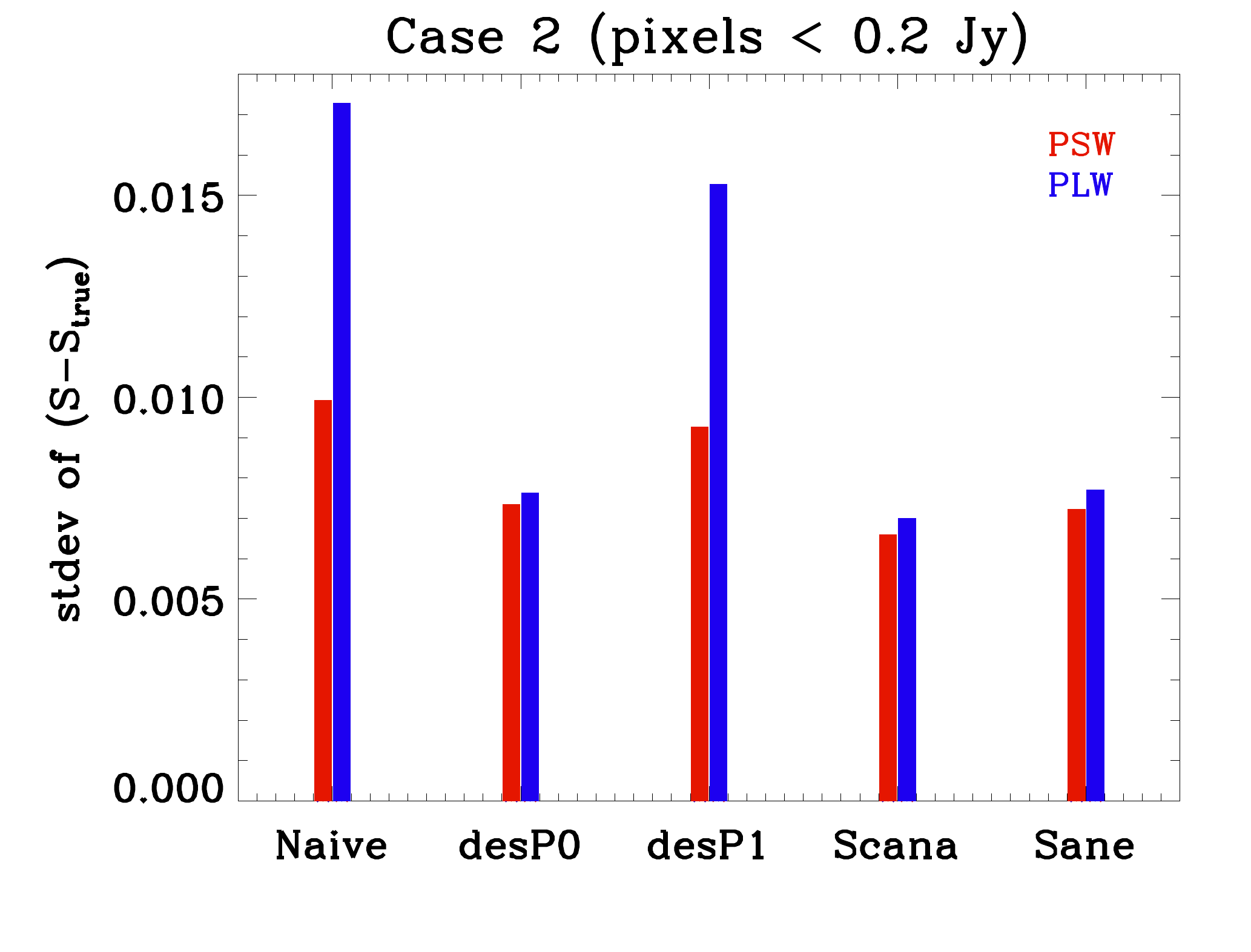}
    \includegraphics[width=8.0cm, angle=0]{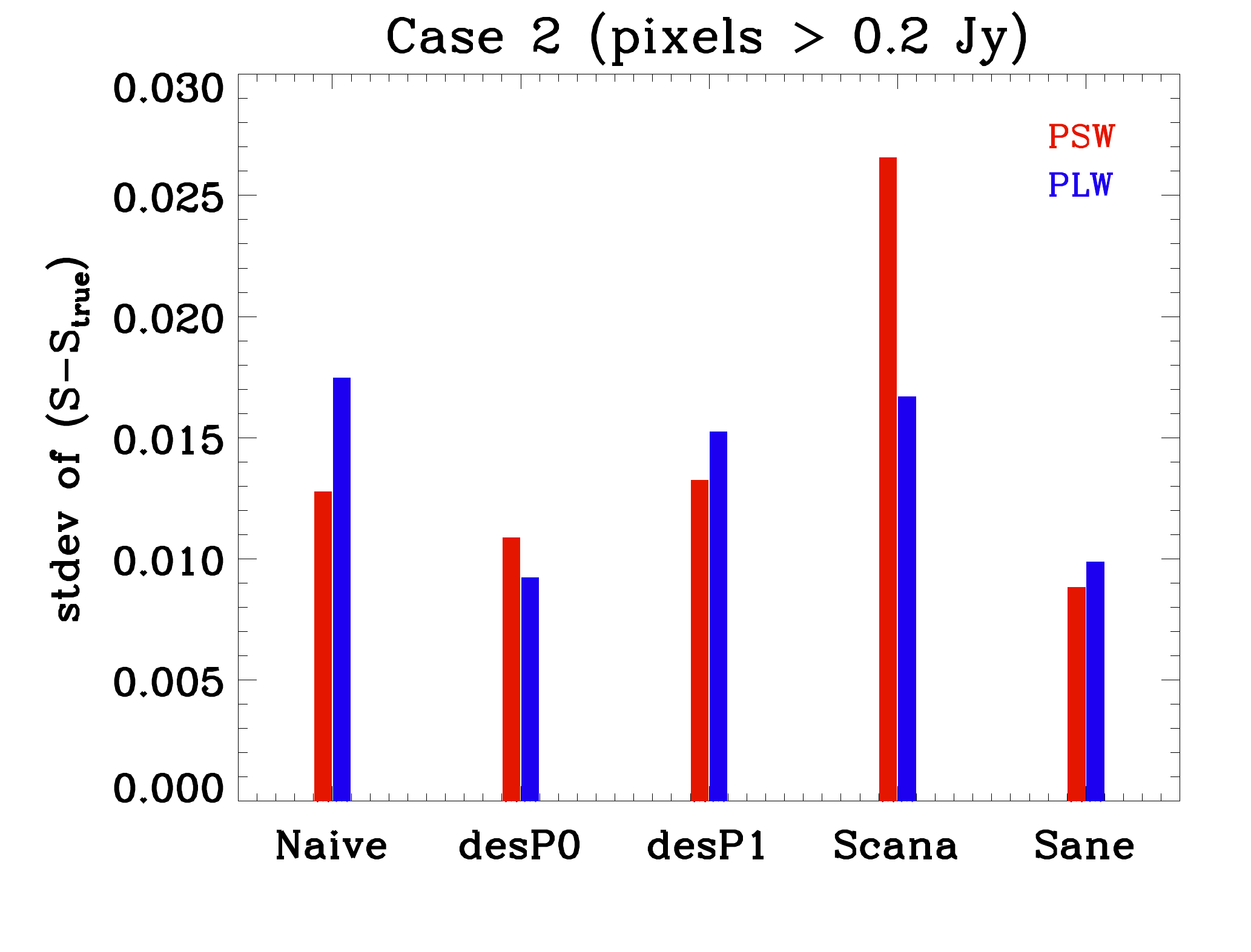}
\caption{Standard deviation of the difference from the truth, for case 2, PSW (red) and PLW (blue) bands, separately for the S$_{\rm true}$ $<$ 0.2~Jy (faint), 
         and S$_{\rm true}$ $>$ 0.2~Jy (bright) emission.}
\end{figure*}
The case 2 difference maps of both bands show a gradient for the 'Naive' and the 'destriperP1' methods which may come from the oversubtraction
of the bright emission of the truth image in the Southern part. In the 'Naive' case a median baseline removal was applied; while a 1$^{st}$-order 
polynomial baseline fit in 'destriperP1'. The Naive map also produces stripy pattern due to a simpler median subtraction of the residual offsets 
from the different detectors.
The gradients are visualized in the scatter plots as anticorrelation of the difference vs truth values. The other methods compensate for this effect. 
For example, with Scanamorphos, it is the /galactic option which takes into account the complexity of the field; the different amount of emission 
at the map edges.

\subsubsection{Case 4 (nominal M51)}

The case 4 difference map scatter plots are not so suggestive, as there are fewer map pixels; and their main feature is the vertical line, 
due to the observed data edges which was embedded in a very faint emission background. 
The median-removed difference maps are very homogeneous for almost all the mappers. The Naive mapper leaves over-subtracted bands in the 
scanning directions which disappears with human intervention, when a mask can be created over the central source to be left out in the 
baseline determination. The human intervention on the other hand introduces an artifact over the high dynamic range central source.
\begin{figure*}[bhhh!]
    \centering
    \includegraphics[width=16.0cm, angle=0]{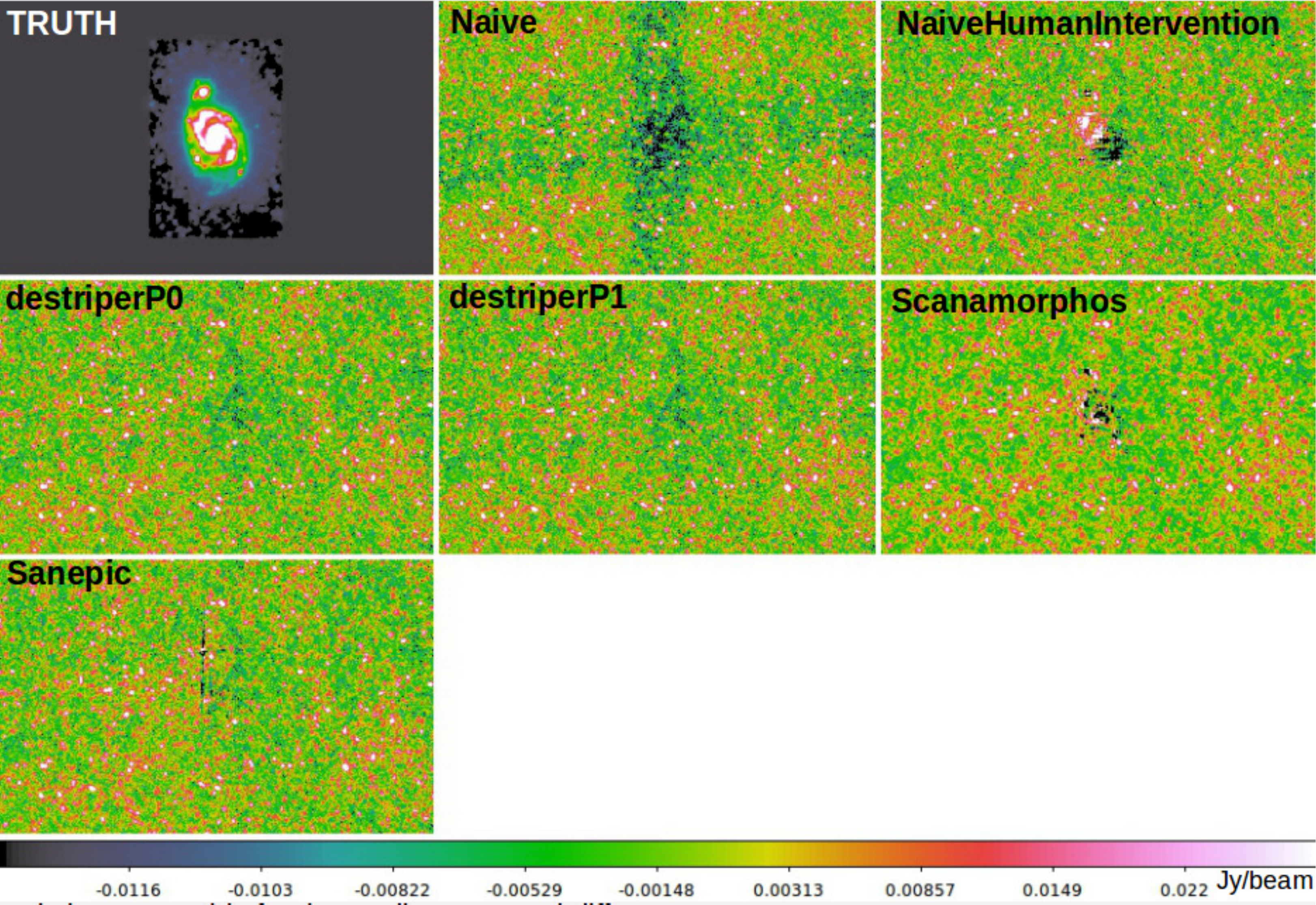}
\caption{Truth and Difference maps for case 4, PSW (nominal M51). For the latter ones, (Diff -- median(Diff)) is shown, 
         therefore the Jy/beam scale is comparable for the median-removed difference maps.}
\end{figure*}
\begin{figure*}
    \centering
    \includegraphics[width=17.0cm, angle=0]{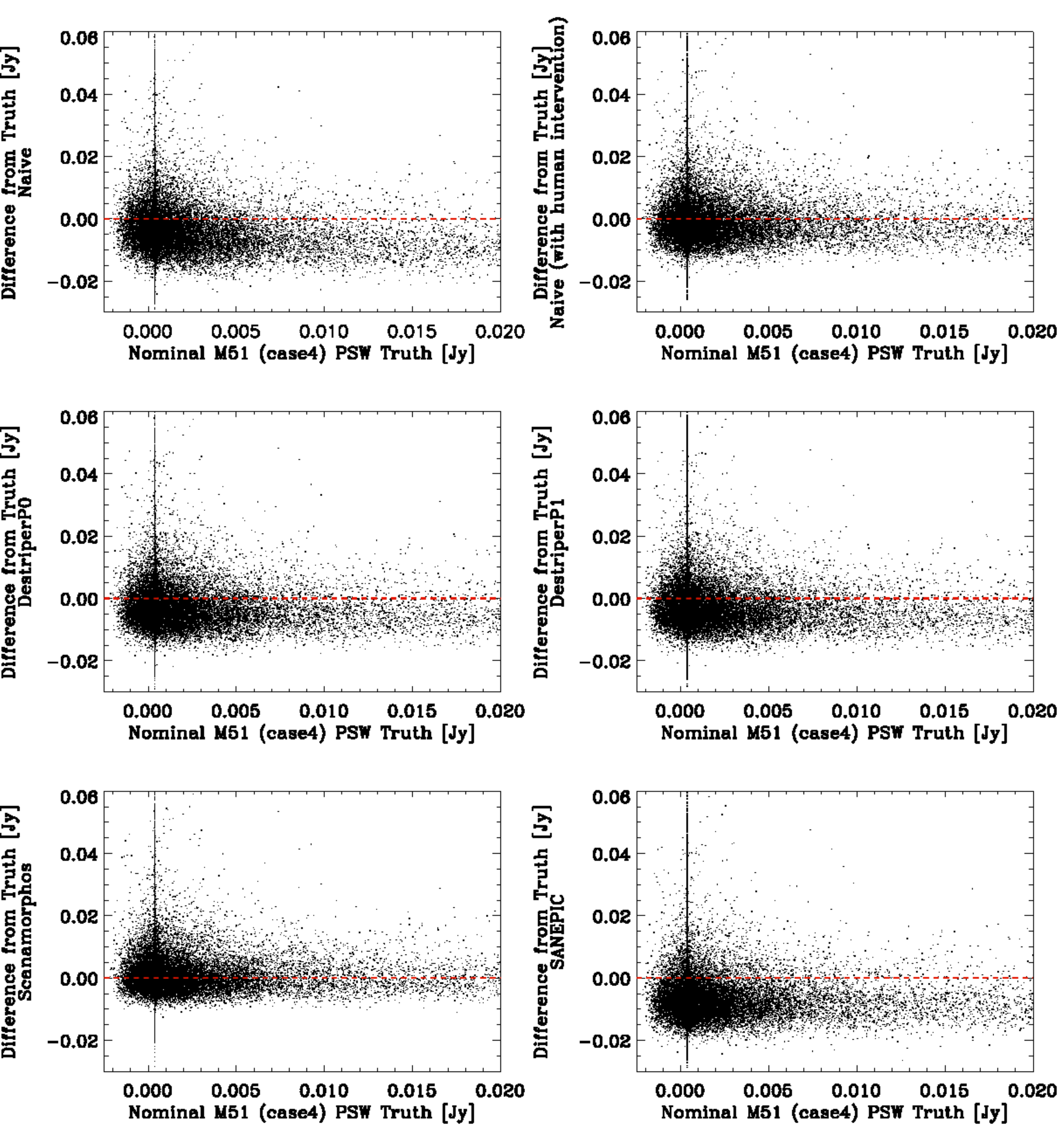}
\caption{Scatter plots of the difference maps for case 4, PSW; (S -- S$_{\rm true}$) vs S$_{\rm true}$.}
\end{figure*}
\begin{figure*}
    \centering
    \includegraphics[width=17.0cm, angle=0]{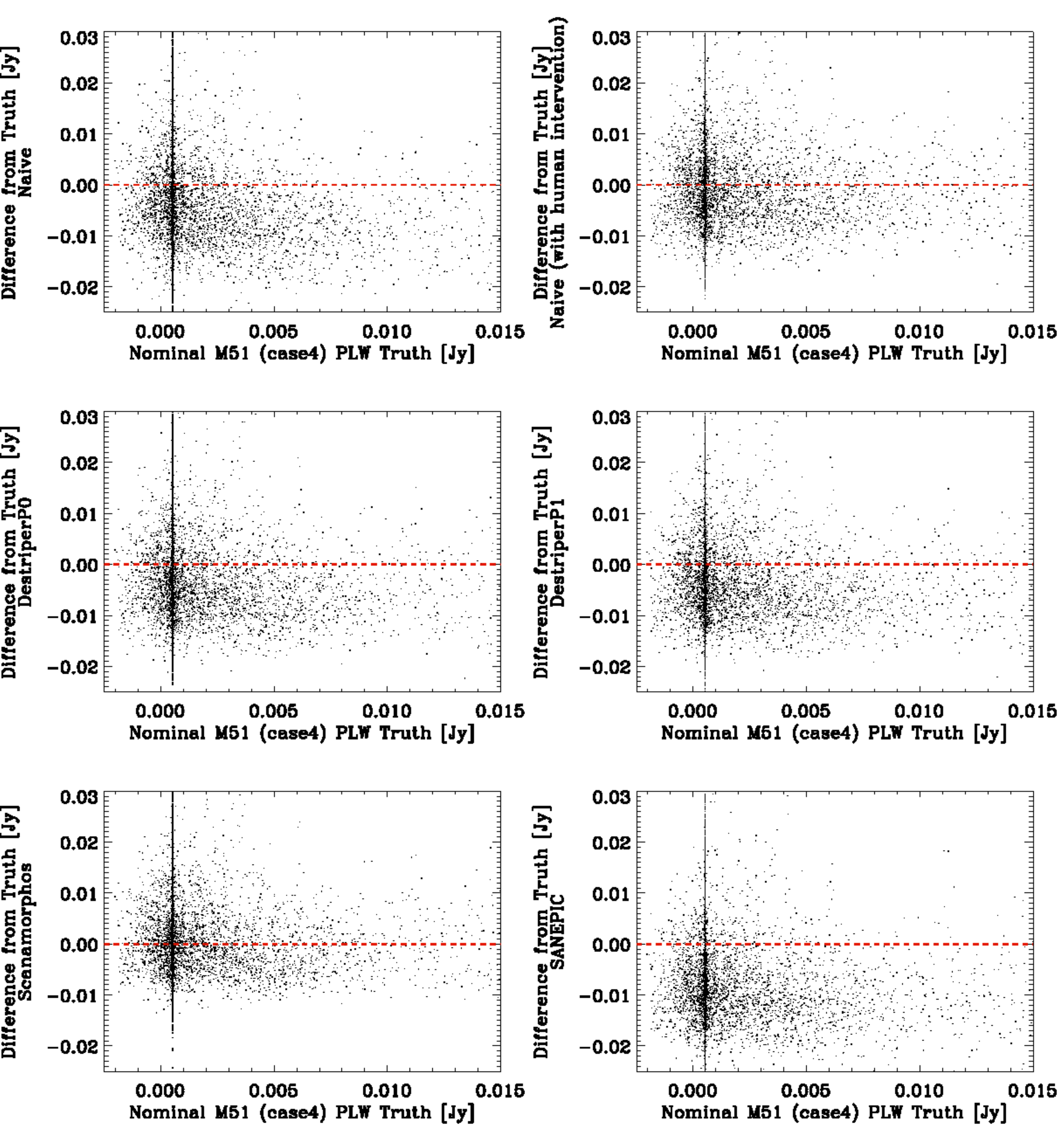}
\caption{Scatter plots of the difference maps for case 4, PLW; (S -- S$_{\rm true}$) vs S$_{\rm true}$.}
\end{figure*}
\begin{figure*}[thh!]
    \centering
    \includegraphics[width=16.0cm, angle=0]{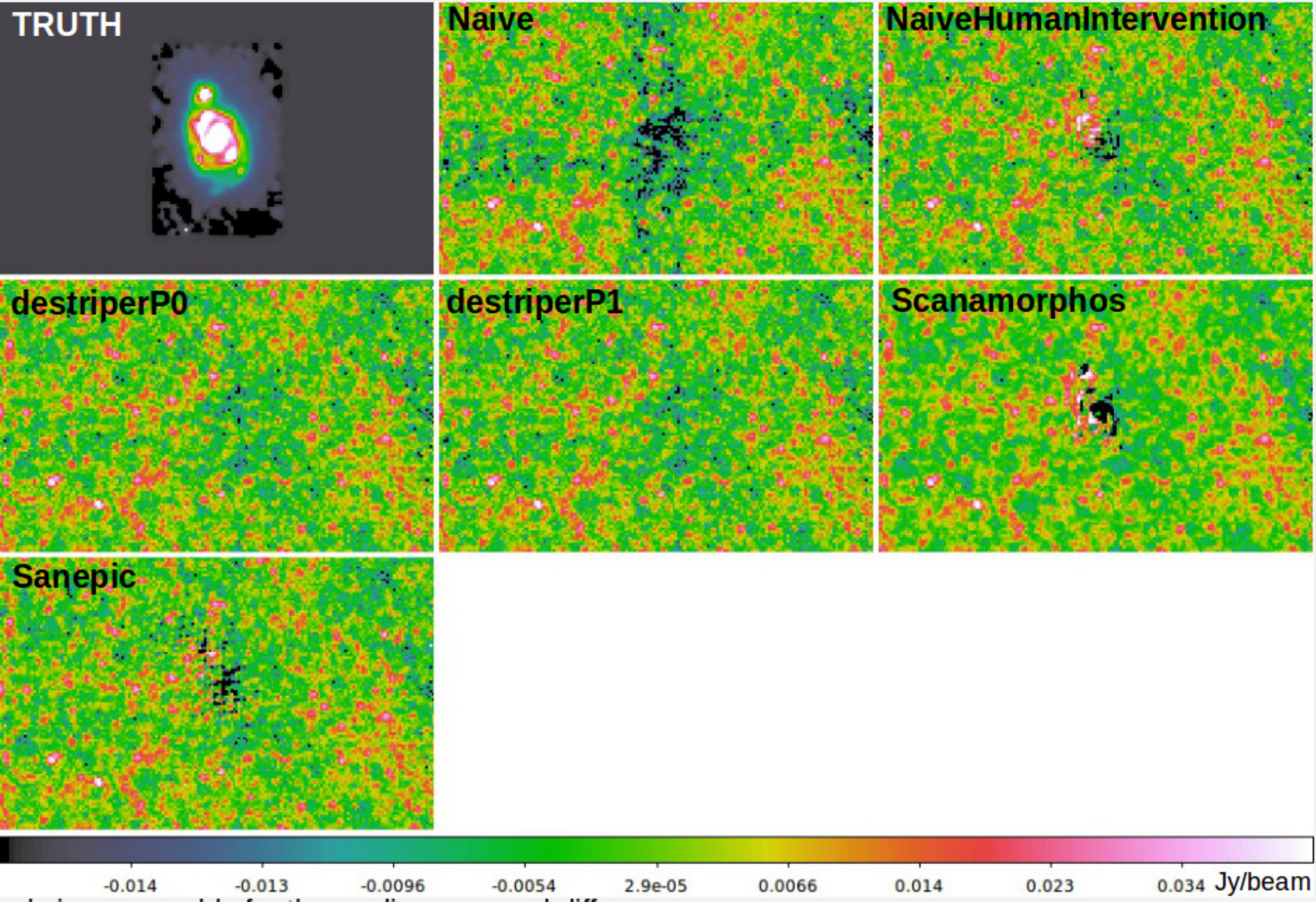}
\caption{Truth and Difference maps for case 4, PLW (nominal M51). For the latter ones, (Diff -- median(Diff)) is shown, 
         therefore the Jy/beam scale is comparable for the median-removed difference maps.}
\end{figure*}
\begin{figure*}[bhhh!]
    \centering
    \includegraphics[width=8.0cm, angle=0]{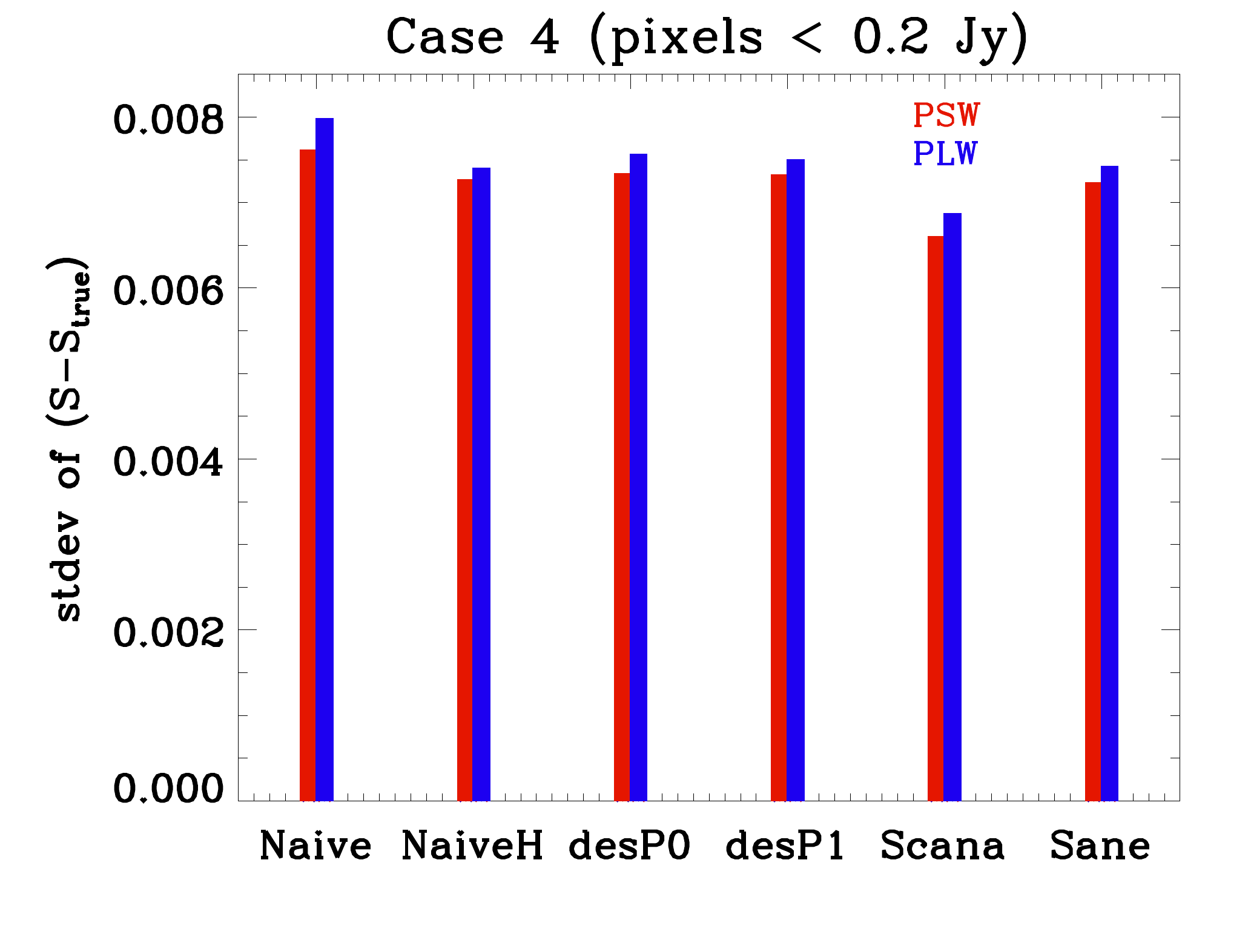}
    \includegraphics[width=8.0cm, angle=0]{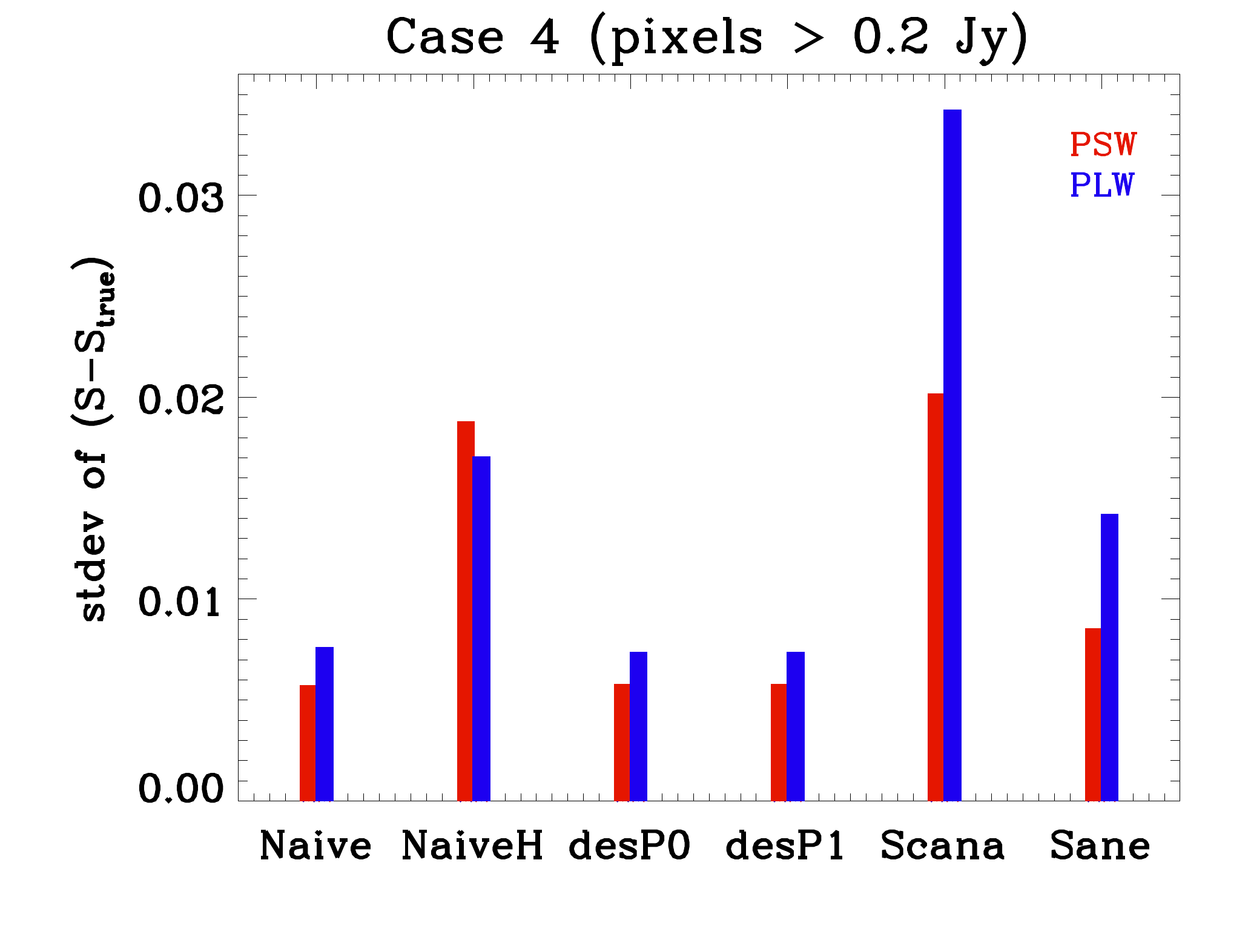}
\caption{Standard deviation of the difference from the truth, for case 4, PSW (red) and PLW (blue) bands, separately for the S$_{\rm true}$ $<$ 0.2~Jy (faint), 
         and S$_{\rm true}$ $>$ 0.2~Jy (bright) emission.}
\end{figure*}

%\clearpage

\subsubsection{Case 6 (fast scan Galactic center)}

Complex fields are more complicated, the difference maps can show various artifacts. 
With Destriper P0 and P1: either erroneous bolometers, scanning through bright data peaks, are leaving outlier scans in the resulting map,
or this outlier baseline removal is due to the inability of the Destripers to deal with the SPIRE "cooler-burps".

\begin{figure*}[bbhh!]
    \centering
    \includegraphics[width=16.0cm, angle=0]{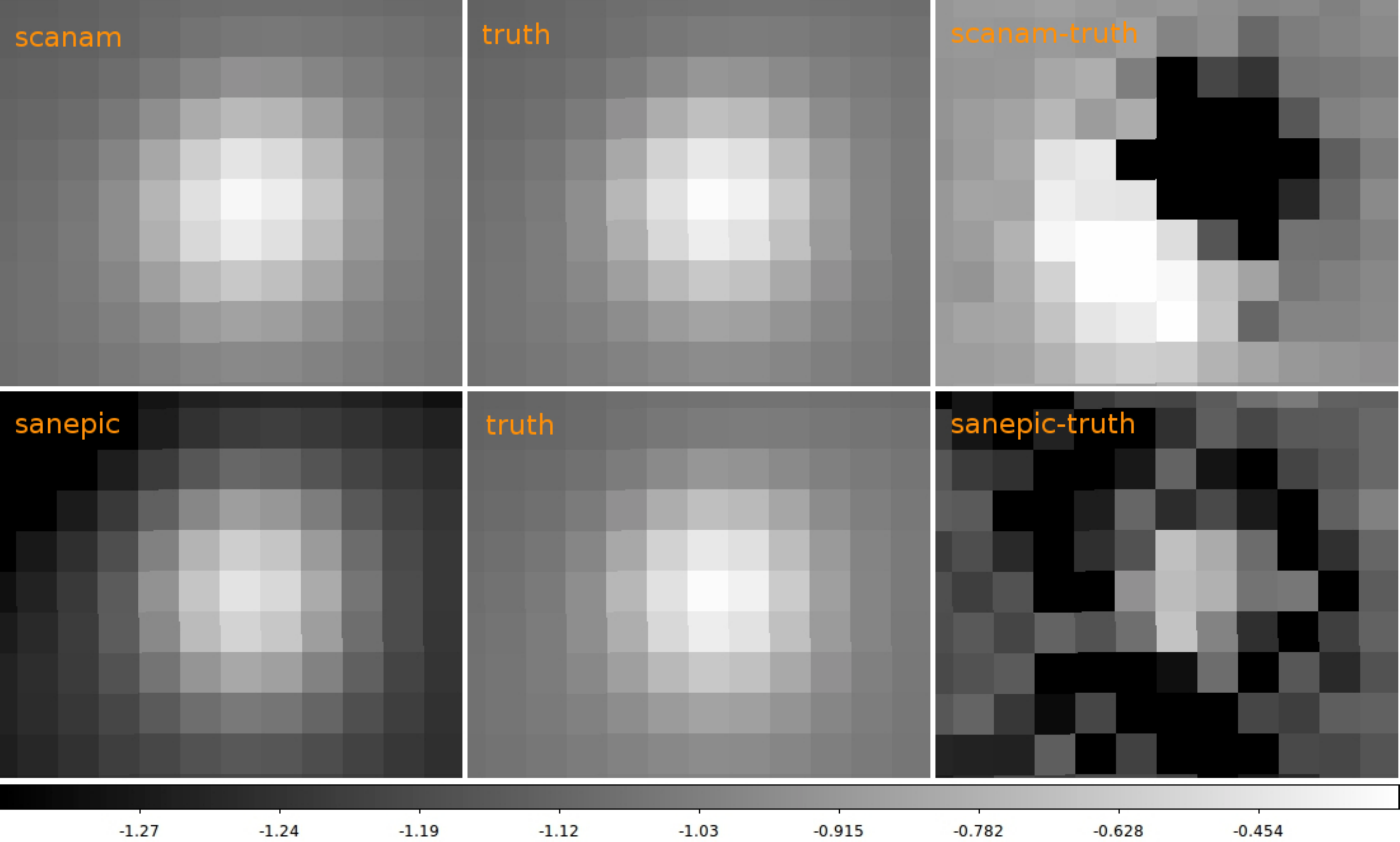}
\caption{A case 6/PSW map detail in the processed Scanamorphos map (upper left), the truth map 
 (middle), and the difference map from the truth (upper right). The same is given for Sanepic 
 (lower panels) for comparison. All the maps are on the same grid.}
\label{fig:offset}
\end{figure*}

The over-/under-subtraction in the Galactic plane, or an apparent positional offset(?) in the Scanamorphos difference map is very likely due to 
the fact that it slightly modifies the PSF. We must emphasize that this apparent positional offset in the difference map of Scanamorphos 
*does not* manifest itself (in the processed Scanamorphos maps) as sources projected off-center from the actual truth positions.  This effect is visualized in Fig.~\ref{fig:offset}.
This was also confirmed by the true position offset checks of the \textit{Point and Extended Sources Photometry} metrics.

Cross-like point source artifacts around bright sources can be well seen in the Sanepic PLW map. This is a typical feature of GLS (Generalized 
Least Square) approaches, possibly due to misalignment of the PSF with respect to the final projected sky grid. One method to remove them
(P/WGLS routines -- e.g., for Unimap) is described in Piazzo (2013), \cite{piazzo2013}, but matching the scan and cross-scan directions with much better 
astrometric accuracy also mitigates this problem (Aussel et al., PACS-ICC investigations).
In the difference map vs truth scatter plots, the Destripers are showing a surprisingly narrow distribution in the Galactic center
 cases (6, 9).   
\begin{figure*}[bbhh!]
    \centering
    \includegraphics[width=16.0cm, angle=0]{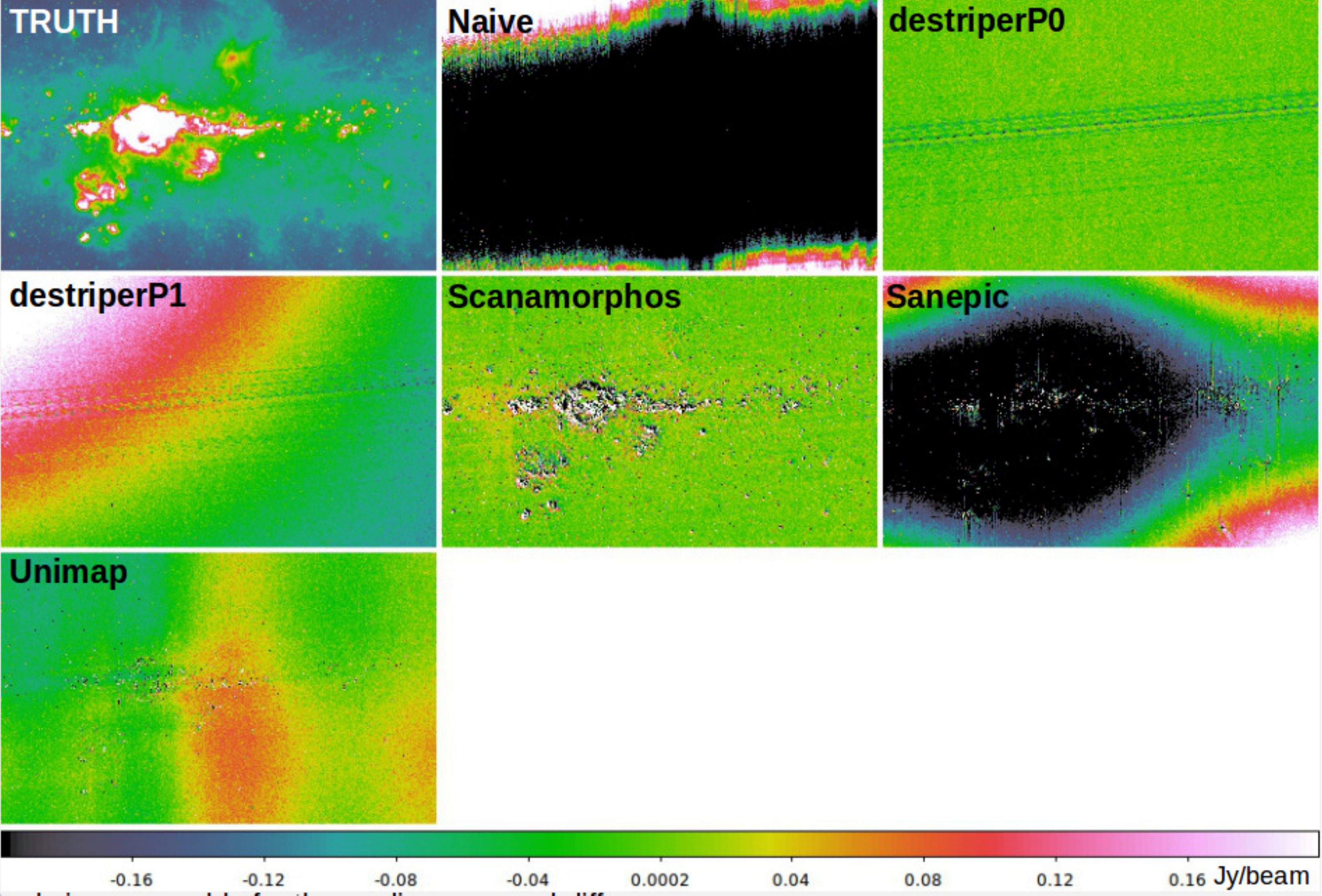}
\caption{Truth and Difference maps for case 6, PSW (fast scan Galactic center). For the latter ones, (Diff -- median(Diff)) is shown, 
         therefore the Jy/beam scale is comparable for the median-removed difference maps.}
\end{figure*}
\newpage
\begin{figure*}[hhh!]
    \centering
    \includegraphics[width=17.0cm, angle=0]{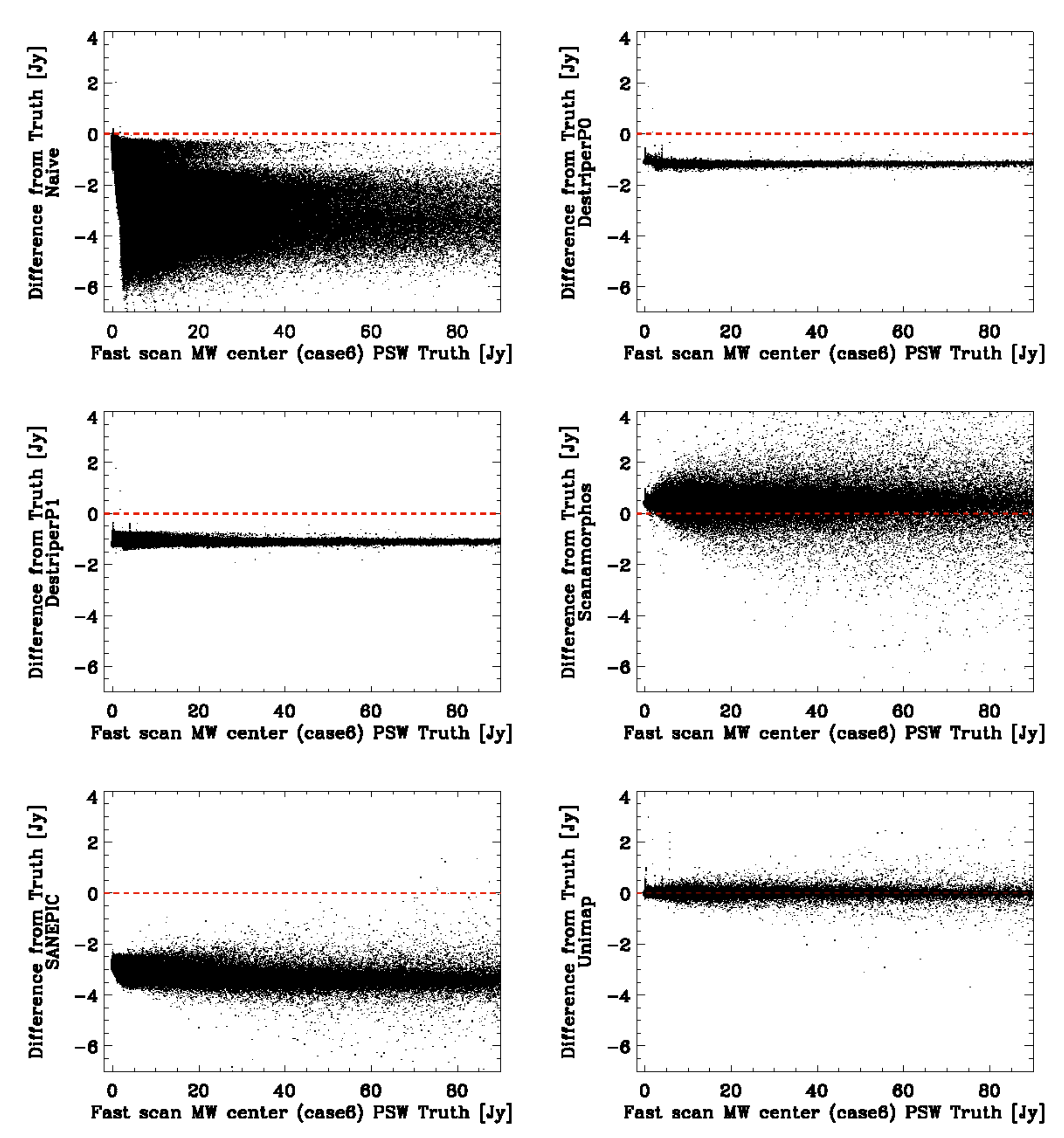}
\caption{Scatter plots of the difference maps for case 6, PSW; (S -- S$_{\rm true}$) vs S$_{\rm true}$.}
\end{figure*}
\clearpage
\begin{figure*}[hhh!]
    \centering
    \includegraphics[width=17.0cm, angle=0]{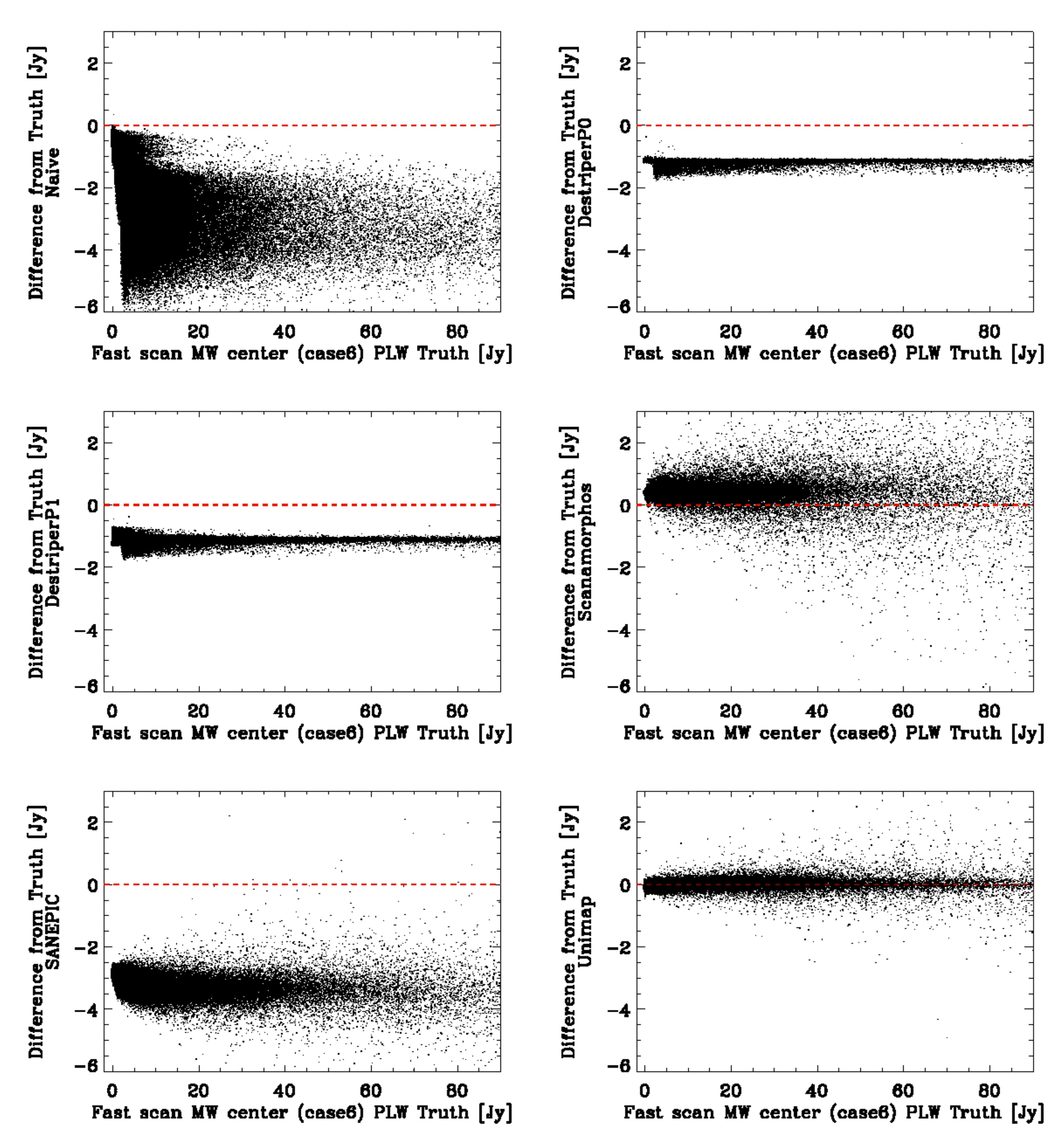}
\caption{Scatter plots of the difference maps for case 6, PLW; (S -- S$_{\rm true}$) vs S$_{\rm true}$.}
\end{figure*}
\clearpage
\begin{figure*}[tthh!]
    \centering
    \includegraphics[width=16.0cm, angle=0]{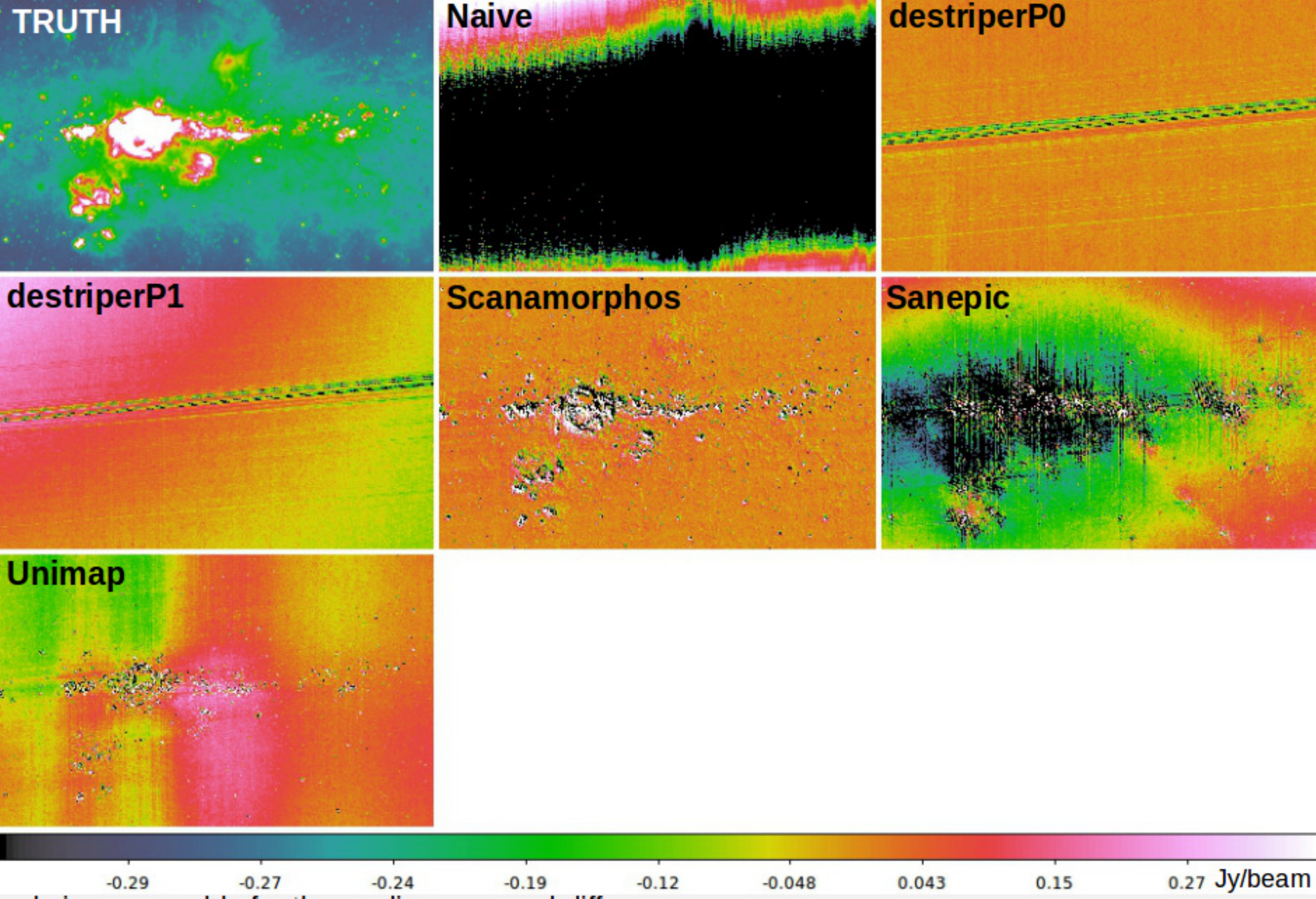}
\caption{Truth and Difference maps for case 6, PLW (fast scan Galactic center). For the latter ones, (Diff -- median(Diff)) is shown, 
         therefore the Jy/beam scale is comparable for the median-removed difference maps.}
\end{figure*}
\begin{figure*}[bbhh!]
    \centering
    \includegraphics[width=8.0cm, angle=0]{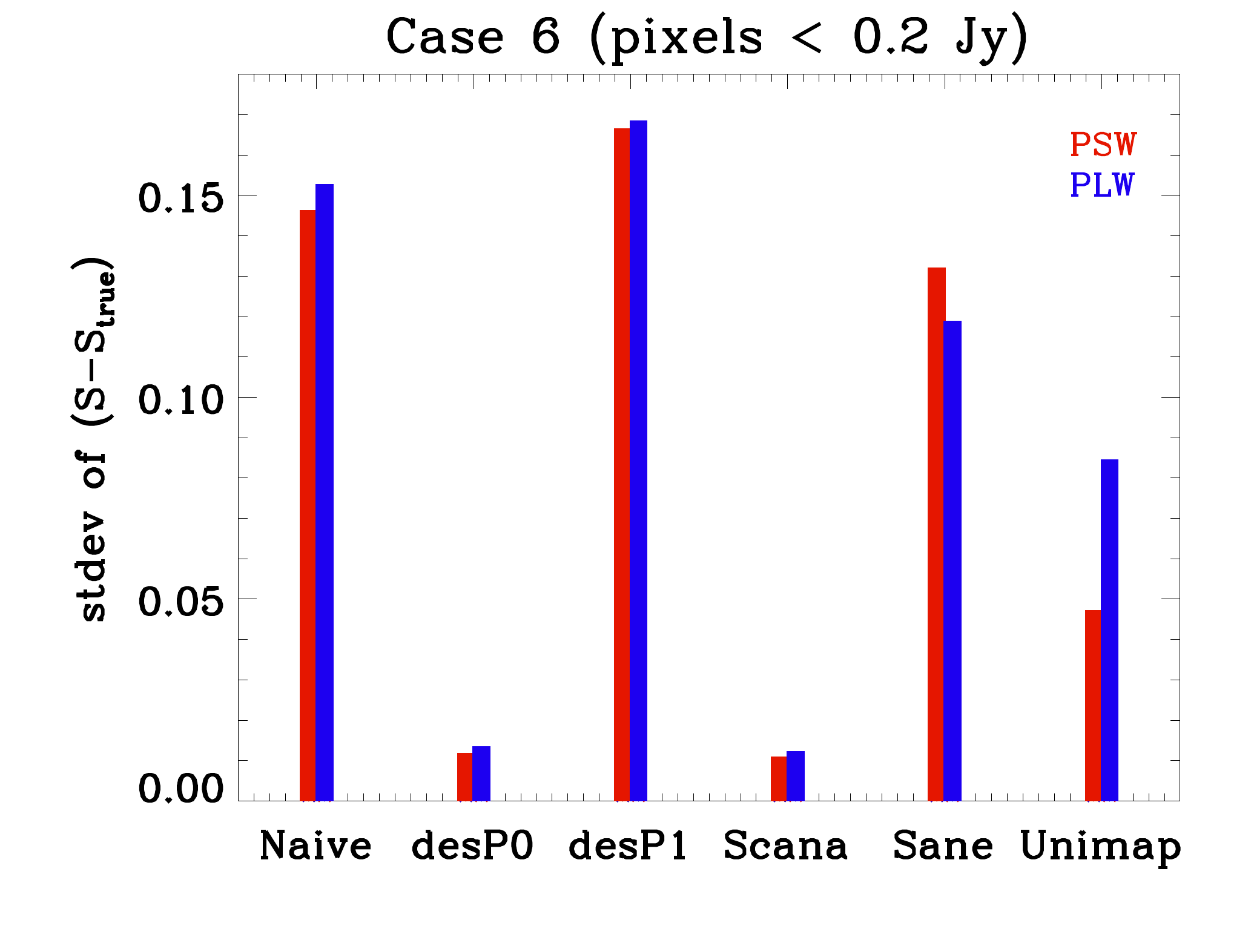}
    \includegraphics[width=8.0cm, angle=0]{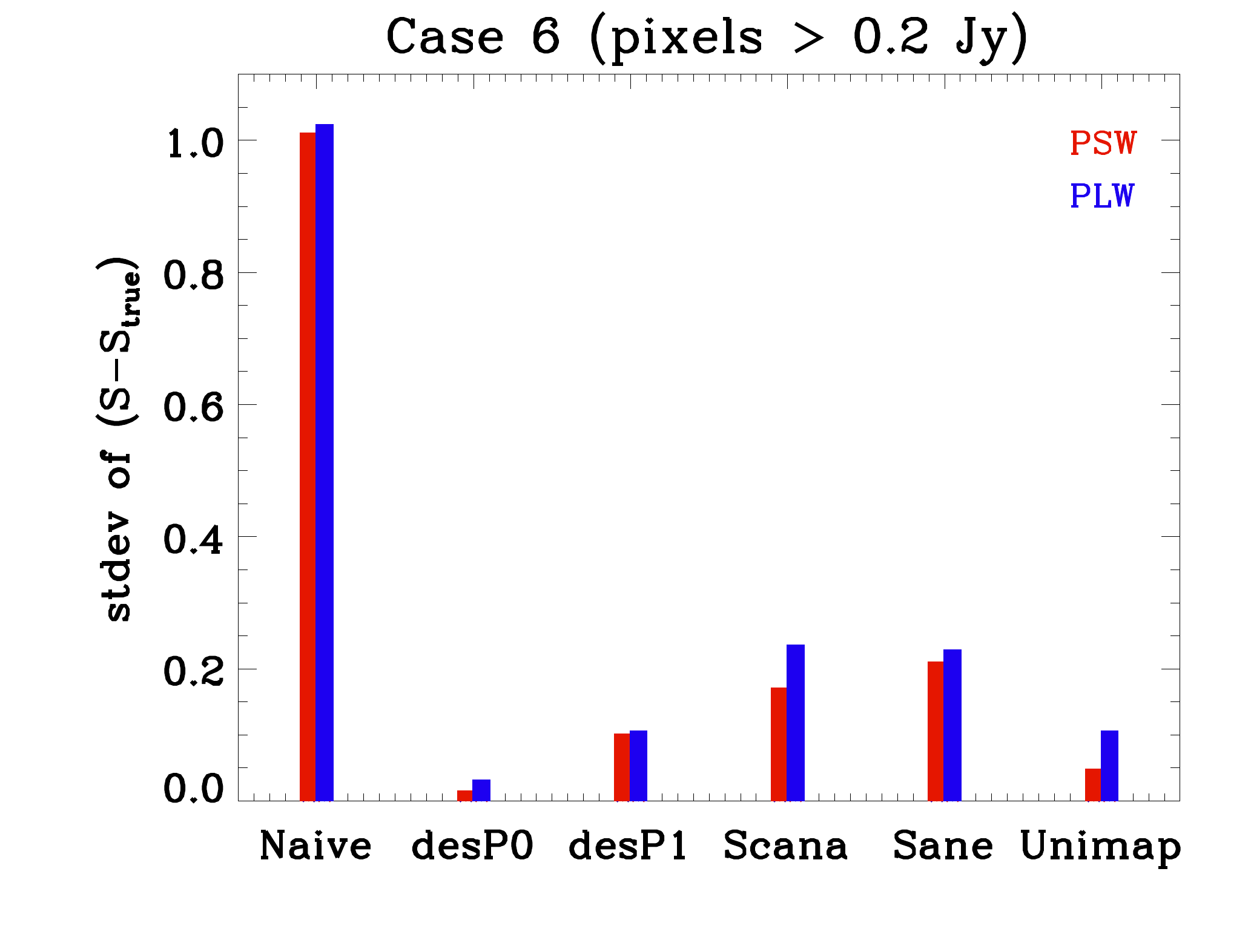}
\caption{Standard deviation of the difference from the truth, for case 6, PSW (red) and PLW (blue) bands, separately for the S$_{\rm true}$ $<$ 0.2~Jy (faint), 
         and S$_{\rm true}$ $>$ 0.2~Jy (bright) domain.}
\end{figure*}
\newpage
\subsubsection{Case 9 (parallel Galactic center)}
\begin{figure*}[hhh!]
    \centering
    \includegraphics[width=17.0cm, angle=0]{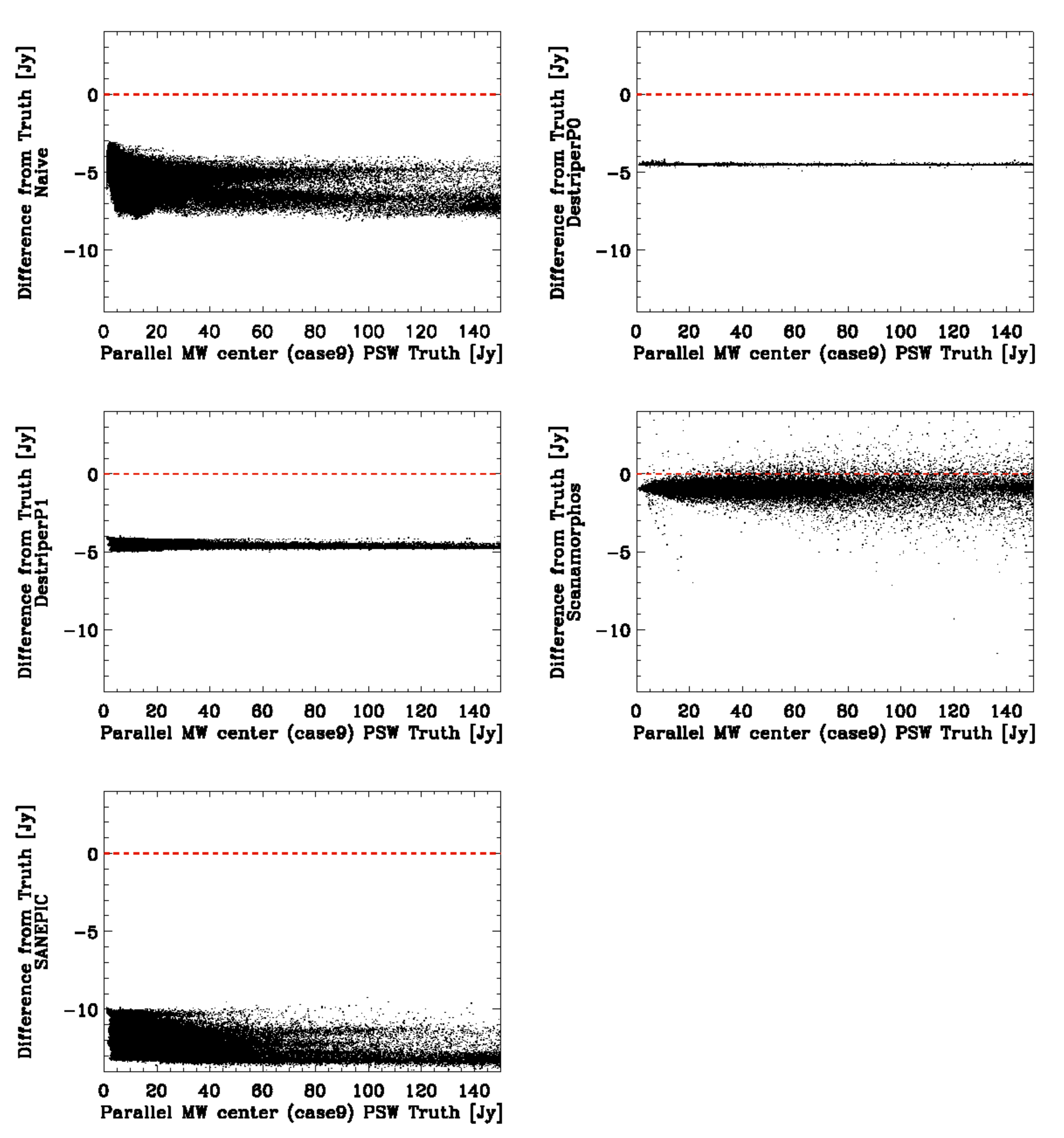}
\caption{Scatter plots of the difference maps for case 9, PSW; (S -- S$_{\rm true}$) vs S$_{\rm true}$.}
\end{figure*}
\newpage
\begin{figure*}[tt!]
    \centering
    \includegraphics[width=16.0cm, angle=0]{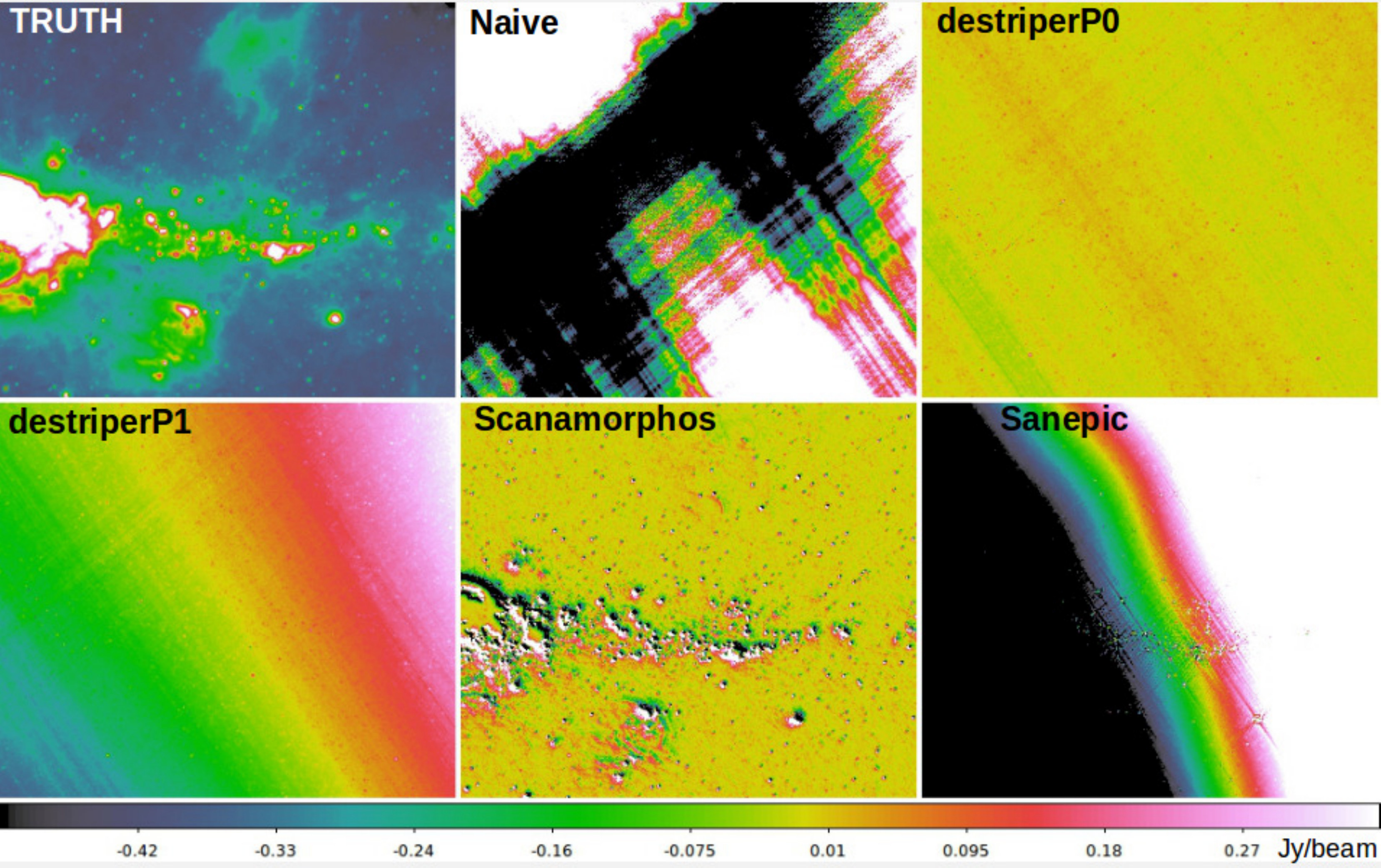}
\caption{Truth and Difference maps for case 9, PSW (parallel Galactic center). For the latter ones, (Diff -- median(Diff)) is shown, 
         therefore the Jy/beam scale is comparable for the median-removed difference maps.}
\end{figure*}
\begin{figure*}[bbb!!!]
    \centering
    \includegraphics[width=8.0cm, angle=0]{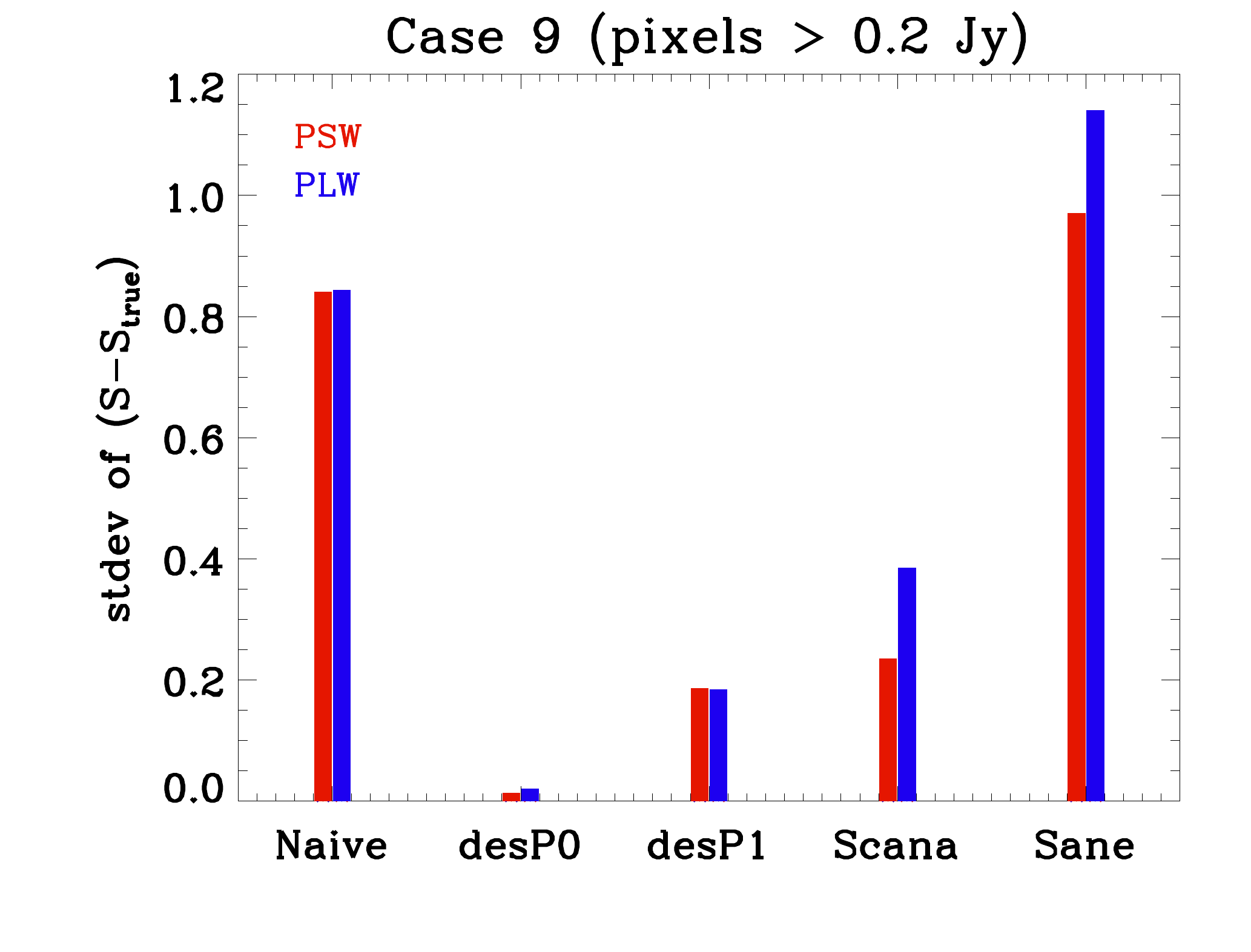}
\caption{Standard deviation of the difference from the truth, for case 9, PSW (red) and PLW (blue) bands, for the S$_{\rm true}$ $>$ 0.2~Jy (bright) emission.}
\end{figure*}
\newpage
\begin{figure*}[hhh!]
    \centering
    \includegraphics[width=17.0cm, angle=0]{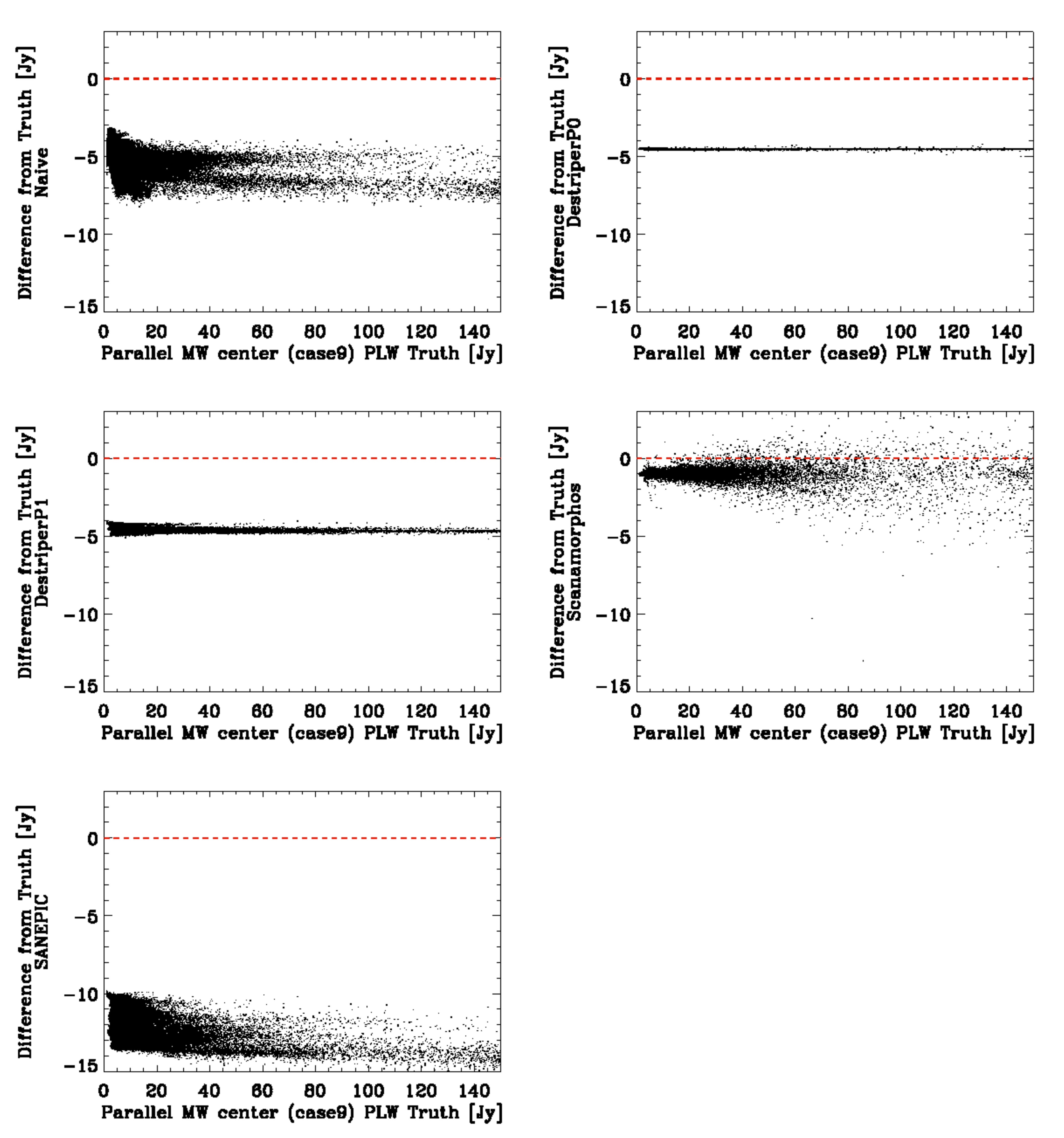}
\caption{Scatter plots of the difference maps for case 9, PLW; (S -- S$_{\rm true}$) vs S$_{\rm true}$.}
\end{figure*}
\clearpage
\begin{figure*}[ttthhh!]
    \centering
    \includegraphics[width=16.0cm, angle=0]{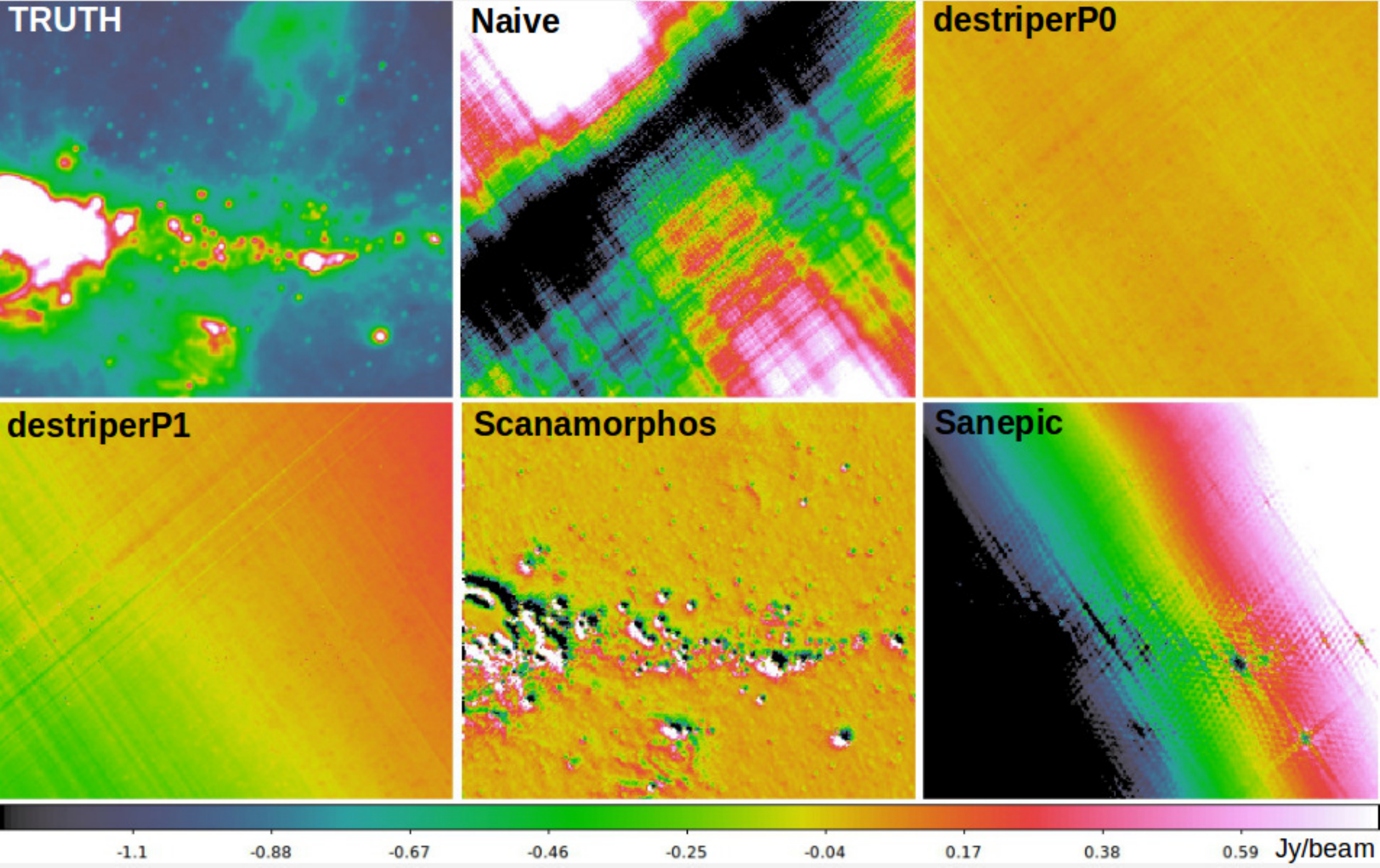}
\caption{Truth and Difference maps for case 9, PLW (parallel Galactic center). For the latter ones, (Diff -- median(Diff)) is shown, 
         therefore the Jy/beam scale is comparable for the median-removed difference maps.}
\label{fig:metrics_report_diff_case9}
\end{figure*}
Observations, statements for case 6 maps are mostly true in this similar case 9 too (e.g., Sanepic, Scanamorphos, Destripers).
In the Sanepic map the strong gradient likely originates from the mapper's assumption that the data is circulant, while in this 
case, it is not. In Scanamorphos, it is the /galactic option which takes into account the non-circulant nature of the data.   
See these details again in the Summary/Conclusions.

\clearpage
\subsubsection{Case 10 (parallel cirrus)}
In this low dynamic range example the map makers produce very similar, and satisfactory results. 
\begin{figure*}[bbbhhh!]
    \centering
    \includegraphics[width=16.0cm, angle=0]{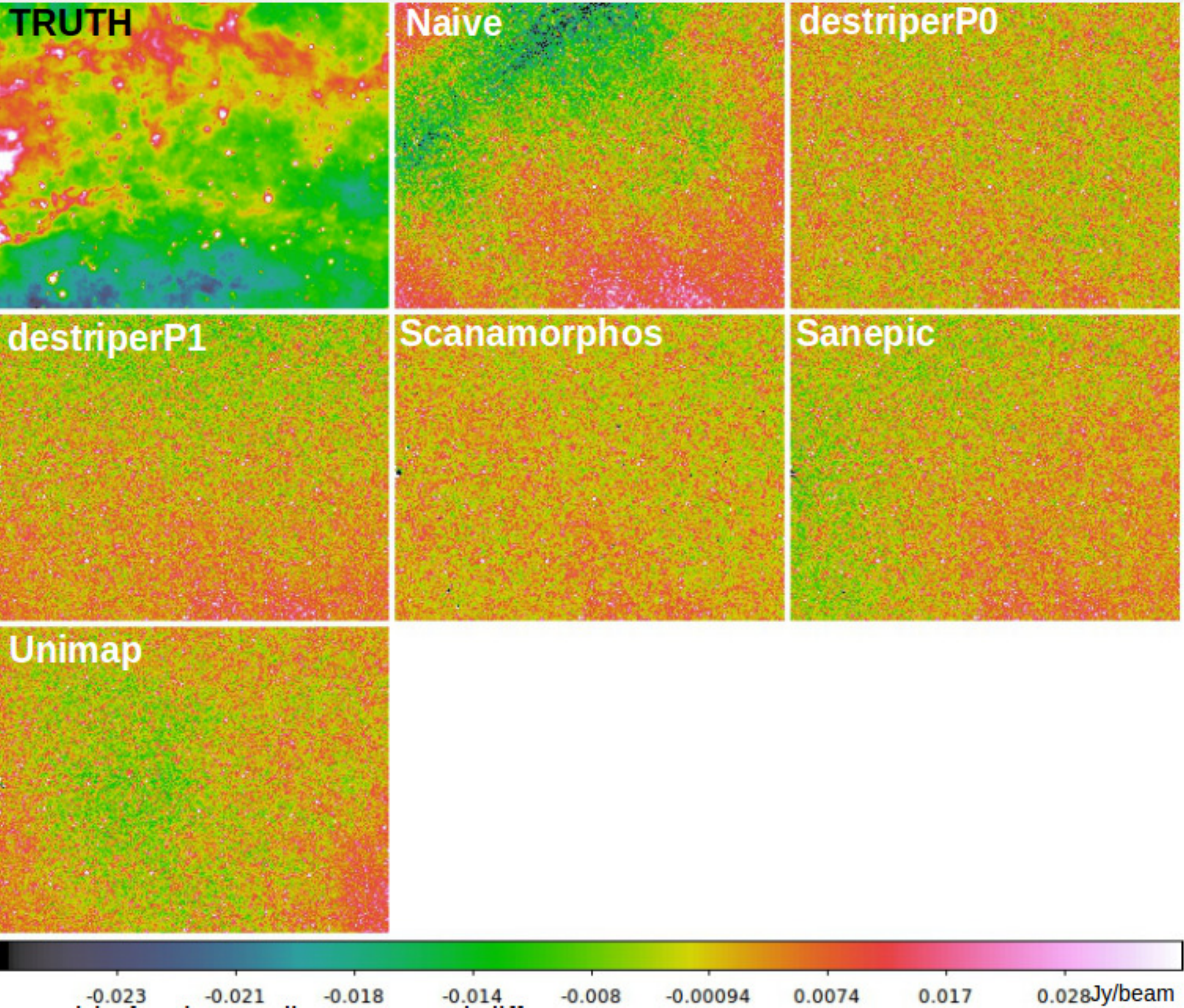}
\caption{Truth and Difference maps for case 10, PSW (parallel cirrus). For the latter ones, (Diff -- median(Diff)) is shown, 
         therefore the Jy/beam scale is comparable for the median-removed difference maps.}
\end{figure*}
\clearpage
\begin{figure*}[hhh!]
    \centering
    \includegraphics[width=17.0cm, angle=0]{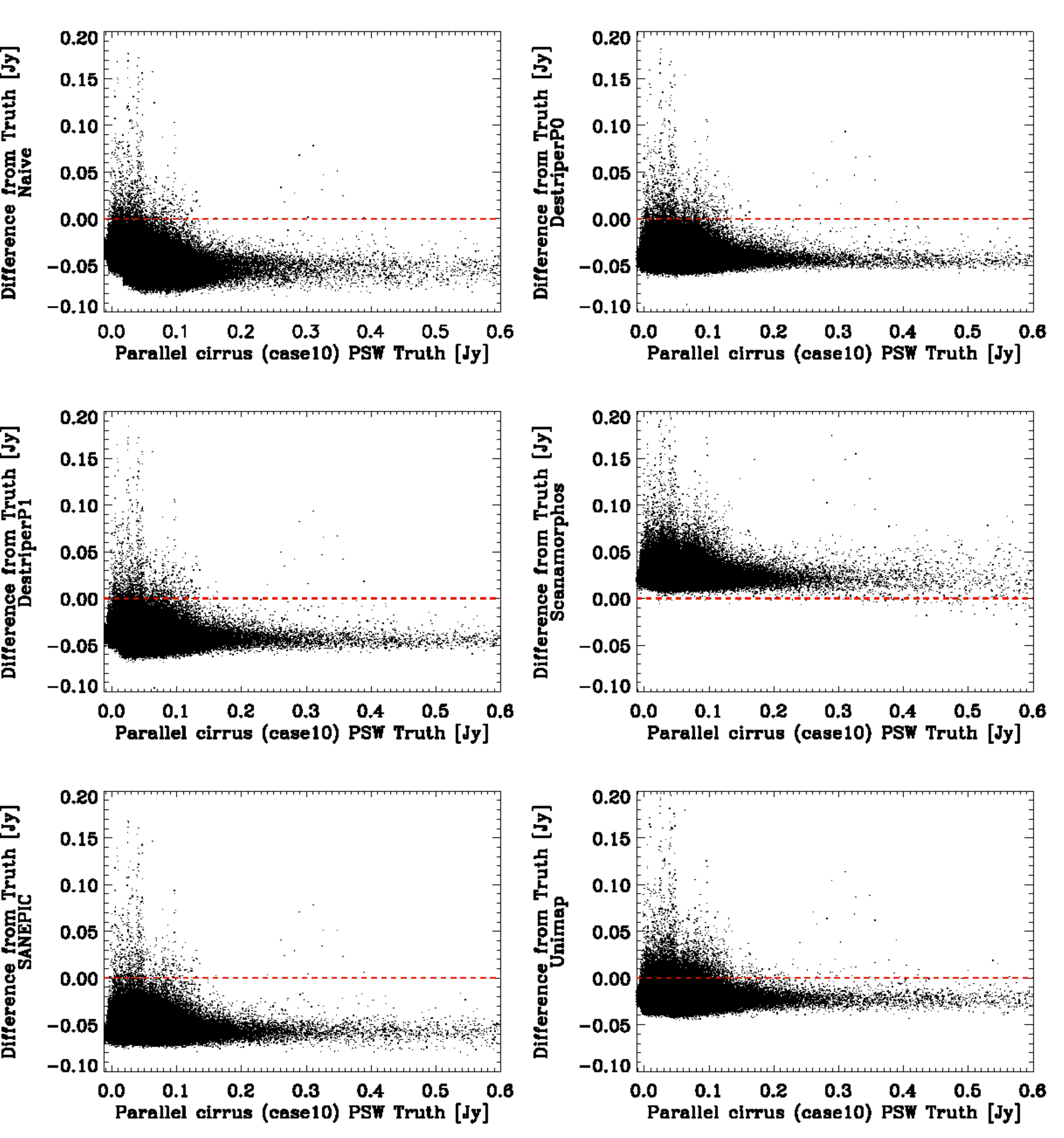}
\caption{Scatter plots of the difference maps for case 10, PSW; (S -- S$_{\rm true}$) vs S$_{\rm true}$.}
\end{figure*}
\clearpage
\begin{figure*}[tttt!]
    \centering
    \includegraphics[width=8.0cm, angle=0]{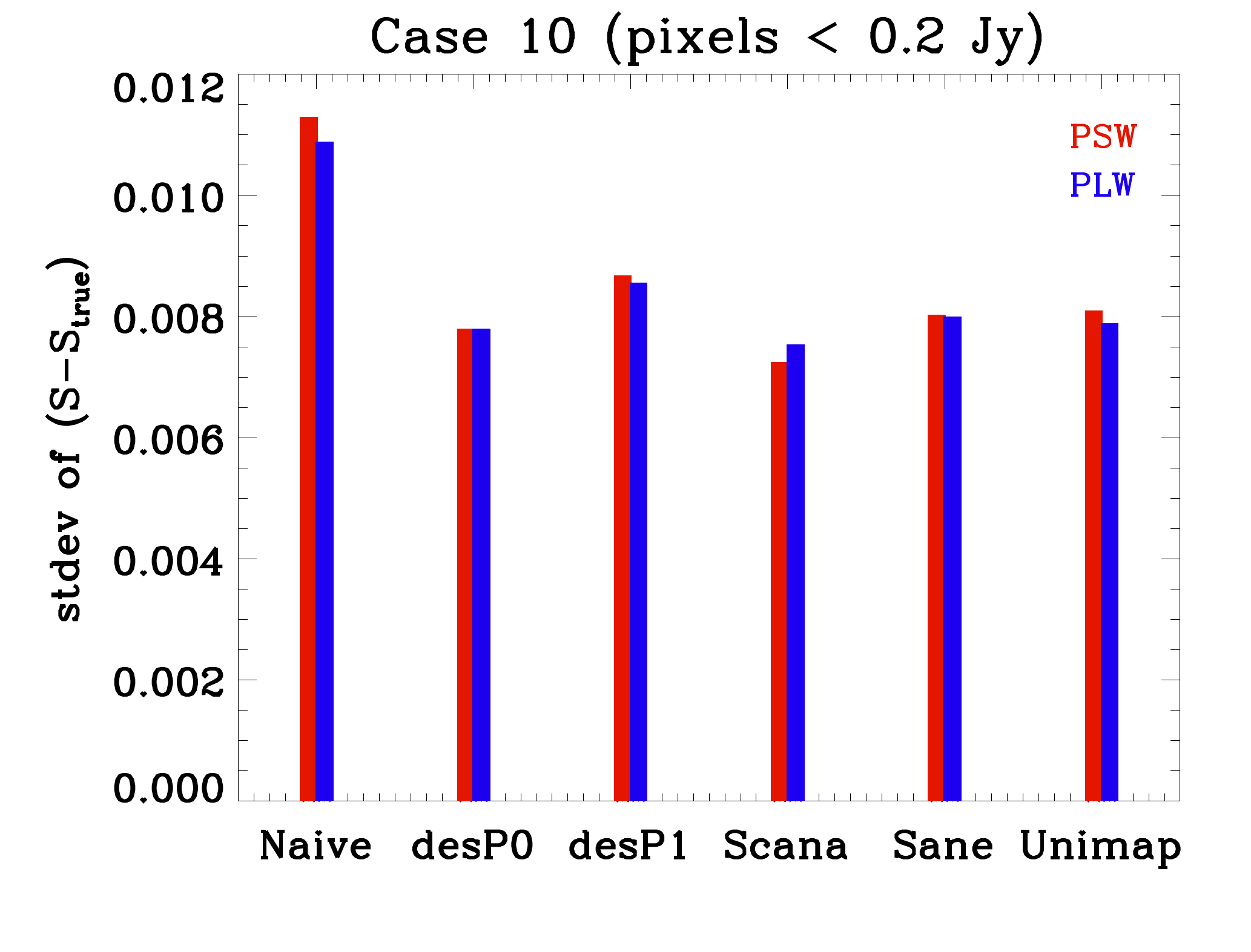}
    \includegraphics[width=8.0cm, angle=0]{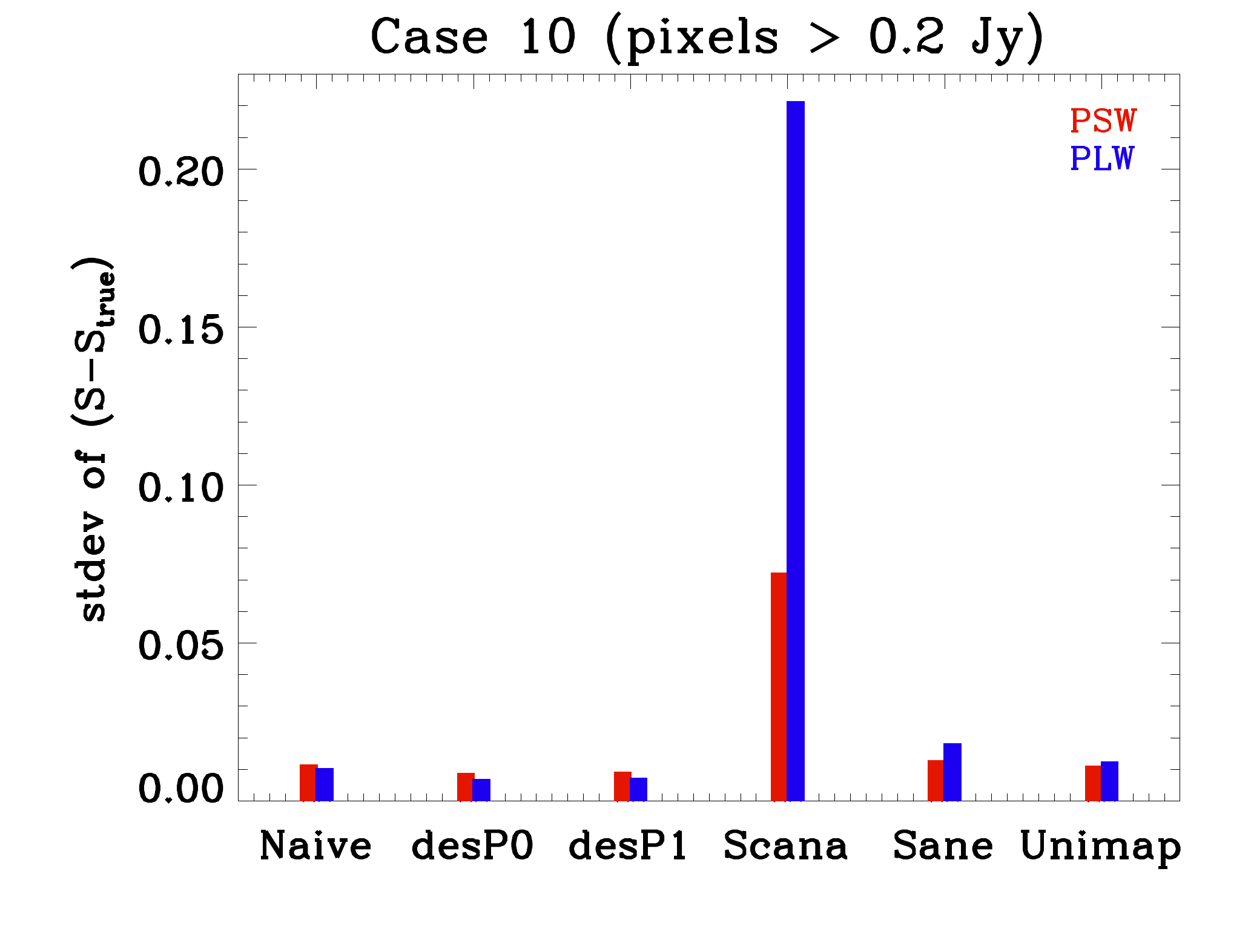}
\caption{Standard deviation of the difference from the truth, for case 10, PSW (red) and PLW (blue) bands, separately for the S$_{\rm true}$ $<$ 0.2~Jy (faint), 
         and S$_{\rm true}$ $>$ 0.2~Jy (bright) domain.}
\end{figure*}
\begin{figure*}[bbbb!]
    \centering
    \includegraphics[width=16.0cm, angle=0]{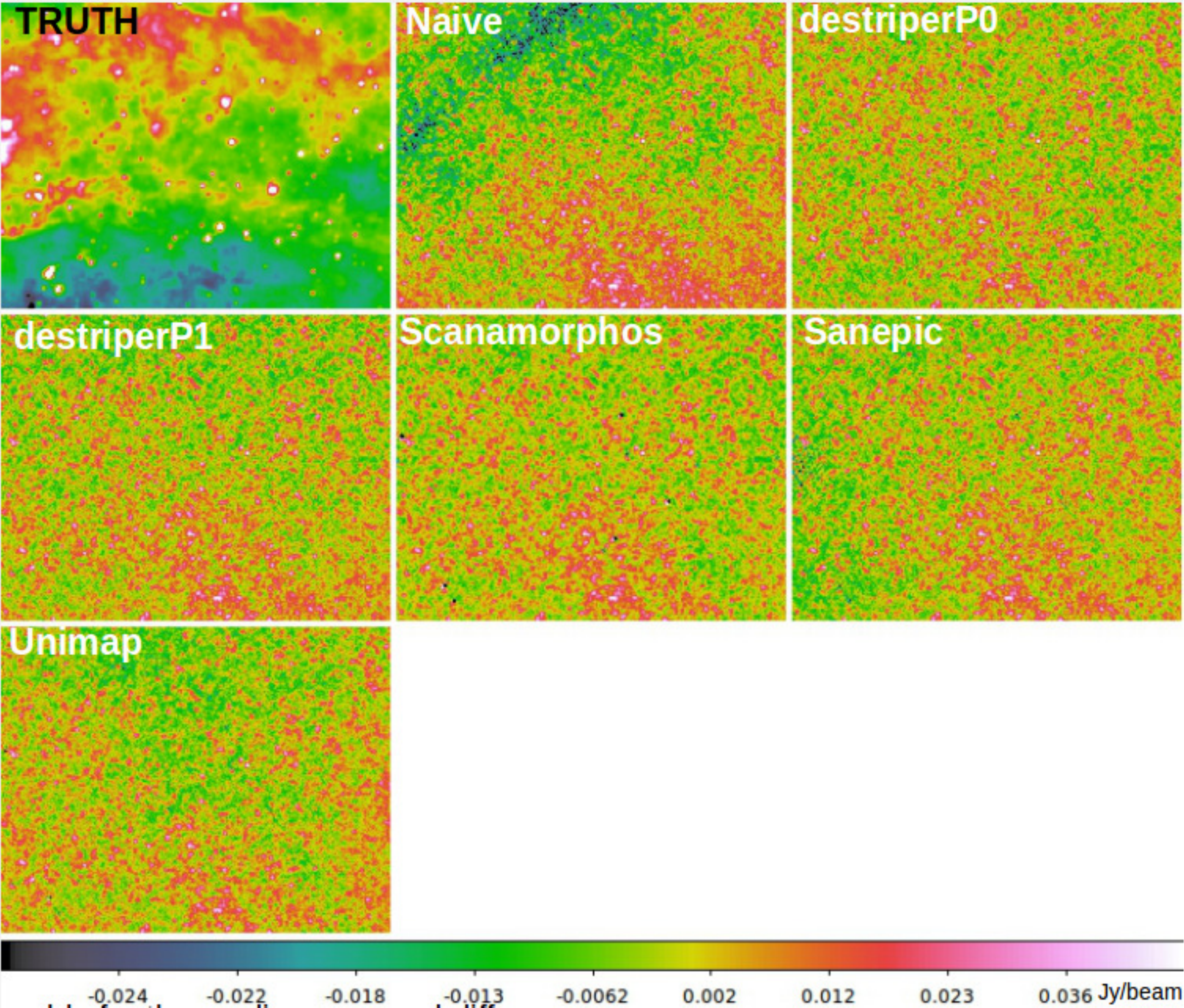}
\caption{Truth and Difference maps for case 10, PLW (parallel cirrus). For the latter ones, (Diff -- median(Diff)) is shown, 
         therefore the Jy/beam scale is comparable for the median-removed difference maps.}
\end{figure*}
\clearpage
\begin{figure*}
    \centering
    \includegraphics[width=17.0cm, angle=0]{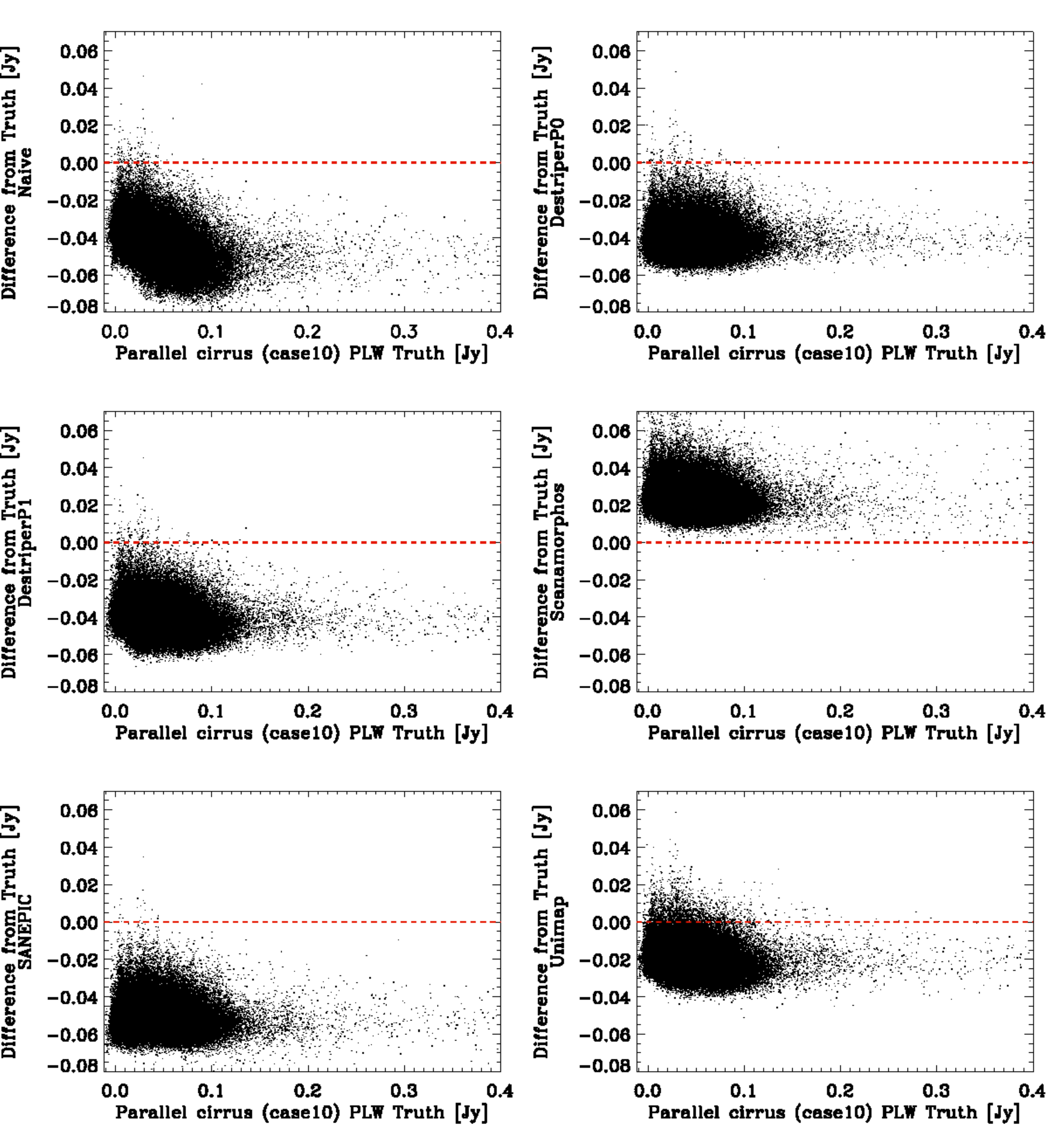}
\caption{Scatter plots of the difference maps for case 10, PLW; (S -- S$_{\rm true}$) vs S$_{\rm true}$.}
\end{figure*}
\clearpage

\subsection{More 'table' results}
\begin{table}[th]
%\begin{center}
\small
\begin{tabular}{l c c c c c c c c}
			&	case 2	&		&		&		&	case 9	&		&		&		\\
\hline
			&	PSW	&		&	PLW	&		&	PSW	&		&	PLW	&		\\
			&	mean	&	stdev	&	mean	&	stdev	&	mean	&	stdev	&	mean	&	stdev	\\
\hline
naive			&  	-0.0429	&	0.0099	&	-0.0996 &	0.0174	&      -5.0302	&	0.8410	&      -5.0599  &	0.8436 \\
destrP0			&  	-0.0427 &	0.0073	&	-0.0991 &	0.0076	&      -4.5131	&	0.0131	&      -4.5184  &	0.0204 \\
destrP1			&  	-0.0429 &	0.0092	&	-0.0994 &	0.0153	&      -4.5192	&	0.1864	&      -4.5231  &	0.1840 \\
scanam			&  	 0.0053 &	0.0066	&	-0.0094 &	0.0072	&      -0.9391 	&	0.2353	&      -1.0083  &	0.3852 \\
sanepic			&  	-0.0445 &	0.0072	&	-0.1023 &	0.0077	&      -11.5947	&	0.9713	&      -11.8134 &	1.1405 \\
\end{tabular}
\vspace{0.3cm}

\begin{tabular}{l c c c c}
			&	case 4	&		&		&		\\
\hline
			&	PSW	&		&	PLW	&		\\
			&	mean	&	stdev	&	mean	&	stdev	\\
\hline
naive			&      -0.0017	&	0.0076	&	-0.0022 &	0.0080  \\
naiveHuman		&      -0.0002	&	0.0074	&	-0.0002 &	0.0076  \\
destrP0			&      -0.0017	&	0.0073	&	-0.0024 &	0.0075  \\
destrP1			&      -0.0017	&	0.0073	&	-0.0025 &	0.0075  \\
scanam			&       0.0017	&	0.0068	&	 0.0013 &	0.0078  \\
sanepic			&      -0.0059	&	0.0072	&	-0.0076 &	0.0075  \\	 
\end{tabular}
\vspace{0.3cm}

\begin{tabular}{l c c c c c c c c}
			&	case 6	&		&		&		&	case 10	&		&		&		\\
\hline
			&	PSW	&		&	PLW	&		&	PSW	&		&	PLW	&		\\
			&	mean	&	stdev	&	mean	&	stdev	&	mean	&	stdev	&	mean	&	stdev	\\
\hline
naive			&  	-1.3883 &      1.0154	&       -1.4105 &      1.0264	&       -0.0463 &      0.0113	&	-0.0441 &	0.0109	\\
destrP0			&  	-1.1215 &      0.0153	&       -1.1357 &      0.0303	&       -0.0420 &      0.0078	&	-0.0403 &	0.0077	\\
destrP1			&  	-1.1149 &      0.1197	&       -1.1292 &      0.1235	&       -0.0421 &      0.0086	&	-0.0403 &	0.0085	\\
scanam			&  	 0.4073 &      0.1546	&	0.3911	&      0.2138	&        0.0221 &      0.0090	&	 0.0235 &	0.0173	\\
sanepic			&  	-3.0149 &      0.2215	&       -3.0339 &      0.2294	&       -0.0545 &      0.0080	&	-0.0507 &	0.0080	\\
unimap			&  	-0.0008 &      0.0493	&       -0.0211 &      0.1042	&       -0.0219 &      0.0081	&	-0.0201 &	0.0079	\\      
\end{tabular}

\caption{Absolute deviations: mean and standard deviation of (S - S$_{\rm true}$). Results are shown for cases 2, 4, 6, 9, and 10,
         for the available map makers.}
\label{tab:absDev}
%\end{center}
\end{table}
\normalsize
The slopes of the difference map scatter plots were obtained as coeff(1) of the IDL's robust\_linefit routine, and their 
errors were normalized by the map pixels (csig(1)*sqrt(N)). This is provided for the S$_{\rm true}$ $>$ 0.2~Jy domain and
shows if a map-maker introduces any bias for the bright pixels. This is visualized in the scatter plot figures
for each case and PSW/PLW bands.
\begin{table}[th!]
%\begin{center}
\small

\begin{tabular}{l c c c c c c c c}
			&	case 2	&		&		&		&	case 9	&		&		&		\\
\hline
			&	PSW	&		&	PLW	&		&	PSW	&		&	PLW	&		\\
			&	mean	&	stdev	&	mean	&	stdev	&	mean	&	stdev	&	mean	&	stdev	\\
\hline
naive			&      -1.0752	&       0.3474	&       -1.0656 &      0.2999	&      -0.9479  &       0.5620	&    -0.9466	&	  0.5611 \\
destrP0			&      -1.1168	&       0.4647	&       -1.1100 &      0.4363	&      -0.9051  &       0.5889	&    -0.8995	&	  0.5852 \\
destrP1			&      -1.0804	&       0.3586	&       -1.0706 &      0.3109	&      -0.8977  &       0.5678	&    -0.8923	&	  0.5646 \\
scanam			&       0.1383	&       0.1841	&       -0.1078 &      0.0957	&      -0.1873  &       0.1220	&    -0.1979	&	  0.1290 \\
sanepic			&      -1.1680	&       0.4943	&       -1.1426 &      0.4435	&      -2.2767  &       1.4251	&    -2.2901	&	  1.4241 \\
\end{tabular}
\vspace{0.3cm}

\begin{tabular}{l c c c c}
			&	case 4	&		&		&		\\
\hline
			&	PSW	&		&	PLW	&		\\
			&	mean	&	stdev	&	mean	&	stdev	\\
\hline
naive			&      -3.1237  &      103.5528 &      -2.3820  &     163.2418  \\
naiveHum		&      -0.5859  &      108.5850 &       0.2562  &     181.8850  \\
destrP0			&      -3.8805  &      107.0133 &      -3.2674  &     152.9565  \\
destrP1			&      -3.9140  &      107.1141 &      -3.4283  &     141.3440  \\
scanam			&       4.1419  &      91.5670  &       2.9724  &     160.8298  \\
sanepic			&      -13.9558 &      118.2347 &      -12.3324 &     94.6355	\\	 
\end{tabular}
\vspace{0.3cm}

\begin{tabular}{l c c c c c c c c}
			&	case 6	&		&		&		&	case 10	&		&		&		\\
\hline
			&	PSW	&		&	PLW	&		&	PSW	&		&	PLW	&		\\
			&	mean	&	stdev	&	mean	&	stdev	&	mean	&	stdev	&	mean	&	stdev	\\
\hline
naive			&      -1.8433	&     4256.5560 &     -7.0235	&     4394.7150 &      -0.9807	&      457.3350 &     -1.8878	&      111.8000 \\
destrP0			&      -0.7295	&     7248.2743 &     -14.1292	&     9185.9890 &      -0.9219	&      489.7473 &     -1.9283	&      125.8154 \\
destrP1			&      -1.0635	&     6151.4683 &     -12.9828	&     8145.7430 &      -0.9116	&      439.8811 &     -1.8479	&      117.9252 \\
scanam			&       0.5090	&     2579.5553 &      5.0161	&     3218.3239 &	0.6074	&      237.1912 &      1.2517 	&      91.1021  \\
sanepic			&      -2.4014	&     18275.315 &     -35.7553	&     22986.732 &      -1.2562	&      588.3852 &     -2.5008 	&      176.5462 \\
unimap			&       0.2322	&     398.8190  &      0.2143	&     435.0549  &      -0.4689	&      247.9541 &     -0.8593	&      48.8528  \\      
\end{tabular}

\caption{Relative deviations: Table of mean and standard deviations of (S - S$_{\rm true}$)/S$_{\rm true}$. Results are shown for cases 2, 4, 6, 9, and 10,
         for the available map makers.}
\label{tab:relDev}
%\end{center}
\end{table}
\normalsize
\clearpage

\begin{table}[th!]
%\begin{center}
\small

\begin{tabular}{l c c c c c c c c}
			&	case 2	&		&		&		&case 9 	&		&		&		\\
\hline
			&	PSW	&		&	PLW	&		&	PSW	&		&	PLW	&		\\
			&	slope	&	slopeErr&	slope	&	slopeErr&	slope	&	slopeErr&	slope	&	slopeErr\\
\hline
naive			&       0.0062	&	0.0331	&        0.0660 &       0.1179	&     -0.007826 &      0.022550 &     -0.007985 &	0.023011 \\
destrP0			&      -0.0018	&	0.0159	&        0.0044 &       0.0574	&     -0.000038 &      0.000333 &     -0.000045 &	0.000550 \\
destrP1			&       0.0044	&	0.0271	&        0.0429 &       0.0975	&     -0.001122 &      0.005092 &     -0.000983 &	0.005104 \\
scanam			&      -0.0409	&	0.0497	&       -0.0265 &       0.0918	&      0.000007 &      0.000723 &     -0.000091 &	0.001560 \\
sanepic			&       0.0030	&	0.0239	&        0.0081 &       0.0679	&     -0.009294 &      0.025961 &     -0.014845 &	0.029227 \\
\end{tabular}
\vspace{0.3cm}

\begin{tabular}{l c c c c}
			&	case 4 	&		&		&		\\
\hline
			&	PSW	&		&	PLW	&		\\
			&	slope	&	slopeErr&	slope	&	slopeErr\\
\hline
naive			&      -0.0010	&       0.0096	&      -0.0030	&        0.0146  \\
naiveHum		&       0.0091	&       0.0305	&       0.0044	&        0.0339  \\
destrP0			&      -0.0007	&       0.0096	&      -0.0023	&        0.0140  \\
destrP1			&      -0.0007	&       0.0096	&      -0.0023	&        0.0140  \\
scanam			&      -0.0065	&       0.0250	&      -0.0131	&        0.0497  \\
sanepic			&      -0.0004	&       0.0144	&      -0.0049	&        0.0264  \\       
\end{tabular}
\vspace{0.3cm}

\begin{tabular}{l c c c c c c c c}\scriptsize
			&	case 6 	&		&		&		&	case 10 &		&		&		\\
\hline
			&	PSW	&		&	PLW	&		&	PSW	&		&	PLW	&		\\
			&	slope	&	slopeErr&	slope	&   slopeErr	&	slope	&	slopeErr&	slope	&    slopeErr	\\
\hline
naive			&     -0.119569 &     0.147264	&     -0.123366 &    0.148822	&    -0.000459  &      0.005589 &     -0.001415 &       0.007617 \\
destrP0			&     -0.000056 &     0.000652	&     -0.000092 &    0.000924	&     0.000046  &      0.003197 &     -0.000215 &       0.004927 \\
destrP1			&      0.000323 &     0.005220	&      0.000114 &    0.005612	&     0.000103  &      0.003505 &     -0.000193 &       0.005307 \\
scanam			&      0.001072 &     0.002506	&      0.001435 &    0.003560	&    -0.009363  &      0.013734 &     -0.019243 &       0.028724 \\
sanepic			&     -0.014426 &     0.025079	&     -0.017577 &    0.022181	&     0.001269  &      0.005870 &     -0.011593 &       0.010348 \\
unimap			&     -0.000036 &     0.002143	&      0.000329 &    0.004832	&    -0.000420  &      0.004908 &     -0.003886 &       0.008665 \\      
\end{tabular}

\caption{Slopes and errors of the difference map scatter plots, calculated only for the S$_{\rm true}$ $>$ 0.2~Jy range.}
\label{tab:slope}
%\end{center}
\end{table}
\normalsize

\subsection{Summary}\label{sect:diff_summary}

We can draw a general conclusion: The data treatment is not optimal if the map-maker's hypothesis does not meet the 
conditions of the simulations. For example, Sanepic in case 9, where the data is not circulant.

\vspace{5mm}
\noindent
Further conclusions in the low dynamic range cases (2, 4, 10):
\begin{itemize}
  \item Destriper/P0, Scanamorphos, and Sanepic produce similar results.
%  \item Scanamorphos produces a larger scatter which is only due to a large scale slope introduced by the mapper. 
  \item By running Scanamorphos with the /galactic option (which preserves large scales) the residual slope is avoided, 
   and the Scanamorphos scatter plots are comparable to the Sanepic and Destriper/P0.
%   and overall display slightly smaller stdev to the other two mappers. 
  \item By running the Destriper with a 1st order polynomial baseline removal, residual slope will be introduced in the map. 
\end{itemize}

\vspace{5mm}
\noindent
In high dynamic range cases (6, 9):
\begin{itemize}
  \item Destriper/P1, Sanepic, Unimap introduce different types of large spatial scale noise. 
  \item Destriper/P0 avoids introducing large spatial scales.

     However, for specific scans the baseline is not properly removed, due to the inability of the Destriper/P0 
     (and P1 too) to deal with the SPIRE "cooler-burps"?
  \item Scanamorphos also avoids introducing any large spatial scale noise. 

     However, the scatter plots still display larger standard deviation, likely due to a slight positional offset 
     introduced by the mapper (e.g., Scanamorphos difference maps, Case 9, PLW), and a slight change in the beam size.  
\end{itemize}

%\bibliographystyle{abbrv}
%\bibliography{metrics_SPIRE_DIFFmap_report_update2013july}
%
%\end{document}

%% file: metric_report_fft_v2.1_pdf.tex
% Template for map-maker reports.
% CKX, April 7, 2013
%\documentclass[letterpaper,11pt]{report}
%\usepackage{titletoc}% http://ctan.org/pkg/titletoc
%\usepackage{amsmath} %Never write a paper without using *amsmath* for
%                     %its many new commands
%\usepackage{amssymb} %Some extra symbols
%\usepackage{makeidx} %If you want to generate an index, automatically
%\usepackage{graphicx} %If you want to include postscript graphics
%\usepackage{caption} %If you want to include postscript graphics
%\usepackage{epsfig}
%\setcounter{chapter}{1}
%  Do not change above lines.
%  Do not play with the margins or fonts.  This annoys the reviewers.
%  Unusual fonts do not render on all systems.  Don't change the
%  font size of the major headers either
% 

%\newcommand{\truth}{\emph{truth}}
%\newcommand{\no}{\emph{noise-only}}
%\usepackage[normalem]{ulem}
%\usepackage{subfigure}

%\begin{document}
\section{Power Spectra (Luca Conversi)}

\subsection{Test Data}\label{lc:data}

Five simulated test cases (cases 2, 4, 6, 9, 10) and five real cases (cases 3, 7, 11, 12 and 13) were examined in this metric. Tab.~\ref{tab:lc:cases} lists the content of the different simulated cases. Cases 3, 7, and 11 are real SPIRE maps of extragalactic dark fields observed in different modes (SPIRE-only or parallel, slow or fast scan speed). Case 12 is a real SPIRE large map observation of NGC-628 taken at nominal scan speed, while case 13 is a Parallel mode observation of a Hi-Gal field taken at fast scan speed.

\begin{table}[htd]
	\centering
	\begin{tabular}{r|l|l}
	Case & Mapping Mode	& Extended Layer\\
	\hline
	2	 & Nominal		& 24 micron cirrus\\
	4	 & Nominal		& M51	\\
	6	 & Fast scan		& 24 micron Galactic Centre\\\
	9	 & Parallel		& 24 micron Galactic Centre\\
	10	 & Parallel		& 24 micron cirrus\\
	\end{tabular}
	\caption{List of the different simulated cases.}\label{tab:lc:cases}
\end{table}

\begin{figure}[htb]
	\centering
	\includegraphics[width=\textwidth,keepaspectratio]{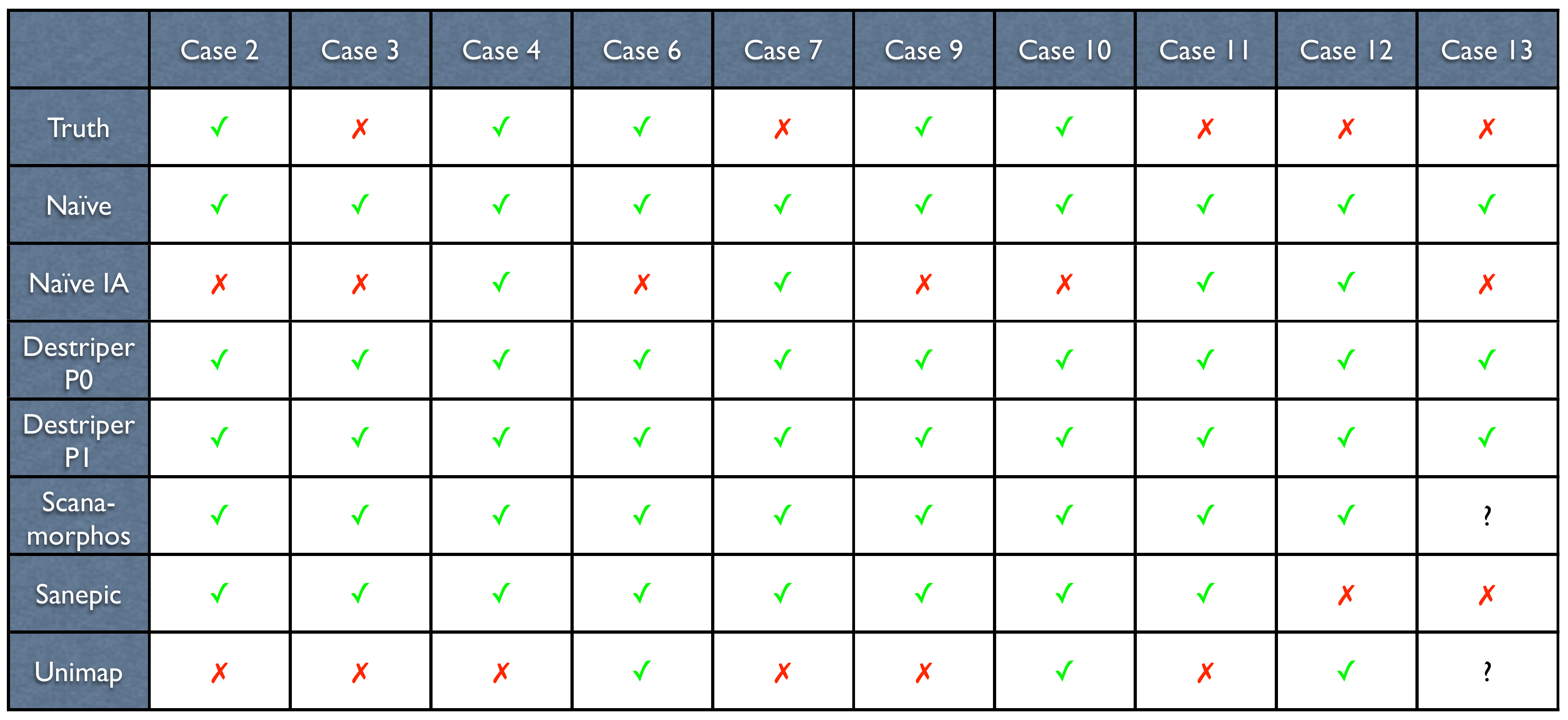}
	\caption{List of cases reduced with different map-makers: green marks mean maps for a given case/map-maker combination was available; red crosses indicate data was not reduced with the selected map-maker; a question mark indicates the cases were maps were available but had a different WCS in respect to the others, making more difficult a direct comparison of the power spectra.}\label{fig:lc:cases}
\end{figure}

Fig.~\ref{fig:lc:cases} shows all the combinations between cases to be analyzed and available maps produced with different map-makers:

\begin{description}
	\item[Truth] means that the input, \emph{noise-free} map was available. Clearly, only cases with simulated extended data had such maps. N.B.: \uline{all truth maps are meant as simulated only data (no noise) but convolved with the instrument's beam} and reprojected on standard SPIRE pixel sizes (6 to 14~arcsec).
	\item[Na\"{\i}ve] are cases reduced with the standard HIPE median baseline subtraction plus na\"{\i}ve map-maker: all cases were available.
	\item[Na\"{\i}ve IA] are cases reduced in interactive analysis, where the user selected a suitable polynomial baseline removal and then ran the na\"{\i}ve map-maker: only cases 4, 7, 11 and 12 were available.
	\item[Destriper P0 and P1] are cases reduced with the HIPE destriper, using either a polynomial order of 0 or 1: all cases were available.
	\item[Scanamorphos] maps were available for all cases; however case 13 had a different WCS (maps were larger) in respect to the other map-makers maps, so a direct comparison of the power spectra was not possible (see Figs.~\ref{fig:lc:case13plw} and \ref{fig:lc:case13psw}).
	\item[Sanepic] maps were available only for the simulations, hence cases 12 and 13 were missing.
	\item[Unimap] maps instead were available only for cases 6, 10 and 12. Case 13 had different WCS (larger maps).
\end{description}

\subsection{Analyses and Results}\label{results}

The analysis is based on an IDL script kindly provided by Jim Ingalls.

\begin{enumerate}
	\item It computes the inverse 2D fast Fourier transform of the input map, normalizing to the total surface brightness of the image, hence the importance of having maps with same WCS.
	\item Then it creates a map of angular scale bins $k$, depending on mapÕs size and resolution.
	\item The average value in each $k$ bin is computed.
	\item Finally, the power spectra $P(k)$ results are fitted between $k = [0.01, 1]$ acrmin$^{-1}$
\end{enumerate}

For the cases where a \truth\ map was available, its power spectrum was used as benchmark. We defined the divergence as:

\begin{equation}
	\Delta (k) = \frac{| P_{\it Map-maker}(k) - P_{\it Truth}(k) |}{P_{\it Truth}(k)}
\end{equation}\label{eq:divergence}

Given the amount of data and plots, we will discuss only some cases. Results obtained on the cases not discussed are reported in Appx~\ref{lc:appendix}.

\subsubsection{Case 2}

\begin{figure}[p!]
	\centering
	\includegraphics[width=10cm,keepaspectratio]{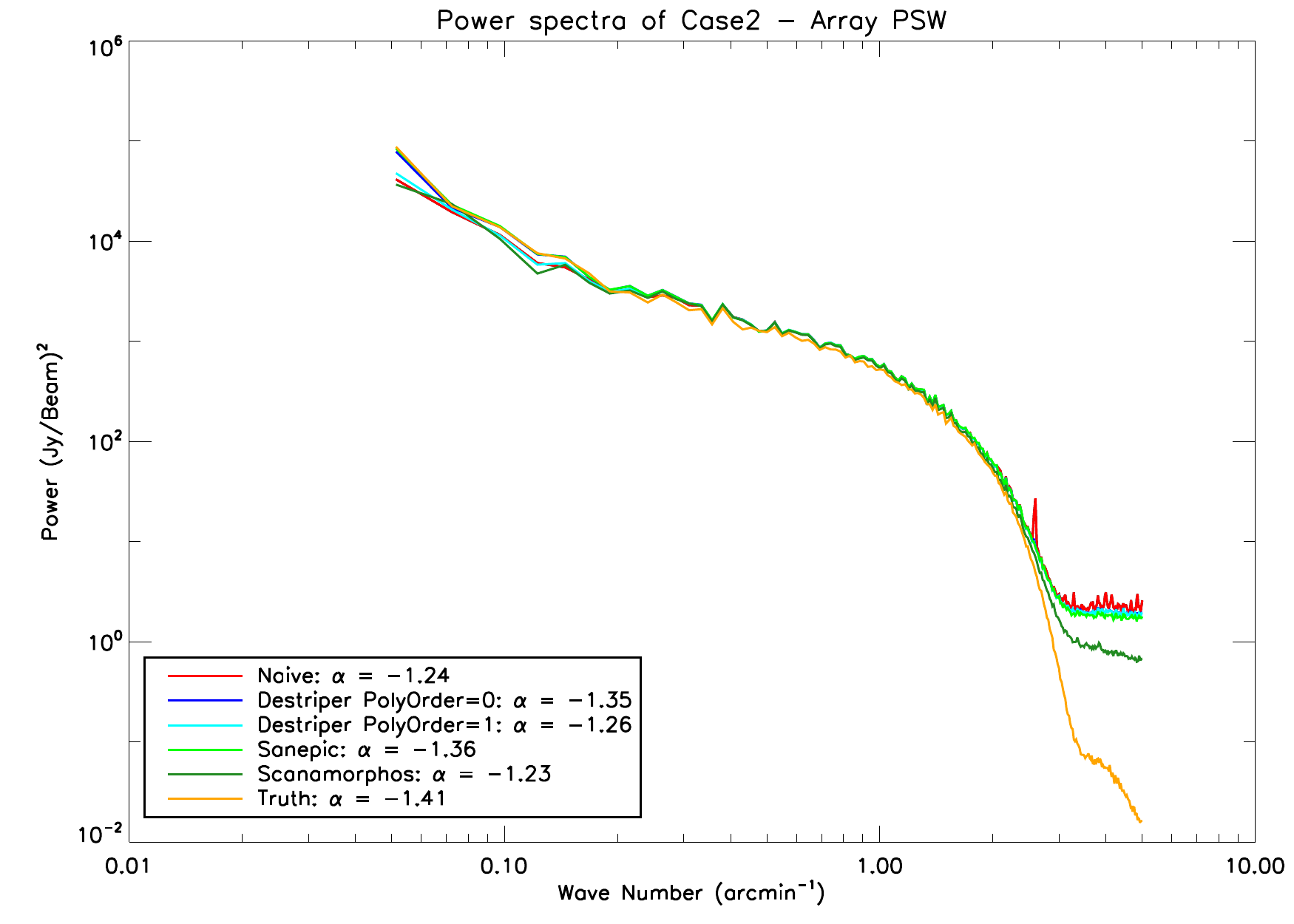}
	\caption{Results of power spectra for case 2 (see Tab.~\ref{tab:lc:cases}), PSW map: nominal speed, SPIRE-only scan map with simulated cirrus emission.}\label{fig:lc:case2psw}
	\vspace{0.5cm}
	\includegraphics[width=10cm,keepaspectratio]{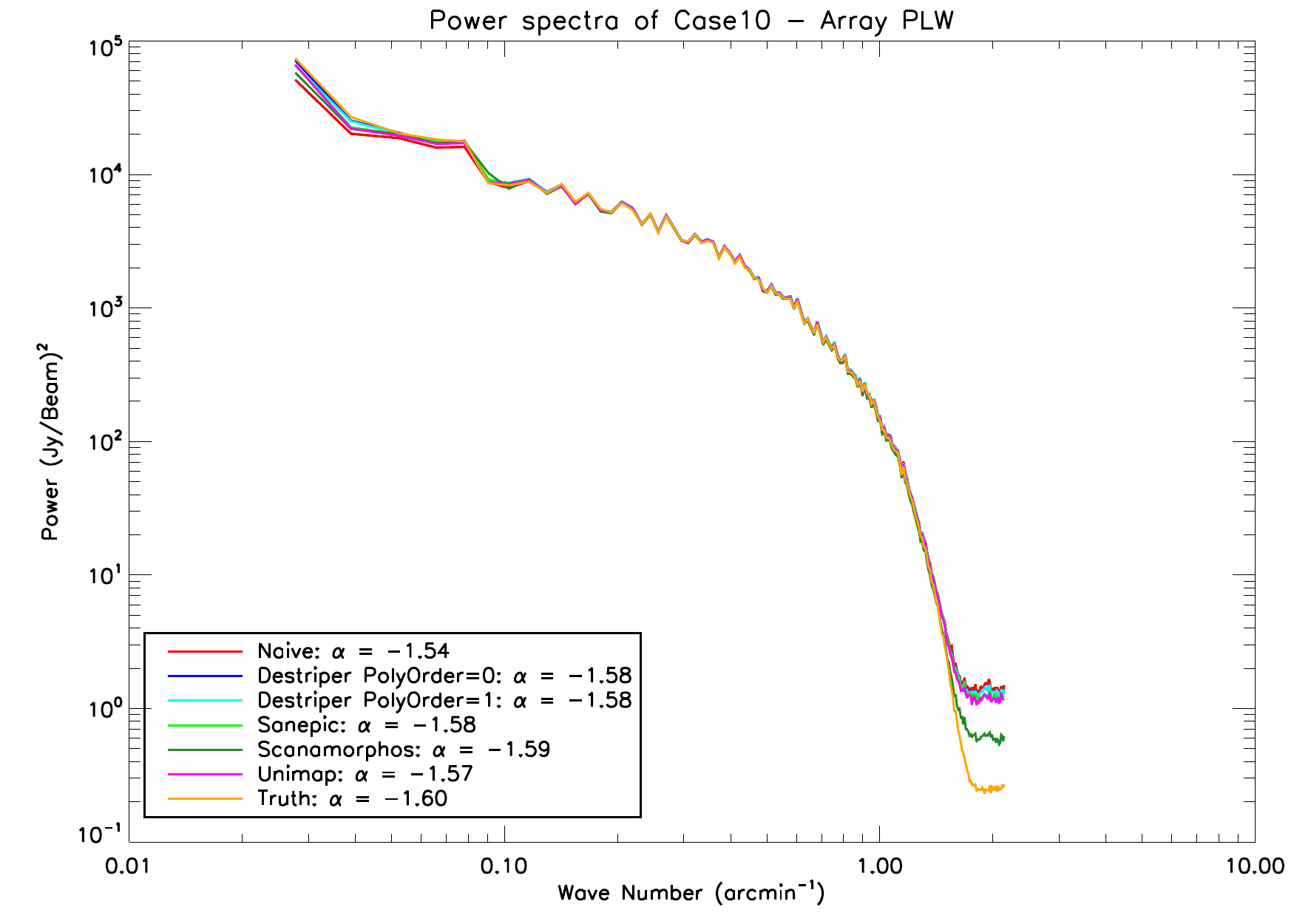}
	\caption{Results of power spectra for case 10 (see Tab.~\ref{tab:lc:cases}), PLW map: parallel mode observation with simulated cirrus emission.}\label{fig:lc:case10plw}
\end{figure}

\begin{figure}[p!]
	\centering
	\subfigure[Power spectra of PLW map]
	{\includegraphics[width=9.5cm,keepaspectratio]{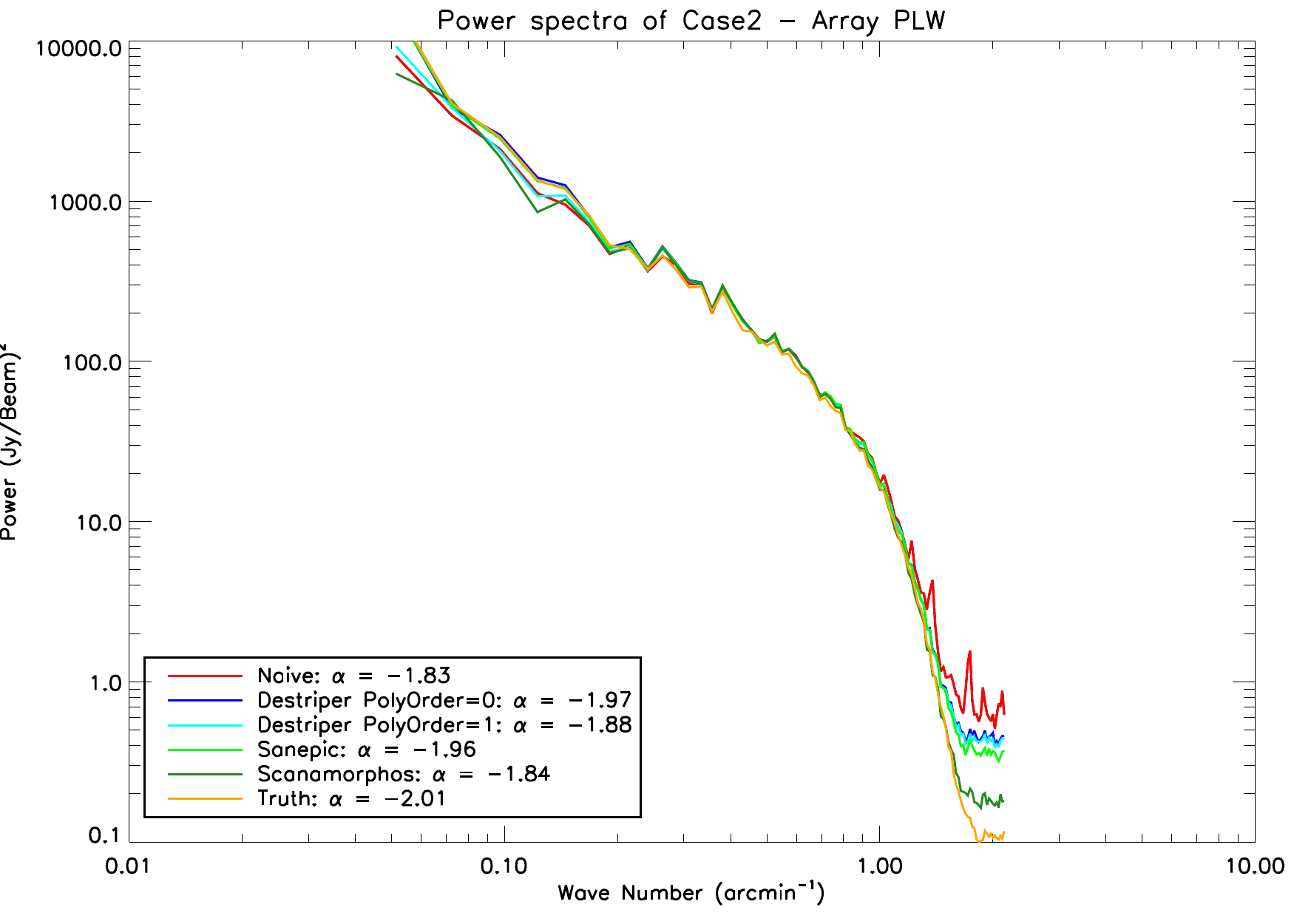}}
	\vspace{0.5cm}
	\subfigure[Divergence from \truth\ of PLW map power spectra]
	{\includegraphics[width=9.5cm,keepaspectratio]{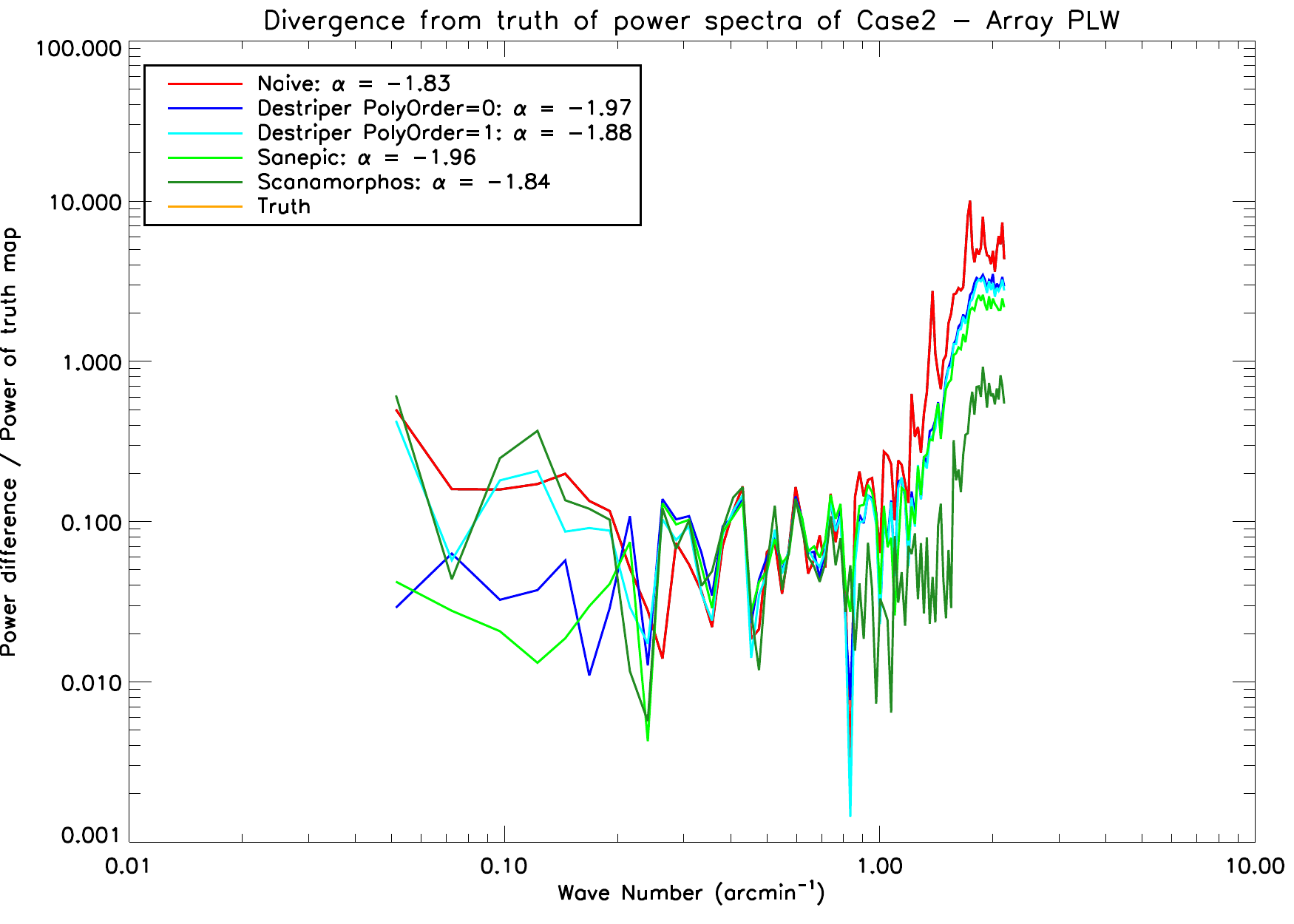}}\\
	\caption{Results of power spectra for case 2 (see Tab.~\ref{tab:lc:cases}), PLW map: nominal speed, SPIRE-only scan map with simulated cirrus emission. In Fig.~\ref{fig:lc:case2plw}a the power spectra computed on available maps are reported, while Fig.~\ref{fig:lc:case2plw}b shows the divergence between the power spectrum computed on the \truth\ map and the others.}\label{fig:lc:case2plw}
\end{figure}

Results of power spectra for case 2 (nominal speed, SPIRE-only scan map with simulated cirrus emission, see Tab.~\ref{tab:lc:cases}) are shown in Fig.~\ref{fig:lc:case2psw} for PSW and in Fig.~\ref{fig:lc:case2plw} for PLW: the power spectra computed on available maps are reported in Fig.~\ref{fig:lc:case2plw}a, while Fig.~\ref{fig:lc:case2plw}b shows the divergence from the \truth\ power spectrum of the various map-makers.

From these plots it is evident how the na\"{\i}ve map-maker is higher than the others at small scales ($k > 1$ arcmin$^{-1}$). Also, Sanepic and the two HIPE destriper \emph{flavors} P0 and P1 agree very nicely, being the closer to the \truth\ power spectrum at large scales ($k < 0.2$ arcmin$^{-1}$). Finally, Scanamorphos is the map-maker giving the lowest power spectrum at small scales (i.e.\ closer to the \truth\ power spectrum).

A more detailed analysis of the PSW power spectra is discussed in later sections.

\subsubsection{Case 10}

\begin{figure}[p!]
	\centering
	\subfigure[Power spectra of PSW map]
	{\includegraphics[width=9.5cm,keepaspectratio]{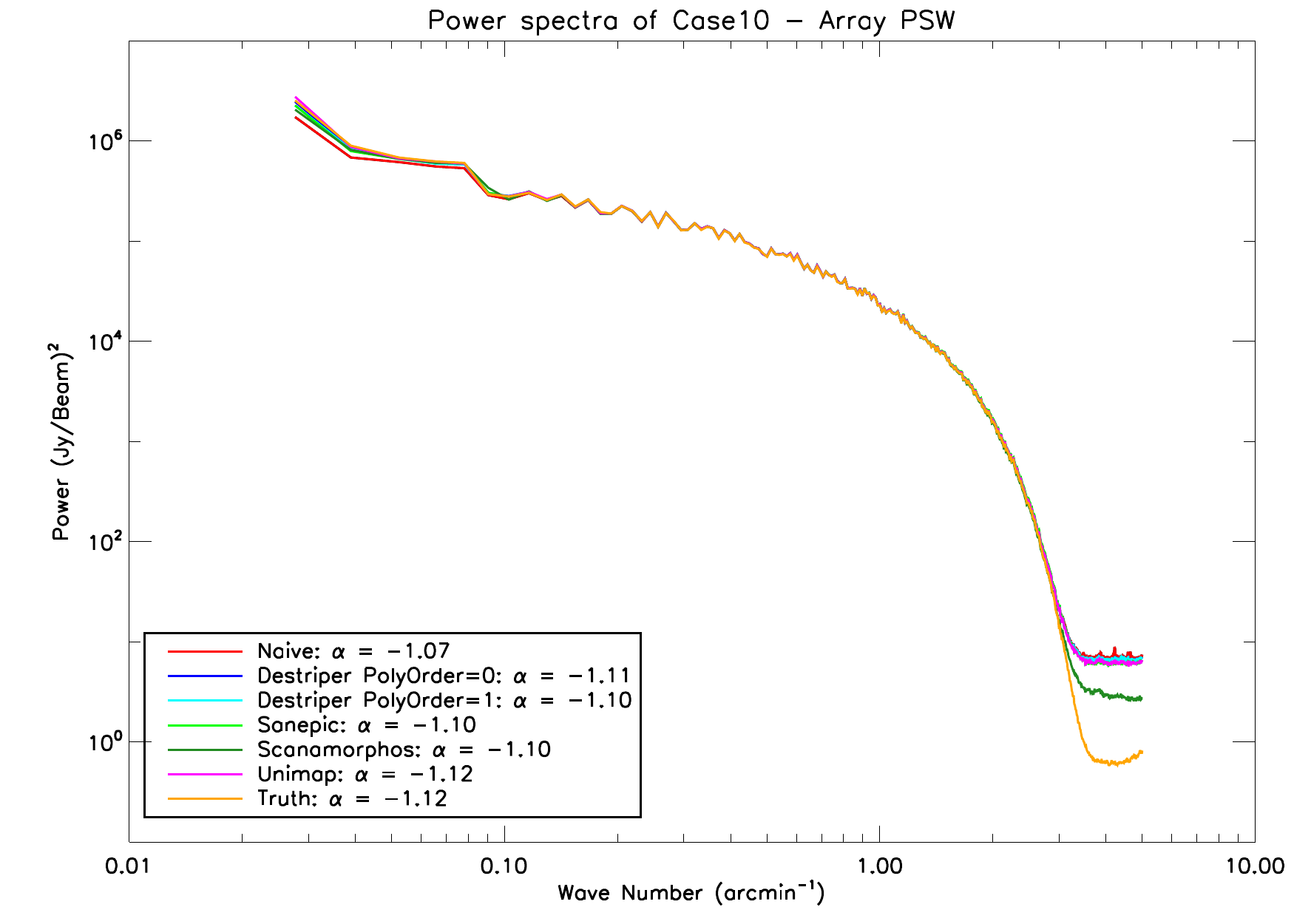}}
	\vspace{0.5cm}
	\subfigure[Divergence from \truth\ of PSW map power spectra]
	{\includegraphics[width=9.5cm,keepaspectratio]{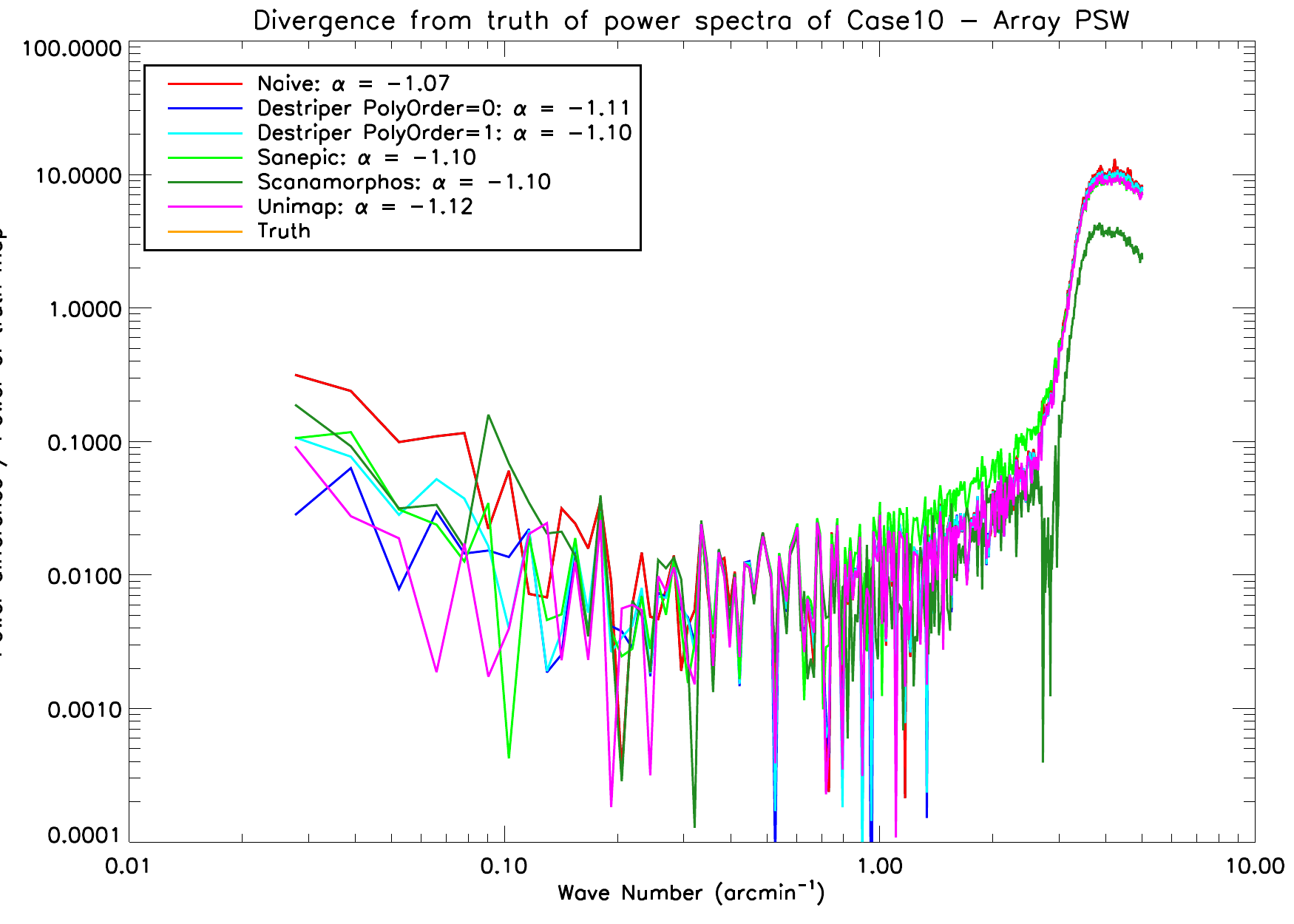}}\\
	\caption{Results of power spectra for case 10 (see Tab.~\ref{tab:lc:cases}), PSW map: nominal speed, SPIRE-only scan map with simulated cirrus emission. In Fig.~\ref{fig:lc:case10psw}a the power spectra computed on available maps are reported, while Fig.~\ref{fig:lc:case10psw}b shows the divergence between the power spectrum computed on the \truth\ map and the others. Also worth noting that Destriper P0 and Unimap are performing best at large scales ($k < 0.2$ arcmin$^{-1}$), being closest to the \truth\ power spectrum.}\label{fig:lc:case10psw}
\end{figure}

Results of power spectra for case 10 (parallel mode observation with simulated cirrus emission, see Tab.~\ref{tab:lc:cases}) are shown in Fig.~\ref{fig:lc:case10plw}b for PLW and in Fig.~\ref{fig:lc:case10psw} for PSW: the power spectra computed on available maps are reported in Fig.~\ref{fig:lc:case10psw}a, while Fig.~\ref{fig:lc:case10psw}b shows the divergence from the \truth\ power spectrum by the others.

Results are similar to the case 2 analysed earlier: the na\"{\i}ve map-maker is slightly higher than the others at small scales, but the general agreement among map-makers is good (within few percent). The only exemption is the Scanamorphos map-maker, which has the lowest power spectrum at small scales and being again closer to the \truth\ power spectrum: this may be due Scanamorphos reproduction algorithm, which effectively smoothes the image (and hence the noise as well).

\subsubsection{Cases 6 \& 9}

Results of power spectra for cases 6 (fast scan speed, SPIRE-only scan map with simulated galactic centre emission) and 9 (parallel mode observation with simulated galactic centre emission, see Tab.~\ref{tab:lc:cases}) are discussed together because they show the same behavior, being the sky signal really strong and much higher than the noise.

\begin{figure}[p!]
	\centering
	\subfigure[Power spectra of PSW map]
	{\includegraphics[width=9.5cm,keepaspectratio]{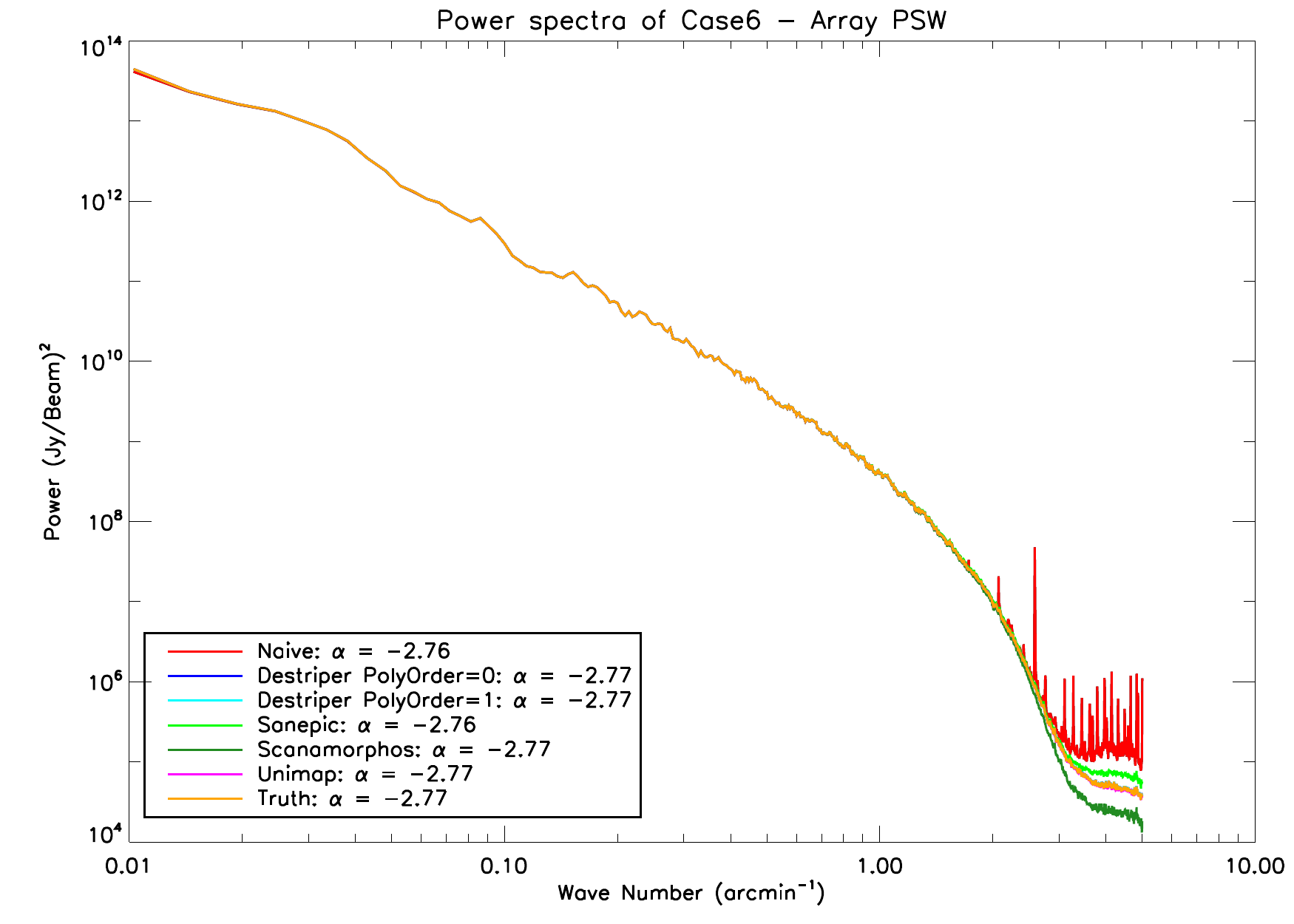}}
	\vspace{0.5cm}
	\subfigure[Divergence from \truth\ of PSW map power spectra]
	{\includegraphics[width=9.5cm,keepaspectratio]{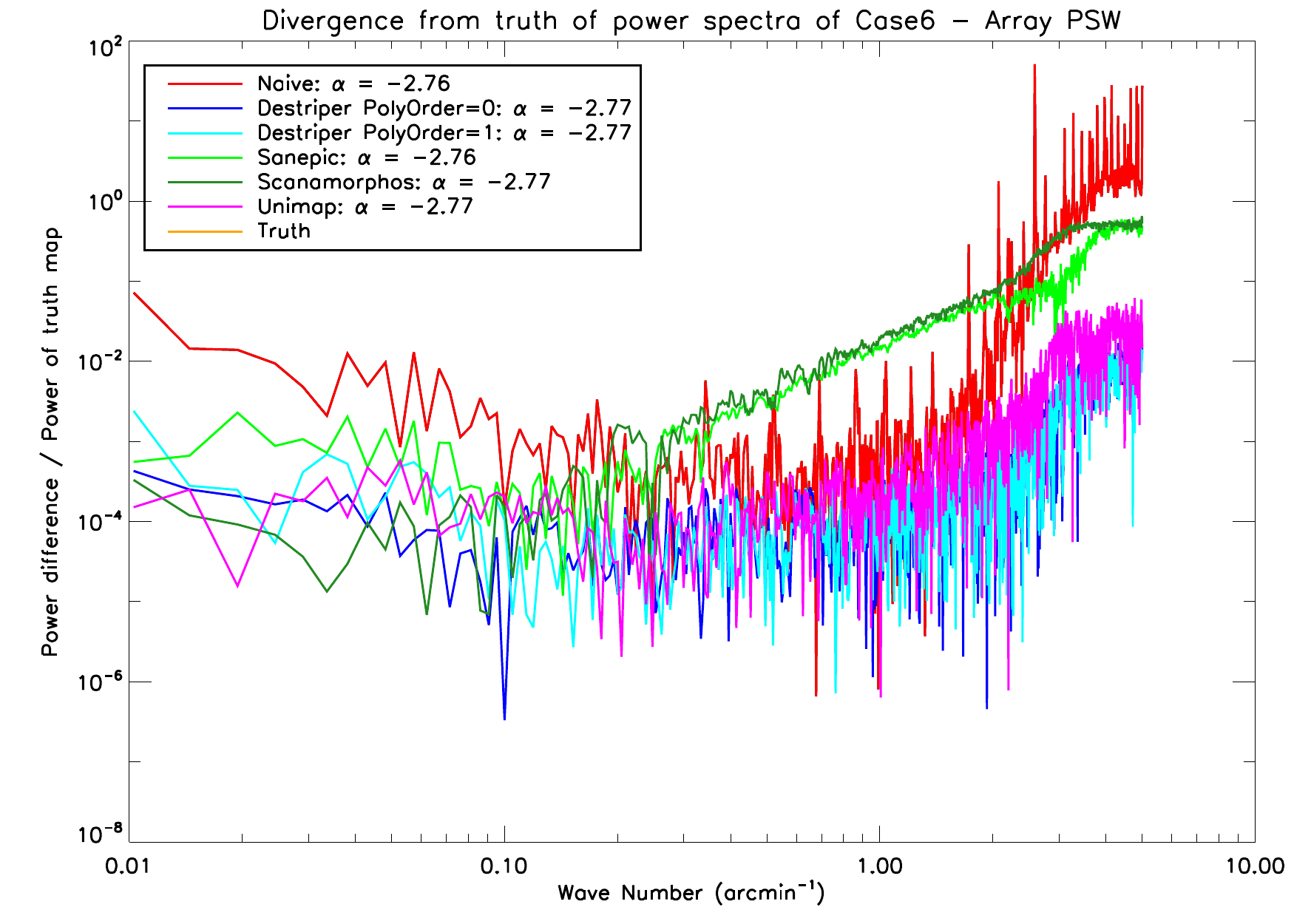}}\\
	\caption{Results of power spectra for case 6 (see Tab.~\ref{tab:lc:cases}), PSW map: fast scan speed, SPIRE-only scan map with simulated galactic centre emission. In Fig.~\ref{fig:lc:case6psw}a the power spectra computed on available maps are reported, while Fig.~\ref{fig:lc:case6psw}b shows the divergence between the power spectrum computed on the \truth\ map and the others.}\label{fig:lc:case6psw}
\end{figure}

\begin{figure}[p!]
	\centering
	\subfigure[Power spectra of PSW map]
	{\includegraphics[width=9.5cm,keepaspectratio]{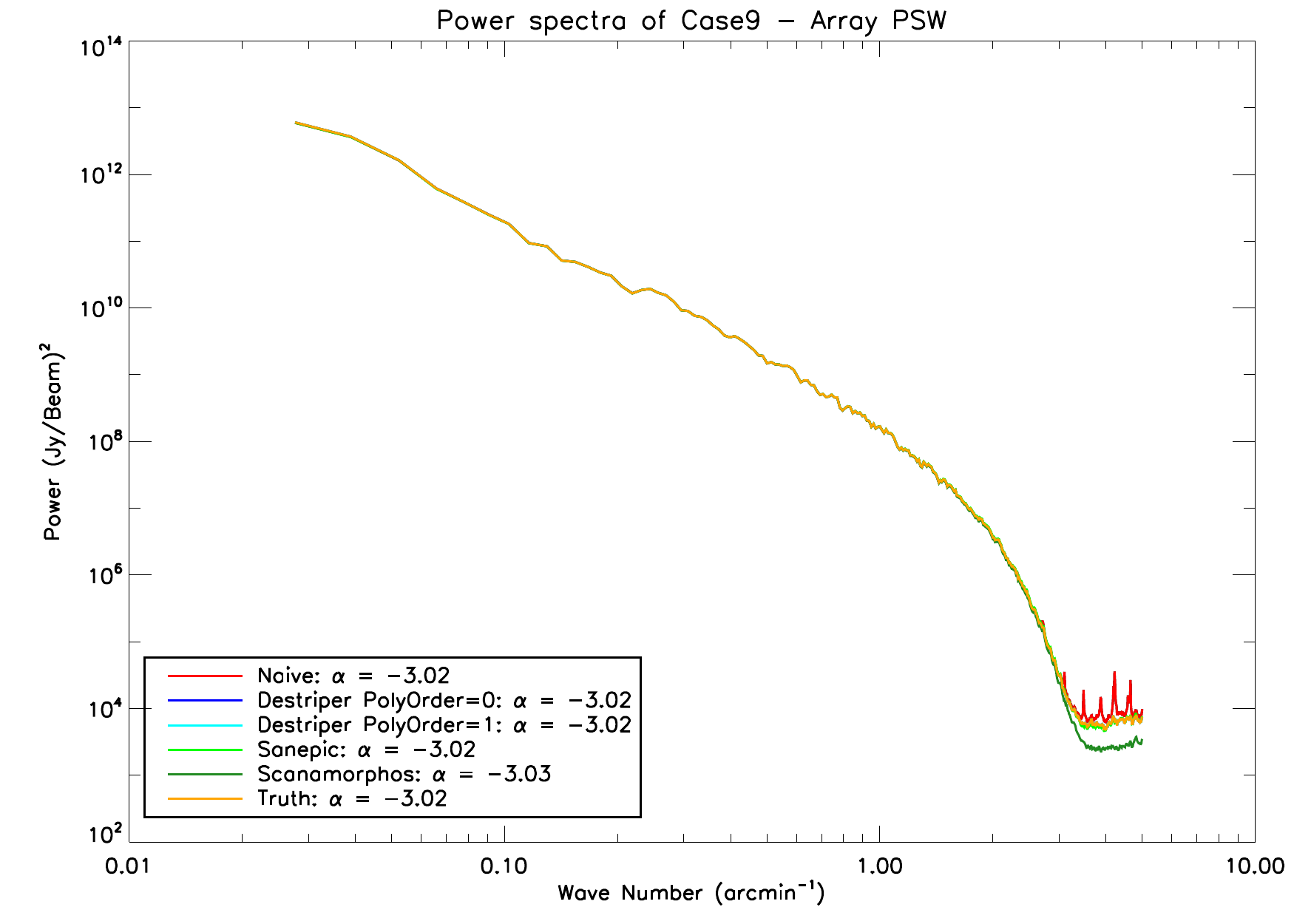}}
	\vspace{0.5cm}
	\subfigure[Divergence from \truth\ of PSW map power spectra]
	{\includegraphics[width=9.5cm,keepaspectratio]{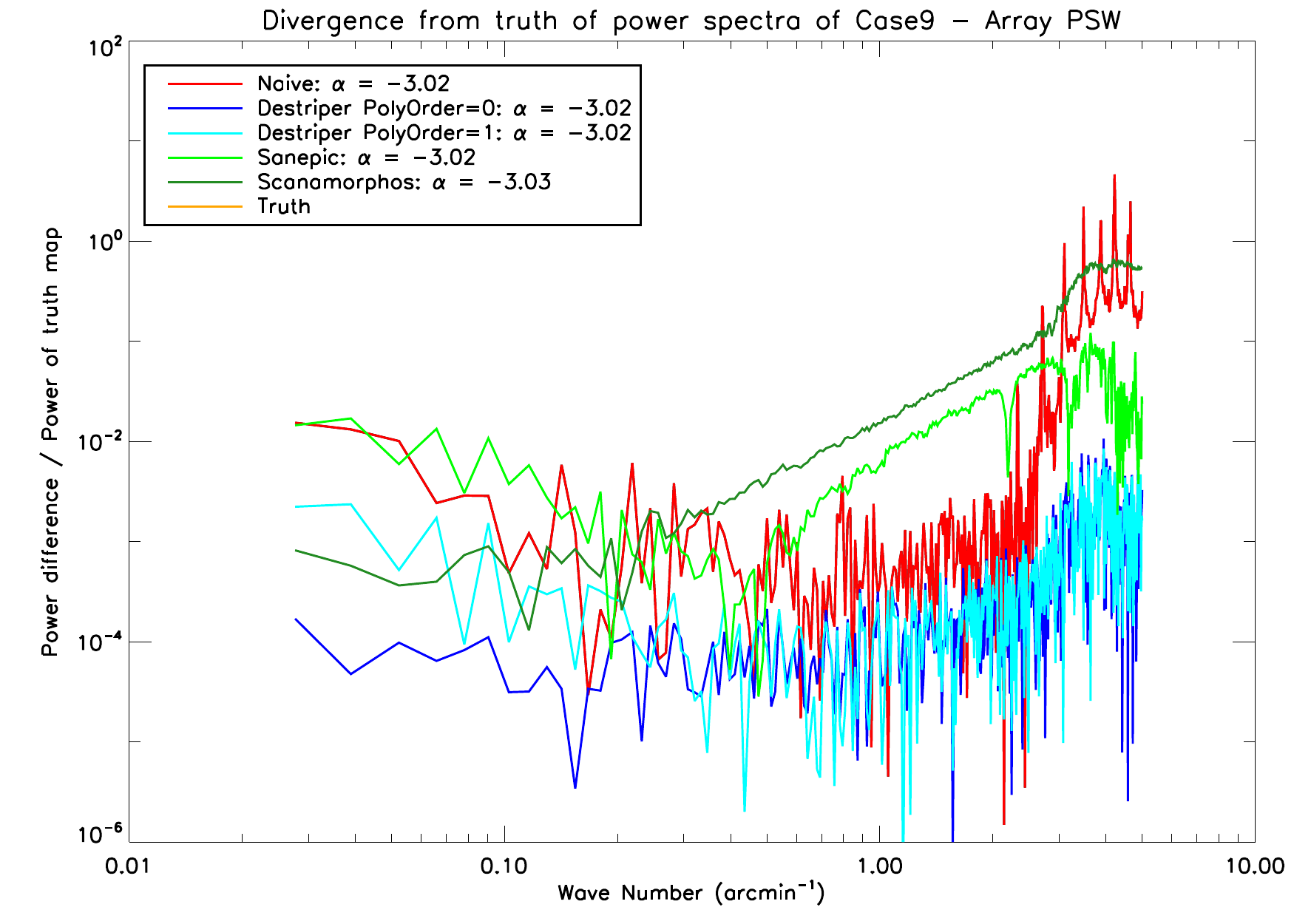}}\\
	\caption{Results of power spectra for case 9 (see Tab.~\ref{tab:lc:cases}), PSW map: parallel mode observation with simulated galactic centre emission. In Fig.~\ref{fig:lc:case9psw}a the power spectra computed on available maps are reported, while Fig.~\ref{fig:lc:case9psw}b shows the divergence between the power spectrum computed on the \truth\ map and the others. Again, Destriper P0 is the closest to the \truth\ power spectrum at large scales ($k < 0.2$ arcmin$^{-1}$).}\label{fig:lc:case9psw}
\end{figure}

Fig.~\ref{fig:lc:case6psw}a and Fig.~\ref{fig:lc:case9psw}a present the power spectra computed on available PSW maps, for cases 6 and 9, respectively. The divergences of map-maker power spectra from the \truth\ are reported in Figs.~\ref{fig:lc:case6psw}b and \ref{fig:lc:case9psw}b. 

As already mentioned, the sky signal is dominated by the galactic centre emission: as a consequence, all map-makers agree really well with the \truth\ at all but small scales ($k > 3$ arcmin$^{-1}$). Here, the na\"{\i}ve map-maker power spectrum is again higher than the others; it also presents a lot of \emph{peaks} due to the leftover stripes: these are produced by the median baseline subtraction, an inadequate method given the high dynamic range of the present in the data. Also the Sanepic power spectrum is higher than the \truth\ but to a lower extent. Again, Scanamorphos is giving a power spectrum which is lower than all the others, but this time even lower then the \truth\ one.

The two HIPE destriper \emph{flavors} P0 and P1 and the Unimap power spectra agree very nicely with the \truth\ power spectrum, having a divergence of $10^{-3} \div 10^{-2}$, i.e.\ $\sim$2 orders of magnitude lower than the other map-maker power-spectra.

\subsubsection{Case 7}

Case 7 is a \no, fast scan speed, SPIRE-only scan map (see Tab.~\ref{tab:lc:cases}). The original SPIRE data contained a so-called \emph{cooler burp}, i.e.\ the observation has been taken in the first 8h after a cooler recycle, where the cooler temperature is particularly unstable and the current pipeline is not capable to correct for the induced effects. Since there is no simulated signal added, the power spectra evidence the capabilities of the different map-makers to cope with such effect. 

\begin{figure}[h!]
	\centering
	\subfigure[Power spectra of PSW map]
	{\includegraphics[width=6cm,keepaspectratio]{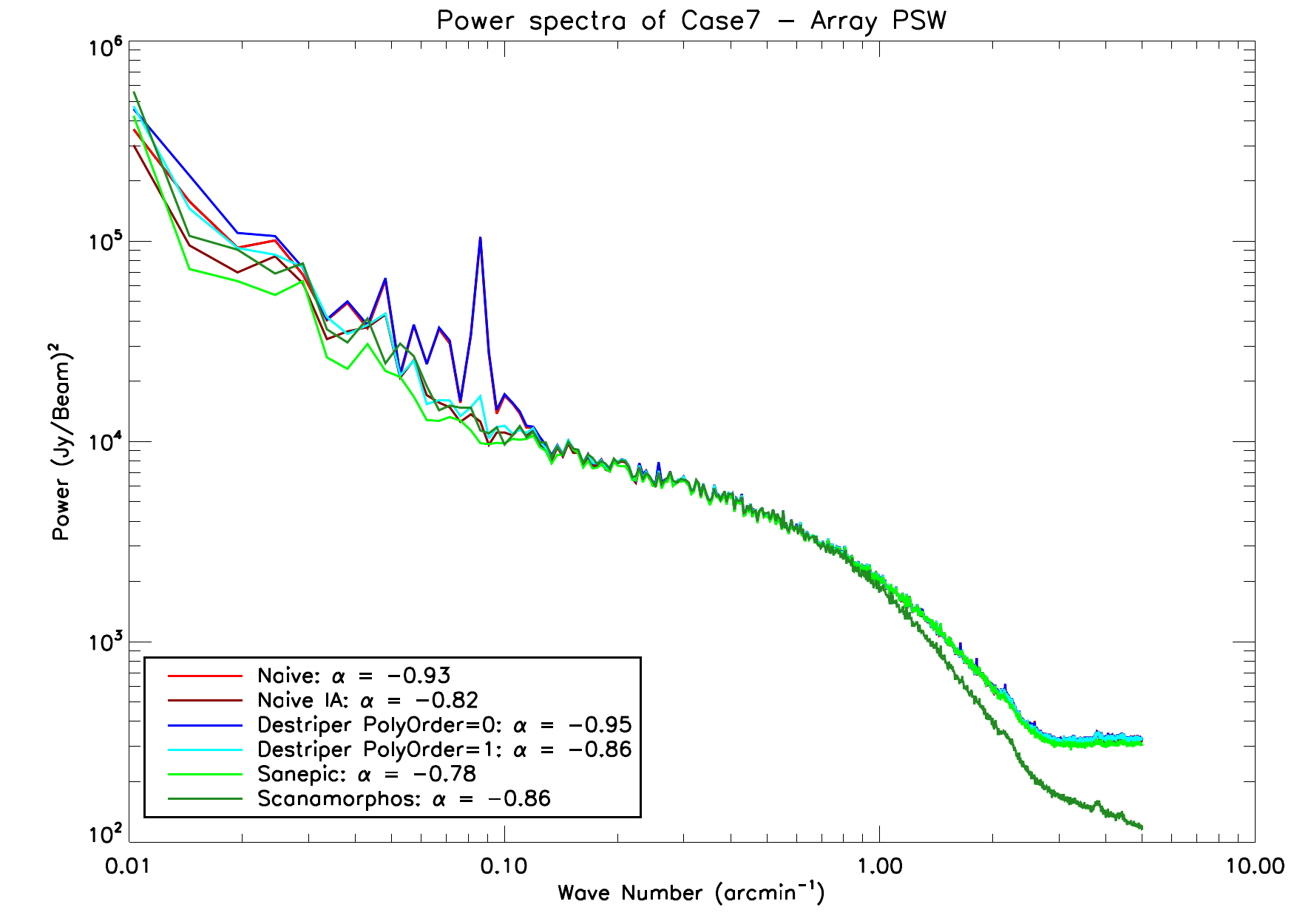}}
	\hspace{0.5cm}
	\subfigure[Zoom-in of PSW maps power spectra]
	{\includegraphics[width=5cm,keepaspectratio]{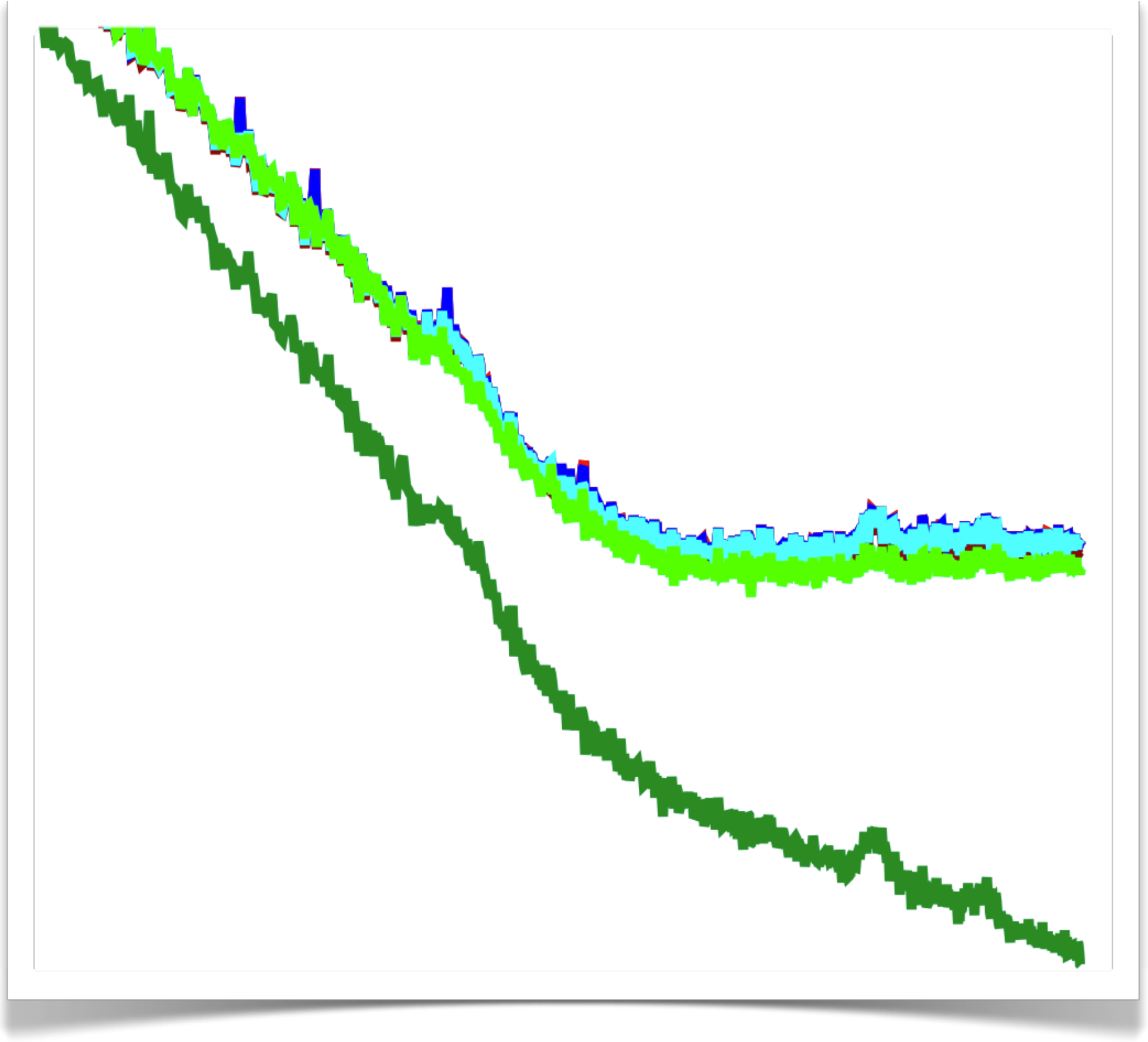}}\\
	\caption{Results of power spectra for case 10 (see Tab.~\ref{tab:lc:cases}), PSW map: \no, fast scan speed, SPIRE-only scan map. In Fig.~\ref{fig:lc:case7psw}a the power spectra computed on available maps are reported, while Fig.~\ref{fig:lc:case7psw}b shows a zoom-in at small scales ($k > 1$ arcmin$^{-1}$).}\label{fig:lc:case7psw}
\end{figure}

Fig.~\ref{fig:lc:case7plw} reports the results for the PLW array: Fig.~\ref{fig:lc:case7plw}a is the full range, while Fig.~\ref{fig:lc:case7plw}b shows a zoom-in at small scales ($k > 1$ arcmin$^{-1}$). Fig.~\ref{fig:lc:case7psw} is the same for the PSW maps. From the plots, it is possible to deduce that:

\begin{figure}[h!]
	\centering
	\subfigure[Power spectra of PLW map]
	{\includegraphics[width=6cm,keepaspectratio]{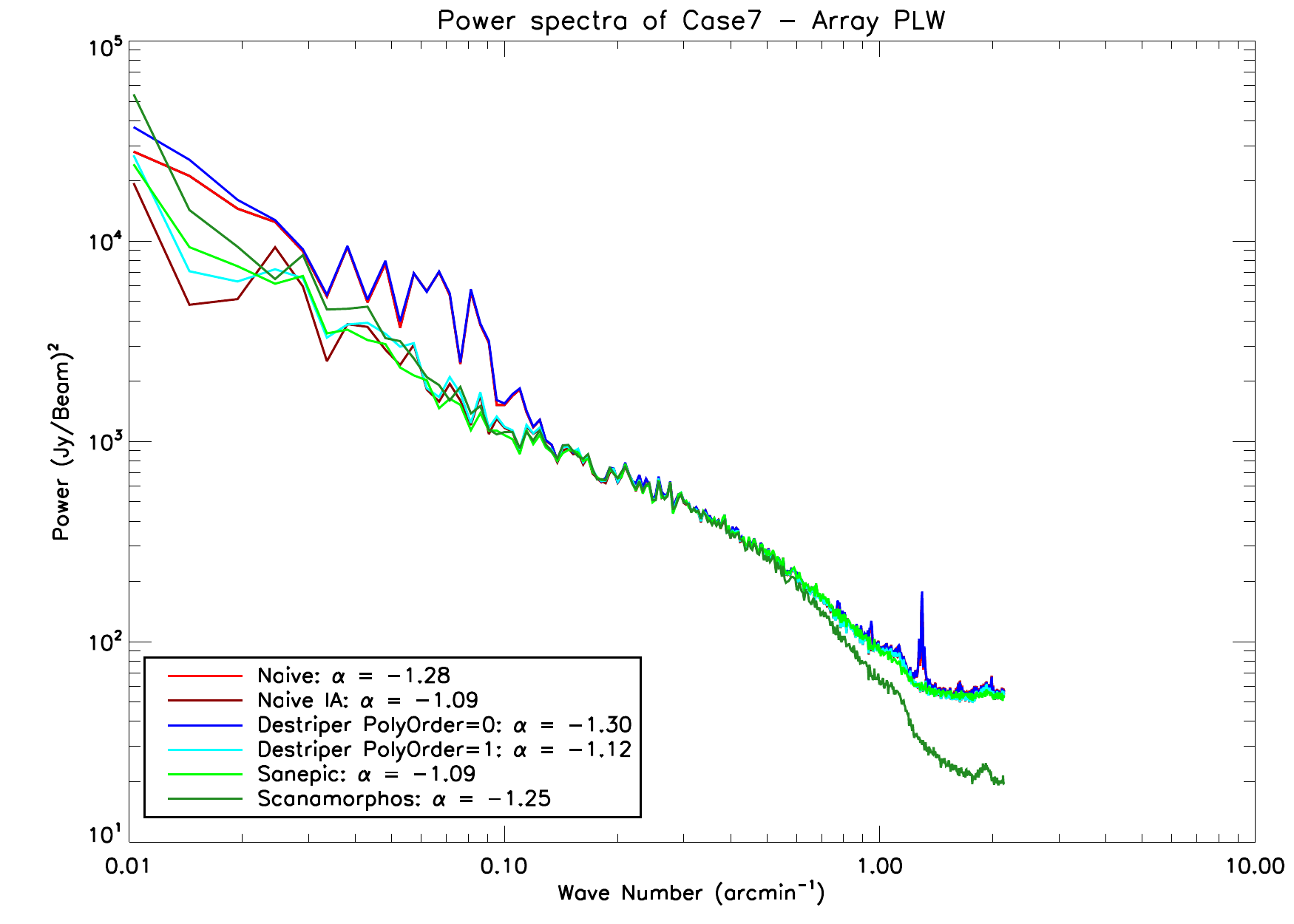}}
	\hspace{0.5cm}
	\subfigure[Zoom-in of PLW maps power spectra]
	{\includegraphics[width=5cm,keepaspectratio]{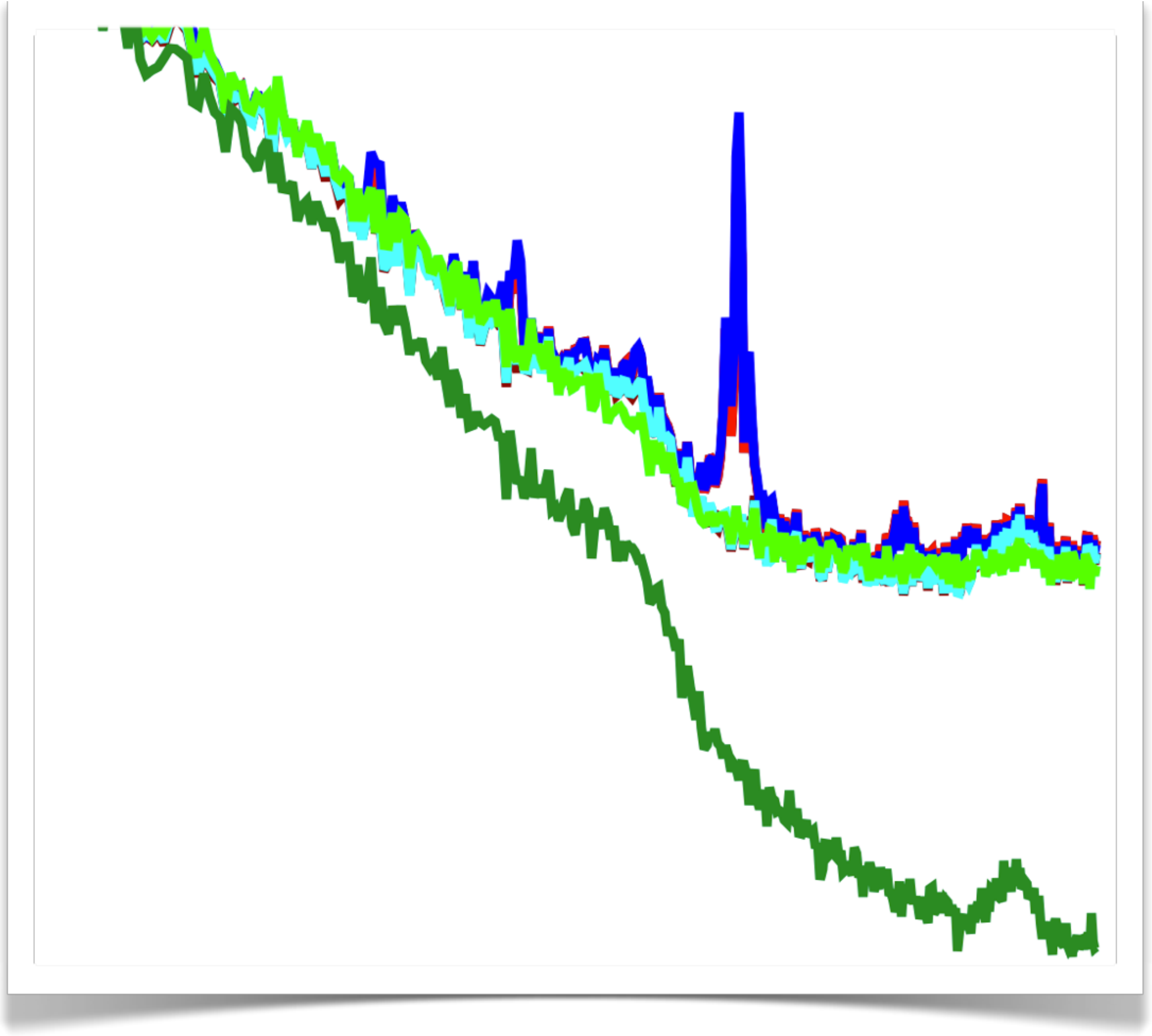}}\\
	\caption{Results of power spectra for case 10 (see Tab.~\ref{tab:lc:cases}), PLW map: \no, fast scan speed, SPIRE-only scan map. In Fig.~\ref{fig:lc:case7plw}a the power spectra computed on available maps are reported, while Fig.~\ref{fig:lc:case7plw}b shows a zoom-in at small scales ($k > 1$ arcmin$^{-1}$).}\label{fig:lc:case7plw}
\end{figure}

\begin{itemize}
	\item The na\"{\i}ve map-maker and destriper with zero-th polynomial order are higher at large scales: this is a clear consequence that none of them was designed to correct for the \emph{cooler burp}. Allowing a higher polynomial order fit solves the problem, as evidenced by the power spectra of Na\"{\i}ve IA and Destriper P1 map-makers, which are in perfect agreement with the others.
	\item PLW maps generated by Na\"{\i}ve and Destriper P0 map-makers also present a \emph{peak} at $k \simeq 1.5$ (see Fig.~\ref{fig:lc:case7plw}b), which is not present in the PSW maps (see Fig.~\ref{fig:lc:case7psw}b): its origin is still unclear, but most probably related to the \emph{cooler burp} too as it is not visible in any other case.
	\item The power spectrum obtained from Scanamorphos maps is  again lower than the other spectra, while Sanepic, Na\"{\i}ve IA and Destriper P1 agree at all scales.
\end{itemize}

\subsubsection{Case 12 \& 13}

Case 12 is the only real observation of which we have power spectra produced with all map-makers but Sanepic (see Fig.~\ref{fig:lc:cases}): in fact, maps of case 13 reduced with Scanamorphos are bigger and have a different WCS, making the power spectra comparison not straightforward.

\begin{figure}[p!]
	\centering
	\subfigure[Power spectra of PMW map]
	{\includegraphics[width=6.5cm,keepaspectratio]{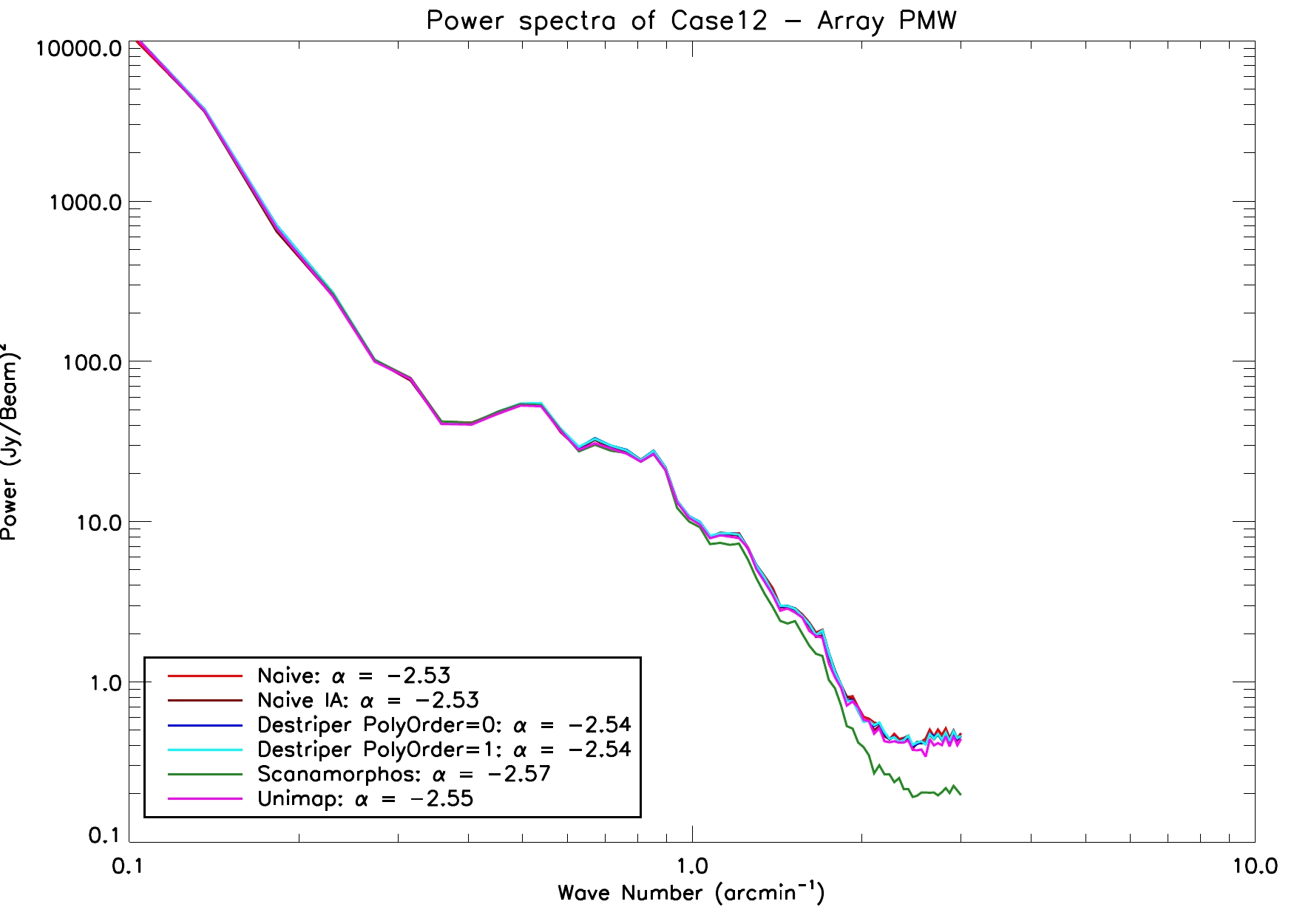}}
	\hspace{0.5cm}
	\subfigure[Zoom-in of PMW maps power spectra]
	{\fbox{\includegraphics[width=5cm,keepaspectratio]{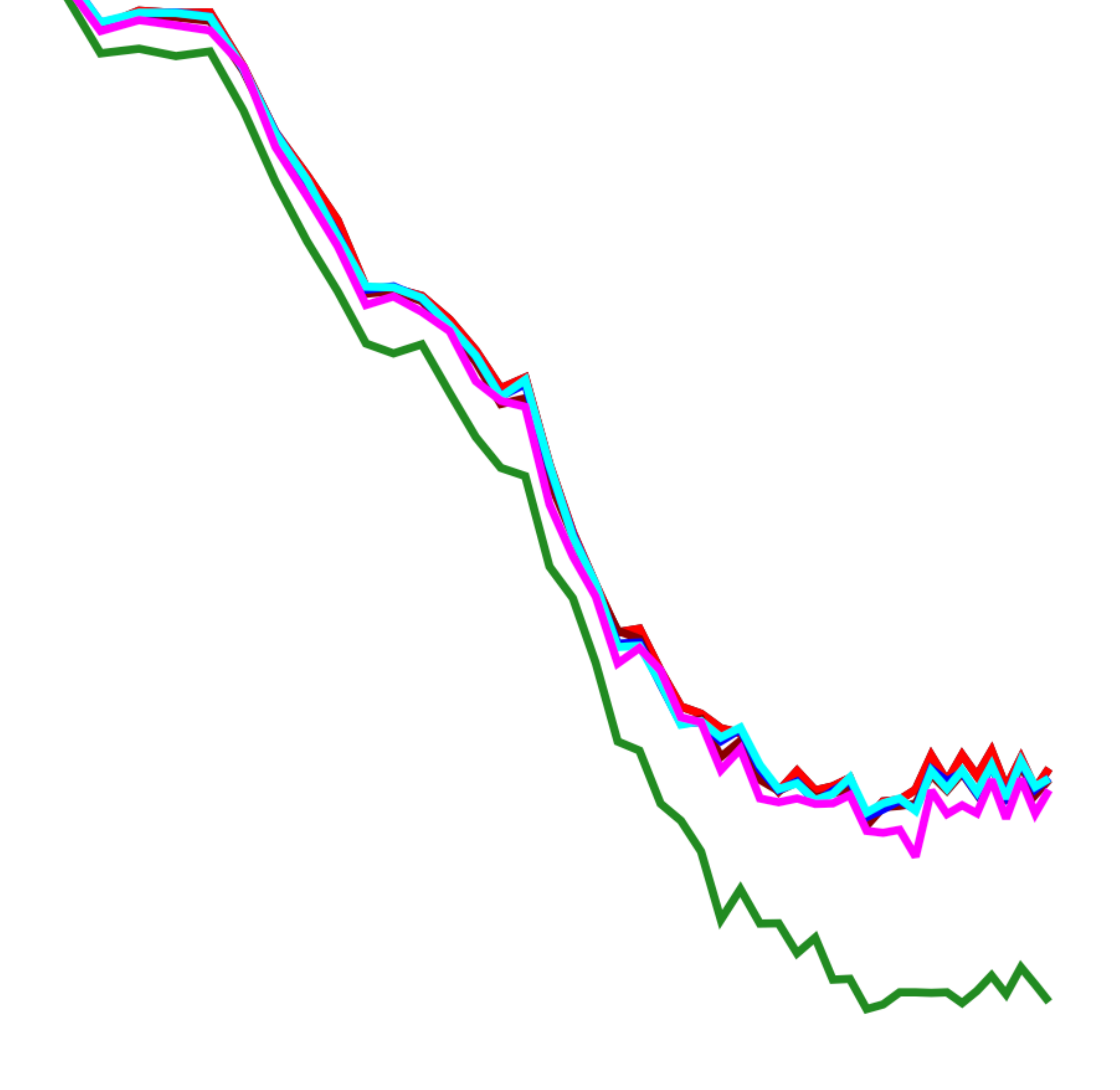}}}\\
	\caption{Results of power spectra for case 12, PMW map: real SPIRE-only large map observation of NGC-628 taken at nominal scan speed. In Fig.~\ref{fig:lc:case12pmw}a the power spectra computed on available maps are reported, while Fig.~\ref{fig:lc:case12pmw}b shows a zoom-in at small scales ($k > 1$ arcmin$^{-1}$).}\label{fig:lc:case12pmw}
	\vspace{0.5cm}
	\subfigure[Power spectra of PLW map]
	{\includegraphics[width=6.5cm,keepaspectratio]{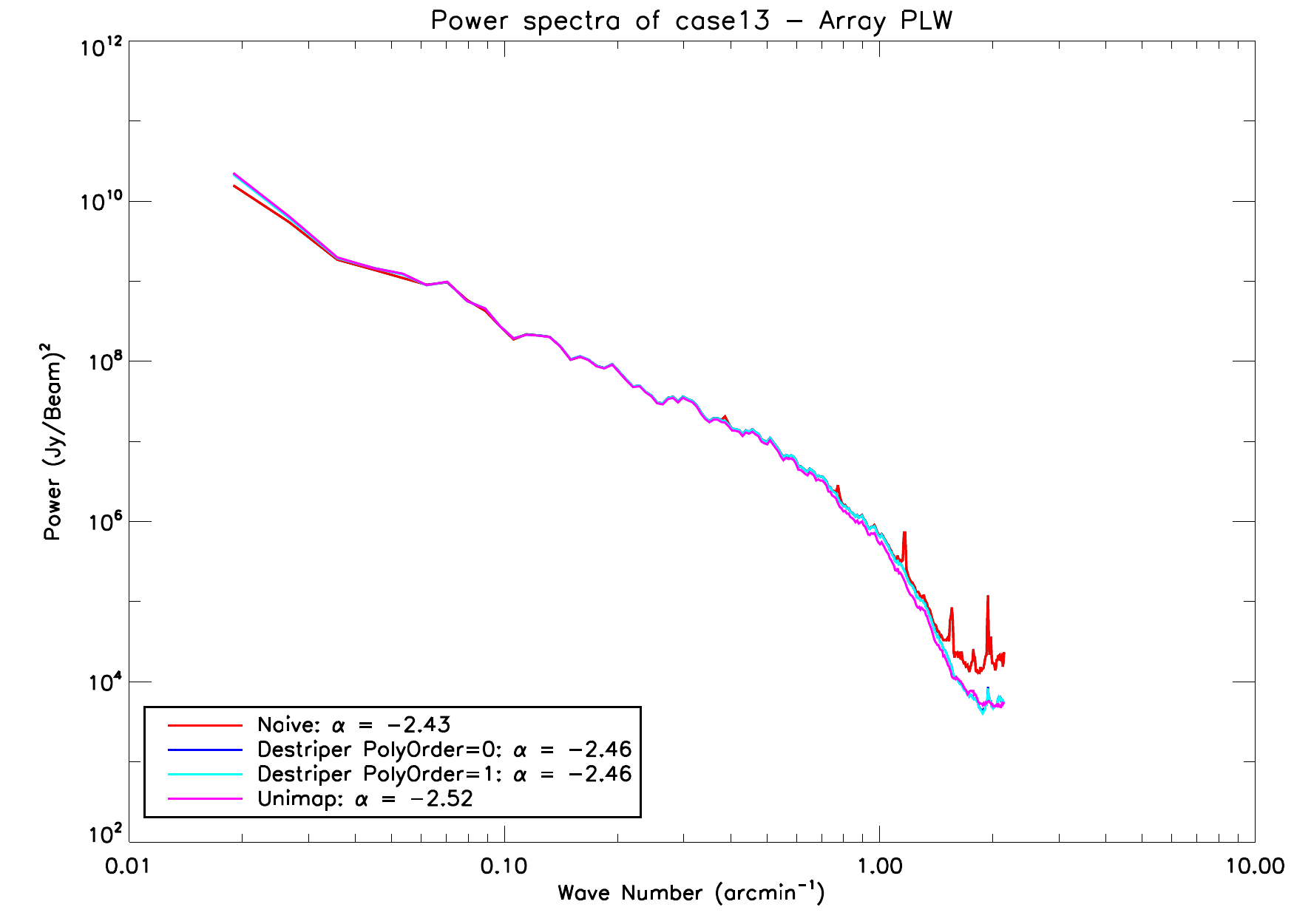}}
	\hspace{0.5cm}
	\subfigure[Zoom-in of PLW maps power spectra]
	{\fbox{\includegraphics[width=5cm,keepaspectratio]{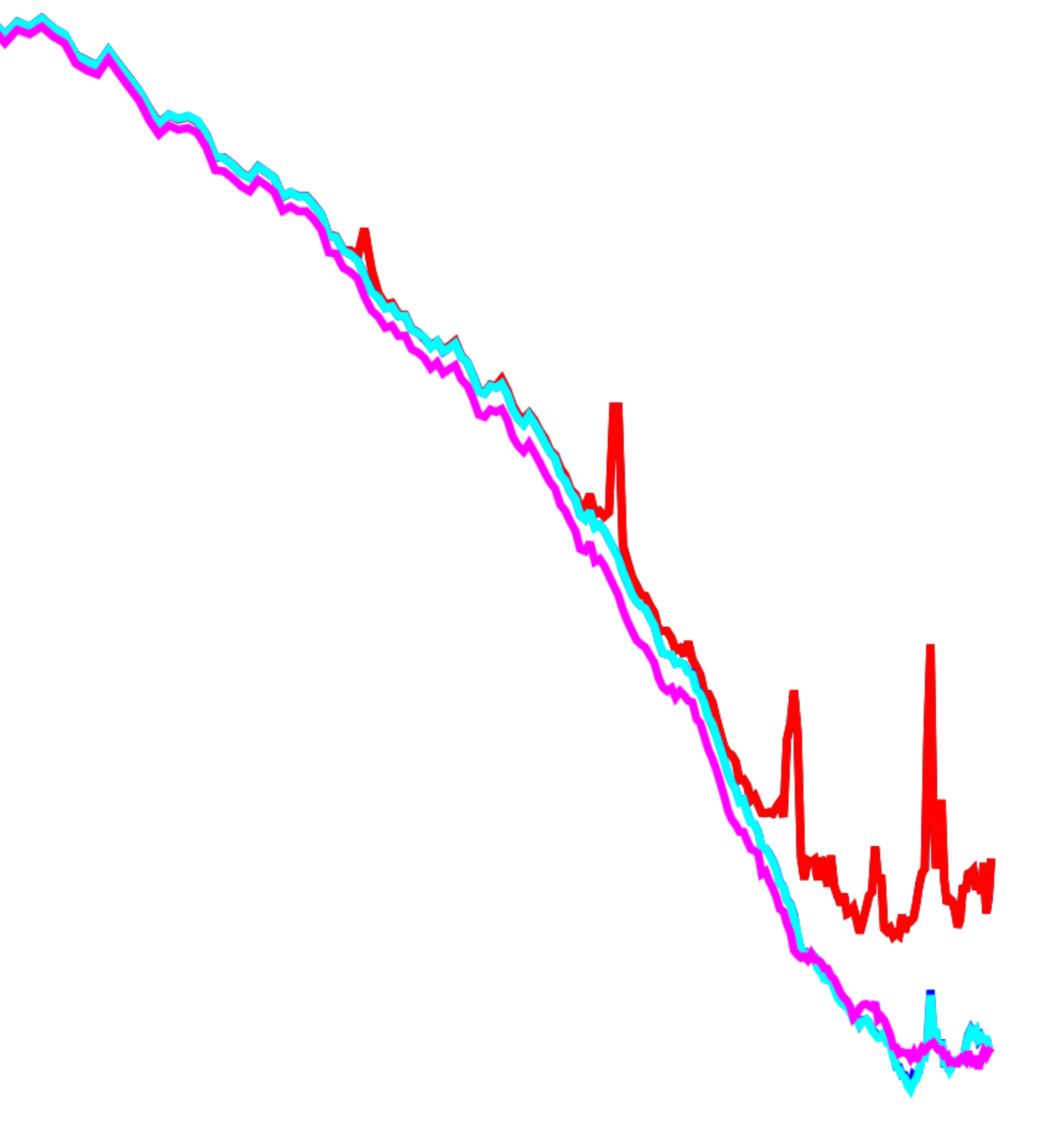}}}\\
	\caption{Results of power spectra for case 13, PLW map: real Parallel mode observation of a Hi-Gal field taken at fast scan speed. In Fig.~\ref{fig:lc:case13plw}a the power spectra computed on available maps are reported, while Fig.~\ref{fig:lc:case13plw}b shows a zoom-in at small scales ($k > 1$ arcmin$^{-1}$).}\label{fig:lc:case13plw}
\end{figure}

\begin{figure}[p!]
	\centering
	\subfigure[Power spectra of PLW map]
	{\includegraphics[width=6cm,keepaspectratio]{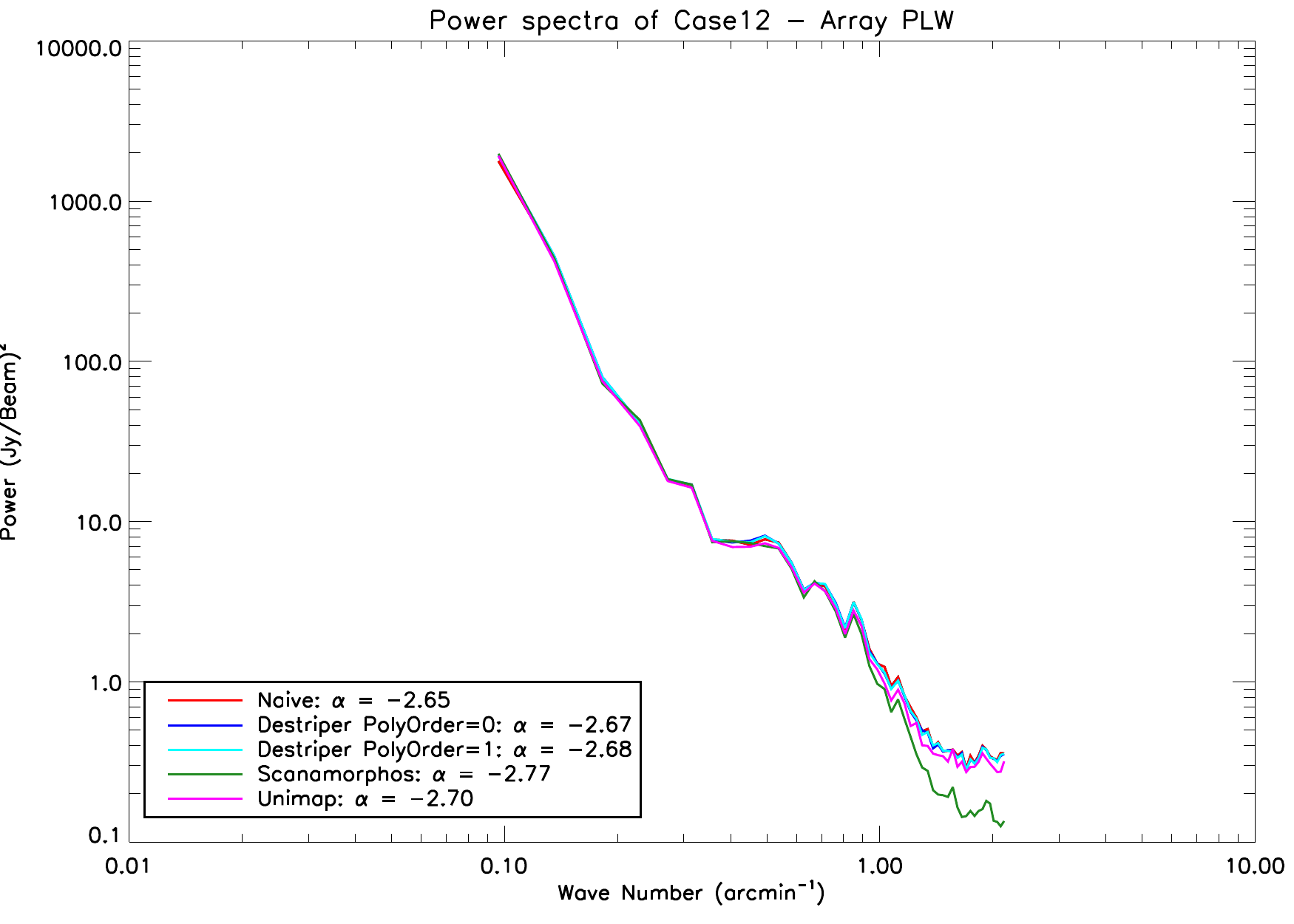}}
	\subfigure[Power spectra of PSW map]
	{\includegraphics[width=6cm,keepaspectratio]{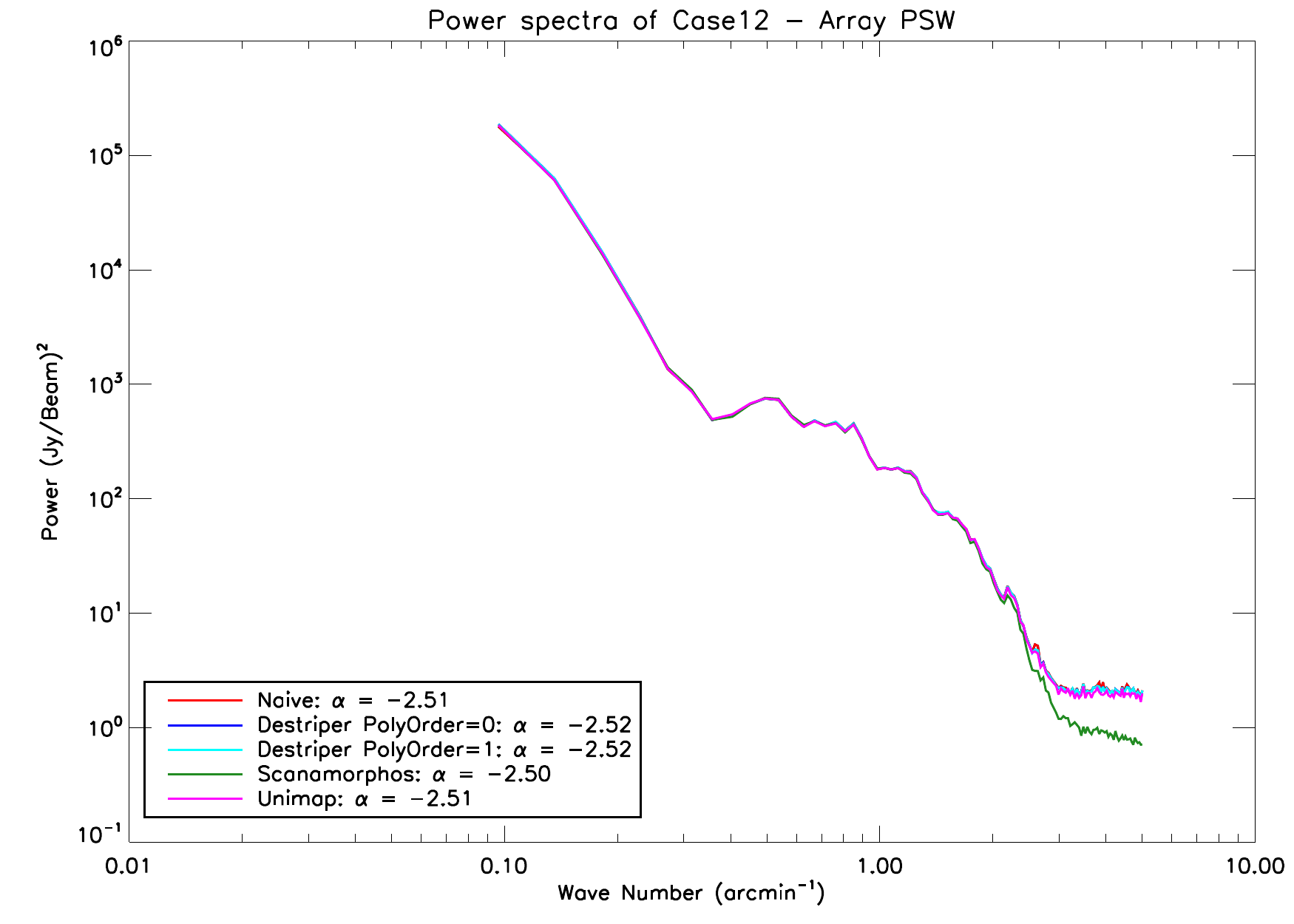}}\\
	\caption{Results of power spectra for case 12.}\label{fig:lc:case12}
	\vspace{1cm}
	\includegraphics[width=\textwidth,keepaspectratio]{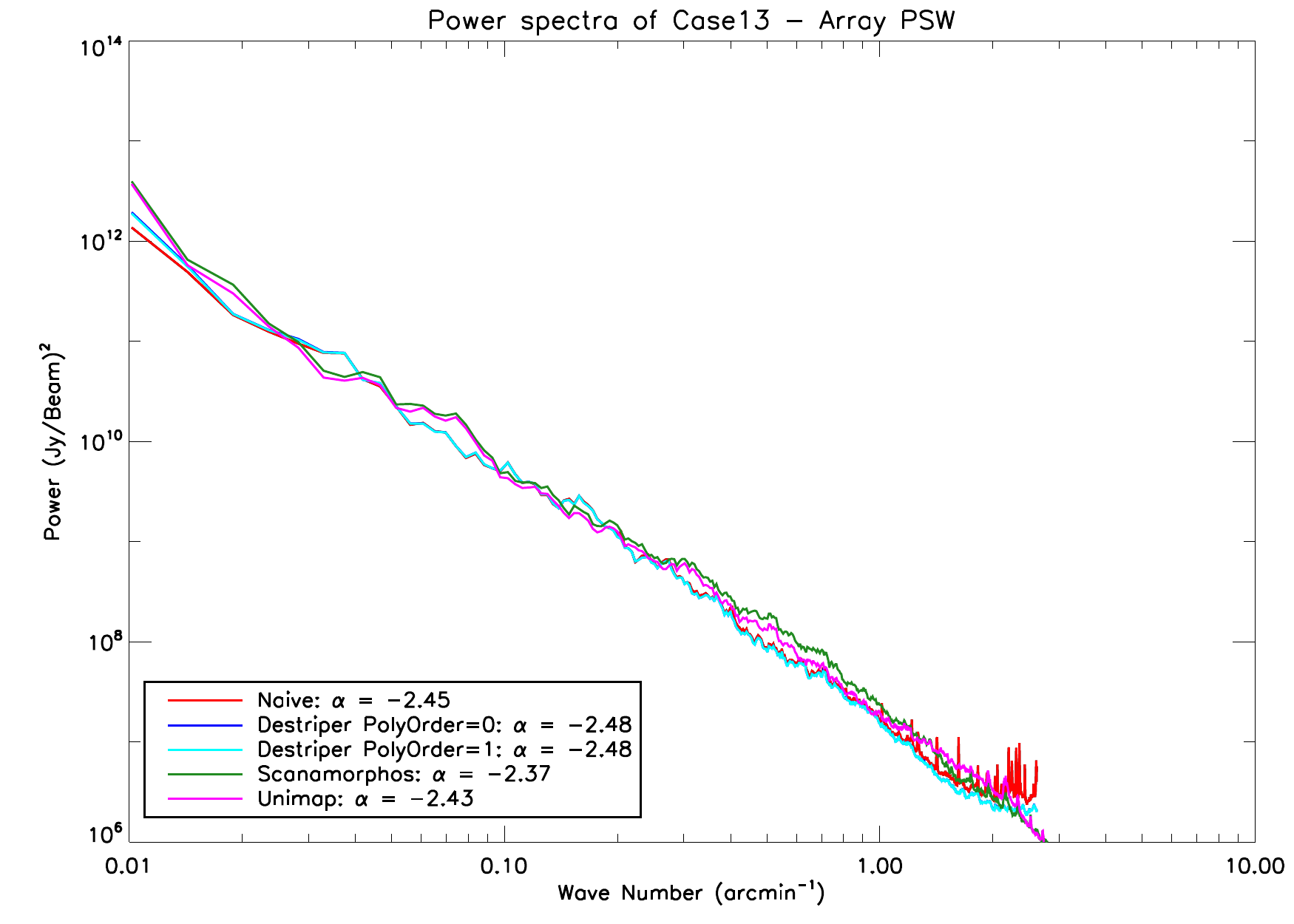}\\
	\caption{Results of power spectra for case 13, PSW map. Note that Unimap and Scanamorphos power spectra have been multiplied by 0.2 to be \emph{aligned} with the others.}\label{fig:lc:case13psw}
\end{figure}

Case 12's power spectra results for the PMW array are shown in Fig.~\ref{fig:lc:case12pmw}a, with a zoom-in at small scales ($k < 1$ arcmin$^{-1}$) reported in Fig.~\ref{fig:lc:case12pmw}b. Results for PLW and PSW are reported in Appx~\ref{lc:appendix} (see Fig.~\ref{fig:lc:case12}). Results are similar to the other cases previously discussed: all map-makers agree very well at large scales, while at small scales Scanamorphos is lower than the others. Among them, Destriper P0, P1 and Na\"{\i}ve IA are almost identical, whereas Unimap and Na\"{\i}ve IA are slightly lower and higher, respectively. 

Fig.~\ref{fig:lc:case13plw}a reports the case 13's power spectra results for the PLW array, with a zoom-in at small scales ($k > 1$ arcmin$^{-1}$) reported in Fig.~\ref{fig:lc:case13plw}b. At small scales, the Na\"{\i}ve map-maker spectrum presents peaks: these are due to the residual stripes in the maps, produced by the inadequate median baseline subtraction for the high dynamic range of the Hi-gal fields.

\subsection{Summary}\label{lc:summary}

Most of the power spectra, either coming from real or simulated data, \no\ or with extended emission, show very similar results.

\begin{itemize}
	\item In the ``central part'' ($k = [0.1, 1]$ arcmin$^{-1}$), results among different map-makers vary little: $\sim 1\%$ for cases where a \truth\ map was available as benchmark.
	\item Differences with reference to \truth\ map increase at $k > 1$ arcmin$^{-1}$:
	\begin{itemize}
		\item HIPE destripers with zero-th and first polynomial order, Sanepic and Unimap results are always very close;
		\item Scanamorphos is usually lower;
		\item The automatic HIPE Na\"{\i}ve mapper produces power spectra higher than the other map-makers, but in most cases (i.e.\ where the field dynamic range is not too high for the median baseline subtraction) it produced reasonable results. Spikes present in some cases at small scales (see e.g. Figs.~\ref{fig:lc:case6psw}, \ref{fig:lc:case9psw} and \ref{fig:lc:case13plw}) are due to leftover stripes in the image. Moreover, once the Na\"{\i}ve mapper is ran in interactive analysis using a higher polynomial order fit for baseline removal, it gave results in line with other map-makers.
	\end{itemize}		
	\item At large scales ($k < 0.2$ arcmin$^{-1}$), the spread among map-makers is larger, being the Destriper P0 and Unimap usually closer to the \truth\ power spectrum.
	\item Case 7 (\no\ map with \emph{cooler burp}, see Tab.~\ref{tab:lc:cases}) is a case apart:
	\begin{itemize}
		\item Na\"{\i}ve and Destriper P0 map-makers perform worse: their power spectra are much higher at $k < 0.1$ and present a \emph{peak} at $k \sim 1.5$ in PLW map. This is to be expected as they cannot correct the drift caused by the \emph{cooler burp}.
		\item The other map-makers are in agreement, with Scanamorphos being lower at small scales.
	\end{itemize}
\end{itemize}

%\newpage
\subsection{Appendix: Power spectra of cases not discussed}\label{lc:appendix}

\begin{figure}[h!]
	\centering
	\includegraphics[width=8cm,keepaspectratio]{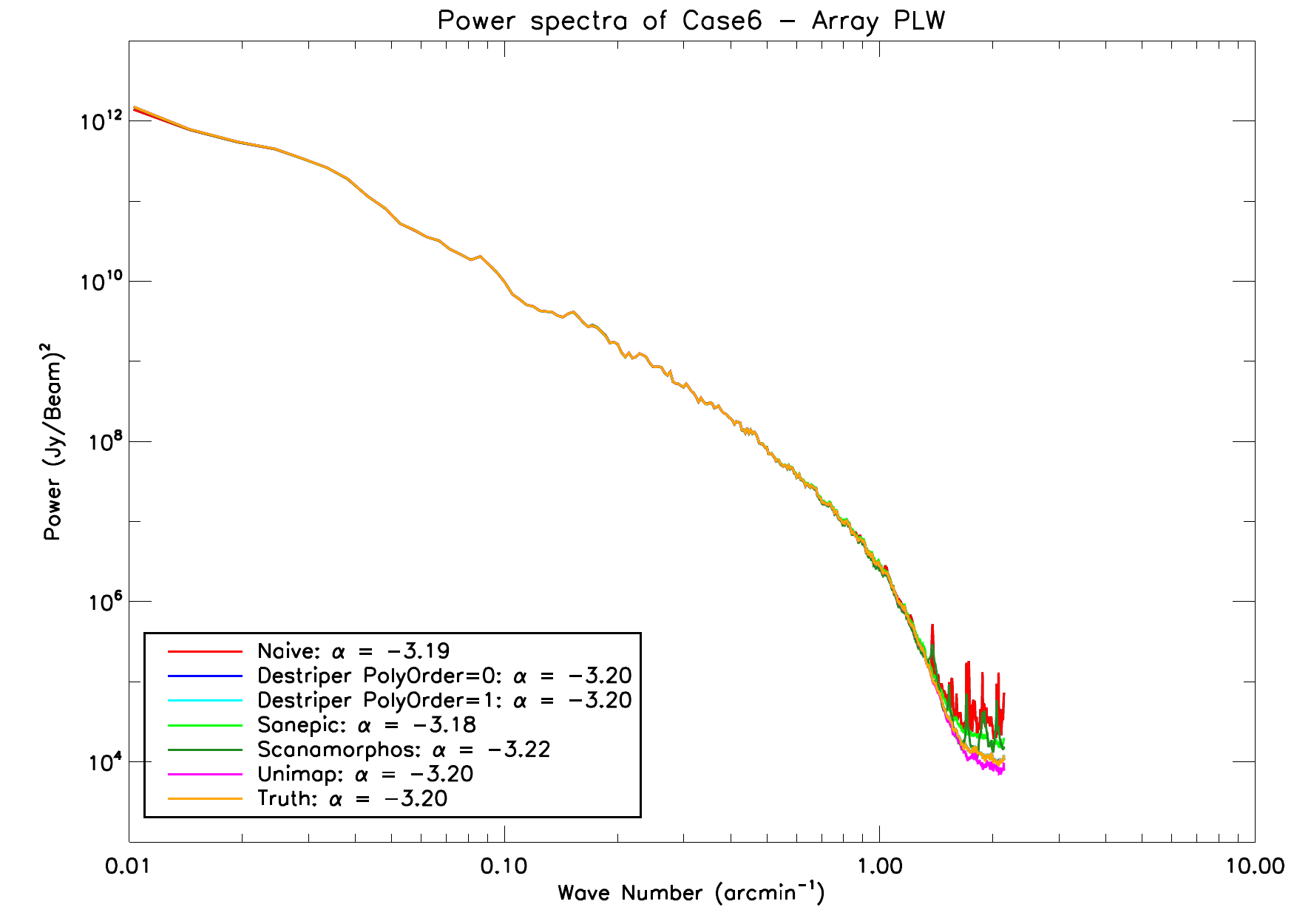}\\
	\caption{Results of power spectra for case 6, PLW map.}\label{fig:lc:case6plw}
\end{figure}

\begin{figure}[p!]
	\centering
	\subfigure[Power spectra of PLW map]
	{\includegraphics[width=6cm,keepaspectratio]{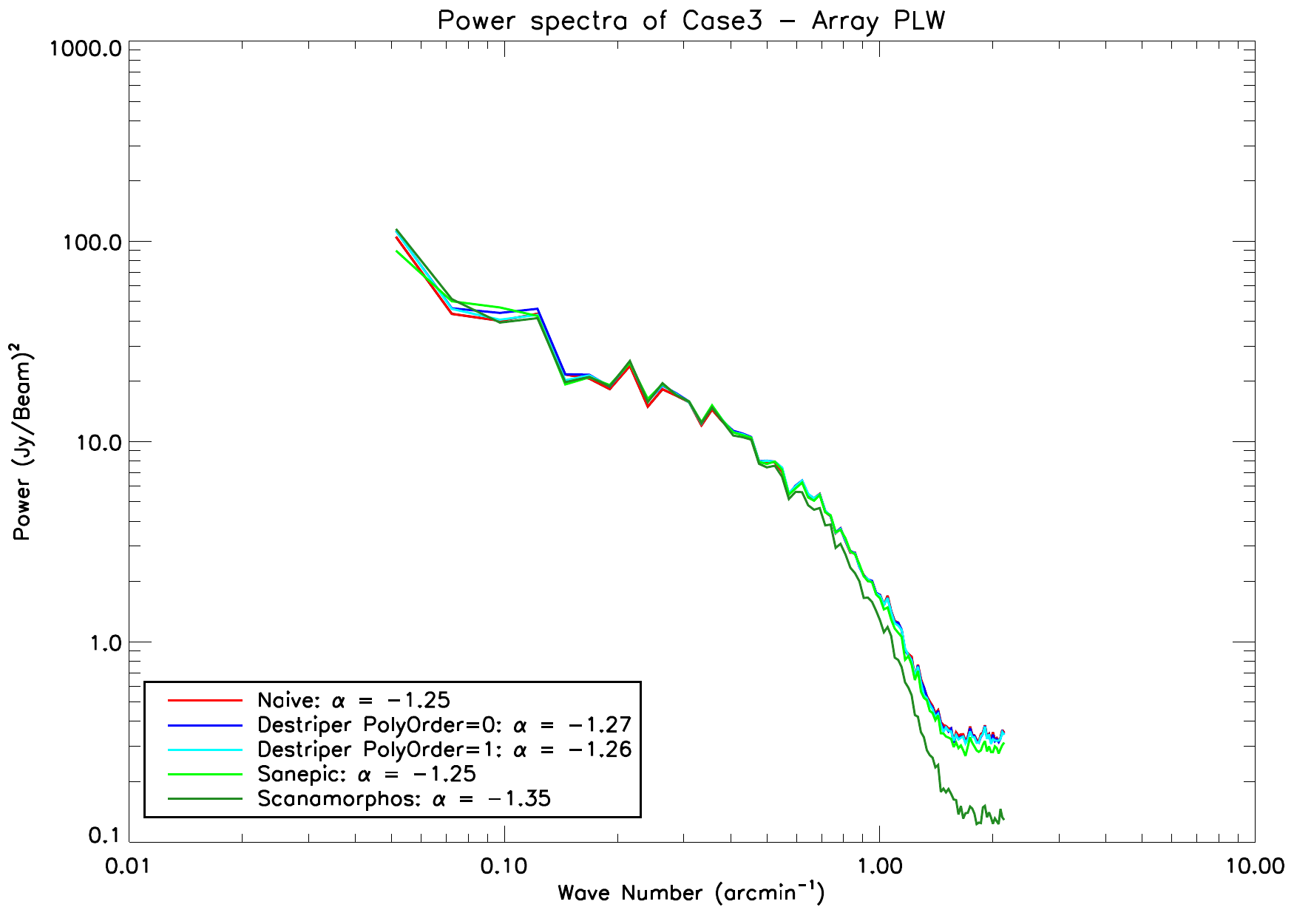}}
	\subfigure[Power spectra of PMW map]
	{\includegraphics[width=6cm,keepaspectratio]{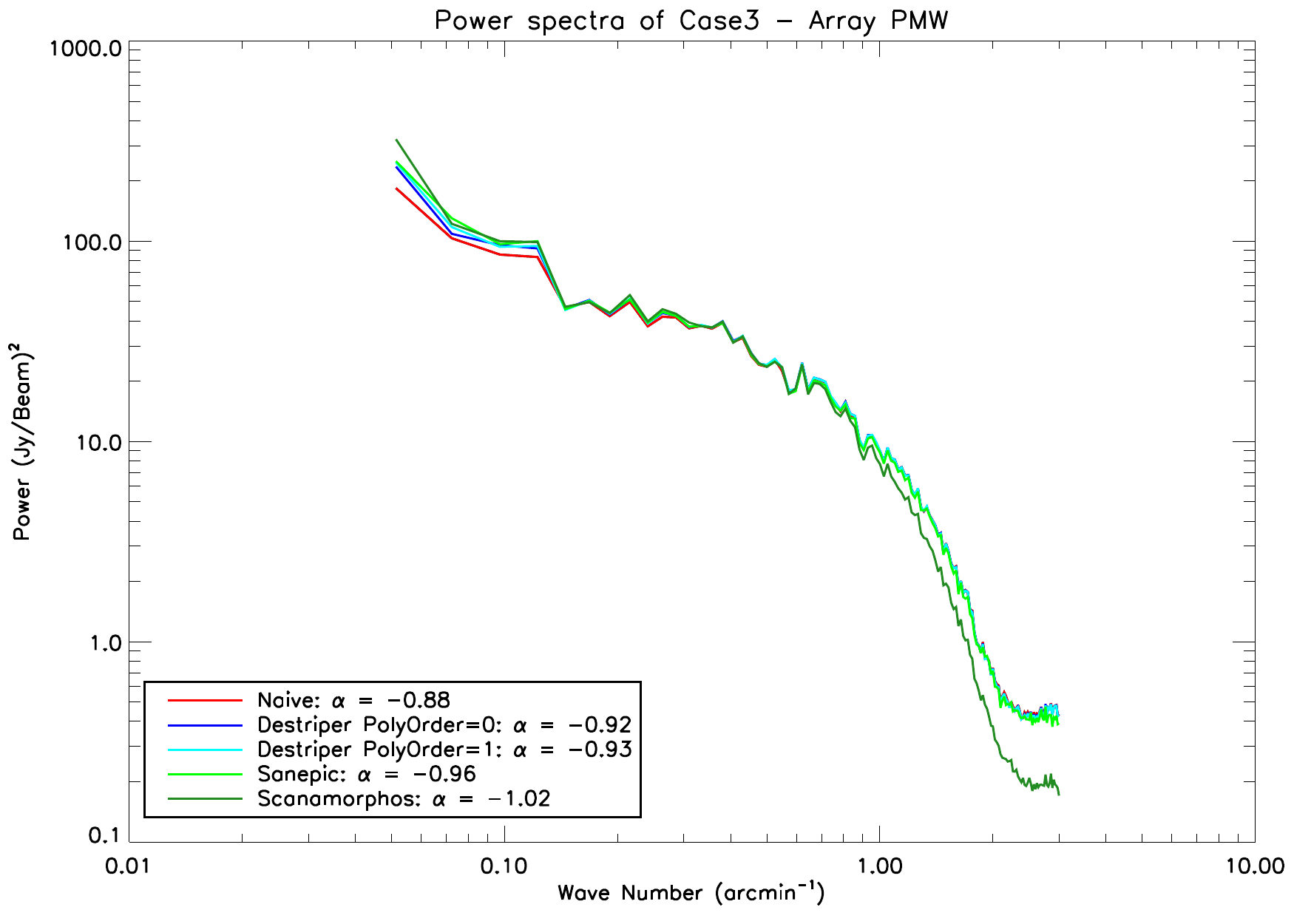}}\\
	\subfigure[Power spectra of PSW map]
	{\includegraphics[width=\textwidth,keepaspectratio]{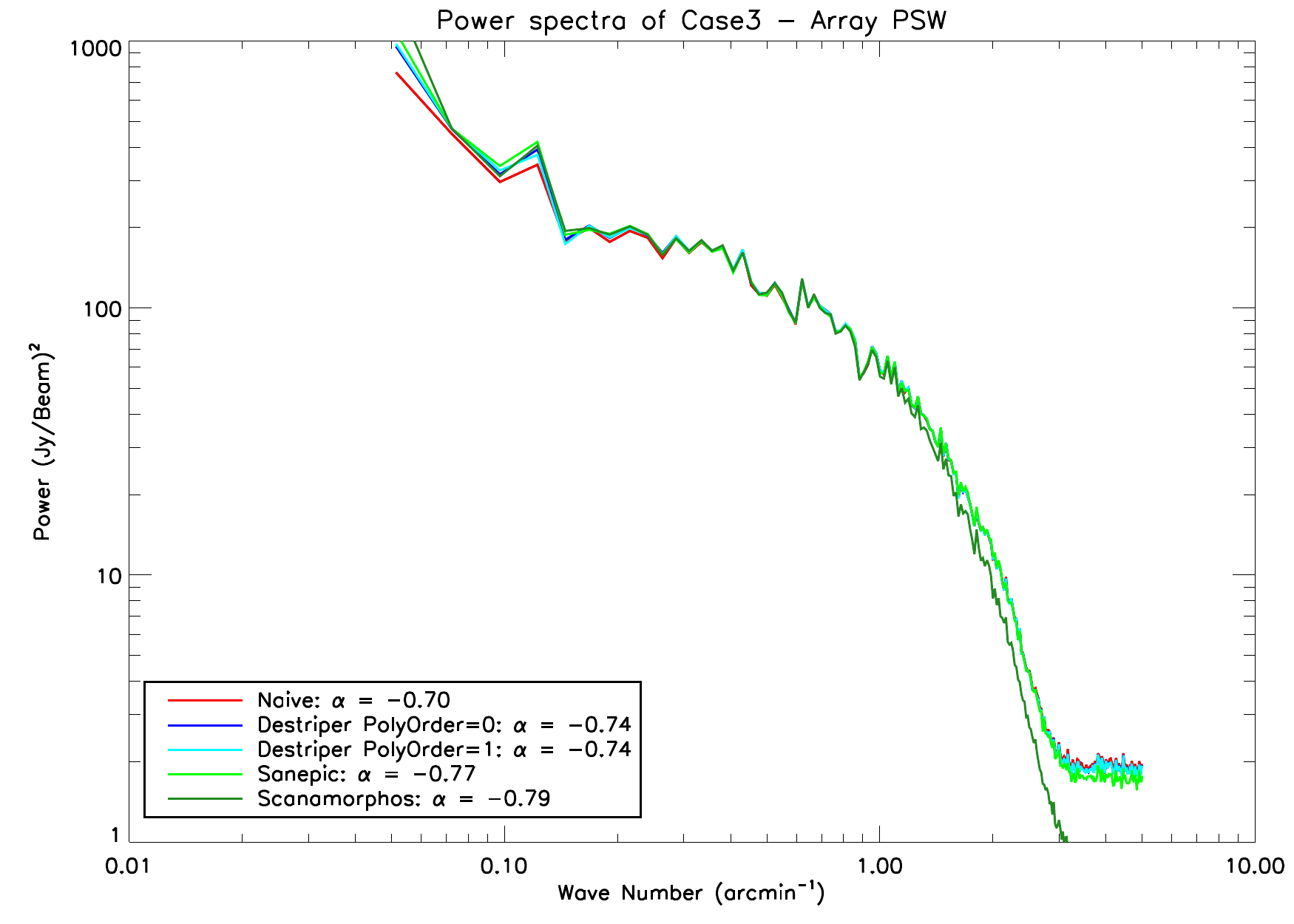}}\\
	\caption{Results of power spectra for case 3.}\label{fig:lc:case3}
\end{figure}

\begin{figure}[p!]
	\centering
	\subfigure[Power spectra of PLW map]
	{\includegraphics[width=6cm,keepaspectratio]{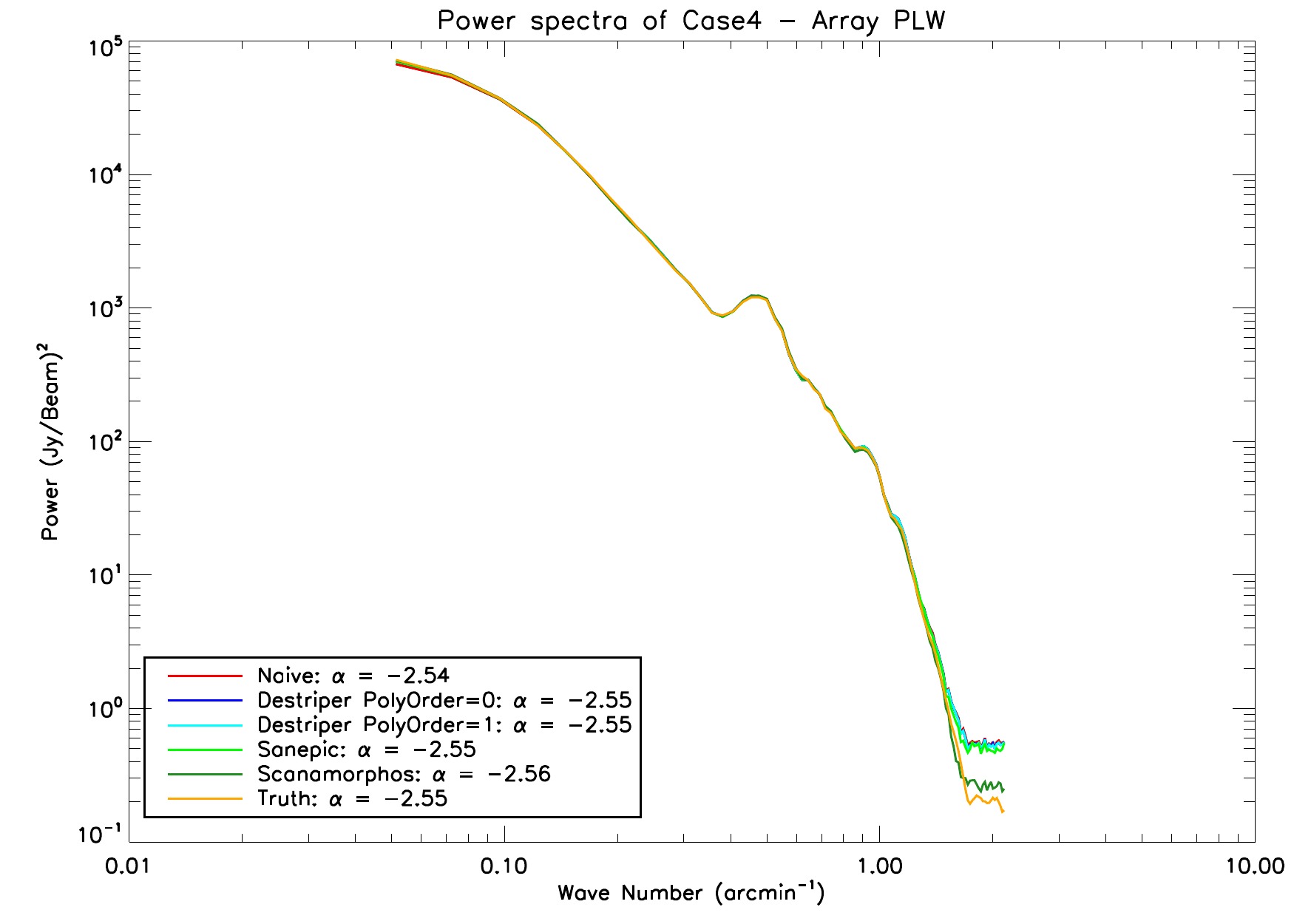}}
	\subfigure[Power spectra of PMW map]
	{\includegraphics[width=6cm,keepaspectratio]{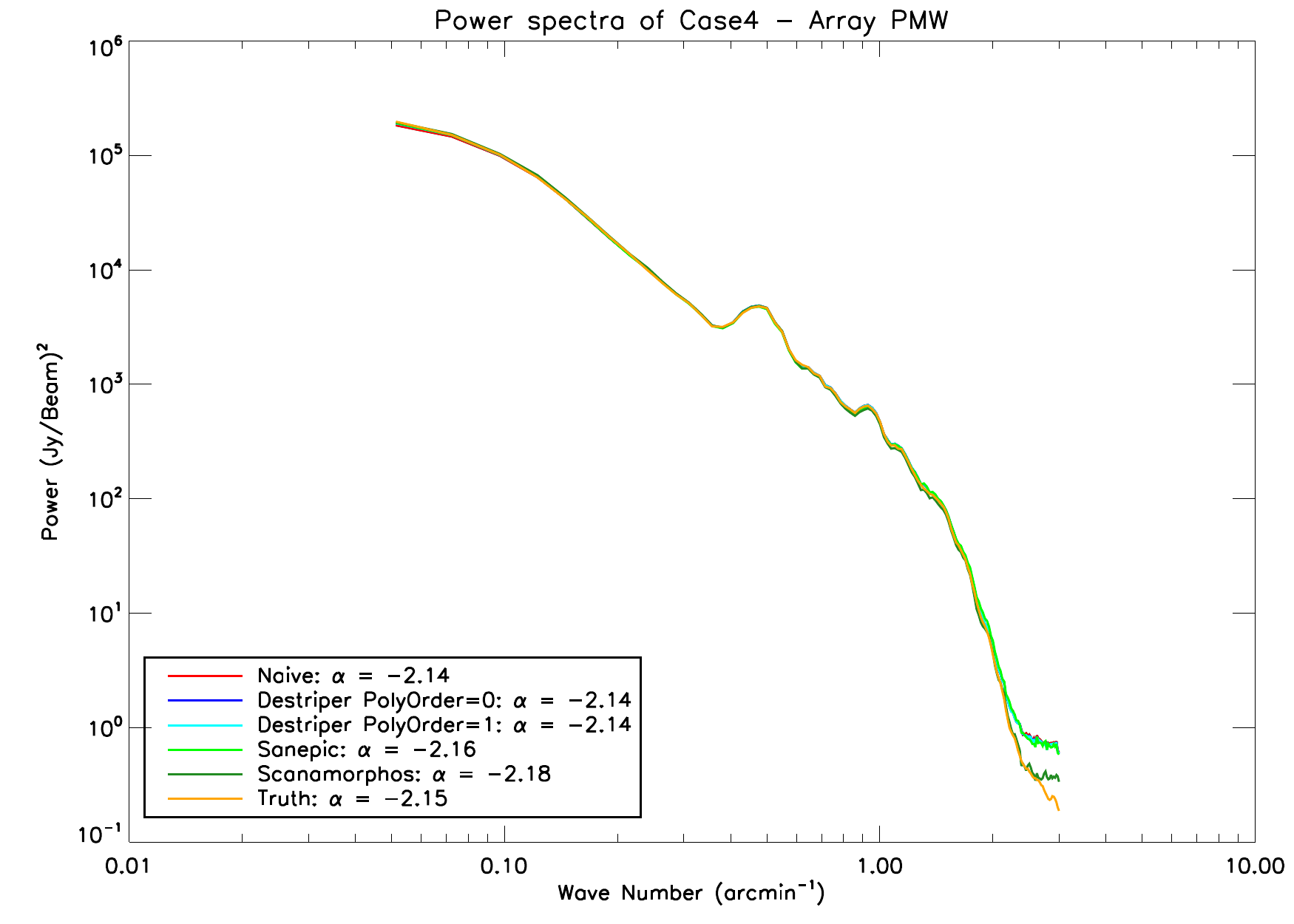}}\\
	\subfigure[Power spectra of PSW map]
	{\includegraphics[width=\textwidth,keepaspectratio]{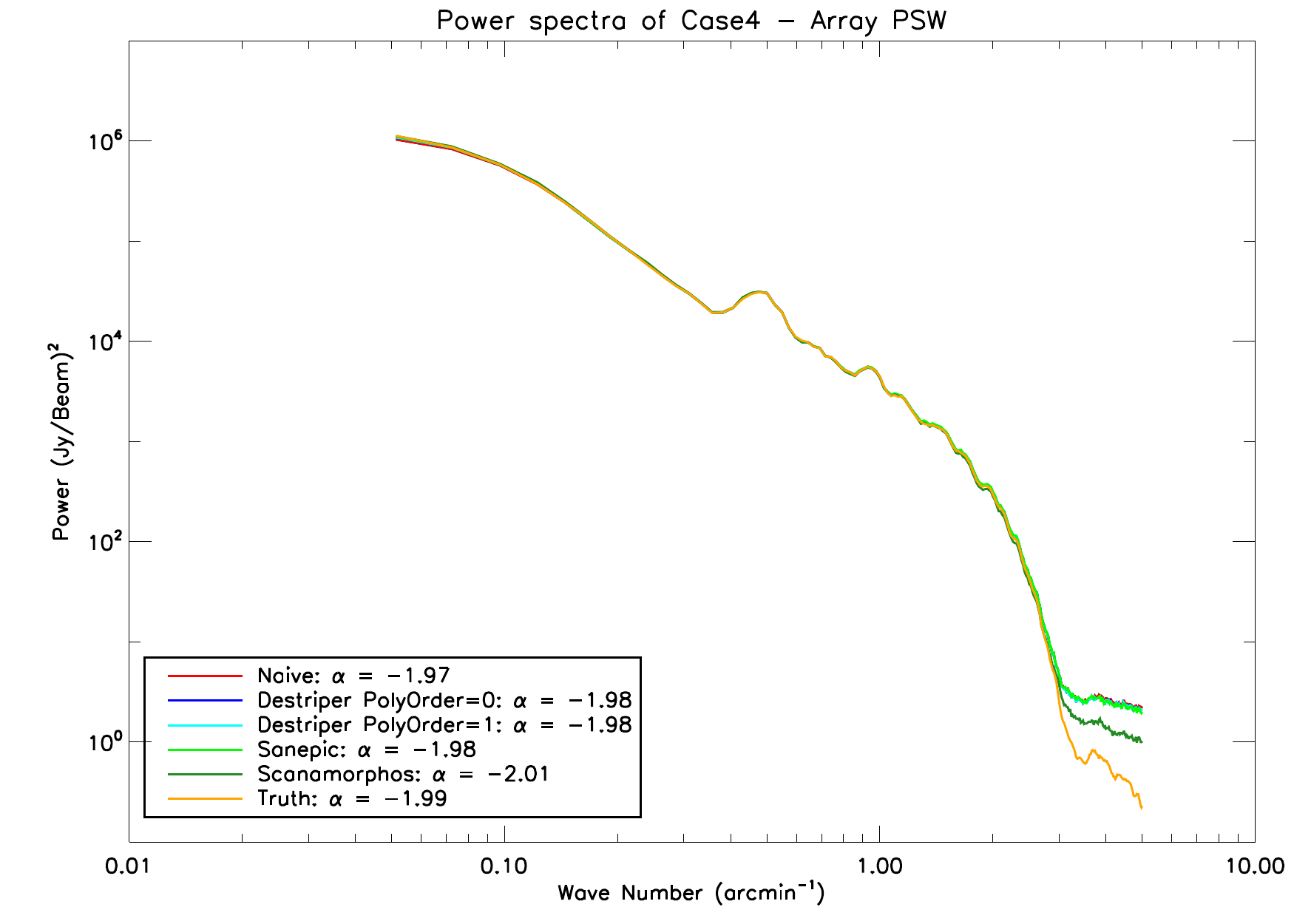}}\\
	\caption{Results of power spectra for case 4.}\label{fig:lc:case4}
\end{figure}

\begin{figure}[p!]
	\centering
	\includegraphics[width=\textwidth,keepaspectratio]{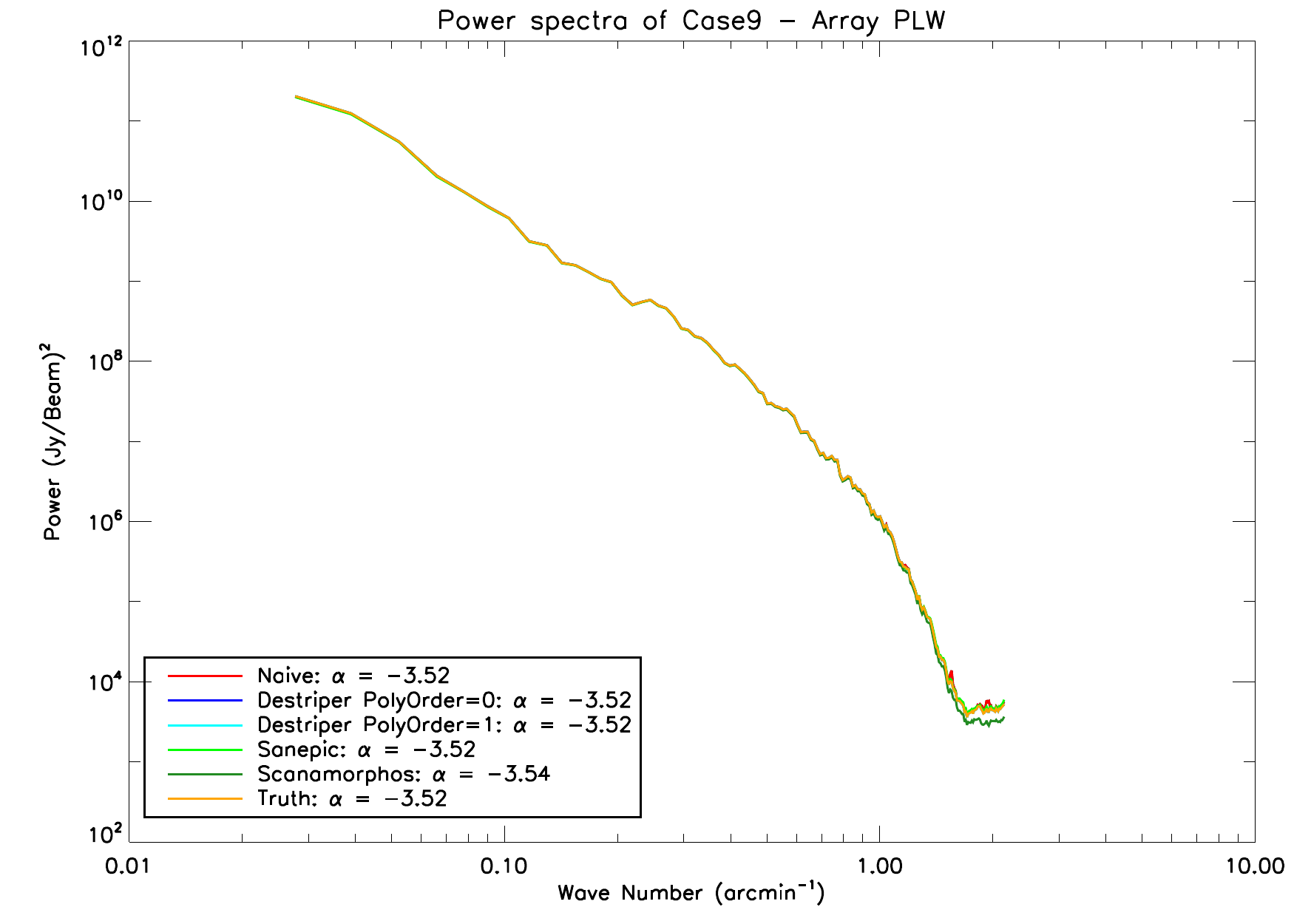}\\
	\caption{Results of power spectra for case 9, PLW map.}\label{fig:lc:case9plw}
	\vspace{1cm}
	\subfigure[Power spectra of PLW map]
	{\includegraphics[width=6cm,keepaspectratio]{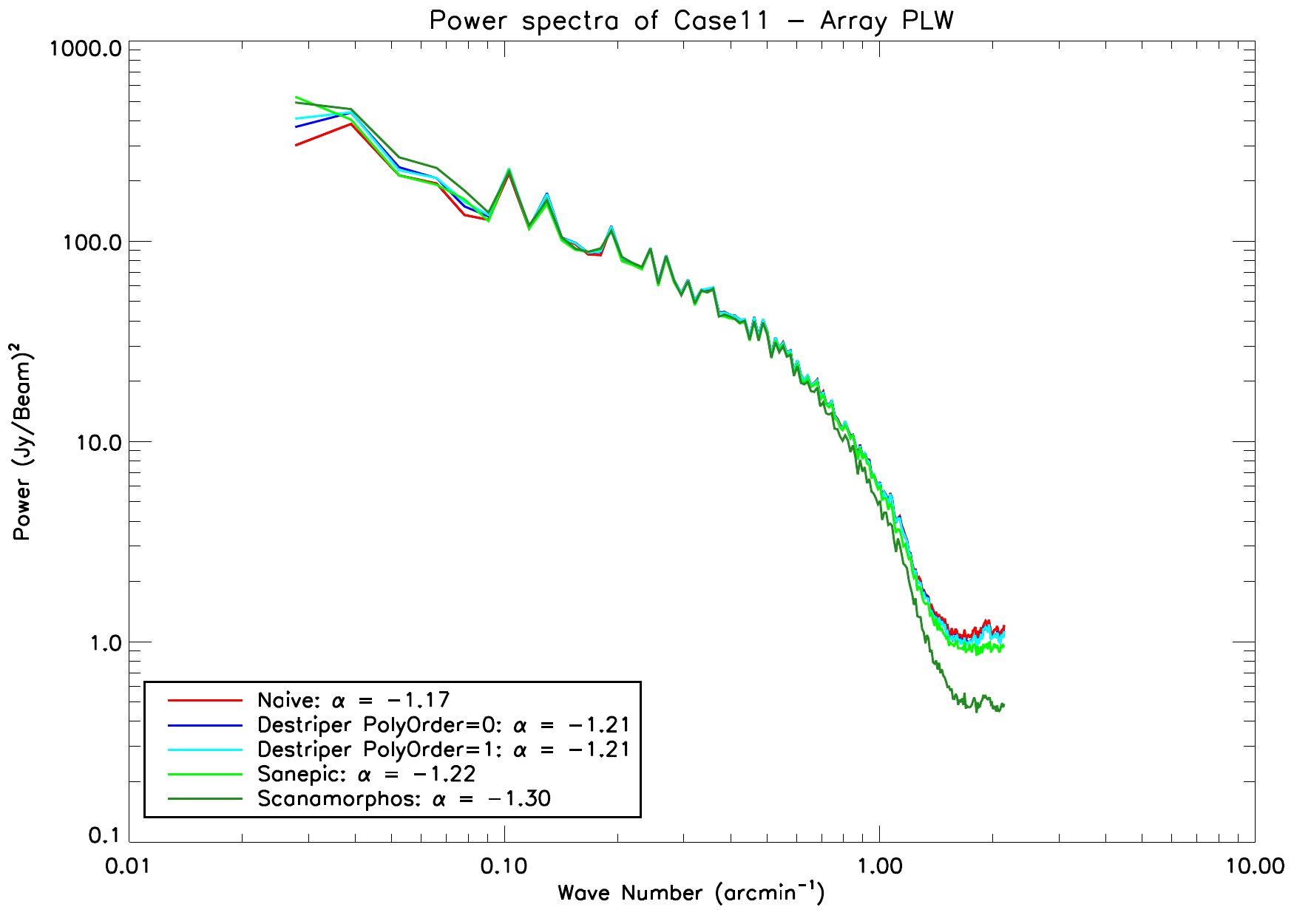}}
	\subfigure[Power spectra of PSW map]
	{\includegraphics[width=6cm,keepaspectratio]{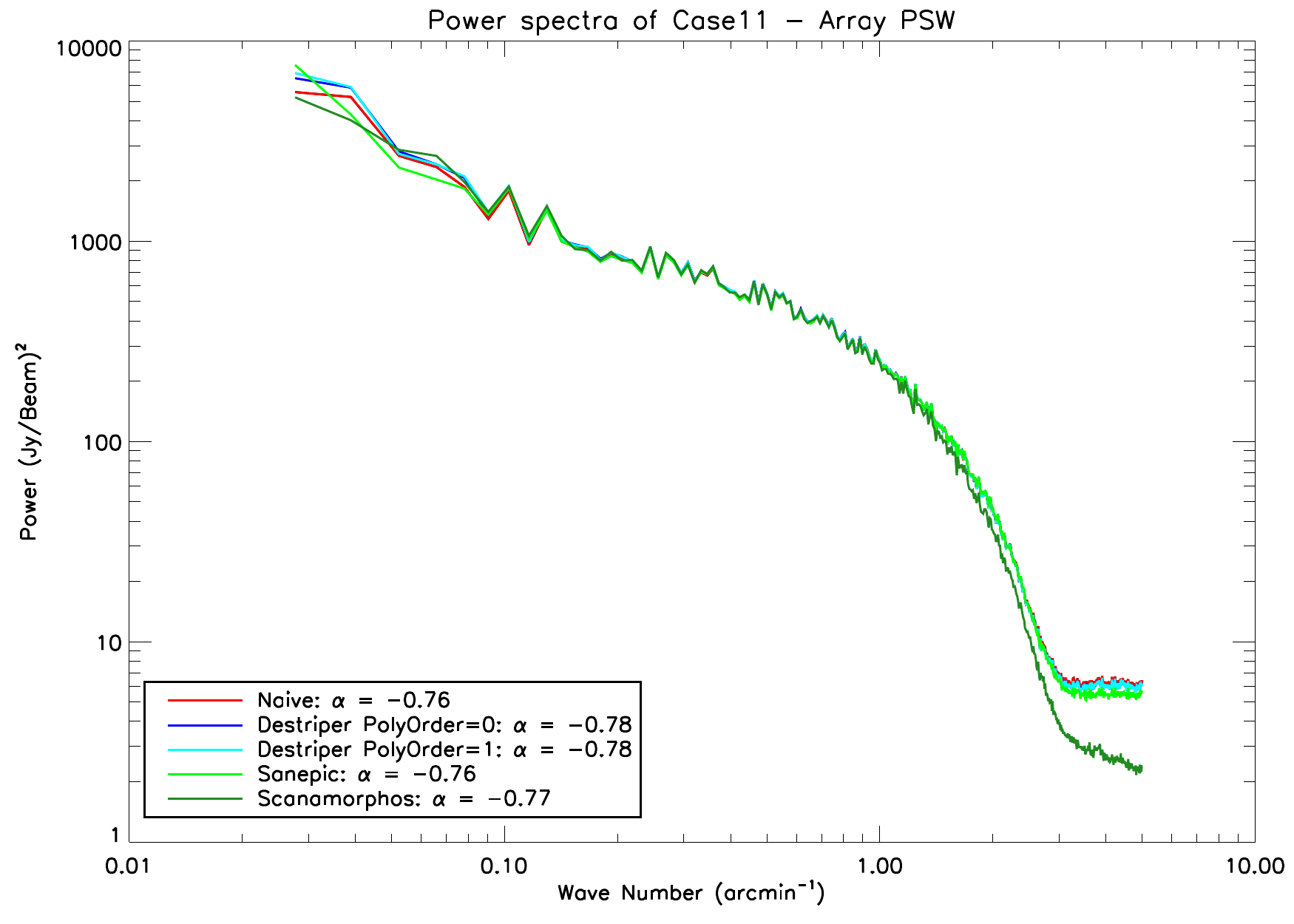}}\\
	\caption{Results of power spectra for case 11.}\label{fig:lc:case11}
\end{figure}

%\end{document}

%% file: chapter5_metrics_sources_pdf.tex
% Update on Aug. 2, 2013. Including the analysis on beam profiles.
% This serves as a template for metrics reports.
% CKX, April 6, 2013

%\documentclass[11pt]{book}
%\usepackage{titletoc}% http://ctan.org/pkg/titletoc
%\usepackage{amsmath} %Never write a paper without using *amsmath* for
                     %its many new commands
%\usepackage{amssymb} %Some extra symbols
%\usepackage{makeidx} %If you want to generate an index, automatically
%\usepackage{graphicx} %If you want to include postscript graphics
%\usepackage{caption} %If you want to include postscript graphics
%\usepackage{epsfig}
%\begin{document}
%%
%\setcounter{chapter}{4}
%\chapter{Metrics and Results}

\section{Point Source and Extended Source Photometry (Kevin Xu)}

\subsection{Test Data}\label{data}

Three simulated test cases (Cases 1, 5 and 8; see Table 4.1) were
examined in this metric. Four map-makers (Naive, Destriper-P0/Destriper-P1,
Scanamorphos, SANEPIC) produced maps for all the three test
cases. Unimap made maps for two test cases (Cases 1 and 5).

\begin{figure}[ht]
\begin{center}
    \includegraphics[width=5cm, angle=0]{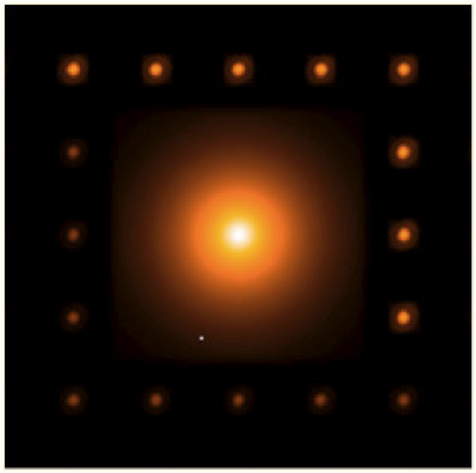}
\end{center}
\caption{Truth map of Case 1 (nominal mode, $\rm 0.7 \times 0.7$ deg$^2$);
in the PLW band.}
\label{metrics_report_sources_fig1}
\end{figure}

The Case 1 simulation (Fig.~\ref{metrics_report_sources_fig1}) 
is for the nominal mode. In the map
of each of the three SPIRE bands, the simulation injected into the
dark field observation (coverage $\rm = 0.7 \times 0.7$ deg$^2$) 
8 bright point
sources (f=300 mJy), 8 faint point sources (f=30 mJy), and one
extended source in the map center . The point sources were simulated
using the observed PSF of the given band. The extended source (fpeak =
1 Jy/beam) is modeled as an exponential disk with the e-folding
radius of 90 arcsec before being convolved with the PSF.

\begin{figure*}
    \centering
    \includegraphics[width=6cm, angle=0]{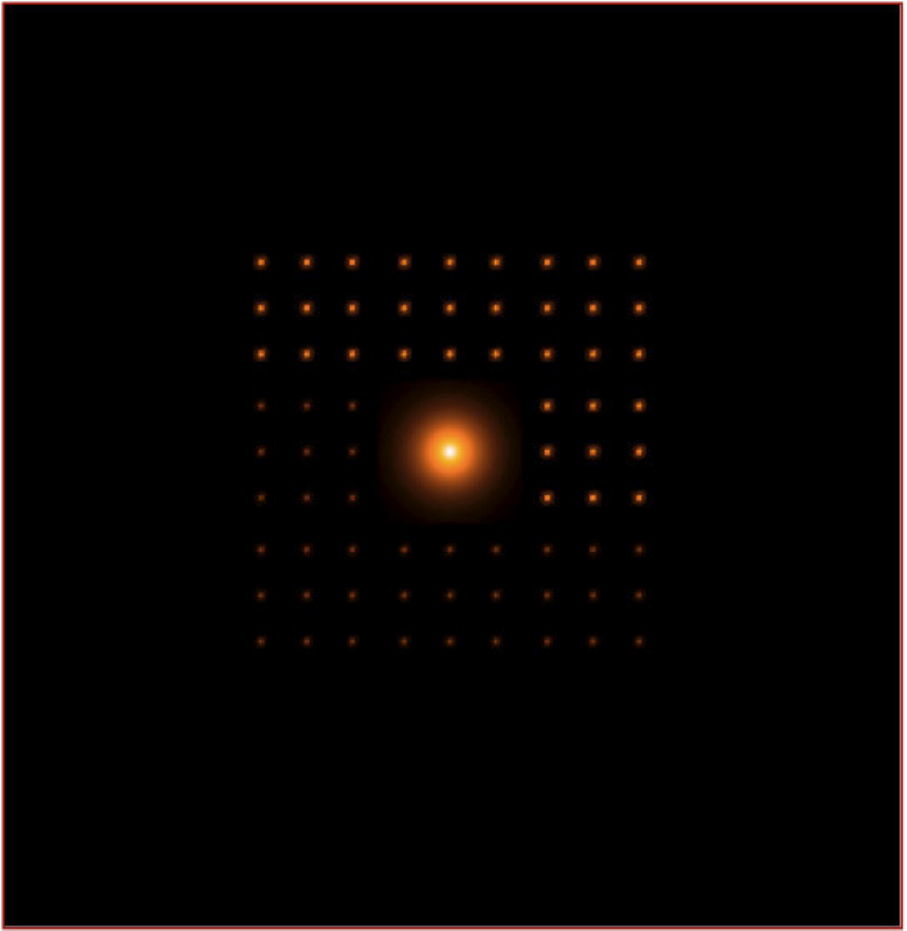}
    \includegraphics[width=4.5cm, angle=0]{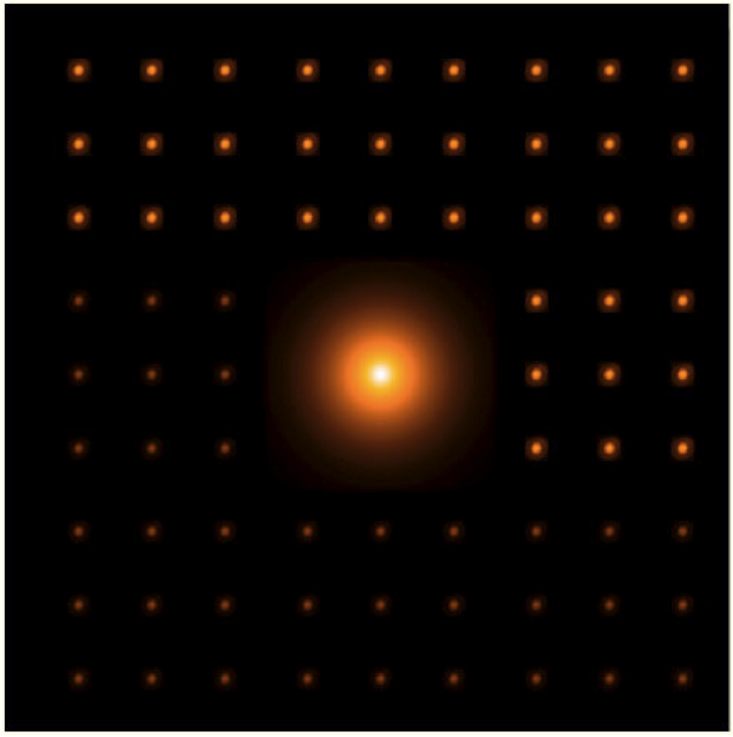}
\caption{{\bf Panel a (left)}: Truth map of Case 5 
(fast-scan mode, $\rm 3.5 \times 3.5$ deg$^2$); in the PLW band.
{\bf Panel b (right)}: Truth map of Case 8 
(parallel mode, $\rm 1.4 \times 1.4$ deg$^2$); in the PLW band.}
\label{metrics_report_sources_fig2}
\end{figure*}

In both Case 5 simulation for the fast-scan mode (coverage=3.5 × 3.5
deg2, Fig.~\ref{metrics_report_sources_fig2}a) 
and Case 8 simulation for the parallel mode
(coverage=1.4 × 1.4 deg2, Fig.~\ref{metrics_report_sources_fig2}b), 
there are 36 bright sources
(f=300 mJy), 36 faint sources (f=30 mJy), and one extended source in
the map of each band (PMW was not simulated).

%Examples of Figures and Tables
%\begin{figure}[ht]
% \begin{center}
% \epsfig{file=file1-eps-converted-to.pdf, width=10cm}
%\end{center}
%\caption{Sample caption for one method of including figures.}
%\end{figure}

%\begin{figure*}
%    \centering
%    \includegraphics[width=7.5cm, angle=0]{file1-eps-converted-to.pdf}
%    \includegraphics[width=7.5cm, angle=0]{file2-eps-converted-to.pdf}
%\caption{Sample caption for another completely different but equivalent method of including figures.}
%\end{figure*}

%\bigskip
%\begin{tabular}{llcc}
%\hline \\
%Source & Coordinates & Flux Density/       & Diameter \\
%       & (J2000)     &   (mJy)             & (arcsec) \\
%\hline \\
%G153.4+0.3 & 17:18:00+59:30:00 & 50 & 5 \\
%G278.3-0.5 & 09:48:38-54:23:7.1 & 3 & 1 \\
%\hline \\
%\end{tabular}

\subsection{Analyses and Results}\label{results}

\subsubsection{Point Sources: PSF Fitting}\label{PSF}

Using PSF fitting method, we checked the following properties of 
point sources in maps made by different map-makers:
\begin{itemize}
\item	Astrometry (position offsets);
\item	flux errors;
\item	faint source detection rate.
\end{itemize}
We chose to use the source extractor Starfinder \cite{Diolaiti2000}
for this task. Starfinder was designed particularly for point sources
in crowded field, matching ideally with our test cases which are point
sources in confusion limited backgrounds. The detection threshold is
set at $\rm 3 \sigma$ 
(local). For a given band in a given case, same PSF (taken
from the truth, including pixelization effect) was used for maps made
by different map-makers.

\begin{figure*}
    \centering
    \includegraphics[width=6cm, angle=0]{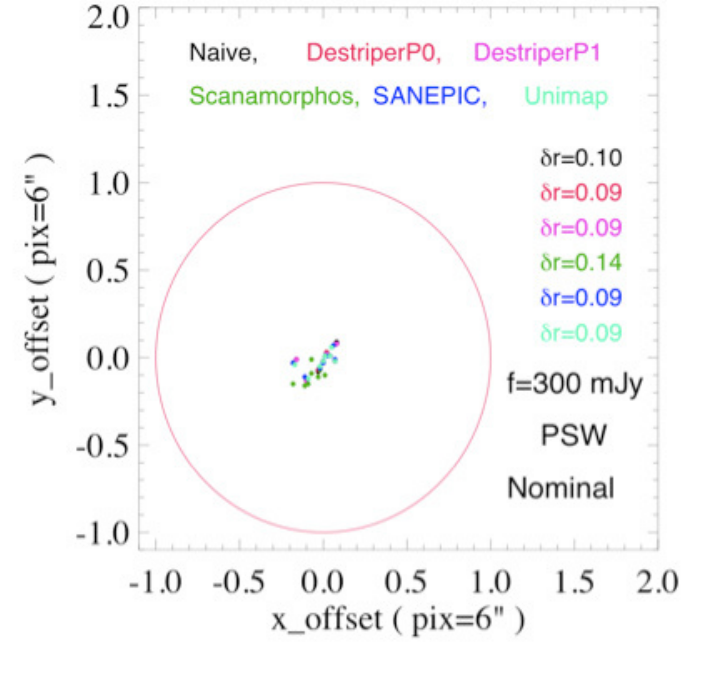}
    \includegraphics[width=6cm, angle=0]{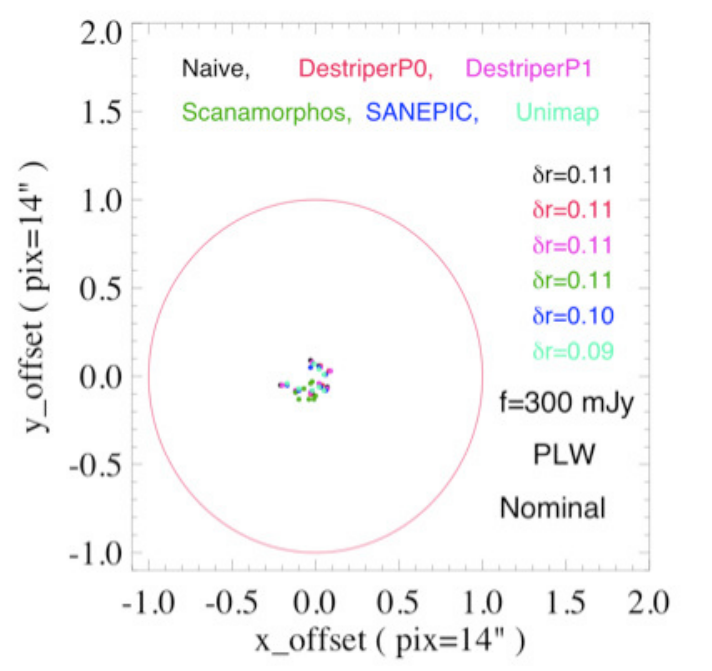}
\caption{
Position offsets (relative to the truth, in units of pixel) of 
bright sources (f=300 mJy) in Case~1 (nominal mode). 
The left panels are for sources in the PSW band, and right
panels for sources in the PLW band. Results on maps made by different
map-makers are color coded (as shown in the figures). $\rm \delta r$ 
is the mean of $\rm \sqrt{x^2_{offset} + y^2_{offset}})$.
}
\label{metrics_report_sources_fig3}
\end{figure*}

\begin{figure*}
    \centering
    \includegraphics[width=6cm, angle=0]{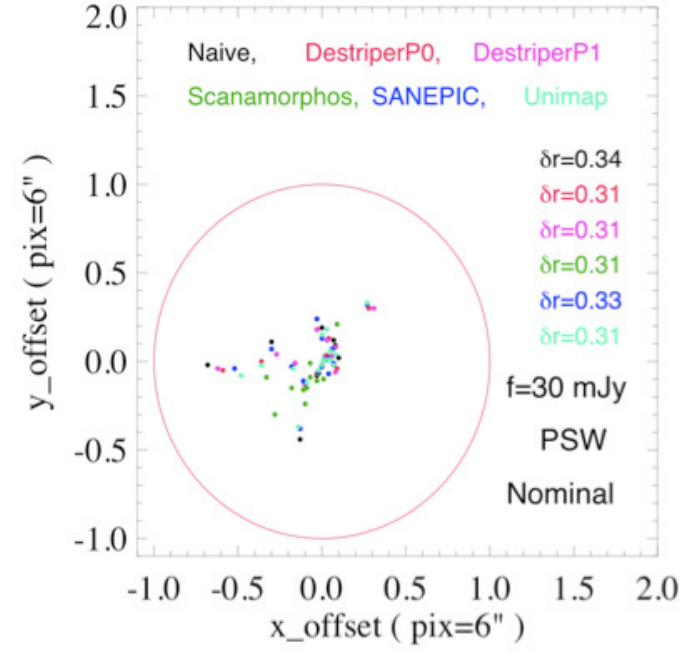}
    \includegraphics[width=6cm, angle=0]{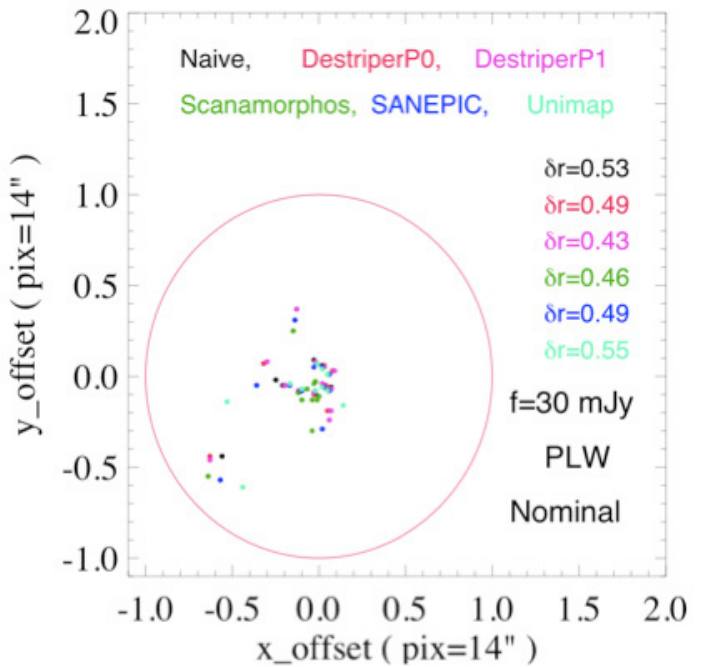}
\caption{Same as Fig.~\ref{metrics_report_sources_fig3}, 
for faint sources (f=30mJy).}
\label{metrics_report_sources_fig4}
\end{figure*}

In Fig.~\ref{metrics_report_sources_fig3}, 
the position offsets of the bright sources 
(f=300 mJy) in Case 1
(nominal mode) are plotted. In a given map, when searching for a
correspondence to a “truth source” (searching radius = 1 pixel), we
always pick the one closest to the truth position. If there is no
detection within the searching radius, the simulated source is deemed
undetected. For the bright sources, the offsets are pretty small in
all maps, on the order of 0.1 pixels. The offsets of the bright
sources in Scanamorphos maps are somewhat larger than in other maps
(e.g., in the PSW band, the former have a mean $\rm \delta r$ 
 of 0.14 pix, while for the same sources in all other maps it is
$≤ 0.10$). There are three
possible causes for this.  First, Scanamorphos applies a relative
gain correction, accounting for the variation of the beam area of
different detectors, to the level-1 data. However, because this effect
was not included in the simulation, other map-makers did not include
correction. Potentially, this correction could change the shape of the
PSF and cause larger point source offsets. In order to test this
hypothesis, Helene Roussel (the author of Scanamorphos) made
additional Case 1 maps using Scanamorphos, this time without the
relative gain correction. However, the tests carried out using these
new maps produced essentially the same results, indicating the effect
due to the relative gain correction is not the cause of the larger
position offset. The second possible cause for the larger position
offsets is the different method used by Scanamorphos to distribute the
signal sampled in a given sky position to adjacent sky pixels. While
in other maps (including the truth) the entire signal is assigned to
the closest sky pixel, Scanamorphos distributes it among adjacent
pixels according to the distance between the sampling point and the
pixel center. This indeed can cause a difference between the PSF in a
Scanamorphos map from that in the corresponding truth map, which in
turn cause the larger offset. In a private correspondence, Helene
Roussel cited a previous test which indicated that this is a minor,
insignificant effect. The third possibility, which was hinted by
results of the metrics on {\it ``Deviation from the truth''} 
(c.f. Section~\ref{sect:diff_summary}), is due to a slight positional offset 
introduced by the mapper. Indeed, in all position-offset plots
(Fig~\label{metrics_report_sources_fig3} -- \ref{metrics_report_sources_fig8})
we see systematic offsets toward the lower-left direction
for sources in Scanamorphos maps, consistant with the trend
found in the difference maps (.e.g. Fig.~\ref{fig:metrics_report_diff_case9}).

For the faint sources (Fig.~\ref{metrics_report_sources_fig4}), 
the position offsets in all maps are significantly larger
than those of bright sources. Because the flux level of the faint
sources is very close to the confusion limit, their positions from PSF
fitting can be significantly affected by the blending with background
sources. In addition to the PSW and PLW bands, for Case 1 we also
simulated and checked the PMW band. The results for the PMW band agree
with those in the PSW and PMW band.
\begin{figure*}
    \centering
    \includegraphics[width=6cm, angle=0]{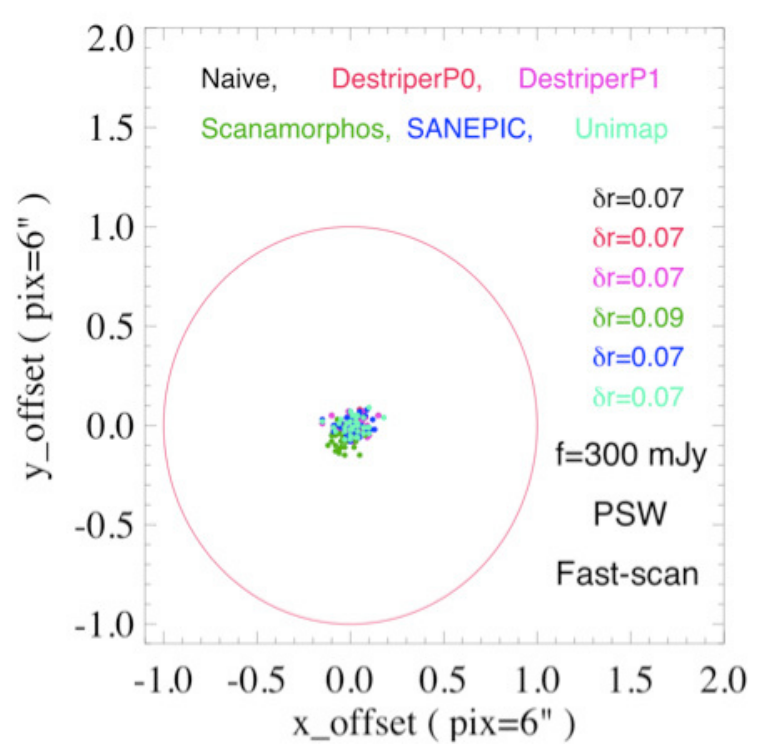}
    \includegraphics[width=6cm, angle=0]{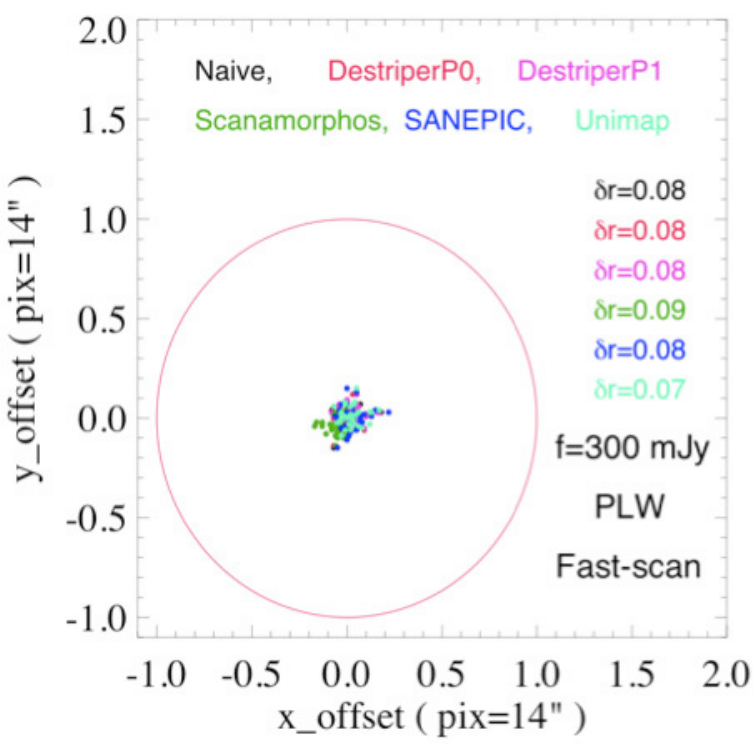}
\caption{Position offsets of bright sources in Case 5 (fast-scan mode).} 
\label{metrics_report_sources_fig5}
\end{figure*}

\begin{figure*}
    \centering
    \includegraphics[width=6cm, angle=0]{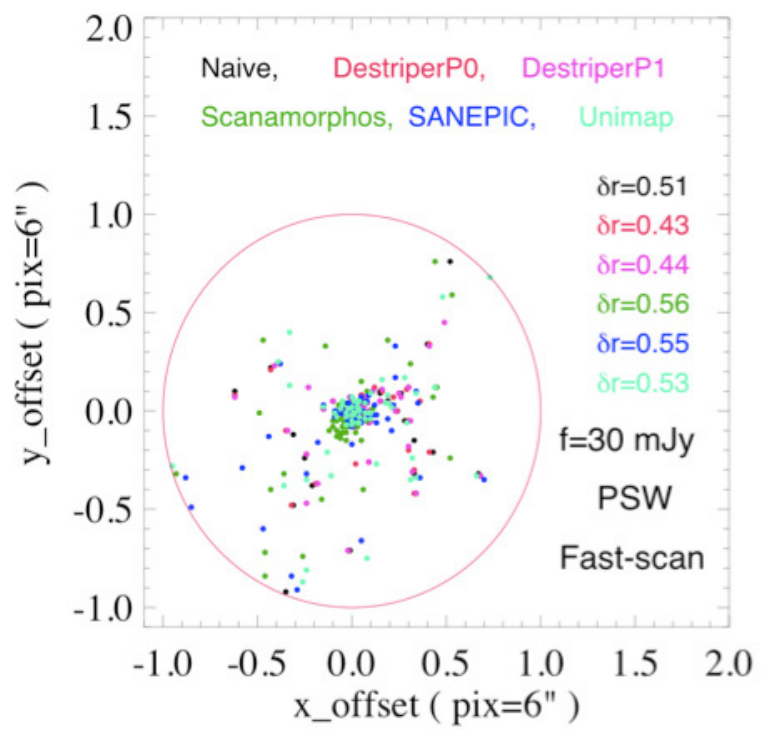}
    \includegraphics[width=6cm, angle=0]{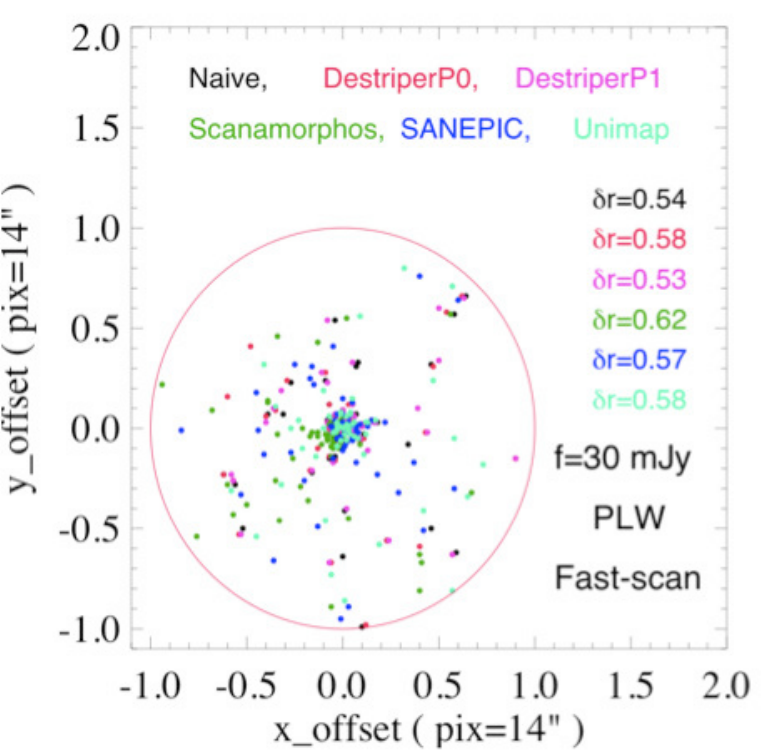}
\caption{Same as Fig.~\ref{metrics_report_sources_fig5}, 
for faint sources (f=30mJy).}
\label{metrics_report_sources_fig6}
\end{figure*}

In Fig.~\ref{metrics_report_sources_fig5}, 
the position offsets of the bright sources in Case 5
(fast-scan mode) are plotted. In this case, the map area simulated is
much larger than that of the Case 1, and therefore more truth sources
were injected (see Section 5.3.1).  Here again we see good position
accuracy for the bright sources, and slightly larger offsets in the
Scanamorphos maps than those in other maps. It is worthwhile noting
that these position offsets are somewhat less than those in Case 1
(Fig.~\ref{metrics_report_sources_fig3}), 
although the simulated survey in Case 5 is much
shallower than that in Case 1 (Table~\ref{tbl:simulations}). 
Indeed, when comparing the
PSW band coverage maps of the Case 5 and of Case 1, we found the
coverage of Case 5 is about a factor of 6 shallower than that of Case
1. It appears that the position accuracy of the bright sources is not
improved with the depth of the coverage. The reason why the bright
sources in nominal observations have poorer position accuracies than
those in fast-scan observations, although only marginally significant,
needs further investigation. For the faint sources 
(Fig.~\ref{metrics_report_sources_fig6}), we
see again larger offsets in all maps, and there are no significant
differences between results for different map-makers.
\begin{figure*}
    \centering
    \includegraphics[width=6cm, angle=0]{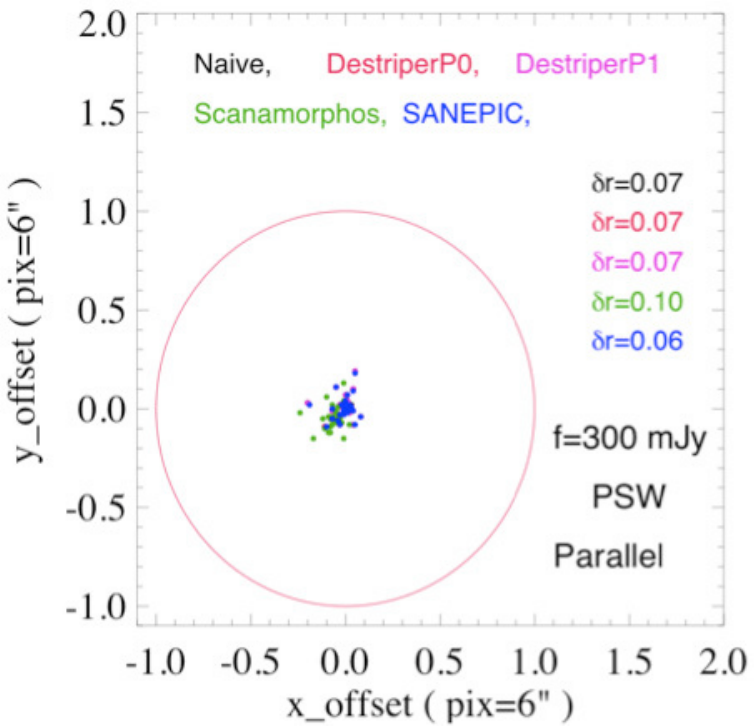}
    \includegraphics[width=6cm, angle=0]{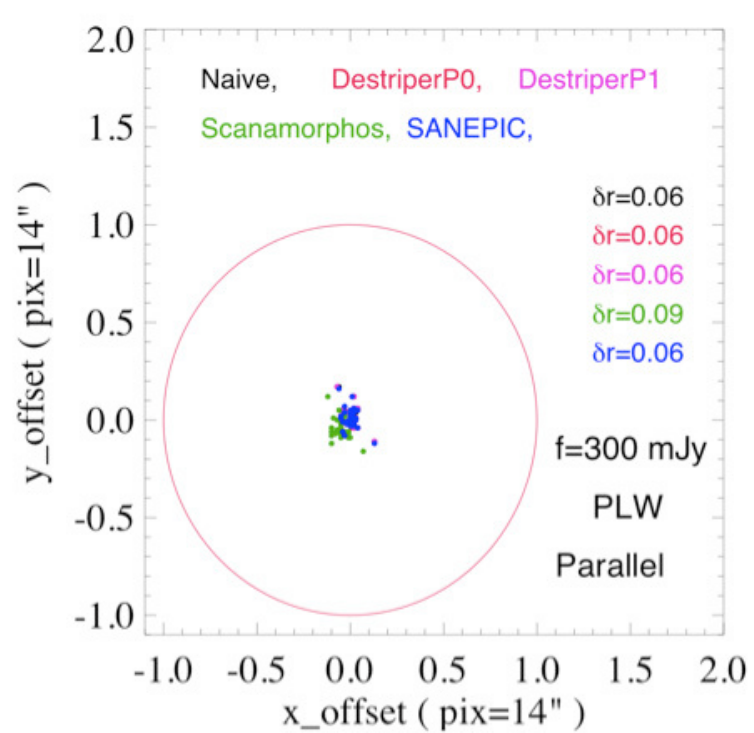}
\caption{Position offsets of bright sources in Case 8 (parallel mode).} 
\label{metrics_report_sources_fig7}
\end{figure*}
\begin{figure*}
    \centering
    \includegraphics[width=6cm, angle=0]{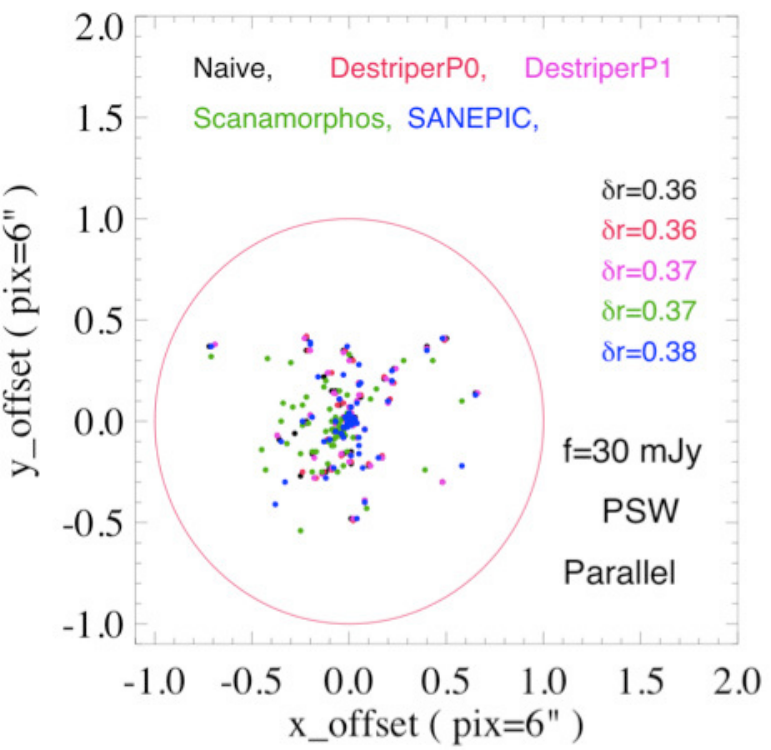}
    \includegraphics[width=6cm, angle=0]{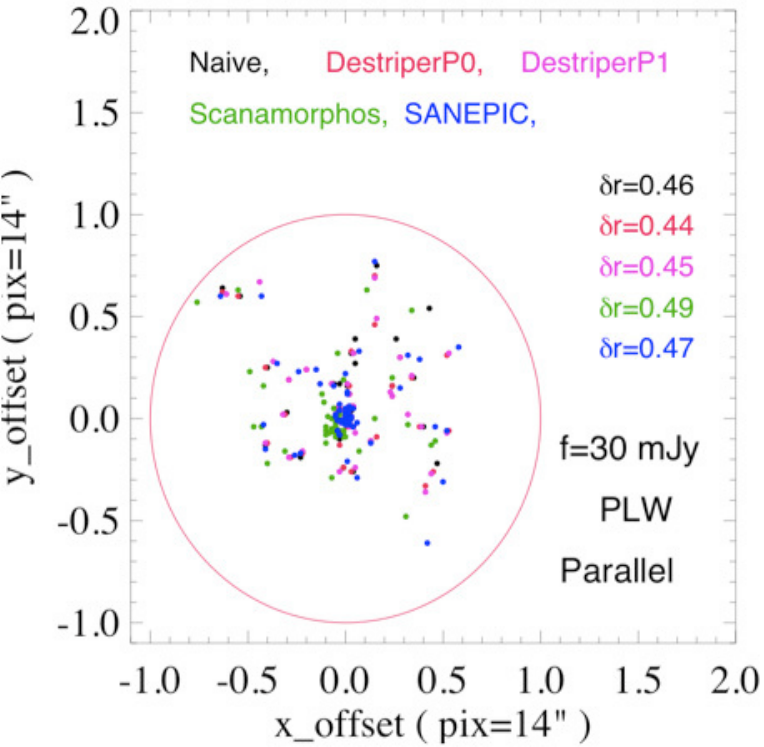}
\caption{Same as Fig.~\ref{metrics_report_sources_fig7}, 
for faint sources (f=30mJy).}
\label{metrics_report_sources_fig8}
\end{figure*}

The position offsets of the point sources in Case 8 (parallel mode)
are plotted in Fig.~\ref{metrics_report_sources_fig7} 
and Fig.~\ref{metrics_report_sources_fig8}. These results are very
similar to those for Case 5.

\begin{table}\scriptsize
%\begin{table}\tiny
\caption{Position offsets (relative to the truth, in units of pixel), defined as the mean of $\rm \delta r 
= \sqrt{x^2_{offset} + y^2_{offset}})$. The bright sources have flux = 300 mJy, and faint sources flux = 30 mJy.}
\begin{tabular}{l c c c c c c c c c c c c}
\hline                                                                                                                
			&Case 1	&	&      &	&Case 5	&	&      &      &Case 8	&	&      &	\\
\hline                                                                                                                
			&PSW	&      & PLW   &	& PSW	&       & PLW  &      &PSW	&      & PLW   &        \\
			& Bright&Faint & Bright& Faint	&Bright	&Faint &Bright & Faint& Bright&Faint & Bright& Faint	\\
\hline                                                                                                                
Naive			&  0.10 &  0.34	& 0.11 &  0.53	&  0.07 &  0.51	& 0.08 & 0.54 &  0.07 &  0.36	& 0.06 &  0.46	  \\
Destriper-P0		&  0.09 &  0.31	& 0.11 &  0.49	&  0.07 &  0.43	& 0.08 & 0.58 &  0.07 &  0.36	& 0.06 &  0.44	  \\
Destriper-P1		&  0.09 &  0.31	& 0.11 &  0.43	&  0.07 &  0.44	& 0.08 & 0.53 &  0.07 &  0.37	& 0.06 &  0.45	  \\
Scanamorphos		&  0.14 &  0.31	& 0.11 &  0.46	&  0.09 &  0.56	& 0.09 & 0.62 &  0.10 &  0.37	& 0.09 &  0.49	  \\
SANEPIC			&  0.09 &  0.33	& 0.10 &  0.49	&  0.07 &  0.55	& 0.08 & 0.57 &  0.06 &  0.38	& 0.06 &  0.47	  \\
Unimap			&  0.09 &  0.31	& 0.09 &  0.56	&  0.07 &  0.53	& 0.07 & 0.58 &       &  	&      &  	  \\      
\hline                                                                                                                
\end{tabular}
\label{tab:offsets}
%\end{center}
\end{table}
\normalsize

\begin{table}\scriptsize
%\begin{table}[t]\tiny
\caption{Dectection rates of bright sources (f = 300 mJy) and faint sources (f = 30 mJy).}
\begin{tabular}{l c c c c c c c c c c c c}
\hline                                                                                                                
			&Case 1	&	&      &	&Case 5	&	&      &      &Case 8	&	&      &	\\
\hline                                                                                                                
			&PSW	&      & PLW   &	& PSW	&       & PLW  &      &PSW	&      & PLW   &        \\
			& Bright&Faint & Bright& Faint	&Bright	&Faint &Bright & Faint& Bright&Faint & Bright& Faint	\\
\hline                                                                                                                
Naive			& 1.0  &  0.87	& 1.0 &  0.25	&1.0   &  0.58	& 1.0 & 0.66 & 1.0  &  0.88	& 1.0 &  0.61	  \\
Destriper-P0		& 1.0  &  0.87	& 1.0 &  0.37	&1.0   &  0.61	& 1.0 & 0.69 & 1.0  &  0.86	& 1.0 &  0.69	  \\
Destriper-P1		& 1.0  &  0.87	& 1.0 &  0.62	&1.0   &  0.55	& 1.0 & 0.69 & 1.0  &  0.83	& 1.0 &  0.69	  \\
Scanamorphos		& 1.0  &  0.50	& 1.0 &  0.50	&1.0   &  0.66	& 1.0 & 0.72 & 1.0  &  0.88	& 1.0 &  0.52	  \\
SANEPIC			& 1.0  &  0.87	& 1.0 &  0.50	&1.0   &  0.63	& 1.0 & 0.63 & 1.0  &  0.83	& 1.0 &  0.66	  \\
Unimap			& 1.0  &  1.00	& 1.0 &  0.37	&1.0   &  0.75	& 1.0 & 0.69 & 1.0  &  	        &     &  	  \\      
\hline                                                                                                                
\end{tabular}
\label{tab:detectionrate}
%\end{center}
\end{table}
\normalsize

\begin{table}\scriptsize
%\begin{table}[t]\tiny
\caption{Fractional flux deviations relative to the truth ($\rm \delta f$) of bright
sources (f = 300 mJy) and faint sources (f = 30 mJy).}
\begin{tabular}{l c c c c c c c c c c c c}
\hline                                                                                                                
			&Case 1	&	&      &	&Case 5	&	&      &      &Case 8	&	&      &	\\
\hline                                                                                                                
			&PSW	&      & PLW   &	& PSW	&       & PLW  &      &PSW	&      & PLW   &        \\
			& Bright&Faint & Bright& Faint	&Bright	&Faint &Bright & Faint& Bright&Faint & Bright& Faint	\\
\hline                                                                                                                
Naive			& 0.02  &  0.08	& 0.03 &  0.23	& 0.02  &  0.32	& 0.03 & 0.29 & 0.02  & 0.20 	& 0.03 & 0.14 	  \\
Destriper-P0		& 0.02  &  0.11	& 0.02 &  0.20	& 0.02  &  0.28	& 0.03 & 0.33 & 0.02  & 0.20 	& 0.03 & 0.16 	  \\
Destriper-P1		& 0.02  &  0.10	& 0.03 &  0.19	& 0.02  &  0.30	& 0.03 & 0.34 & 0.02  & 0.18 	& 0.03 & 0.17 	  \\
Scanamorphos		& 0.01  &  0.17	& 0.02 &  0.17	& 0.02  &  0.21	& 0.03 & 0.24 & 0.02  & 0.21 	& 0.02 & 0.11 	  \\
SANEPIC			& 0.02  &  0.11	& 0.03 &  0.14	& 0.03  &  0.29	& 0.04 & 0.31 & 0.02  & 0.22 	& 0.03 & 0.15 	  \\
Unimap			& 0.02  &  0.12	& 0.03 &  0.14	& 0.03  &  0.29	& 0.03 & 0.24 &       &         &      &          \\      
\hline                                                                                                                
\end{tabular}
\label{tab:fluxdeviation}
%\end{center}
\end{table}
\normalsize

\begin{figure*}
    \centering
    \includegraphics[width=6cm, angle=0]{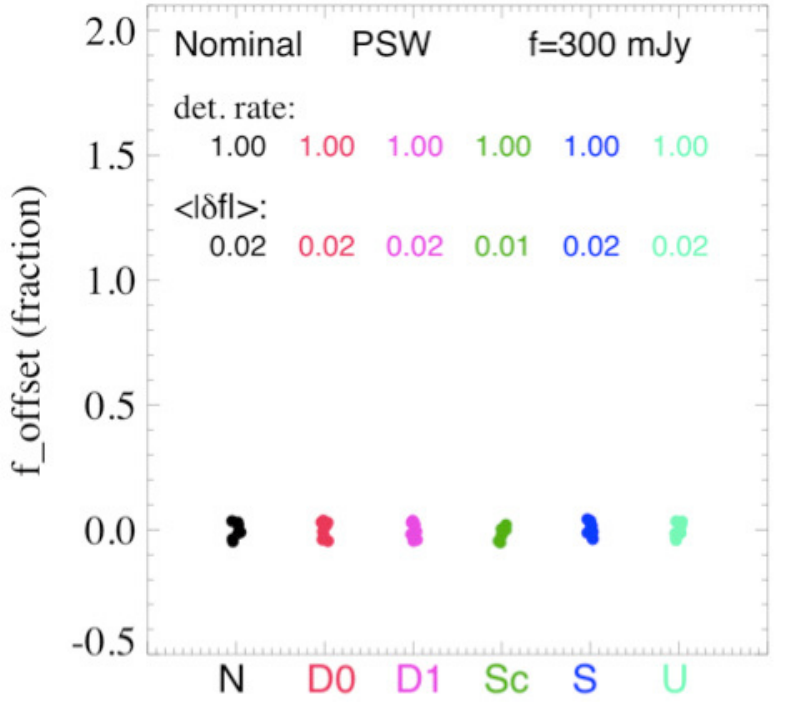}
    \includegraphics[width=6cm, angle=0]{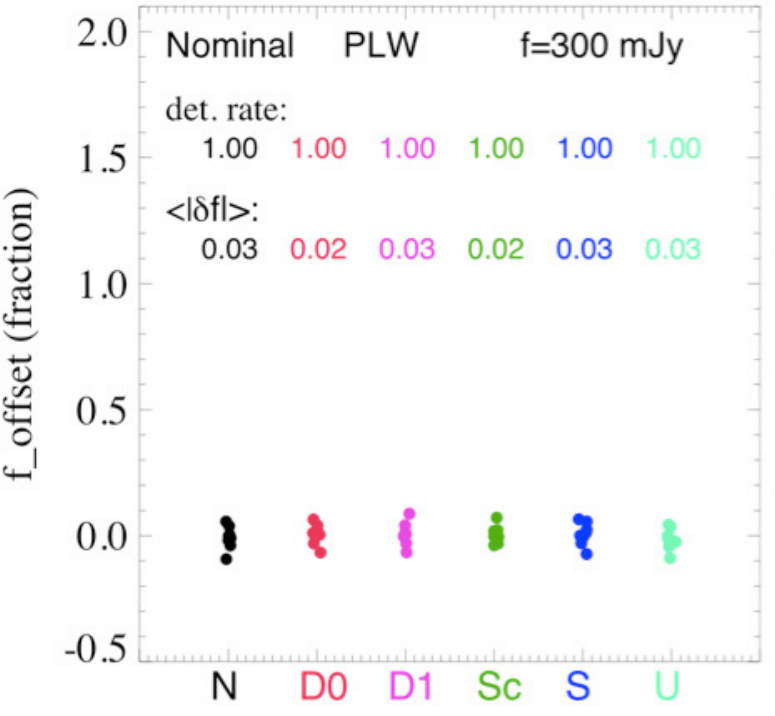}
\caption{Fractional flux deviations (relative to the truth) of bright sources
(f = 300 mJy)
in Case 1 (nominal mode). Each column is for a given map-maker, color coded 
in the same way as in Fig.~\ref{metrics_report_sources_fig3}. 
The left panels are for sources in 
the PSW band, and right panels for sources in the PLW band}
\label{metrics_report_sources_fig9}
\end{figure*}
\begin{figure*}
    \centering
    \includegraphics[width=6cm, angle=0]{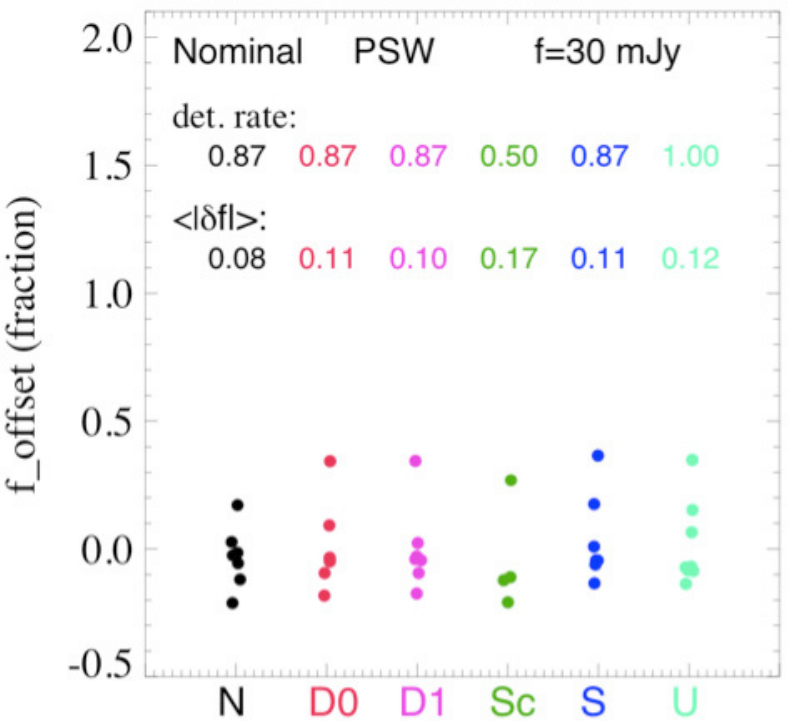}
    \includegraphics[width=6cm, angle=0]{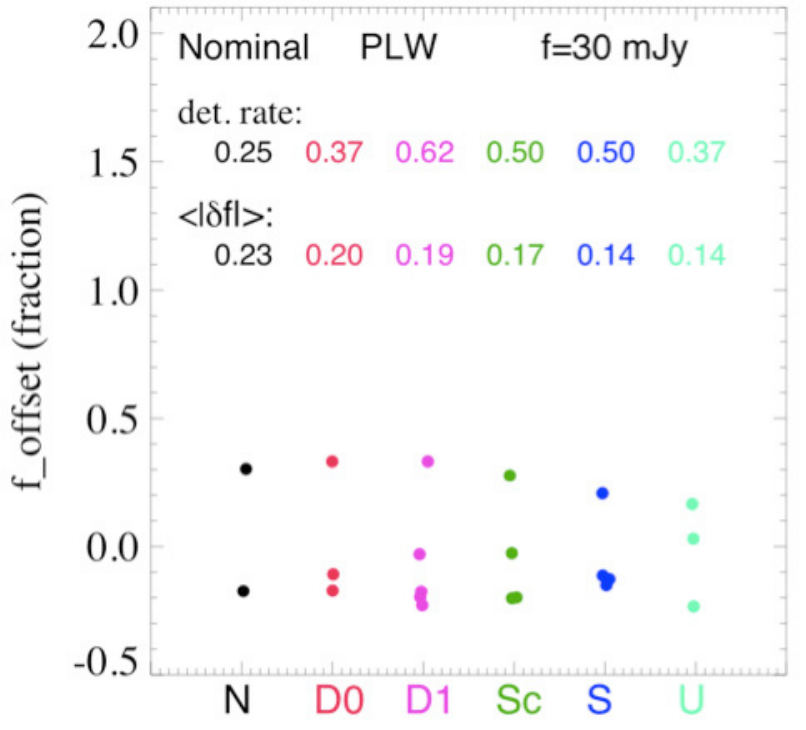}
\caption{Same as Fig.~\ref{metrics_report_sources_fig9}, 
for faint sources (f = 30 mJy).}
\label{metrics_report_sources_fig10}
\end{figure*}
Fraction flux deviations from the truth, as indicators of the flux
errors, are plotted in Fig.~\ref{metrics_report_sources_fig9} 
for the bright sources in Case 1
(nominal mode). In each panel, color coded for the different
map-makers, the detection rates and the mean flux deviations are also
listed. For the bright sources, the detection rates are always 100
percent, and the flux deviations are as low as a couple of
percent. There is no significant difference between results for
different map-makers. For the faint sources 
(Fig.~\ref{metrics_report_sources_fig10}), 
the detection rate is
lower. In particular, in the PLW maps, the detection rates are less or
equal to 50\%.
There are some differences between results for different
map-makers. However, combining the results in both PSW and PLW 
bands, we don’t see any
clear pattern. For example, Unimap gives the highest detection rate in
the PSW band, but the 2nd lowest detection rate in the PLW band.
Then, the detection rate in the PSW map made by Scanamorphos is the
lowest among all PSW maps, while in the PLW band, the detection rate
in the Scanamorphos map is the 2nd highest. The flux deviations are
plotted for the detected faint sources, and the means are listed in
the corresponding panels. The deviations are much higher than for
those the bright sources, because of the lower signal/noise ratios for
the fainter sources.

\begin{figure*}
    \centering
    \includegraphics[width=6cm, angle=0]{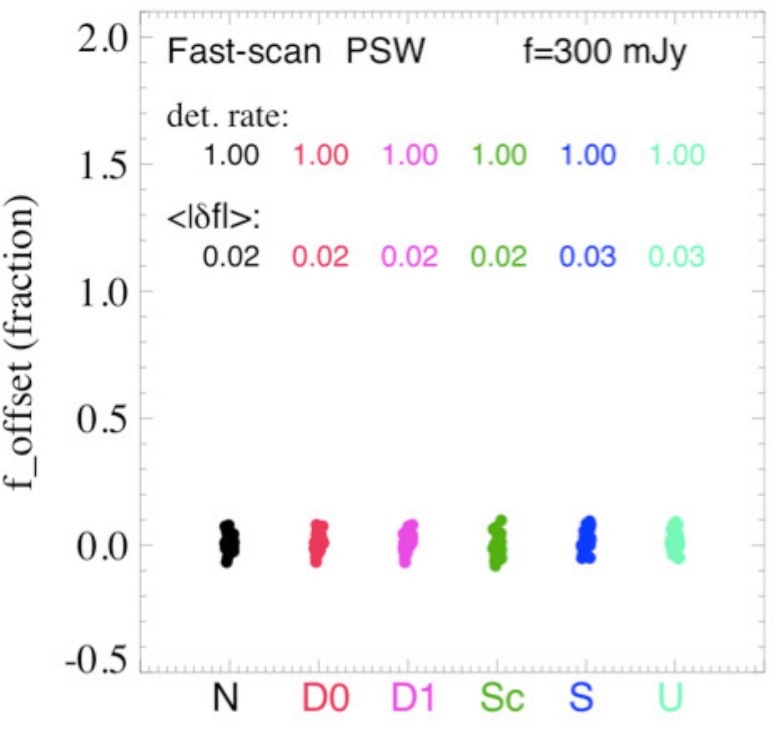}
    \includegraphics[width=6cm, angle=0]{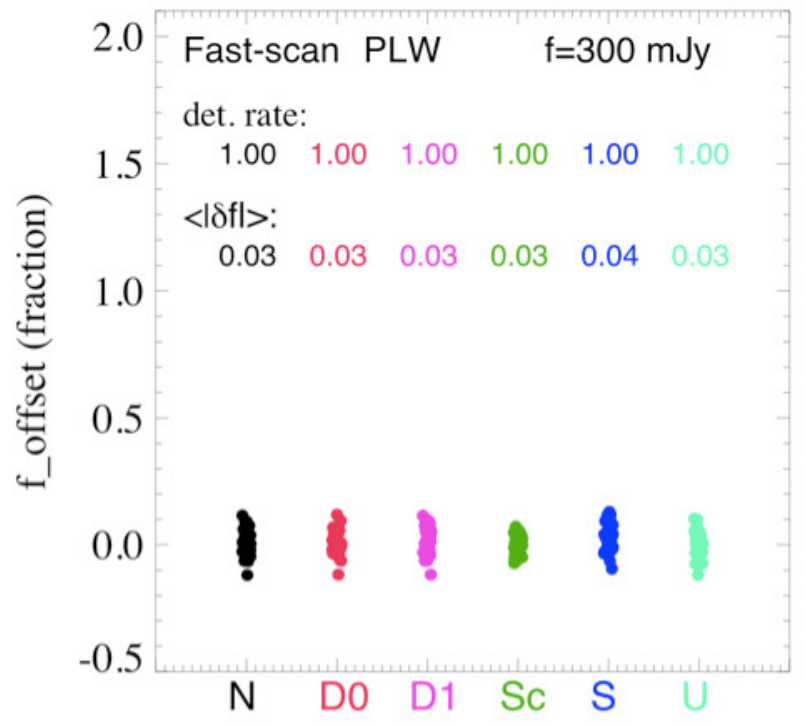}
\caption{Fractional flux deviations of bright sources (f = 300 mJy)
in Case 5 (fast-scan mode). }
\label{metrics_report_sources_fig11}
\end{figure*}
\begin{figure*}
    \centering
    \includegraphics[width=6cm, angle=0]{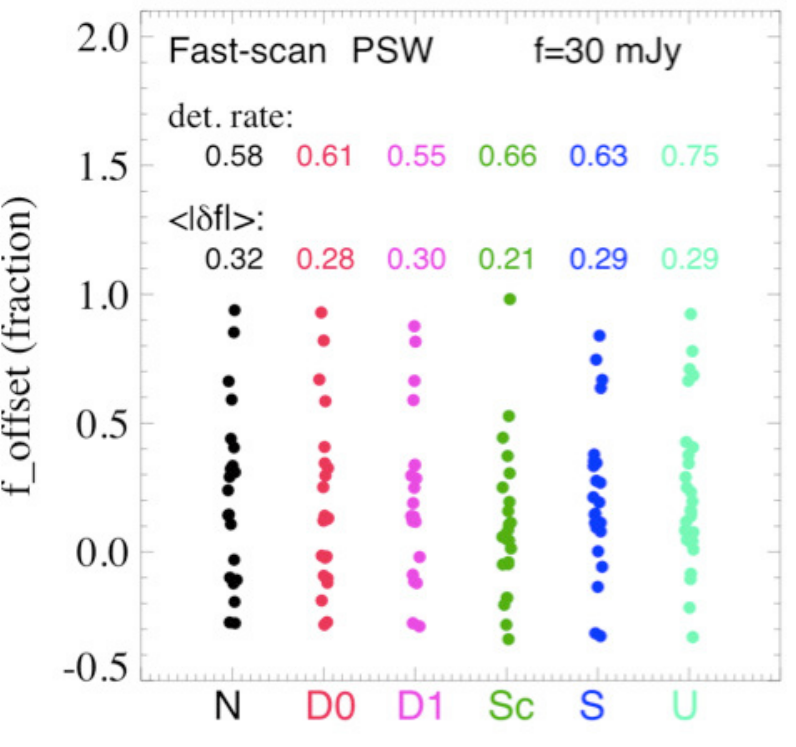}
    \includegraphics[width=6cm, angle=0]{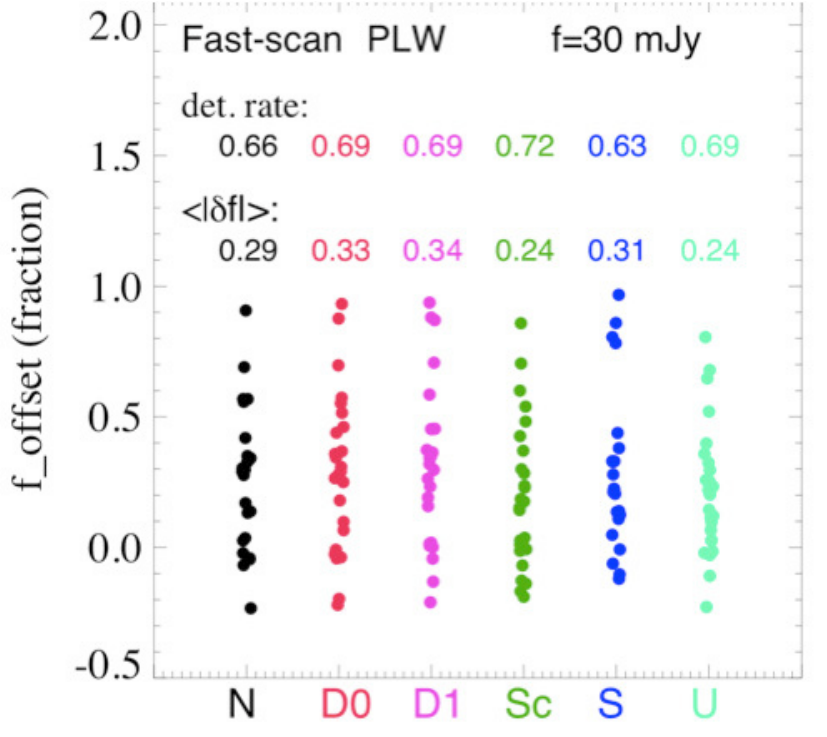}
\caption{Same as Fig.~\ref{metrics_report_sources_fig11}, 
for faint sources (f = 30 mJy).}
\label{metrics_report_sources_fig12}
\end{figure*}

In Fig.~\ref{metrics_report_sources_fig11}, 
fractional flux deviations of the bright sources in Case
5 (fast-scan mode) are plotted. The results are very similar to those
in Fig.~\ref{metrics_report_sources_fig9} 
(Case 1, nominal mode). However, for the faint sources
(Fig.~\ref{metrics_report_sources_fig12}), there is no significant difference between results in
the PSW band and those in the PLW band. Again, there is no significant
difference between results for different map-makers.

\begin{figure*}
    \centering
    \includegraphics[width=6cm, angle=0]{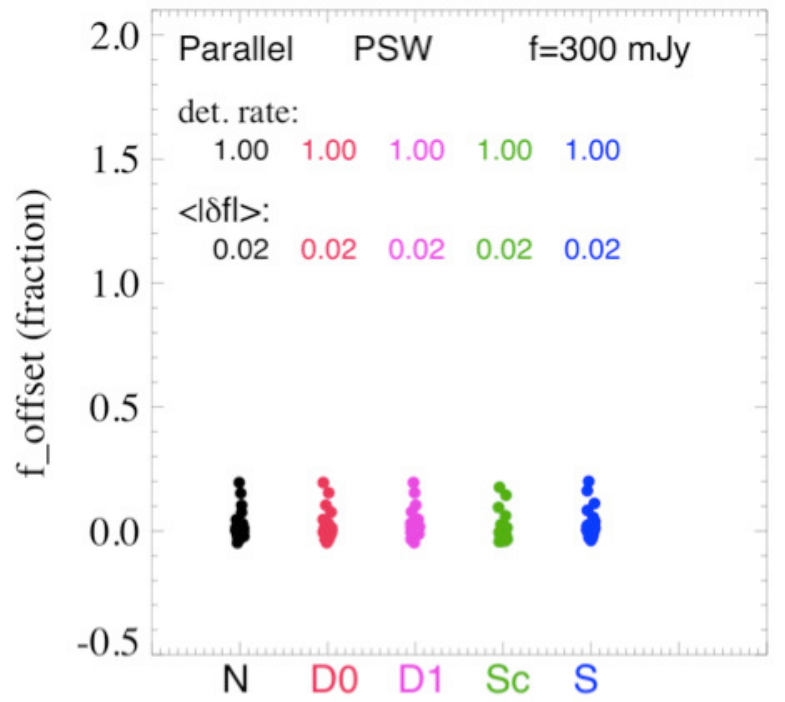}
    \includegraphics[width=6cm, angle=0]{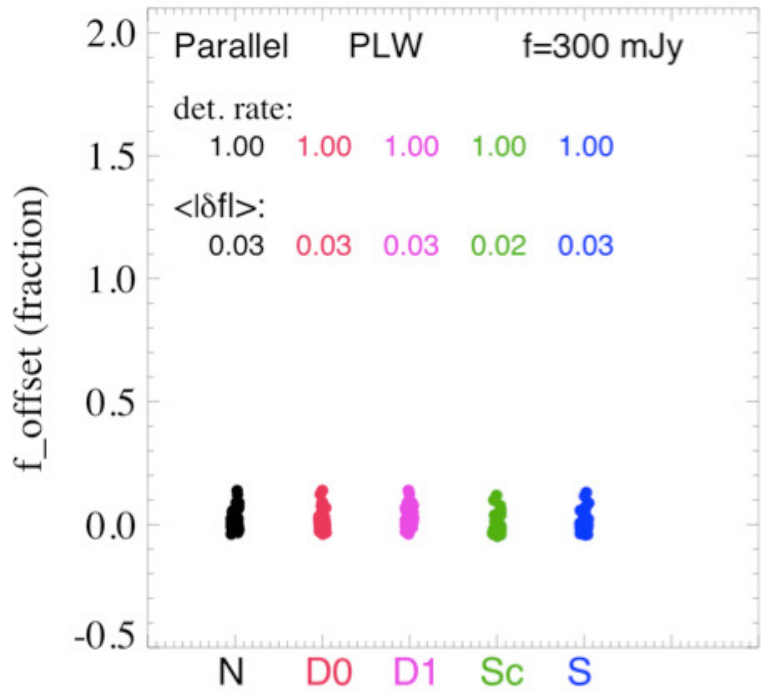}
\caption{Fractional flux deviations of bright sources (f = 300 mJy)
in Case 8 (parallel mode). }
\label{metrics_report_sources_fig13}
\end{figure*}

\begin{figure*}
    \centering
    \includegraphics[width=6cm, angle=0]{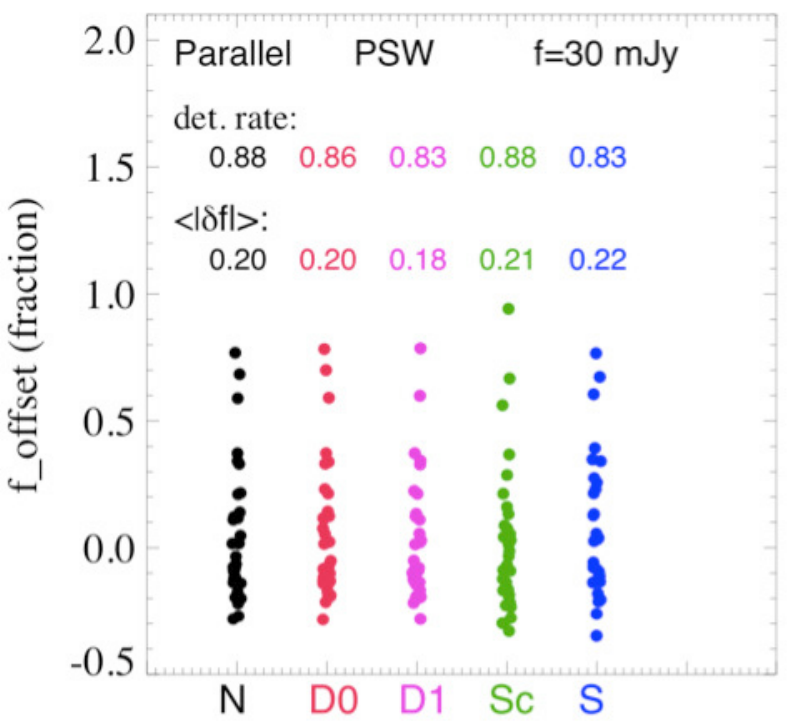}
    \includegraphics[width=6cm, angle=0]{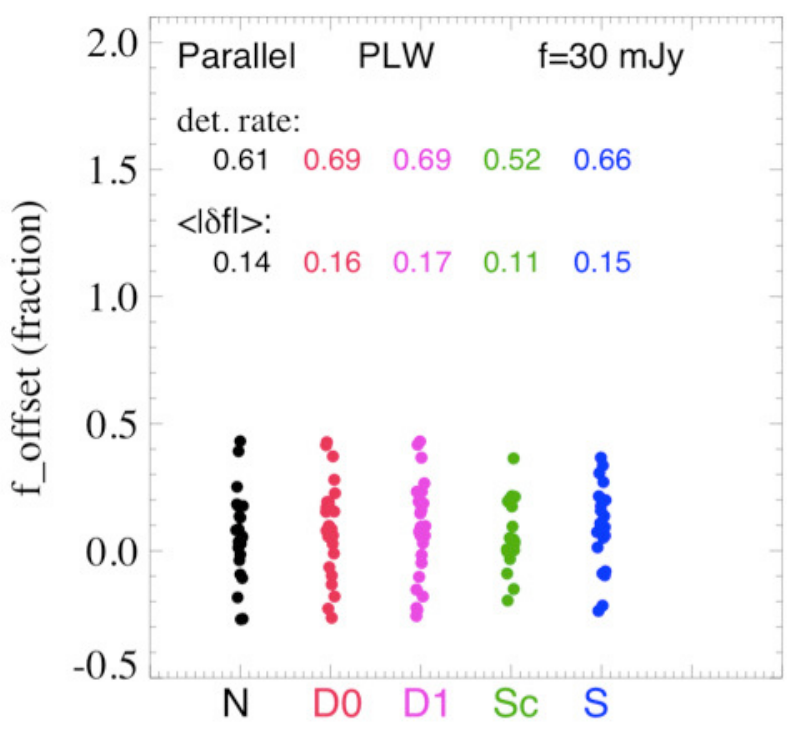}
\caption{Same as Fig.~\ref{metrics_report_sources_fig13}, 
for faint sources (f = 30 mJy).}
\label{metrics_report_sources_fig14}
\end{figure*}

In Fig.~\ref{metrics_report_sources_fig13} 
and  Fig.~\ref{metrics_report_sources_fig14}, 
fractional flux deviations of the point sources in Case 8 
(parallel mode) are plotted. The results are very similar to those 
for Case 5 (fast-scan mode). 

\subsubsection{Point Sources: Aperture Photometry}\label{Aperture}

We used two different apertures for the point source aperture
photometry. The small aperture has a radius of $\rm 0.5 \times FWHM$
of the PSF, and the large aperture a radius of $\rm 2 \times
FWHM$. The background annulus is between two radii of 3 and 4 times of
the FWHM, respectively. The small aperture photometry is sensitive to
the position offset, and large aperture photometry checks the energy
conservation in the map-making.

\begin{figure*}
    \centering
    \includegraphics[width=6cm, angle=0]{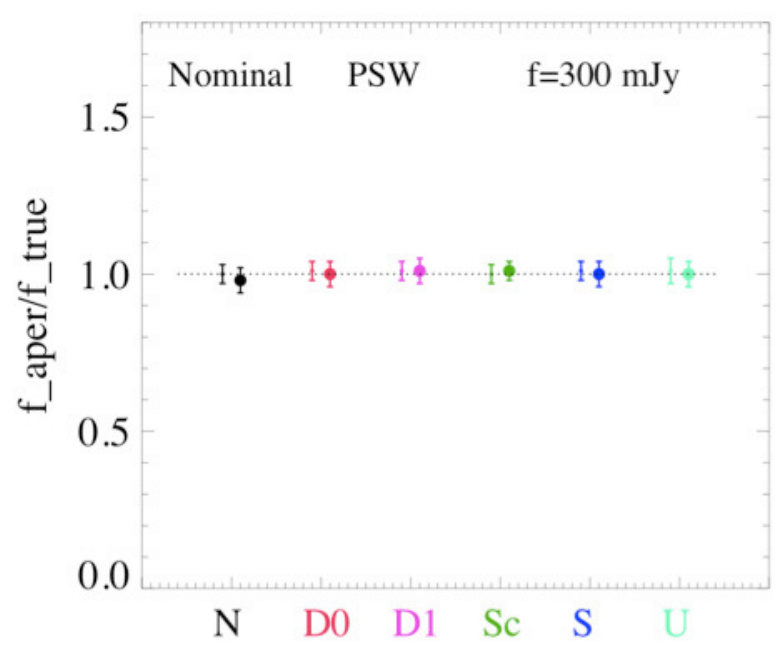}
    \includegraphics[width=6cm, angle=0]{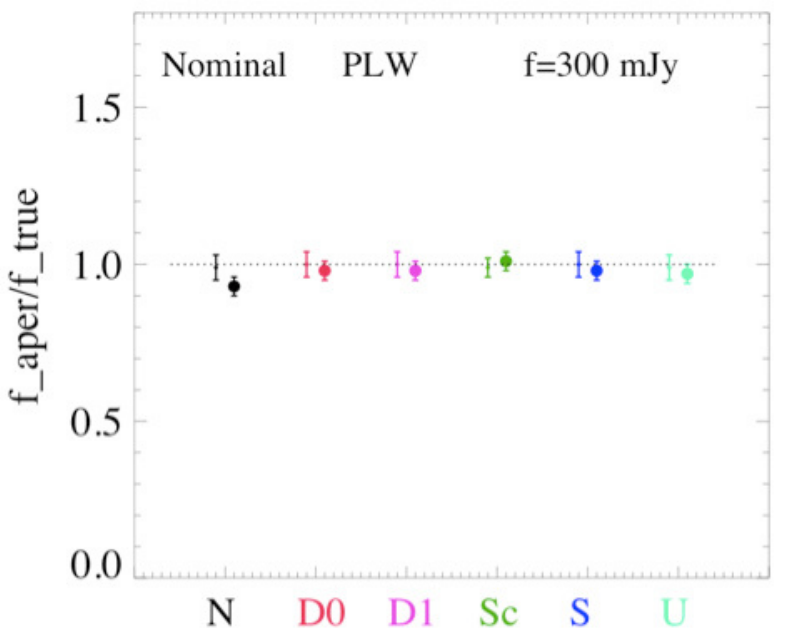}
\caption{Plots of mean $\rm f_{aper}/f_{true}$ 
ratios of bright sources (f = 300 mJy) 
in Case 1 (nominal mode); $\rm f_{aper} $ is
the flux measured in a simulated map, and $\rm f_{true}$ the flux
measured in the corresponding truth map using the same aperture (and
background) parameters. In each panel, for a given map-maker, the
small point (on the left) with error bars presents the result obtained
using the small aperture (r=0.5 FWHM), and the large point (on the
right) with error bars the result obtained using the large aperture
(r=2 FWHM).  }
\label{metrics_report_sources_fig15}
\end{figure*}
\begin{figure*}
    \centering
    \includegraphics[width=6cm, angle=0]{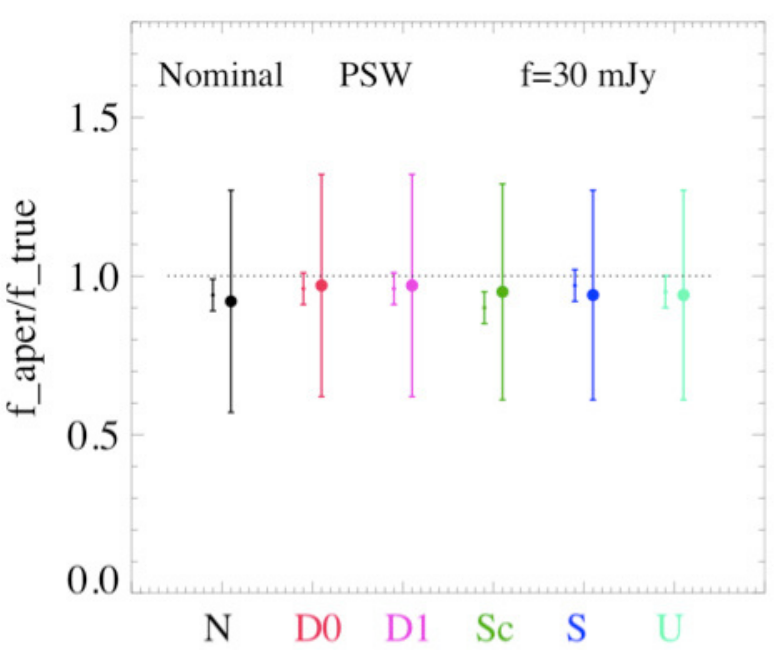}
    \includegraphics[width=6cm, angle=0]{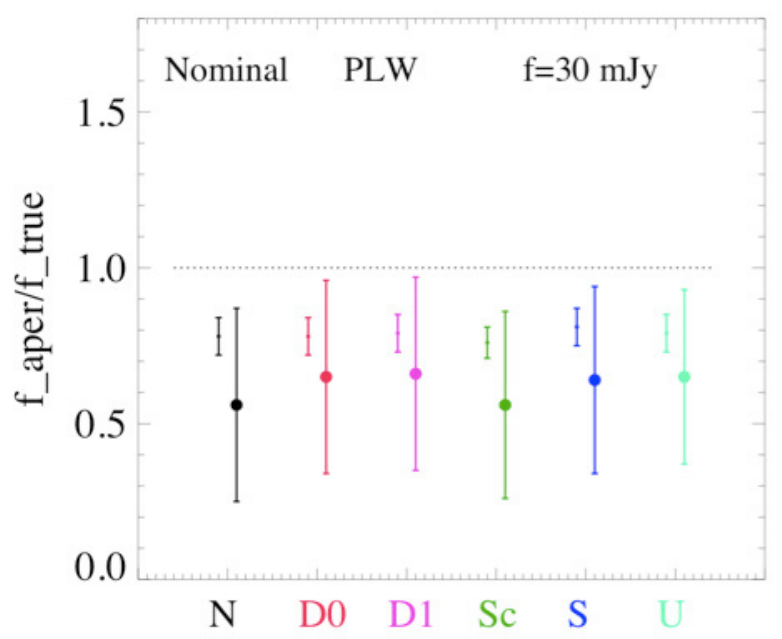}
\caption{Same as Fig.~\ref{metrics_report_sources_fig15},
 for faint sources (f = 30 mJy).}
\label{metrics_report_sources_fig16}
\end{figure*}

In Fig.~\ref{metrics_report_sources_fig15}, 
plots of mean $\rm f_{aper}/f_{true}$ ratios of bright 
sources in
Case 1 (nominal mode) are presented. Here $\rm f_{aper}$ is the flux measured
using the aperture photometry in a given map, and $\rm f_{true}$ the flux
measured in the corresponding truth map using the same aperture. Using
$\rm f_{aper}/f_{true}$ as an indicator for the flux deviation, we are free from
uncertainties due to any aperture correction. For the bright sources,
both the small and larger aperture photometry agree very well with the
truth. There is a slightly under estimation for the large aperture
photometry for the naive map in the PLW band. For the faint sources 
(Fig.~\ref{metrics_report_sources_fig16}),
we see larger deviations and larger error bars because of the
noise. In the PLW maps, there is a systematic underestimation of
fluxes measured by both apertures, which is apparently due to large
confusion noise. Detailed inspections
of the maps showed that 5 out of the 8 faint sources are severely
affected by blending. The low detection rates shown in 
Fig.~\ref{metrics_report_sources_fig4} are due to the same reason.

\begin{figure*}
    \centering
    \includegraphics[width=6cm, angle=0]{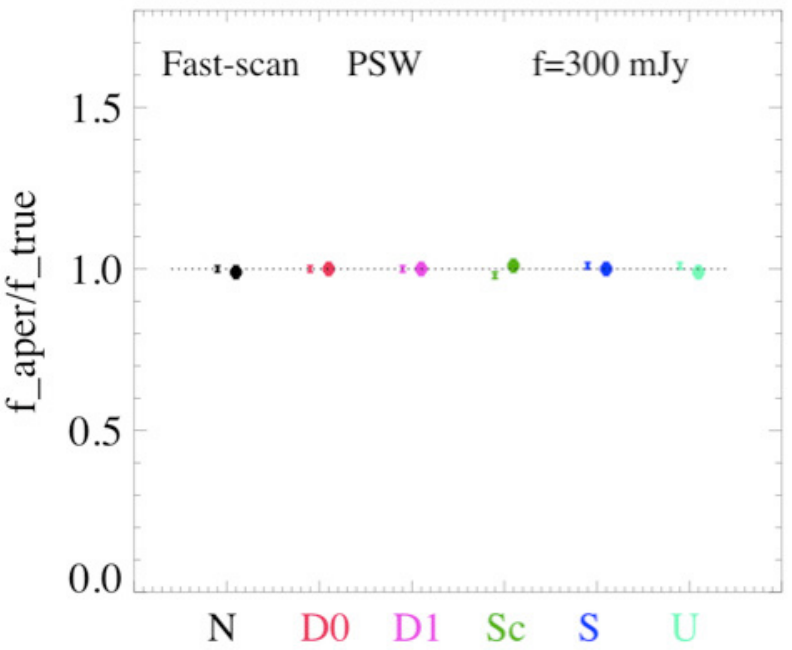}
    \includegraphics[width=6cm, angle=0]{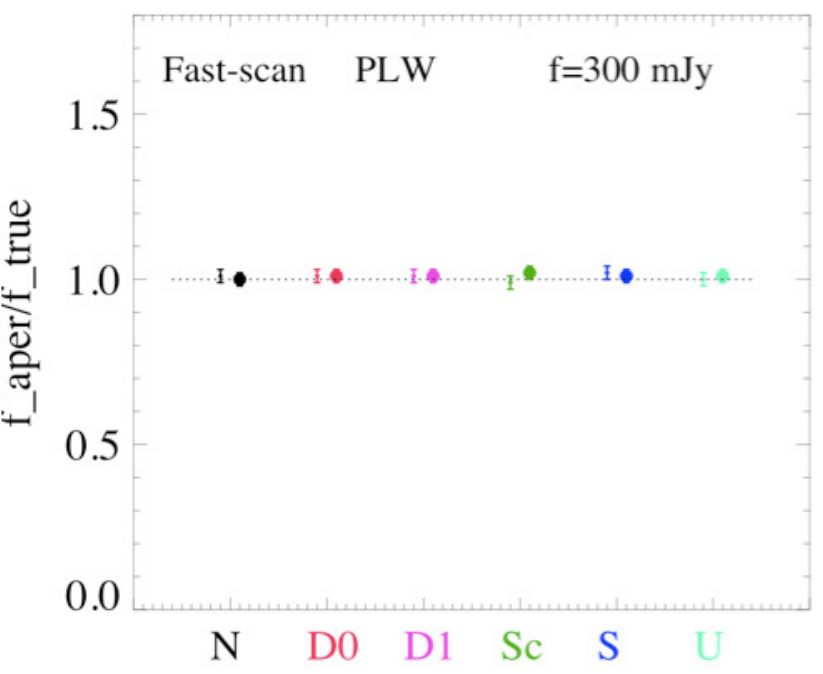}
\caption{Plots of mean $\rm f_{aper}/f_{true}$ ratios of bright sources 
(f = 300 mJy) in Case 5 (fast-scan mode).}
\label{metrics_report_sources_fig17}
\end{figure*}
\begin{figure*}
    \centering
    \includegraphics[width=6cm, angle=0]{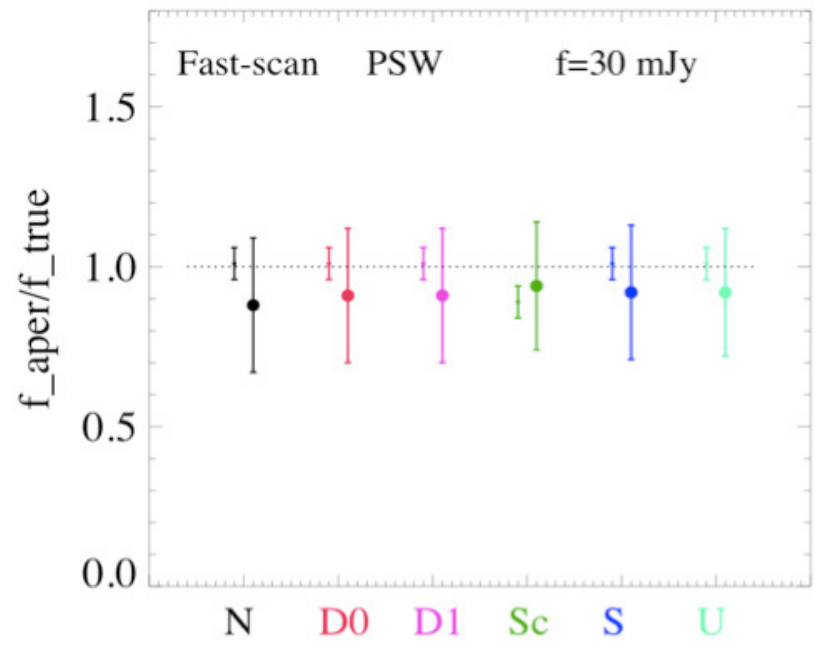}
    \includegraphics[width=6cm, angle=0]{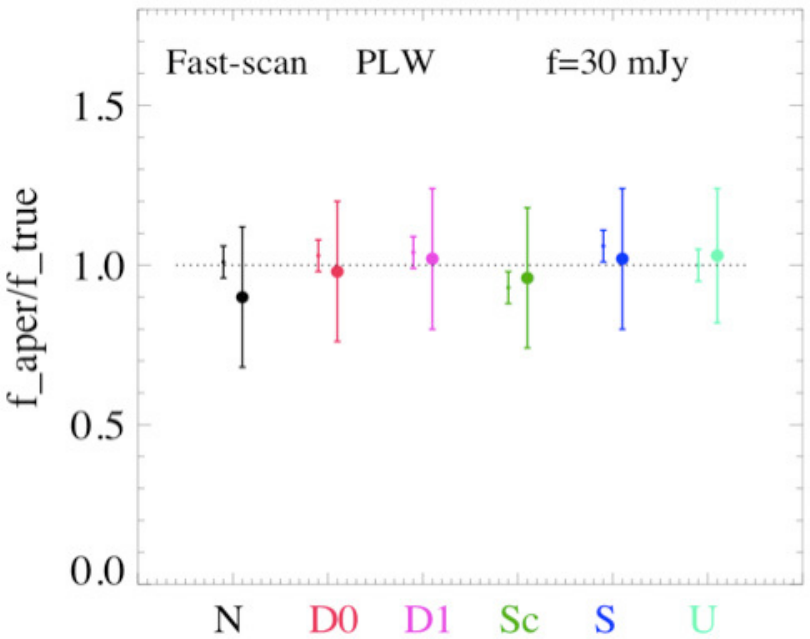}
\caption{Same as Fig.~\ref{metrics_report_sources_fig17},
 for faint sources (f = 30 mJy).}
\label{metrics_report_sources_fig18}
\end{figure*}

The results for Case 5 (fast-scan mode) are presented in Fig.~17
and Fig.~18. Again we see good photometric accuracy for the bright
sources, and large errors for the faint sources. No significant
difference is found between results for different map-makers.

\begin{figure*}
    \centering
    \includegraphics[width=6cm, angle=0]{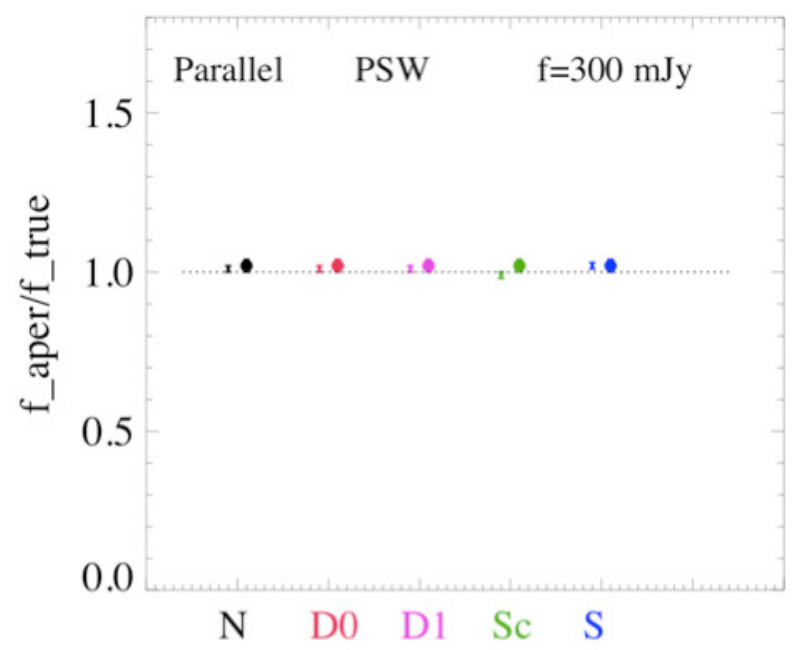}
    \includegraphics[width=6cm, angle=0]{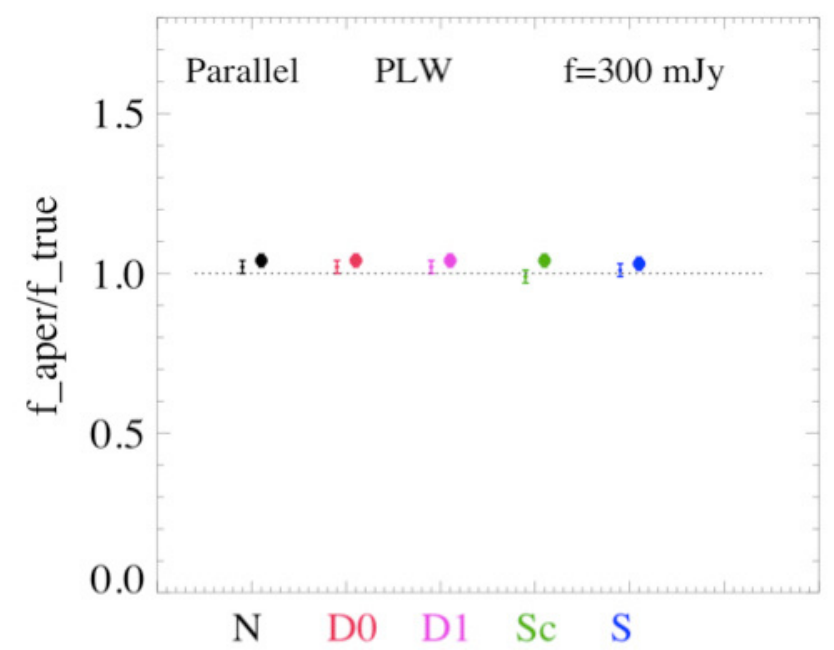}
\caption{Plots of mean $\rm f_{aper}/f_{true}$ ratios of bright sources 
(f = 300 mJy) in Case 8 (parallel mode).}
\label{metrics_report_sources_fig19}
\end{figure*}

\begin{figure*}
    \centering
    \includegraphics[width=6cm, angle=0]{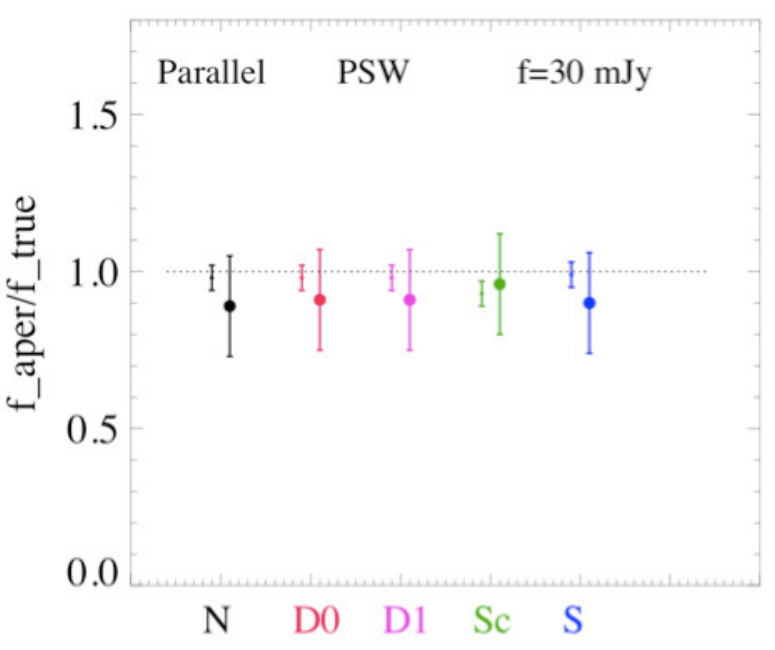}
    \includegraphics[width=6cm, angle=0]{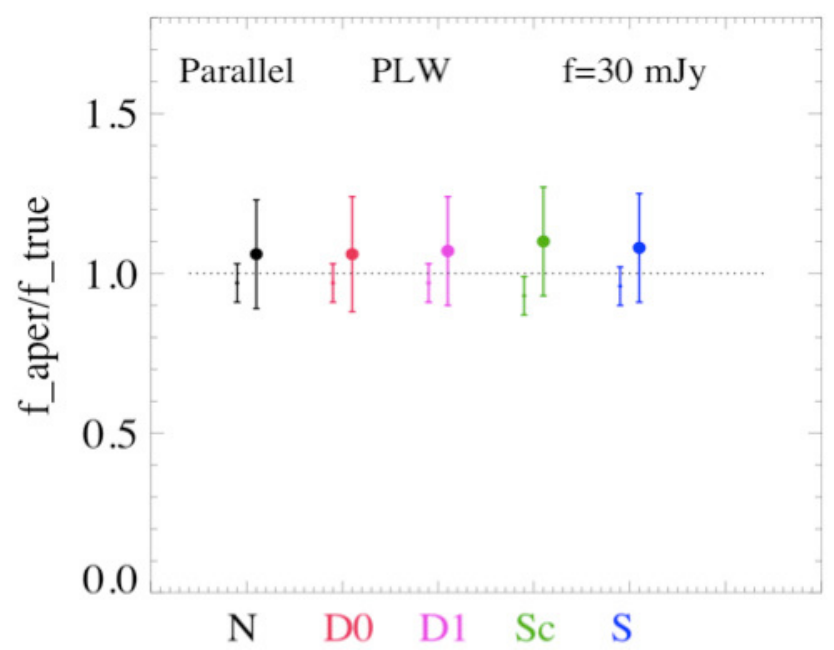}
\caption{Same as Fig.~\ref{metrics_report_sources_fig19}, 
for faint sources (f = 30 mJy).}
\label{metrics_report_sources_fig20}
\end{figure*}

Very similar results are found in Fig.~\ref{metrics_report_sources_fig19}
and Fig.~\ref{metrics_report_sources_fig20}
 for Case 8 (parallel mode).

\subsubsection{Point Sources: Difference in Beam Profile}\label{Aperture}

\begin{figure*}
    \centering
    \includegraphics[width=6cm, angle=0]{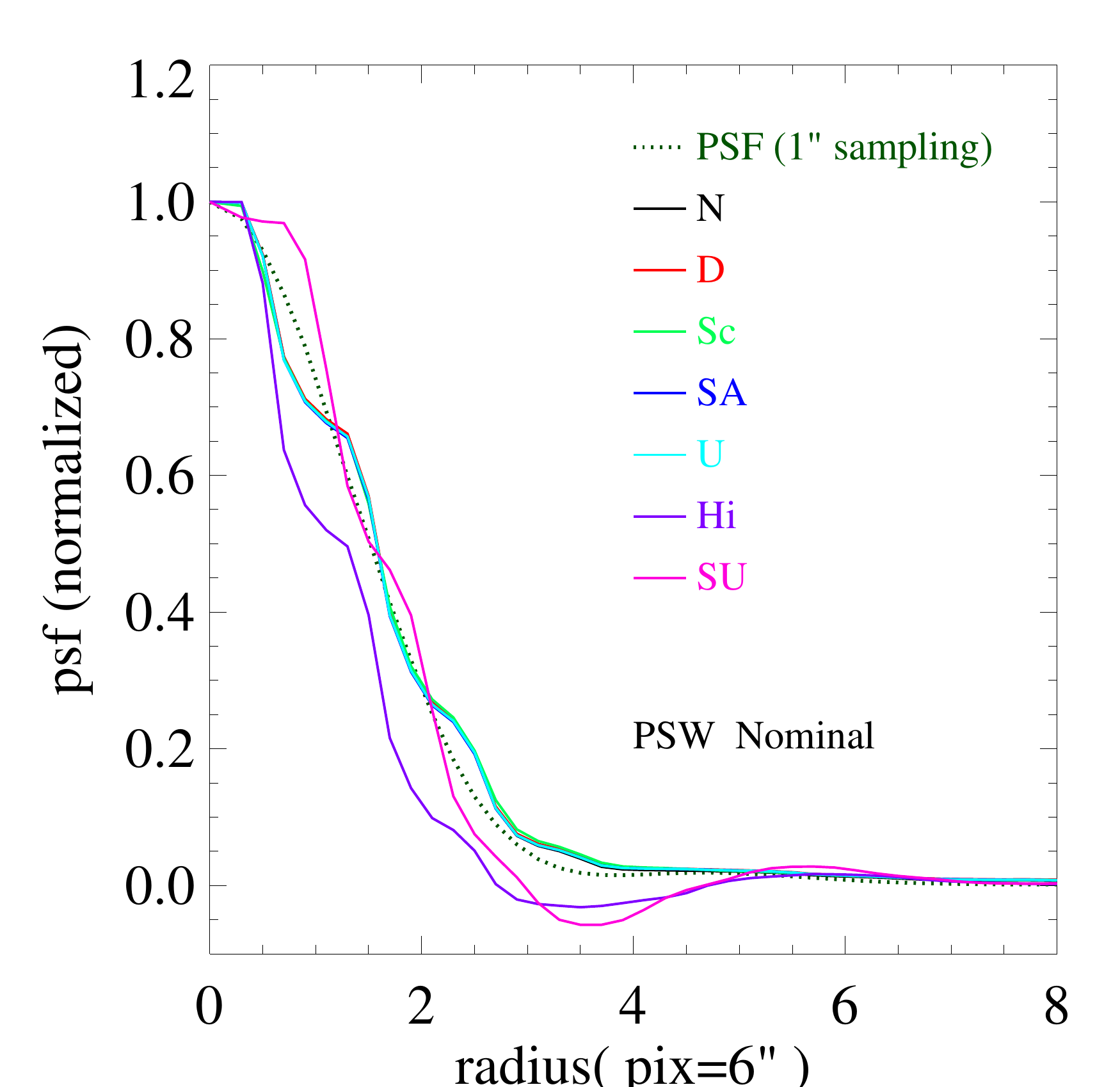}
    \includegraphics[width=6cm, angle=0]{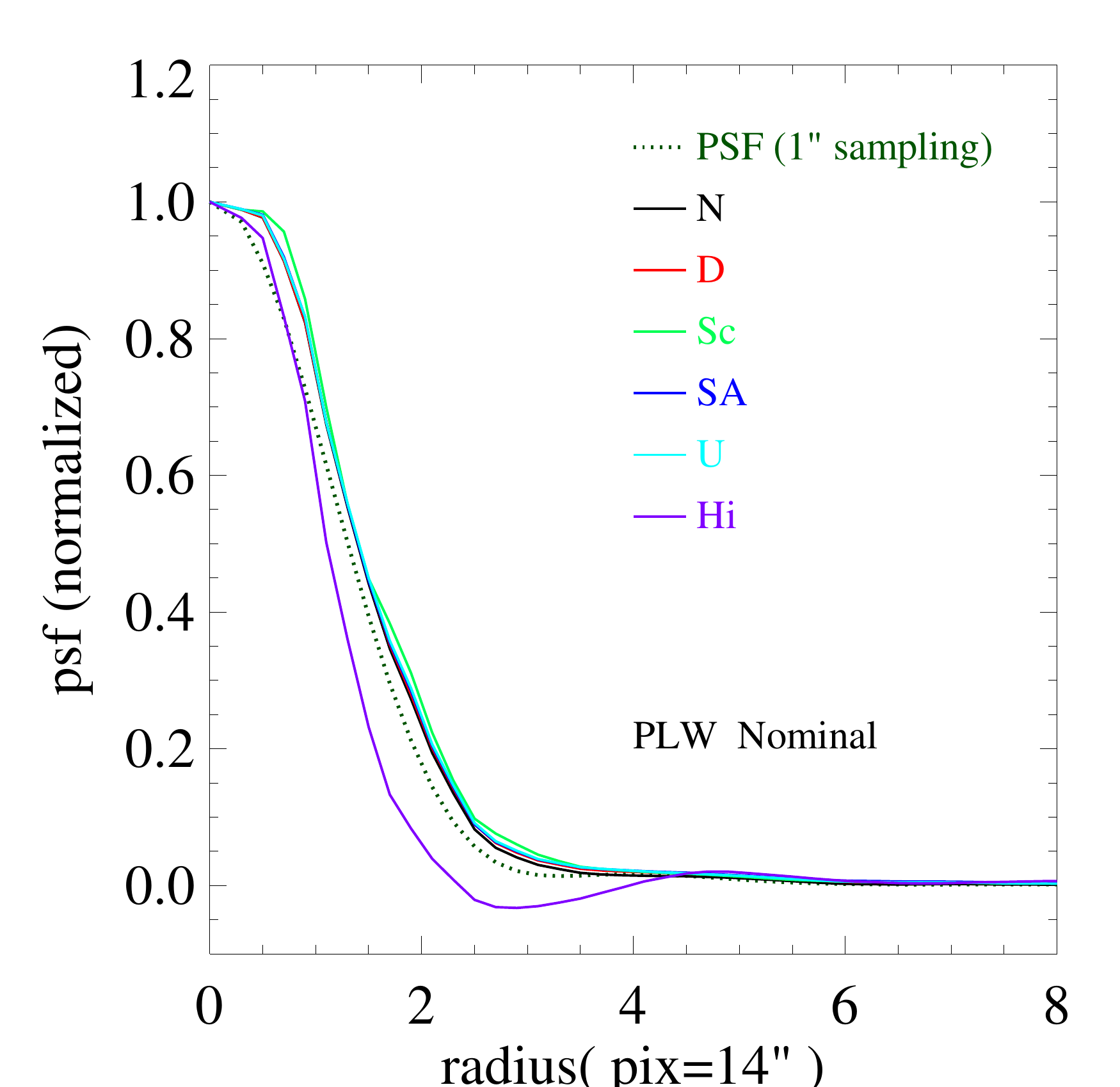}
\caption{Beam profiles for maps in Case 1 (nominal mode), derived using the 
mean radial distribution of the eight bright sources. The PSF data are 
taken from the SPIRE Photometer beam-profiles release note
published on 21 March 2011 by the SPIRE ICC.} 
\label{case1_profiles}
\end{figure*}

\begin{figure*}
    \centering
    \includegraphics[width=6cm, angle=0]{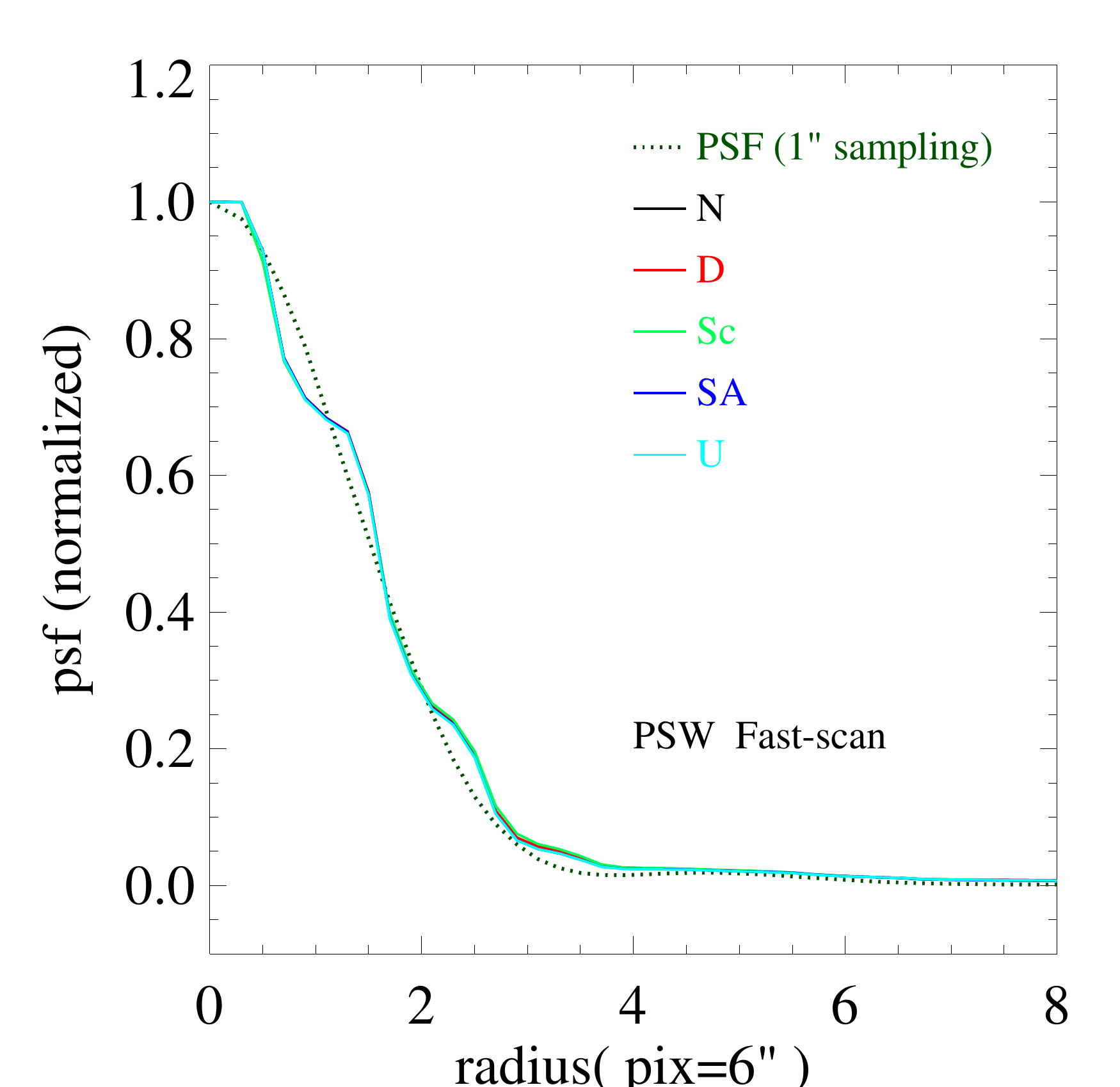}
    \includegraphics[width=6cm, angle=0]{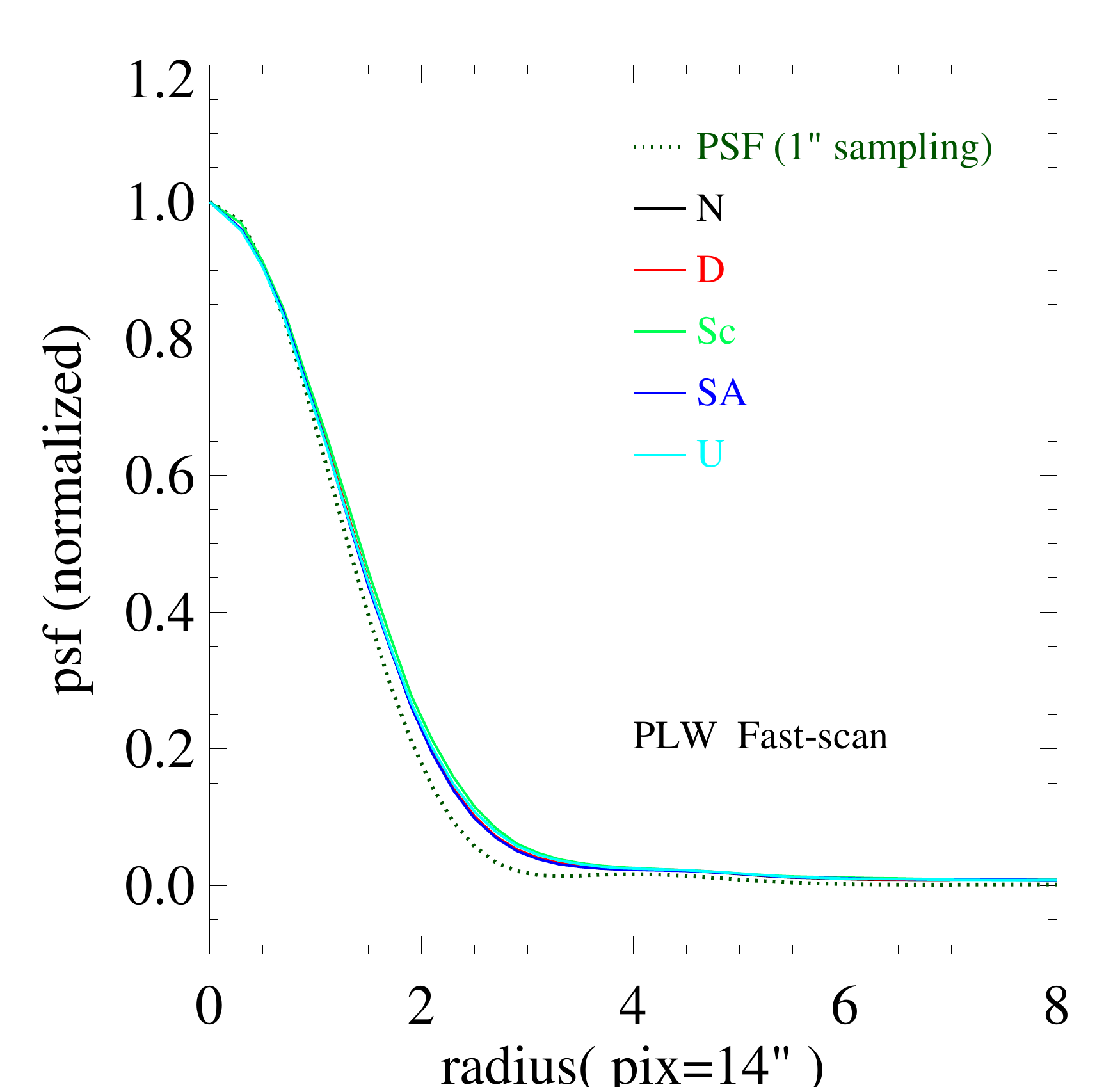}
\caption{Same as Fig.~\ref{case1_profiles}, 
for Case 5 (fast-scan mode).}
\label{case5_profiles}
\end{figure*}

\begin{figure*}
    \centering
    \includegraphics[width=6cm, angle=0]{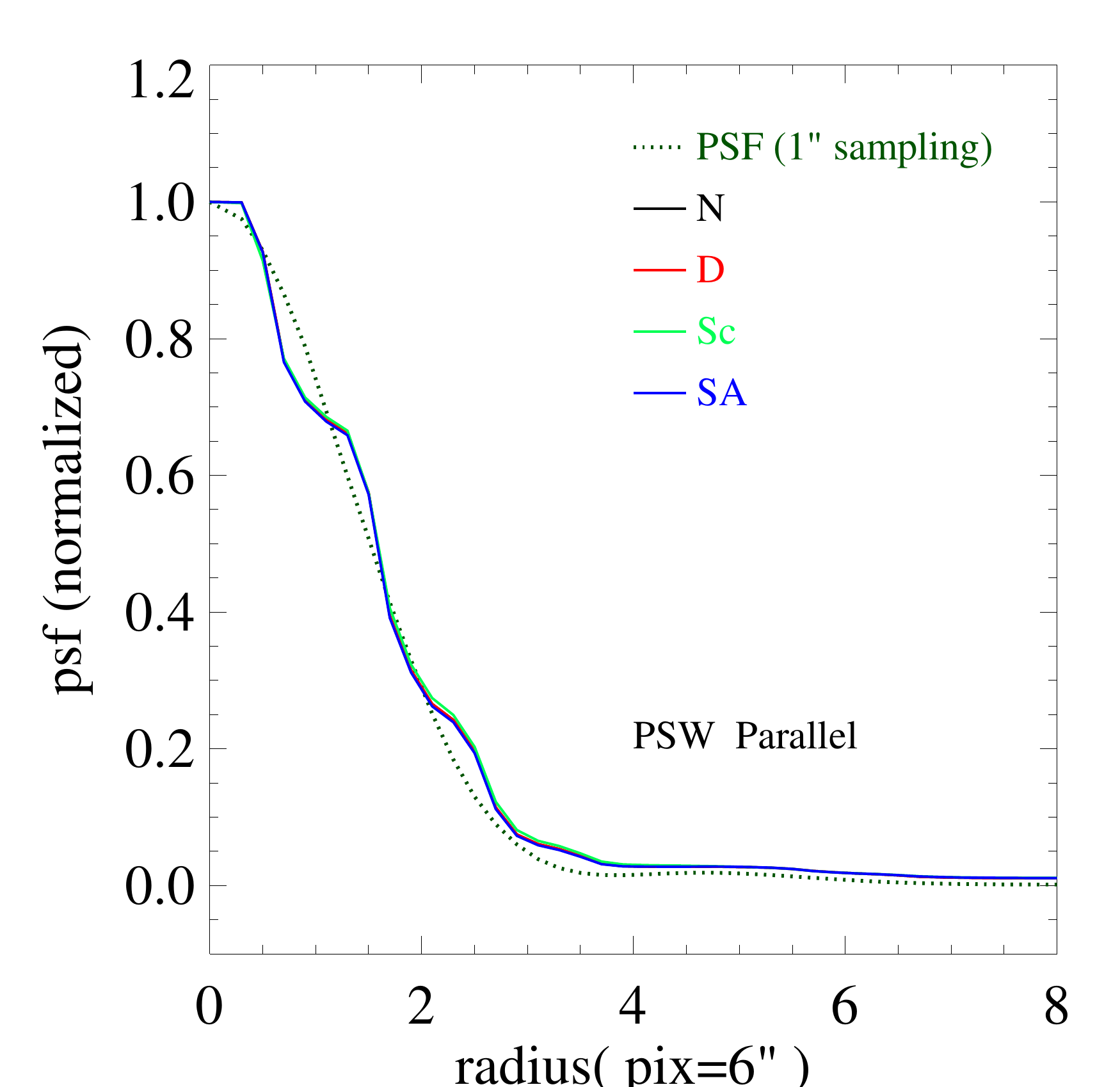}
    \includegraphics[width=6cm, angle=0]{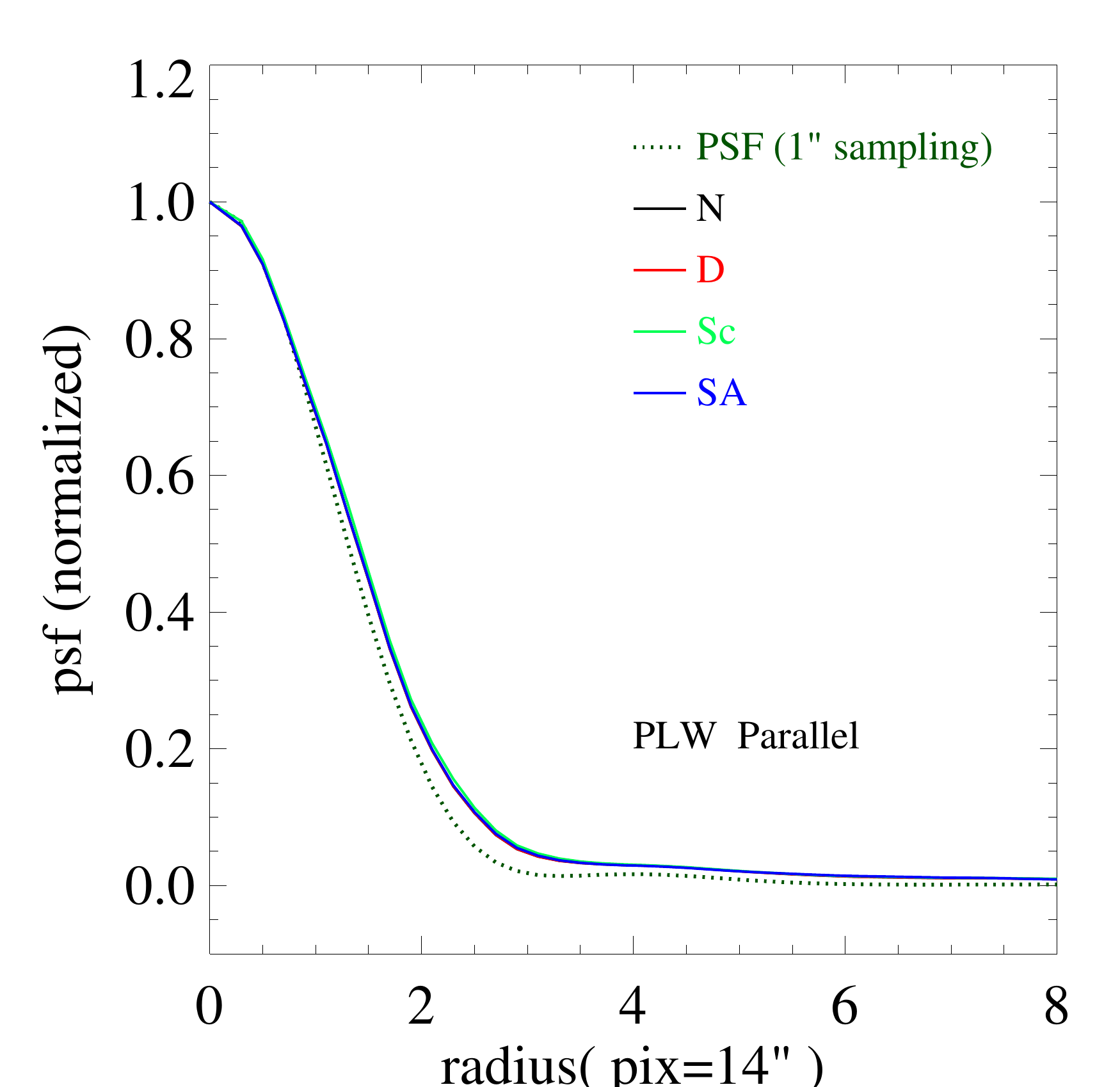}
\caption{Same as Fig.~\ref{case1_profiles}, 
for Case 8 (parallel mode).}
\label{case8_profiles}
\end{figure*}

In Fig.~\ref{case1_profiles}, Fig.~\ref{case5_profiles} and
Fig.~\ref{case8_profiles}, we compare between the beam profiles of
sources in maps made by different map-makers, and the profile of the
standard SPIRE PSFs (taken from the SPIRE Photometer beam-profiles
release note published on 21 March 2011 by the SPIRE ICC).  It appears
that, except for the super-resolution map-makers, the beam profiles of
sources in maps made by different map-makers do not differ
significantly with each other in all cases and in both the PSW and PLW
bands. It is interesting to note that, in these figures, the beam
profiles of sources in maps made by Scanamorphos look nearly identical
as those made by other map-makers. Therefore, although the PSFs in
Scanamorphos maps may differ slightly from those in other maps (c.f.
Section~\ref{sect:scanamorphos}),
as found in the analysis on the deviation from the truth 
(Section~\ref{sect:metrics_diff}), the difference is insignificant on
the scale of $\sim 0.2$ pixels (resolution of the beam profile plots
in this section).

\begin{figure*}
    \centering
    \includegraphics[width=6cm, angle=0]{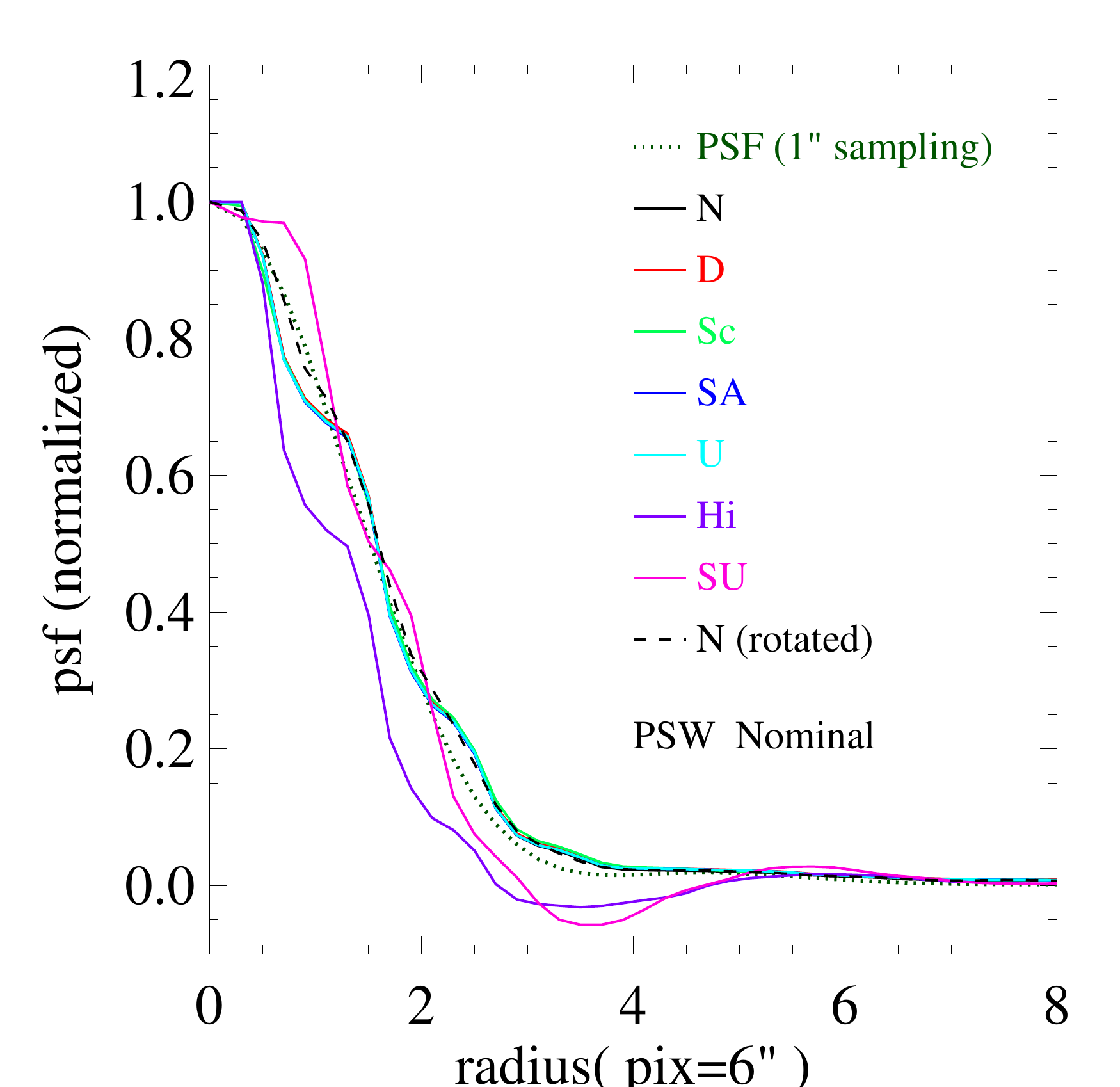}
    \includegraphics[width=6cm, angle=0]{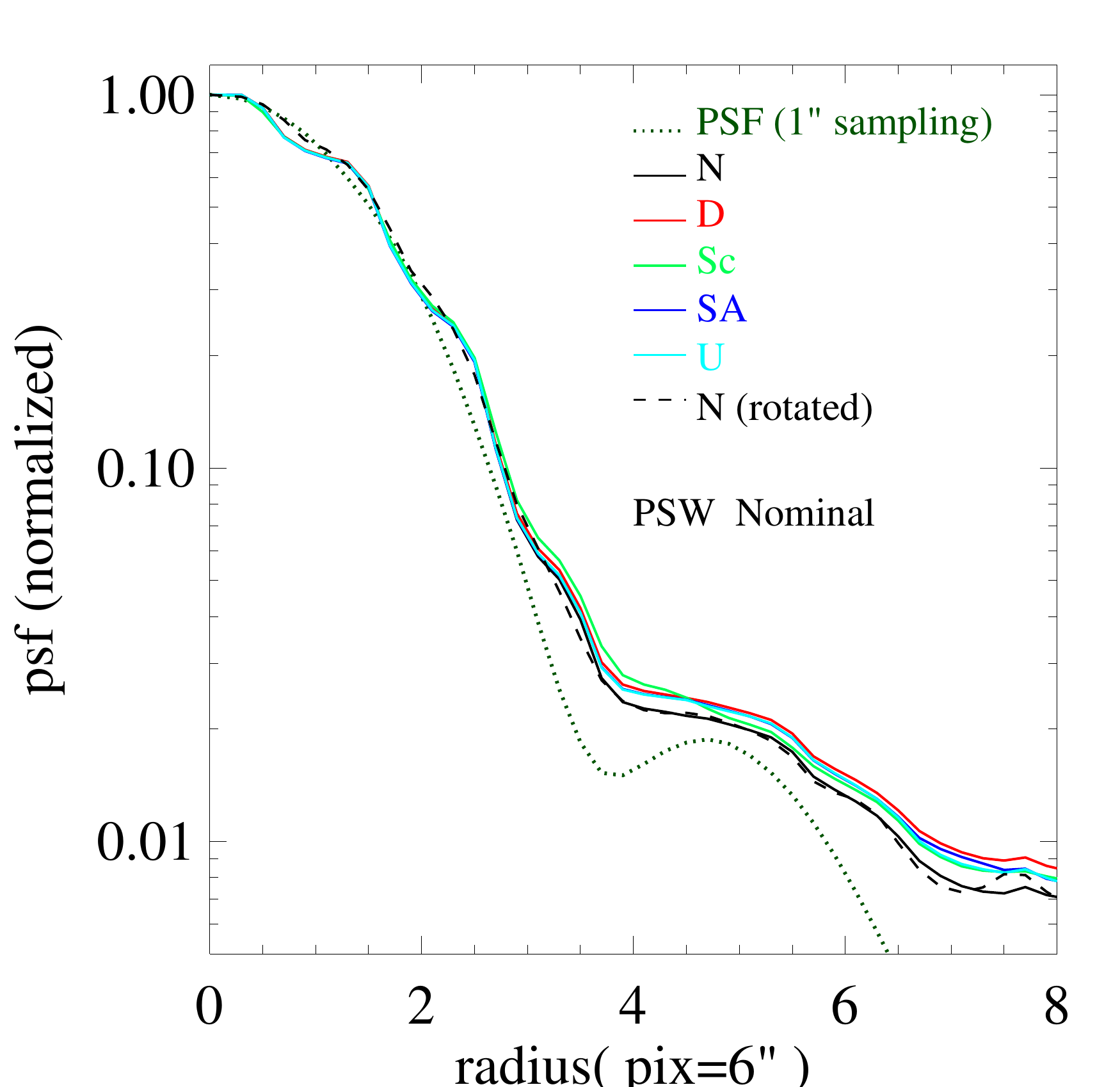}
\caption{{\it Left Panel}: 
Beam profiles for the PSW maps in Case 1. The plot is the same as
that in Fig.~\ref{case1_profiles}, except for an additional beam profile
for sources in a map made by NAIVE mapper with crota=45 (the north direction 
is rotated by 45 degrees clockwise relative to the up direction).
{\it Right Panel}: The same plot in the logarithmic scale. The beam profiles 
for sources in the HiRes map and in the SUPREME map are excluded because of
the negative values around the minimum. At large radii ($>6$ pix), the noise
dominates the signal in the 
measurements for source profiles in all maps, therefore the values of
the measurements are significantly higher than that of the standard PSF.
}
\label{case1_profiles_rotate}
\end{figure*}

Also, again in all cases, the beam profiles in the PSW band
show a much stronger effect of pixelization than those in the
PLW band. A possible explanation for this could be due to a
difference between sources in the PSW maps and those in the PLW maps 
in the simulation: all simulated sources in the PSW maps have
their peaks centered at the pixel centers, while for the PLW sources
this is true only in Case 1. In Case 5 and Case 8, the source centers
in the PLW maps are placed irregularly relative to the pixel centers. So,
for the PLW maps in Case 5 and Case 8, we have much better samplings
of the PSF phase, therefore much reduced effect of the pixelization.
In order to test whether this can indeed be the reason for the
pixelization effect, we carried out the following experiment. We made
a new Naive PSW map whose orientation is rotated by 45$^\circ$
clockwise relative to the default (north-up) orientation. The rotation
of source positions relative to the map grids puts the source centers
at different locations relative to the pixel center. In
Fig~\ref{case1_profiles_rotate}, we compared the beam profile of the
sources in this new map with other results. Indeed the pixelization
effects is weakened, in particular for small radii ($\rm < 1.5 pixel$).
However, the pixelization effect remains about the same at larger radii. 

The second reason for the PSW beam profiles having stronger pixelization
effect could be due to the scan sampling steps, which are
~2 arcsec for Case 1 and case 8, and ~4 arcsec for Case 5. Given the different
sizes of the PLW and PSW beams, these steps may be fine enough for the
PSF sampling in the PLW band, but too coarse for that in the PSW band.
This may explain the difference between the beem profiles of sources 
in the PLW map and of those in the PSW map (without rotation) in Case 1,
both have centers located at the pixel centers. This may also explain the
residual pixelization effect seen in the rotated PSW map in Case 1.

In Fig~\ref{case1_profiles} and Fig~\ref{case1_profiles_rotate},
results from HiRes and SUPREME are also included. In the PSW band, the
PSF in the HiRes map is indeed much narrower.  The PSF in the SUPREME
map has about the same FWHM as other map-makers, but has a rather flat
center peak. This is perhaps why it has more power in the higher
frequency modes in the power-spectra analysis. More detailed analysis
on the beam profiles of sources in super-resolution maps can be found
in Section~\ref{sect:superresolution}.

\subsubsection{Extended Source: Aperture Photometry}\label{Extended}

In all three test cases analyzed here, the central extended source has
the same shape of an exponential disk with the e-folding radius of
$90”$.  The aperture used for its flux measurement has a radius of $\rm r =
270”$, namely $\rm 3\times$ e-folding radius. The background annulus is between
two radii of 5 and 6 times of the e-folding radius, respectively.
\begin{figure*}
    \centering
    \includegraphics[width=6cm, angle=0]{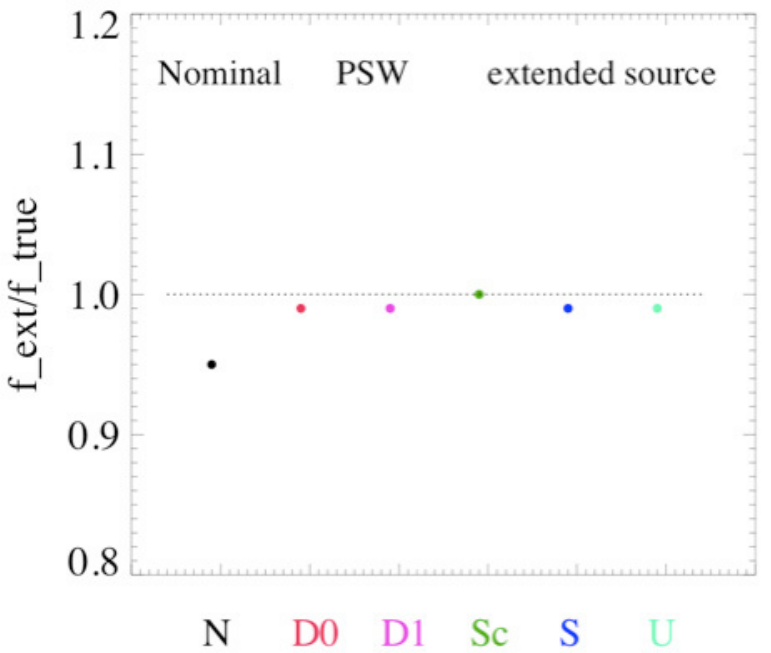}
    \includegraphics[width=6cm, angle=0]{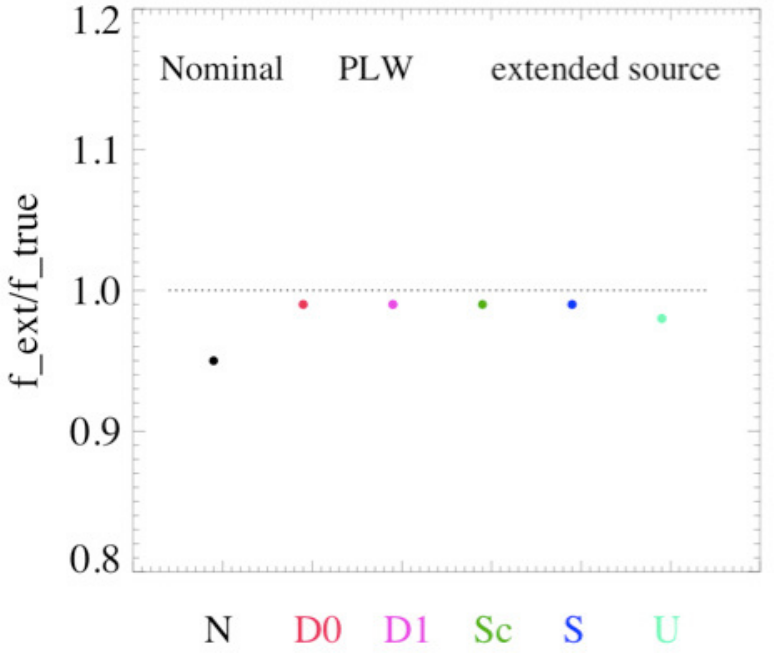}
\caption{Plots of $\rm f_{ext}/f_{true}$ 
ratios of the extended source in Case 1 maps (nominal mode); $\rm f_{ext}$ 
is the flux measured in a map made by a given map-maker, and $\rm f_{true}$ 
the flux measured in the corresponding truth map using the same aperture. 
}1
\label{metrics_report_sources_fig21}
\end{figure*}
\begin{figure*}
    \centering
    \includegraphics[width=6cm, angle=0]{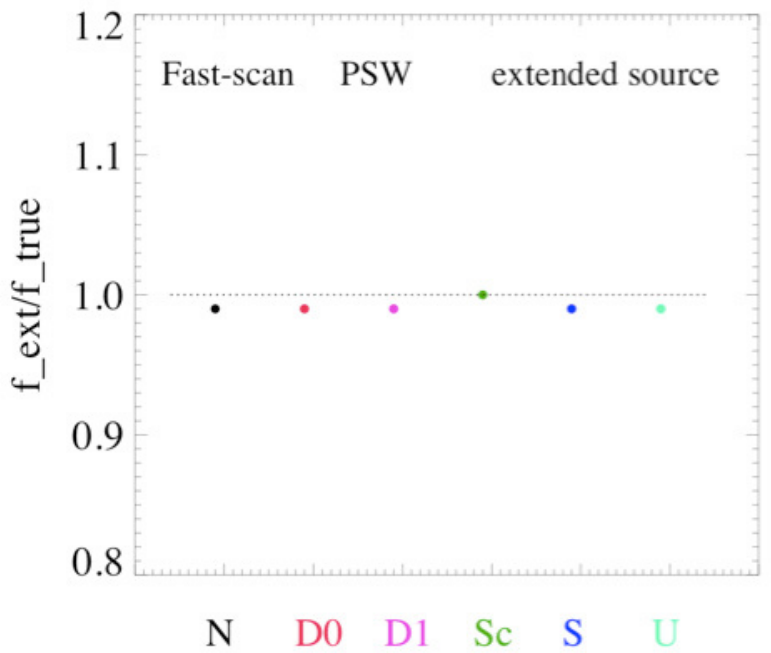}
    \includegraphics[width=6cm, angle=0]{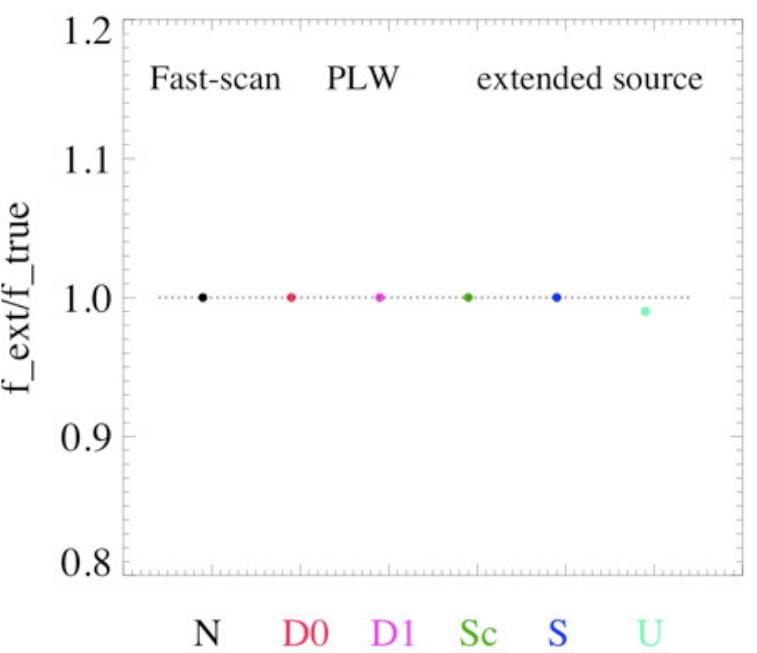}
\caption{Same as Fig.~\ref{metrics_report_sources_fig21}, 
for Case 5 (fast-scan mode).}
\label{metrics_report_sources_fig22}
\end{figure*}

In Fig.~\ref{metrics_report_sources_fig21}, plots of of $\rm f_{ext}/f_{true}$ 
ratios of the extended source
in Case 1 (nominal mode) are presented. Here $\rm f_{ext}$ is the flux
measured in a map made by a given map-maker, and $\rm f_{true}$ the flux
measured in the corresponding truth map using the same aperture. We
see a significant underestimation in maps made by the naive mapper, in
both the PSW and PLW bands. In maps made by other map-makers, the
$\rm f_{ext}/f_{true}$ ratio is consistent with unity. For the Case 5 maps in
the fast-scan mode (Fig.~\ref{metrics_report_sources_fig22}),
 we do not see this difference: the
results obtained from the naive maps are as good as those from other
maps. The underestimation in the naive maps of Case 1 is due to a
known bias in the median baseline removal carried out in the naive
map-making, which over-subtracts the background around bright extended
sources. The bias is much less significant in Case 5 maps because of
the much larger size of the maps (the area occupied by the extended
source is much less than the background area). Results for Case 8 maps
are similar to those for Case 5 maps.

\subsection{Summary}\label{sammary}

\begin{description}
\item{1)} Both PSF fitting and small aperture photometry show small position errors ($< 0.1$ pix) for bright point sources. In some cases, maps made  by Scanamorphos show slightly larger errors.
\item{2)} Photometry for bright point sources in maps made by all map-makers have small errors, indicating good energy conservation in the map making.
\item{3)} The position and the flux errors for the faint sources are significantly affected by the confusion noise. It appears that no map-maker stands out in detecting the faint sources and minimizing the errors.
\item{4)} In both PSW and PLW bands, there is no significant difference between beam profiles of sources in maps made by different map-makers. The beam profiles in the PSW band are affected significantly by the pixelization effect.  
\item{5)} For the photometry of extended sources: in some cases, maps made by Naive mapper are significantly affected by a known bias due to the over-subtraction of baselines. Other maps have no such issue.
\end{description}

%\bibliographystyle{abbrv}
%\bibliography{main}

%\end{document}

%% file: metrics_report_superresolution_v4_pdf.tex
% Template for metrics reports.
% CKX, April 6, 2013

%\documentclass[11pt]{book}
%\usepackage{titletoc}% http://ctan.org/pkg/titletoc
%\usepackage{amsmath} %Never write a paper without using *amsmath* for
                     %its many new commands
%\usepackage{amssymb} %Some extra symbols
%\usepackage{makeidx} %If you want to generate an index, automatically
%\usepackage{graphicx} %If you want to include postscript graphics
%\usepackage{caption} %If you want to include postscript graphics
%\usepackage{epsfig}
%\begin{document}
%
%\setcounter{chapter}{4}
%\chapter{Metrics and Results}

%\setcounter{section}{3}

\section{Metrics for Super-resolution Maps (David Shupe)}
\label{sect:superresolution}

Since the super-resolution mapmakers for SPIRE are intended for uses different than
the other mapmakers, metrics were computed separately for HiRes and SUPREME.
The comparisons in this section were made with the Destriped maps (polynomial order=0)
as the main reference.

\subsection{Test Data}\label{sr-data}

A subset of the input data cases were considered for the super-resolution mapmakers.
\begin{itemize}
    \item Case 1 (sources): simulated data with artificial sources
    \item Case 10 (cirrus): simulated data with the truth image from a 24 $\mu$m map
    \item Case 12 (ngc628) : real data of the face-on spiral galaxy NGC 628
    \item Case 13 (higal\_l30): real data of the HiGal $ l=30^\circ $ field
\end{itemize}

The maps were made with the standard pixel sizes of [6, 10, 14] arcseconds for [250, 350,
500] $\mu$m. For the sources data (case 1), maps were also made with pixels half the size
of the standard ones.

The maps delivered for HiRes were taken from the 20th iteration map and were trimmed
to match the area covered by the Destriped map. The delivered SUPREME maps covered
a wider area, including all the input data, and had errors in the world coordinates ranging from
a few to several dozen pixels. For the analysis reported below, the WCS's of the SUPREME
maps were adjusted based on reference point sources in the images, and the maps were
trimmed to cover the same area as the HiRes and Destriped maps. The registration differs
by a fraction of a pixel since the projection centers of the SUPREME maps do not fall
in the center of a pixel. Since the images are not exactly registered, difference maps are
not a useful analysis tool and are not included here.

The analysis focuses on the PSW (250 $\mu$m channel) as SUPREME maps were available
for this band only.

\begin{figure*}
\centering
\includegraphics[width=5.0in]{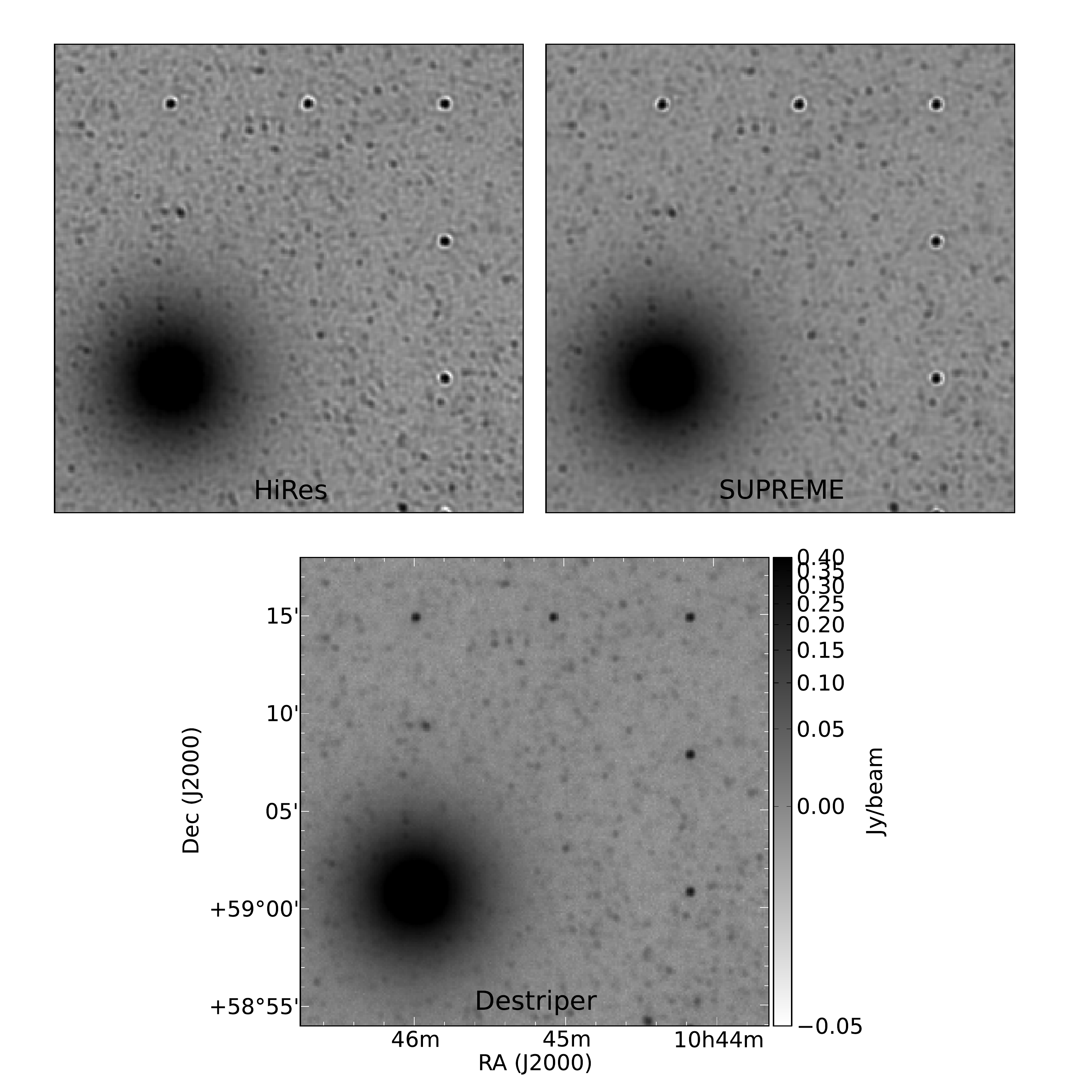}
\caption{Images produced by HiRes, SUPREME, and Destriper, 
for the 250 $\mu$m band for Case 1 (sources). Part of the field is shown.}
\label{fig:sr-pspec-sources}
\end{figure*}

\begin{figure*}
\centering
\includegraphics[width=5.0in]{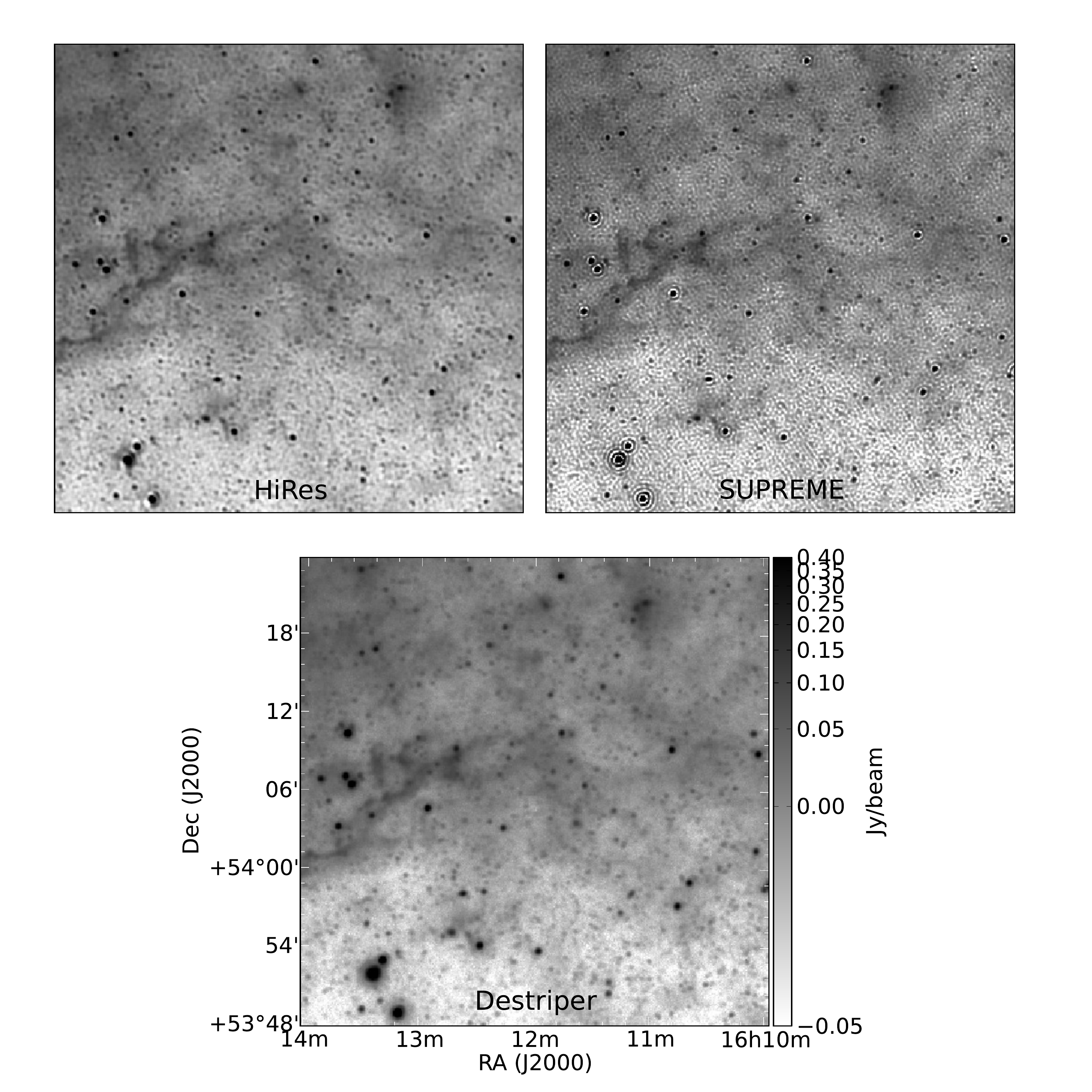}
\caption{Images produced by HiRes, SUPREME, and Destriper, 
for the 250 $\mu$m band for Case 10 (cirrus), zoomed to part of the field.
 The ringing
around point sources is considered normal, as SUPREME has been designed to
enhance the resolution of extended features. The simulation contains many bright
point sources because it is based on a 24 $\mu$m map, and is not representative
of extended emission observations made with SPIRE.}
\label{fig:sr-pspec-cirrus}
\end{figure*}

\begin{figure*}
\centering
\includegraphics[width=5.0in]{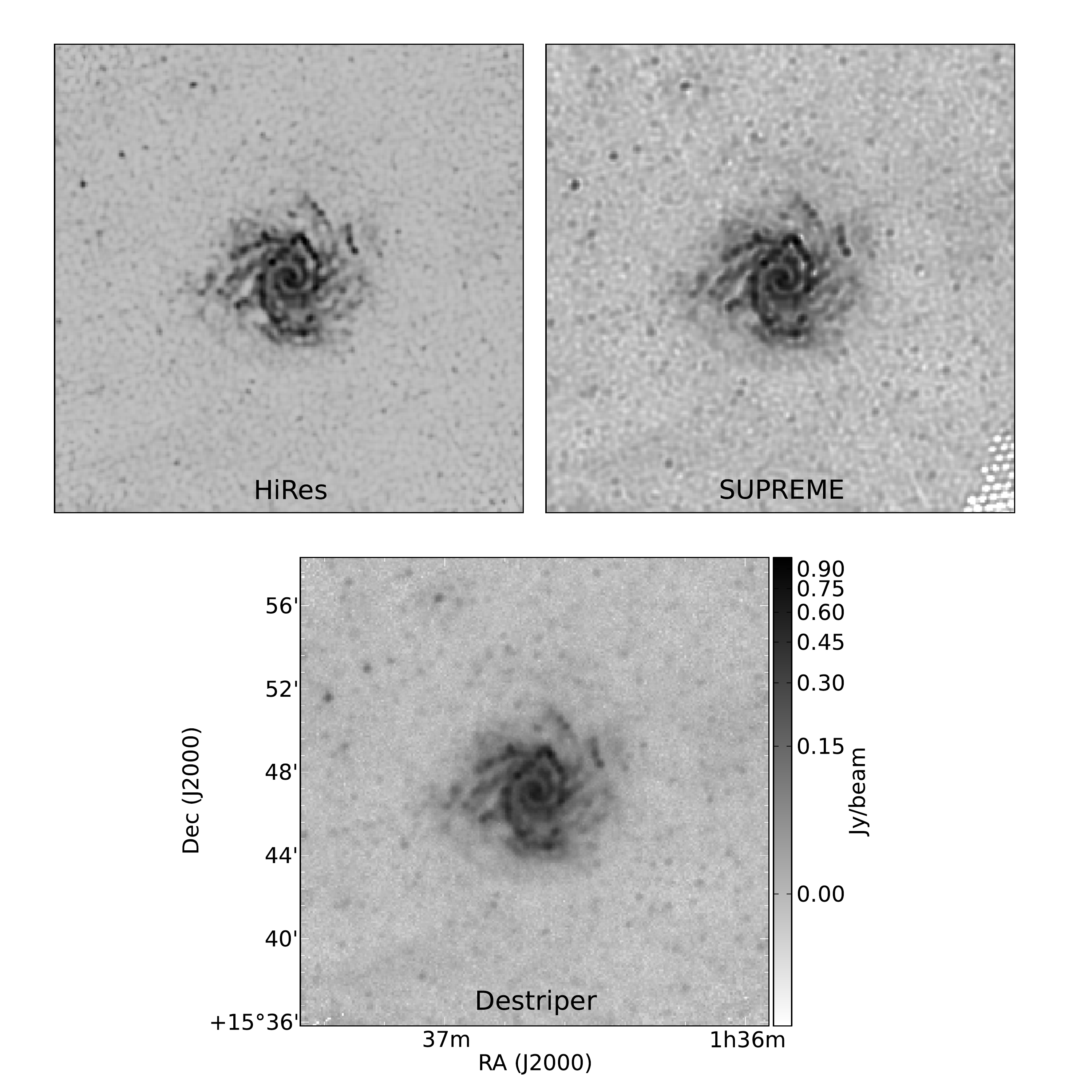}
\caption{Images produced by HiRes, SUPREME, and Destriper, 
for the 250 $\mu$m band for Case 12 (NGC 628).}
\label{fig:sr-pspec-ngc628}
\end{figure*}

\begin{figure*}
\centering
\includegraphics[width=5.0in]{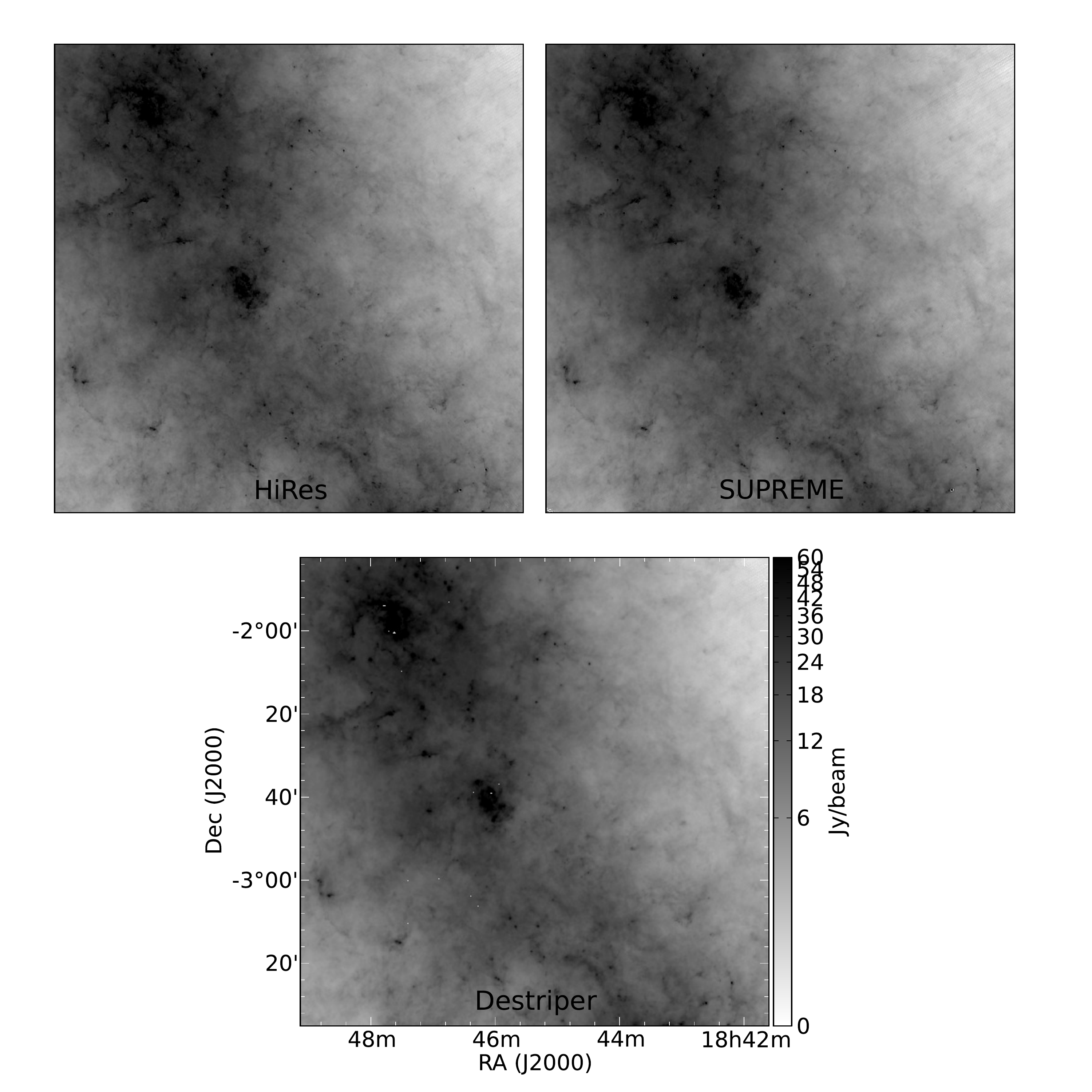}
\caption{Images produced by HiRes, SUPREME, and Destriper, 
for the 250 $\mu$m band for Case 13 (higal\_l30).}
\label{fig:sr-pspec-l30}
\end{figure*}

%Examples of Figures and Tables
%\begin{figure}[ht]
% \begin{center}
% \epsfig{file=file1-eps-converted-to.pdf, width=10cm}
%\end{center}
%\caption{Sample caption for one method of including figures.}
%\end{figure}

%\begin{figure*}
%    \centering
%    \includegraphics[width=7.5cm, angle=0]{file1.ps}
%    \includegraphics[width=7.5cm, angle=0]{file2.ps}
%\caption{Sample caption for another completely different but equivalent method of including figures.}
%\end{figure*}

%\bigskip
%\begin{tabular}{llcc}
%\hline \\
%Source & Coordinates & Flux Density/       & Diameter \\
%       & (J2000)     &   (mJy)             & (arcsec) \\
%\hline \\
%G153.4+0.3 & 17:18:00+59:30:00 & 50 & 5 \\
%G278.3-0.5 & 09:48:38-54:23:7.1 & 3 & 1 \\
%\hline \\
%\end{tabular}

\subsection{Analyses and Results}\label{sr-results}

\subsubsection{Power Spectra}

Power spectra were computed as an azimuthal average of the 2-D power
spectrum. The spectra are normalized to the summed square surface brightness
of the maps. The computations were performed with a Python translation of the
IDL script provided by Jim Ingalls and described in the preceding chapter on power
spectra.

\begin{figure*}
\centering
\includegraphics[width=6.0in]{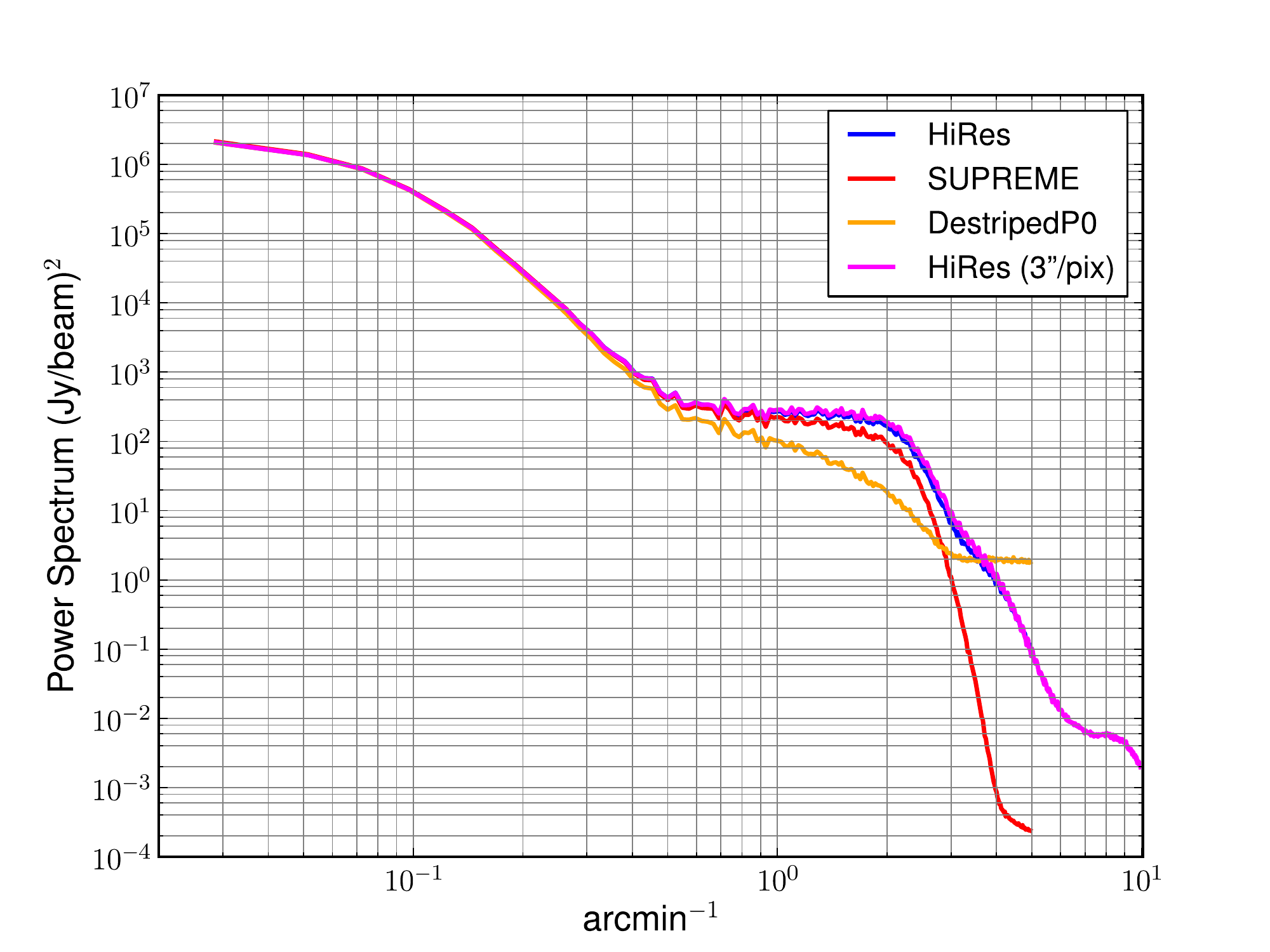}
\caption{Power spectra for the 250 $\mu$m band for Case 1 (sources), including
a 3-arcsecond-pixel image for HiRes.}
\label{fig:sr-pspec-sources}
\end{figure*}

\begin{figure*}
\centering
\includegraphics[width=6.0in]{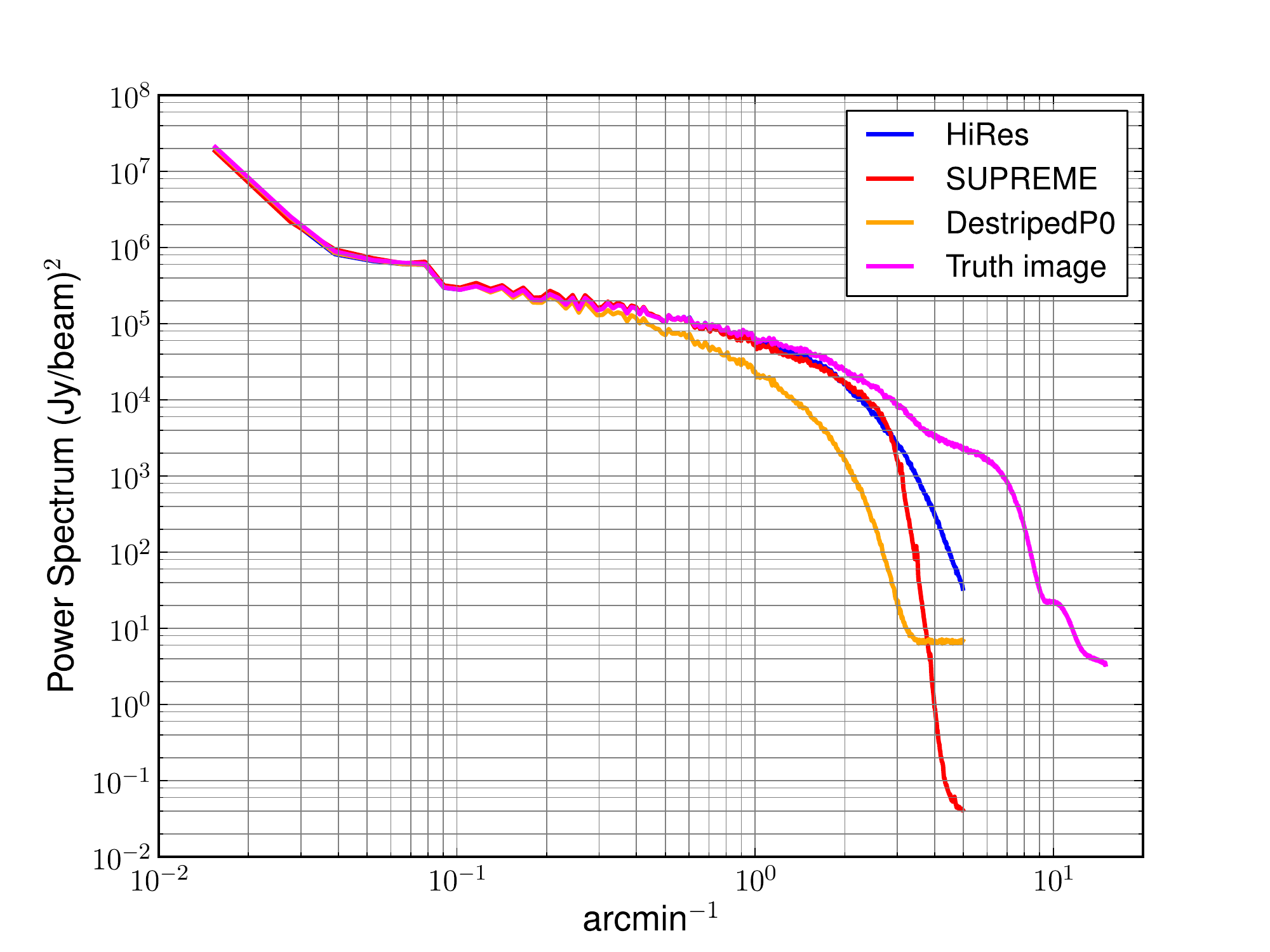}
\caption{Power spectra for the 250 $\mu$m band for Case 10 (cirrus).}
\label{fig:sr-pspec-cirrus}
\end{figure*}

\begin{figure*}
\centering
\includegraphics[width=6.0in]{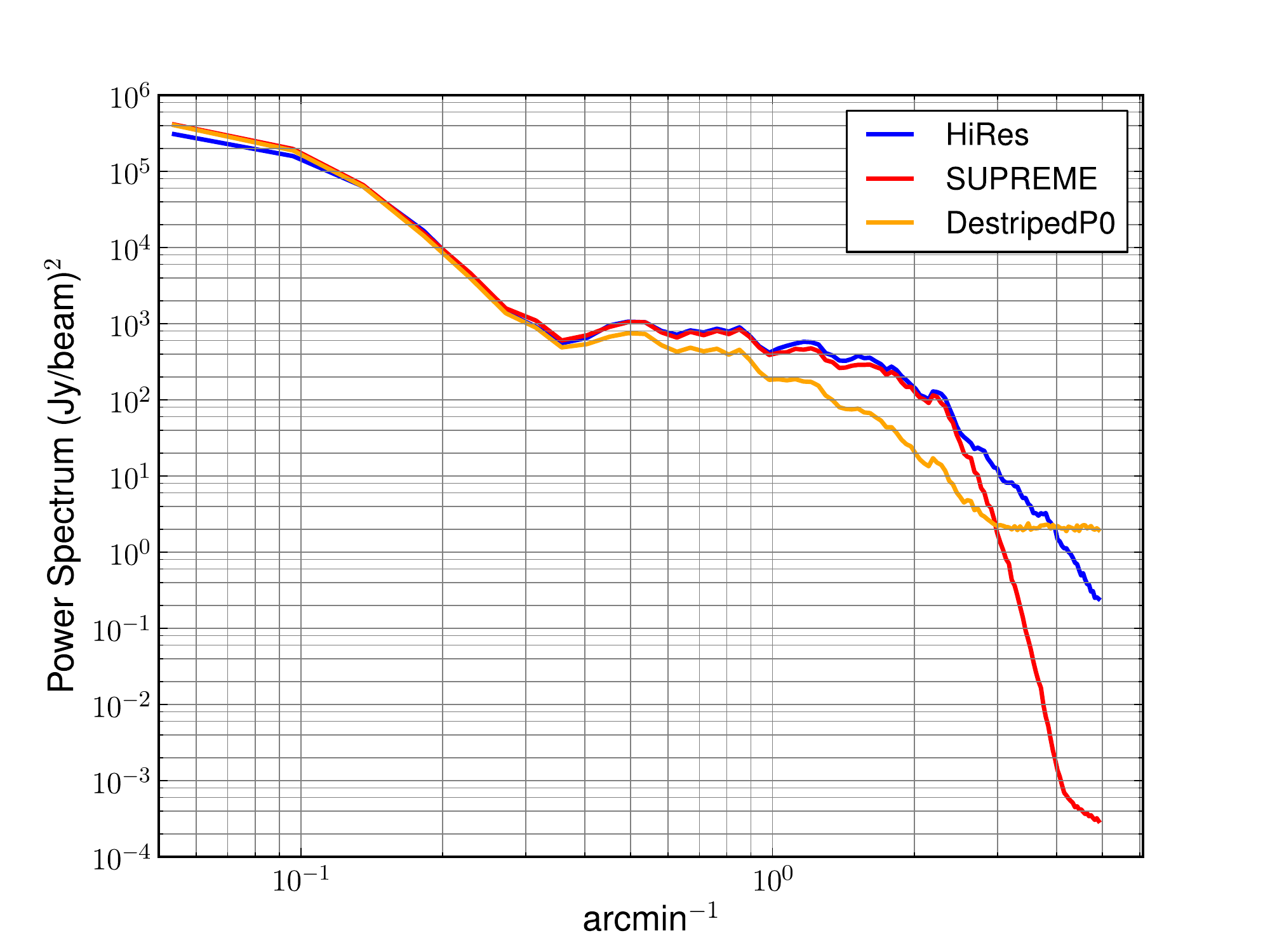}
\caption{Power spectra for the 250 $\mu$m band for Case 12 (NGC 628)}
\label{fig:sr-pspec-sources}
\end{figure*}

\begin{figure*}
\centering
\includegraphics[width=6.0in]{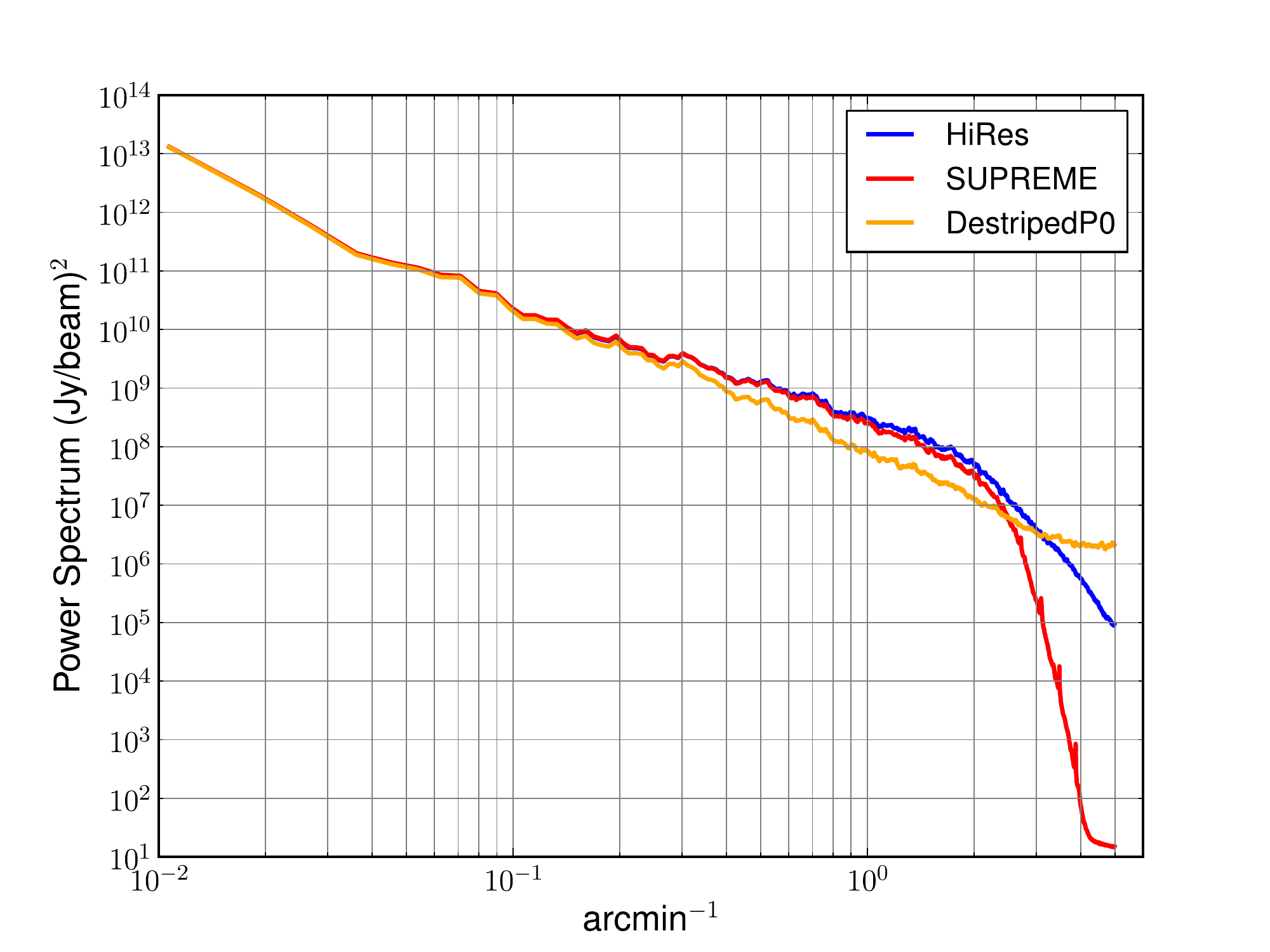}
\caption{Power spectra for the 250 $\mu$m band for Case 13 (higal\_l30)}
\label{fig:sr-pspec-sources}
\end{figure*}

In general, both HiRes and SUPREME perform well, with about a factor of 2-3 improvement
in resolution over the destriper at spatial frequencies of about 2 arcmin$^{-1}$. 
The power spectra show that, as designed, SUPREME
filters the noise below the beam resolution. HiRes contains more power in the
higher frequencies. In more quantitative terms,
SUPREME and HiRes power spectra begin diverging at spatial scales between 20 and 24 arcseconds, a size comparable to the FWHM of a Gaussian approximation to the central SPIRE beam at 250 microns. HiRes contains more power than the DestripedP0 map for spatial scales larger than about 15 arcseconds; at smaller scales (higher frequencies) the  DestripedP0 map contains more power, visible in the images as noise (not unlike "snow" seen in old television sets).

\subsubsection{Point source profiles}

The widths of point sources was estimated by fitting Gaussians to a number of
sources in each field. A drawback of this test is that the beams are not Gaussian
so the results can be misleading. Therefore these results do not give as complete
a picture as the power spectrum analysis provides--the latter analysis is to be preferred
in assessing the resolution enhancements provided by these mapmakers.

The results are shown in Table \ref{tab:sr-fwhm}. The results
show improvement in resolution for both map-makers. HiRes FWHM values are reduced
by about 25\% to 50\%, and SUPREME widths are reduced by about 15\% to 30\%.
These figures should be treated with caution, especially since SUPREME is tuned to
enhance extended emission, and the SUPREMEX method will provide better separation
of point sources.

\bigskip
%\begin{table}\tiny
%\begin{table}\scriptsize
\begin{table}
\centering
\begin{tabular}{llrrr}
\hline \\
Case & Map-maker &   Number & Mean & Std. Dev \\
     &           & of sources   & FWHM &  (FWHM) \\
 &        &   & (arcsec) & (arcsec) \\
\hline \\
1 (sources) & Destriper & 10 & 18.46 & 1.33 \\
1 (sources) & HiRes & 10 & 13.63 & 1.22 \\
1 (sources) & HiRes (3 arcsec) & 10 & 13.11 & 1.15 \\
1 (sources) & SUPREME & 10 & 16.21 & 0.96 \\
 & & & & \\
 10 (cirrus) & Truthimage & 12 & 6.22 & 0.20 \\
 10 (cirrus) & Destriper & 5 & 22.90 & 9.1 \\
 10 (cirrus) & HiRes & 6 & 11.37 & 0.42 \\
 10 (cirrus) & SUPREME & 6 & 15.43 & 1.11 \\
  & & & & \\
12 (NGC 628) & Destriper & 12 & 21.48 & 0.56 \\
12 (NGC 628) & HiRes & 12 & 13.55 & 0.74 \\
12 (NGC 628) & SUPREME & 12 & 14.48 & 0.22 \\
\hline \\
\end{tabular}
\caption{Results from fitting Gaussians to point sources on the 250 $\mu$m images.}
\label{tab:sr-fwhm}
\end{table}

\subsection{Summary}\label{sr-summary}

Overall, both SUPREME and HiRes achieve their goals of resolution enhancement of SPIRE images. For spatial scales larger than about 25 arcseconds in the 250 micron band (comparabale to the beam size), the resolution enhancement of both mapmakers is similar and reaches a
maximum of about 2-3 over destriped maps, at spatial frequencies around 2 arcmin$^{-1}$. The main difference between SUPREME and HiRes is seen for spatial scales smaller than the beam size. SUPREME is tuned to smooth and reduce the noise at scales smaller than the beam size. The HiRes maps contain more power on these scales, effectively turning noise variations into beam-sized blobs. Fitting Gaussians to point sources shows improvement for both SUPREME and HiRes, with slightly more improvement (25\% to 50\%) shown by the latter. However it should be noted that SUPREME is tuned to enhance extended emission, and that the power spectrum analysis is to be preferred when assessing gains in resolution. 

%\bibliographystyle{abbrv}
%\bibliography{main}
%
%\end{document}

%% file: chapter_summary.tex
% NOTE: to be included in the report_main.tex
%\documentclass[11pt]{book}

%\begin{document}

\chapter{Summary}

In this report, we present results of a comprehensive test on SPIRE map-making.
Seven map-makers participated in this test, including Naive-mapper, Destriper
(used in two different configurations),
Scanamorphos, SANEPIC, Unimap, HiRes, and SUPREME (the last two being 
super-resolution mappers). The 13 test cases (8 simulated
observations and 5 real observations) include data sets 
obtained in different observational modes and scan speeds, with different
map sizes, source brightness, and levels of complexity of 
the extended emission. They also include observations suffering from the
``cooler burp'' effect, and those having strong large-scale gradients in the
background radiation. The major goal is to identify strengths and
limitations of different map-makers in dealing with different issues
in SPIRE maps. The results can be summarized as following:

\begin{itemize}
\item The Destriper with the polynomial order of 0 (Destriper-P0), the
default map-maker in the SPIRE scanmap pipeline since HIPE 9, 
performed remarkably well and compared
favorably among all map-makers in all test cases except for those suffering
from the ``cooler burp'' effect. In particular, it can handle observations
with complex extended emission structures and 
with large scale background gradient very well. 
\item In contrast, the Destriper with 
the polynomial order of 1 (Destriper-P1) compared poorly among its peers,
introducing significant artificial large scale gradient in many cases. 
\item Scanamorphos showed noticeable differences in all comparisons. 
On the positive side,
its maps have the smallest deviation from the truth for faint pixels 
($\rm f < 0.2 Jy$) in nearly all cases.
Particularly, as shown in both the difference maps and in
the power-spectra, it can handle the ``cooler burp'' effect very well.
On the negative side, for bright pixels ($\rm f > 0.2 Jy$), 
its maps show significant deviations from the truth, likely due to a 
slight positional offset introduced by the mapper as well as
a slight change in the beam size. This effect is also seen
in the astrometric errors of the bright sources.
However the offset is very small ($\sim 0.1$ pexel), therefore it
does not affect the photometry of
both point sources and extended sources, and does not show up
in the comparison between beam profiles (resolution: 0.2 pixels).
The power spectrum analysis indicates smoothing of the data compared 
to the other mapmakers.
\item The GLS mapper SANEPIC can also minimize the ``cooler burp'' 
effect. It performed quite well in most cases. However, for those cases
with strong variations in very large scales (i.e. comparable to the map size),
its maps show significant deviations from the truth. This
 is because some of its assumptions (e.g. TODs are circulant) are 
invalid for the data.
\item Unimap, another participating GLS mapper, is among the best
performers in most cases. However, because it does not include a mechanism for
handling the ``cooler burp'', its maps show significant
deviations from the truth in the cases affected by the artifact.
\item The Naive-mapper (with simple median background removal) is inferior 
among its peers in general. The most severe bias it introduces is the
over-subtraction of the background when extended emission is present. In the 
cases where the extended emission is in complex structures, this bias
cannot be avoided by simple masks in the background removal.
\item The two super-resolution mapmakers, SUPREME and HiRes, yield
  similar resolution enhancements (factors of 2-3) at spatial scales
  around 2 arcmin$^{-1}$ for the limited datasets tested at 250
  microns. At higher spatial frequencies corresponding to spatial
  scales smaller than the beam size, there is less power in the
  SUPREME maps (intentionally, to smooth and reduce the noise at
  scales smaller than the beam). HiRes contains more power than either
  SUPREME or Destriper-P0 maps between spatial scales of 15-20
  arcseconds. The differences in SUPREME and HiRes arise mainly
  because SUPREME is tuned to enhance extended emission features, and
  HiRes is essentially performing a deconvolution in image space.
\end{itemize}

%\end{document}